\documentclass[12pt,a4paper]{article}
\usepackage[margin=1in]{geometry}
\usepackage{setspace}
\usepackage{ragged2e}
\usepackage[left]{lineno}

\usepackage{amsfonts}
\usepackage{amssymb}
\usepackage{bm}
\usepackage{mathtools}   

\usepackage{array}
\usepackage{booktabs}  
\usepackage[table]{xcolor}
\usepackage{longtable}
\usepackage{tabularx}
\usepackage{threeparttablex}
\usepackage{adjustbox}
\usepackage{siunitx} 
\usepackage{makecell}
\usepackage{placeins}

\usepackage{graphicx}
\usepackage{rotating}
\usepackage{lscape}
\usepackage{float}

\usepackage{subcaption}
\usepackage{caption}
\usepackage{pdflscape}

\usepackage{algorithm2e}
\usepackage{algpseudocode}
\usepackage{algorithmicx}
\usepackage{algpseudocode}

\definecolor{winered}{rgb}{0.5,0,0}
\usepackage[bookmarks=true, bookmarksnumbered=true, allbordercolors={1 1 1}]{hyperref}
\hypersetup{
	colorlinks=true,
	linkcolor=winered,
	urlcolor=blue,
	citecolor=winered
}
\usepackage[open=true,numbered=true]{bookmark}
\usepackage[colorinlistoftodos]{todonotes}
\usepackage{natbib}
\usepackage{varioref}
\usepackage{cleveref}

\usepackage{authblk}
\usepackage{fancyhdr}
\usepackage{afterpage}
\usepackage{verbatim}
\usepackage{tablefootnote}
\usepackage{multirow}

\makeatletter
\def\blfootnote{\xdef\@thefnmark{}\@footnotetext}
\makeatother

\emergencystretch=\maxdimen
\begin{document}
	
	\title{\LARGE{Learning from crises: A new class of time-varying parameter VARs with observable adaptation}\blfootnote{
			Correspondence: Dimitris Korobilis, Professor of Econometrics, Adam Smith Business School, 2 Discovery Place, Glasgow, G11 6EY, United Kingdom, email: \href{mailto:Dimitris.Korobilis@glasgow.ac.uk}{Dimitris.Korobilis@glasgow.ac.uk}. }}
	
	\author[1,2]{Nicolas Hardy}
	\author[2,3]{Dimitris Korobilis} 
	\affil[1]{{\footnotesize Facultad de Administracion y Economia, Universidad Diego Portales}}
	\affil[2]{{\footnotesize Adam Smith Business School, University of Glasgow}}
	\affil[3]{{\footnotesize CAMP, BI Norwegian Business School}}	
	\date{\today}
	
	\maketitle
	\begin{abstract}
\noindent We revisit macroeconomic time-varying parameter vector autoregressions (TVP-VARs), whose persistent coefficients may adapt too slowly to large, abrupt shifts such as those during major crises. We explore the performance of an adaptively-varying parameter (AVP) VAR that incorporates deterministic adjustments driven by observable exogenous variables, replacing latent state innovations with linear combinations of macroeconomic and financial indicators. This reformulation collapses the state equation into the measurement equation, enabling simple linear estimation of the model. Simulations show that adaptive parameters are substantially more parsimonious than conventional TVPs, effectively disciplining parameter dynamics without sacrificing flexibility. Using macroeconomic datasets for both the U.S. and the euro area, we demonstrate that AVP-VAR consistently improves out-of-sample forecasts, especially during periods of heightened volatility.
		\bigskip
		
		\bigskip \medskip
		
	\end{abstract}
	\thispagestyle{empty} 
	
	\newpage
	\onehalfspacing
	\setcounter{page}{1}
	\section{Introduction}	
	The use of time-varying parameter (TVP) models in macroeconomics and finance dates back over half a century \citep{CooleyPrescott1973,HarveyPhillips1982}. Seminal work by \cite{Doanetal1984} and later by \cite{Sims1993} introduced vector autoregressions (VARs) with coefficients that evolve over time to capture gradual changes in macroeconomic dynamics. Since then, TVP regressions and TVP-VARs have become central tools for structural analysis, forecasting, and policy evaluation, particularly in periods of economic instability. A dominant assumption in this literature is that parameters follow random walks and this specification has barely been disputed over several decades.\footnote{As \cite{Sims1993} elegantly puts it: ``Litterman and I have both experimented with specifications where [the time-varying parameters are autoregressions of order one with parameter $\Theta$]... The best choice of $\Theta$ has always turned out to be close to one.''} As a result, the modern canonical TVP-VAR model, formalized in a Bayesian setting by \citet{CogleySargent2005} and extended by \citet{Primiceri2005}, relies on the random walk evolution of autoregressive coefficients and volatilities and inference typically proceeds via state-space methods.\footnote{Alternative approaches include kernel smoothing \citep{Kapetaniosetal2019} and generalized autoregressive score models \citep{Crealetal2013}, but these methods also generate persistent parameter dynamics that effectively mimic the gradual evolution implied by random walk TVP specifications.} Despite their popularity, random walk TVPs impose smoothness and gradual drift on the parameters, which may poorly capture abrupt shocks, such as those experienced during the global financial crisis or the COVID-19 pandemic. 
	
	In this paper, we propose the adaptively-varying parameter VAR (AVP-VAR), a new class of TVP-VARs where autoregressive coefficients exhibit persistent time variation driven by observed exogenous variables, rather than stochastic random walks. These observable drivers can include forward-looking macroeconomic and financial indicators such as measures of uncertainty, stress and volatility, or consumer, business, and investor expectations; that is, variables that are publicly available, interpretable, and often directly linked to policy or economic conditions. In traditional TVP-VARs, time variation is inferred from the likelihood via latent state-space models, which can be sensitive to prior choices and prone to overfitting, especially in high-dimensional settings. By contrast, AVP-VARs tie parameter shifts to observable series, offering a more transparent and economically interpretable mechanism for capturing structural change. Our proposed formulation eliminates the need for state filtering and reduces estimation to standard linear regression techniques, which greatly simplifies computation and makes real-time monitoring and forecasting computationally feasible. Linear estimation also helps manage the high-dimensional parameter space, enhancing stationarity and stability without complex sampling schemes. Finally, by replacing the unobserved state error in the random walk with a linear combination of exogenous variables, we reduce the model’s reliance on prior tuning. In standard TVP-VARs, the covariance of the state error governs how freely parameters evolve over time, and small changes in its prior can lead to dramatically different results. Our approach shifts the source of time variation to observable series, allowing shrinkage methods to determine which variables matter most. This makes the AVP-VAR more robust to prior choices, while preserving flexibility and improving interpretability, replicability, and computational efficiency.
	
	Our primary contribution is to demonstrate the effectiveness of this simple yet powerful framework for macroeconomic TVP-VARs. Recent studies have explored flexible approaches that utilize observable variables to drive time variation in parameters, such as regression trees and random forests \citep{Deshpandeetal2020,Hauzenbergeretal2022,Goulet2024}, neural networks \citep{Guetal2021,LiYuan2024}, and Gaussian processes \citep{FoxDunson2015}.\footnote{\cite{KapetaniosTzavalis2010} and \cite{Uzeda2021} represent a different strand that links time variation in parameters to model residuals instead of exogenous drivers. Despite the fact that \cite{Uzeda2021} show important implications of their specification for structural inference, making parameters a function of residuals limits interpretability for forecasting because parameter shifts reflect model fit rather than economic conditions.} All these machine learning inspired approaches can be valuable for univariate or low-dimensional time-varying parameter inference, however, they may be less suitable for VARs, where the parameter space is large and unrestricted flexibility can lead to unstable time series dynamics and unreliable forecasts. In multivariate settings, even small perturbations in priors or likelihood assumptions can produce dramatically different parameter paths, making results fragile and difficult to replicate.\footnote{Prior sensitivity in TVP-VARs is well documented. For example, \citet{CogleySargent2005} propose a ``business as usual'' prior that effectively shrinks time variation towards recursive least squares. Related approaches include hybrid TVP-VARs by \citet{Chan2023}, and hierarchical shrinkage methods by \citet{Amiretal2020}, which aim to stabilize estimation by controlling the degree of time variation.} 
	
	In contrast, we show that the AVP-VAR consistently improves upon the TVP-VAR benchmark across a range of scenarios. We perform three real-data forecasting applications, focusing on episodes with large structural shifts such as the global financial crisis of 2007-2009 and COVID-19. First, in a monthly U.S. VAR with output, inflation, and the short-term interest rate, we incorporate uncertainty measures and forward-looking indicators as exogenous drivers of autoregressive coefficients and correlations, yielding substantial forecasting gains overall. Second, using quarterly euro area data, we confirm that these gains extend beyond U.S. data, underscoring the general applicability of the approach. Finally, in an online supplement, we repeat the forecasting exercise using a default quarterly U.S. dataset (FRED-QD) and we show once more that the AVP-VAR delivers robust improvements. In all three exercises, the AVP-VAR is not only better than the TVP-VAR in many configurations, but also fairs well against some hard to beat benchmarks such as a heteroskedastic VAR \citep{ChanEisenstat2018} and an outlier-robust stochastic volatility model with Student-t innovations \citep{Carrieroetal2023}.
	
	To understand the source of these forecasting gains, we use real and synthetic data to assess the model's ability to capture parameter variation. Specifically, we establish two stylized facts through simulation. First, when the true parameter dynamics are characterized by random walk or near-random walk behavior, the adaptive parameter approach performs comparably to the canonical TVP estimator, when the exogenous drivers are simple Gaussian noise. This result suggests that our proposed adaptive parameter variation does not suffer meaningful efficiency losses under conventional dynamics. Second, when the random walk evolution is contaminated with random jumps/outliers or regime shifts, the advantages of AVP-VAR become evident. In the presence of even mildly informative observables, the adaptive specification yields more accurate estimates of time variation than standard TVP methods.	
	
	 Next, we document an additional stylized fact using the datasets from our forecasting exercises. In-sample parameter estimates reveal that the canonical TVP-VAR, faced with a large parameter space and a high-dimensional state covariance matrix, tends to heavily regularize the state innovations, leaving most parameters virtually time-invariant.\footnote{This limitation is evident in the estimates of \citet{Primiceri2005}, who shows via marginal likelihoods that the preferred TVP-VAR prior heavily regularizes parameters, making them effectively identical across three representative periods. Yet, simple least squares estimates of VARs on subsamples, such as pre- and post-Great Moderation, reveal parameter shifts that are far more pronounced than his results suggest.} We illustrate this point by comparing the estimated time-varying intercepts from the univariate Unobserved Components Stochastic Volatility (UC-SV) model -- a flexible local-level benchmark known for strong forecasting performance \citep{StockWatson2007} -- with those from the TVP-VAR: while the UC-SV intercepts exhibit substantial variation, the TVP-VAR counterparts remain comparatively more flat. Because VARs are more heavily parametrized than the UC-SV, conventional prior shrinkage tends to overpenalize coefficient variation, potentially muting important dynamics in parameters that should vary more than others, such as intercepts that approximate time-varying trends. In contrast, the AVP-VAR produces intercept dynamics that closely track the UC-SV estimates, while allowing the autoregressive coefficients to display heterogeneous patterns of variation: sometimes nearly constant, sometimes shifting more strongly. This ability to capture realistic time variation without excessive parametrization explains why AVP-VAR delivers sharper estimates and consistently better forecasts.
	
	The remainder of the paper is organized as follows. Section~2 describes our model and outlines the Gibbs sampler. Section 3 establishes the excellent out-of-sample properties of the proposed methodology using different datasets. In Section 4 we delve deeper into the ability of our algorithm to estimate time variation correctly, and we establish the stylized facts. Section 5 concludes the paper.
	
	\section{Methodology} \label{sec:model}
	\subsection{Baseline TVP-VAR Framework}
	
	Our starting point is the $p$-lag time-varying parameter VAR in \cite{Primiceri2005}. For an $n \times 1$ endogenous variable $y_{t}$ this is of the form
	\begin{equation}
		y_{t} = B_{0t} + \sum_{i=1}^{p}B_{it}y_{t-i} + A_{t}H_{t}\varepsilon_{t}, \qquad \varepsilon_{t} \sim \mathcal{N}\left( 0, I_n \right),
	\end{equation}
	where $B_{0t}$ is the $n \times 1$ vector of time-varying intercepts, $B_{it}$ is the $n \times n$ matrix of time-varying autoregressive coefficients for lag $i$, $A_{t}$ is a lower unitriangular $n \times n$ matrix of contemporaneous relations, and $H_{t}$ is a diagonal $n \times n$ matrix of time-varying standard deviations. As a consequence, the VAR covariance matrix of the reduced-form errors follows the implied decomposition of the form
	\begin{equation}
		\Omega_t = A_t H_{t} H_{t} A_t'.
	\end{equation}
	For inference, it is convenient to separate stochastic volatilities from contemporaneous interactions and do inference in $A_{t}$ and $H_{t}$ separately, instead of having to deal with the whole time-varying covariance matrix $\Omega_{t}$. However, the use of lower-triangular $A_t$ imposes a dependence on the ordering of variables in $y_t$, a limitation criticized heavily by \cite{Bognanni2018,Ariasetal2023,Ariasetal2024,Chanetal2024}. The ordering dependence arises because the lower-triangular structure of $A_t$ imposes a recursive identification scheme: the first variable in $y_t$ is assumed to affect all others contemporaneously, the second affects all but the first, and so on. This means that inference about the magnitude and evolution of contemporaneous relationships depends critically on how variables are ordered, which is a choice that is often arbitrary and lacks economic justification.
	
	To address this shortcoming, we adopt a VAR model with time-varying parameters and a factor stochastic volatility (FSV) decomposition of the reduced-form innovations, extending the constant parameter specifications in \cite{Korobilis2022a}, \cite{StockWatson2005JEEA} and others. The model becomes:
	\begin{equation}
		y_t = B_{0t} + \sum_{i=1}^{p}B_{it}y_{t-i} + \Lambda_t f_t + \Sigma_{t}^{1/2} \varepsilon_t, \qquad \varepsilon_t \sim \mathcal{N}(0, I_n),
	\end{equation}
	where $f_t = [f_{1t}, \dots, f_{rt}]^{\prime}$ is an $r \times 1$ vector of latent factors, $f_t \sim \mathcal{N}(0, I_r)$. The matrix $\Lambda_t$ is an $n \times r$ matrix of time-varying factor loadings. The matrix $\Sigma_t^{1/2}$ is an $n \times n$ diagonal matrix of time-varying idiosyncratic standard deviations. In our framework, the term $\Lambda_t f_t$ captures the common time-varying covariance structure across equations. This assumption implies the following factor structure for the time-varying covariance matrix:
	\begin{equation}
		\Omega_t = \underbrace{\Lambda_t \Lambda_t^{\prime}}_{\text{common component}} + \underbrace{\Sigma_t}_{\text{idiosyncratic shocks}}.
	\end{equation}
	That is, innovations to $y_t$ are composed of common latent factors with time-varying loadings and idiosyncratic disturbances with stochastic volatility. Unless we want to uniquely identify the factors $f_{t}$, no restrictions are required in $\Lambda_t$ as the common component is always identified. Since $\Lambda_t f_t$ captures co-movement in the reduced-form shocks without imposing a causal hierarchy, the estimated covariance structure $\Lambda_t \Lambda_t^{\prime}$ remains the same regardless of how we permute the elements of $y_t$, which means that this VAR formulation is order invariant.
	
	We can write this model more compactly in regression form. Let $x_t = [1, y_{t-1}^{\prime}, \dots, y_{t-p}^{\prime}]^{\prime}$ be a $k \times 1$ vector of regressors, where $k = np + 1$. Given the $r \times 1$ vector of latent factors $f_t$, define the regressor matrix $X_t = I_n \otimes [x_t^{\prime}, f_t^{\prime}]$, which is an $n \times n(k + r)$ matrix that stacks $x_t$ and $f_t$ per equation. Define the time-varying coefficient vector $\theta_t = [\beta_t^{\prime}, \lambda_t^{\prime}]^{\prime}$ of dimension $n(k + r) \times 1$, where $\beta_t = \left(B_{0t}^{\prime},\text{vec}(B_{1t})^{\prime},...,\text{vec}(B_{pt})^{\prime} \right)^{\prime}$ contains the time-varying autoregressive coefficients and $\lambda_t = \text{vec}(\Lambda_{t})^{\prime}$ the time-varying factor loadings, where $\text{vec}(A)$ is the operator that stacks the columns of a matrix $A$ into a single column vector. Let $v_t = \Sigma_t^{1/2} \varepsilon_t$, then the model can be rewritten as:
	\begin{equation}
		y_t = X_t \theta_t + v_t, \qquad v_t \sim \mathcal{N}(0, \Sigma_t). \label{eq:VAR}
	\end{equation}
	This formulation nests the standard time-varying parameter VAR and introduces a flexible factor structure in the error term. Cross-equation dependence in the shocks is governed by the latent factors and the time-varying loading matrix $\lambda_t$, while idiosyncratic uncertainty is governed by the stochastic volatilities $\Sigma_t$.
	
	\paragraph{Standard time variation specification.} In the canonical TVP-VAR literature, time variation in the coefficients and volatilities is modeled via independent random walks:
	\begin{align}
		\theta_t &= \theta_{t-1} + \eta_t, \qquad \eta_t \sim \mathcal{N}(0, Q), \label{eq:RW_theta} \\
		\log \sigma_{i,t} &= \log \sigma_{i,t-1} + \zeta_{i,t}, \qquad \zeta_{i,t} \sim \mathcal{N}(0, \omega_i^2), \quad i = 1, \ldots, n, \label{eq:RW_sigma}
	\end{align}
	where $\sigma_{i,t}$ denotes the $i$-th diagonal element of $\Sigma_t^{1/2}$, and $Q$ is the covariance matrix of the state innovations. This random walk specification allows parameters to drift gradually over time, capturing smooth structural changes. However, this approach has several limitations. First, it requires estimating the high-dimensional covariance matrix $Q$, which is computationally demanding and often requires restrictive assumptions such as diagonal or reduced-rank structures \citep{Chanetal2020}. Second, the random walk does not directly link parameter changes to observable economic drivers or regime shifts, making the source of time variation difficult to interpret. Third, the flexibility of the random walk can lead to overfitting, especially in short samples or when structural breaks are sparse \citep{Amiretal2020}. Our proposed framework addresses these limitations by explicitly modeling time variation through observable drivers.

	\subsection{Adapting the evolution of parameters to match changing economic conditions: The AVP-VAR} \label{subsec:instruments}
	
	We depart from the standard TVP specification in equations \eqref{eq:RW_theta}--\eqref{eq:RW_sigma} by replacing the unobserved state innovation with an observable drift. Let $Z_t\in\mathbb{R}^m$ denote a vector of exogenous drivers (e.g., forward-looking indicators of uncertainty, stress, volatility, or expectations), and define the cumulative drivers $C_t := \sum_{s=1}^{t-1} Z_s$. The time-varying coefficients evolve according to
	\begin{equation}
		\theta_t \;=\; \theta_{t-1} + \gamma Z_{t-1},
		\qquad\Rightarrow\qquad
		\theta_t \;=\; \theta + \gamma C_t,
		\label{eq:theta_evolve}
	\end{equation}
	where $\theta\in\mathbb{R}^{n(k+r)}$ collects baseline coefficients and $\gamma\in\mathbb{R}^{n(k+r)\times m}$ maps drivers to the drift of the coefficients. Unlike \citet{Fischeretal2023}, whose time-varying parameters combine pre-specified effect modifiers with stochastic innovations, AVP-VAR replaces state errors with linear combinations of observable economic drivers. This specification preserves the persistent dynamics of a random walk while ensuring all parameter drifts are explicitly determined by economic conditions.
	
	Let $x_t=[1,y_{t-1}^\prime,\ldots,y_{t-p}^\prime]^\prime\in\mathbb{R}^{k}$ with $k=np+1$, and $f_t\in\mathbb{R}^r$. Define $X_t := I_n\otimes[x_t^\prime,\,f_t^\prime]\in\mathbb{R}^{n\times n(k+r)}$ and $\theta_t=[\beta_t^\prime,\lambda_t^\prime]^\prime\in\mathbb{R}^{n(k+r)}$, where $\beta_t$ stacks the VAR coefficients and $\lambda_t=\mathrm{vec}(\Lambda_t)$ stacks the factor loadings. The measurement equation \eqref{eq:VAR} can then be written as
	\begin{equation}
		y_t \;=\; X_t \theta_t + v_t \;=\; X_t\theta + X_t\gamma C_t + v_t,
		\qquad v_t\sim\mathcal{N}(0,\Sigma_t),\quad \Sigma_t=\mathrm{diag}(\sigma_{1t}^2,\ldots,\sigma_{nt}^2).
		\label{eq:meas_avp}
	\end{equation}
	Using the mixed-product rule, $X_t\gamma C_t=(C_t^\prime\otimes X_t)\mathrm{vec}(\gamma)$, so that stacking $Y=[y_1^\prime,\ldots,y_T^\prime]^\prime$ and $V=[v_1^\prime,\ldots,v_T^\prime]^\prime$ yields the static linear regression
	\begin{equation}
		Y \;=\; \tilde{X}(1_T\otimes \theta) +
		\tilde W \mathrm{vec}(\gamma) \;+\; V,
		\label{eq:stacked_model}
	\end{equation}
	with $\tilde X = I \otimes X \in\mathbb{R}^{Tn\times Tn(k+r)}$ and $\tilde W= \left[ C_1^\prime\!\otimes X_1, \hdots, C_T^\prime\!\otimes X_T \right]^{\prime} \in\mathbb{R}^{Tn\times mn(k+r)}$. This representation collapses the state equation into the measurement equation once $\{C_t\}_{t=1}^{T}$ are precomputed; no Kalman filtering is required.
	
	\paragraph{Connections to existing frameworks.} Our specification relates to several strands of the literature. If we place a spherical prior on \(\mathrm{vec}(\gamma)\) of the form \(\mathrm{vec}(\gamma)\sim\mathcal N(0,g^{-1}I)\), then
	\[
	\theta_t \mid \theta_{t-1}, Z_t \;\sim\; \mathcal N\!\Big(\theta_{t-1},\; g^{-1}\,(Z_{t-1}'Z_{t-1})\, I_{n(k+r)}\Big).
	\]
	This implied prior for $\theta_{t}$ represents a reformulation of Zellner's $g$-prior \citep{Zellner1986}, where the prior covariance is scaled by the outer product of the observable drivers $Z_t$. Unlike standard $g$-priors that scale the prior by the design matrix $X_t^{\prime} X_t$, our approach uses external information $Z_t$ to inform the degree of parameter variability, thereby incorporating economic knowledge into the prior structure. This connection highlights how our model embeds prior beliefs about the relationship between parameter shifts and observable economic conditions. The key advantage of this reformulation is that it allows the prior variance of $\theta_t$ to adapt endogenously to economic conditions: when $Z_t$ signals heightened uncertainty or structural stress, the prior automatically permits larger parameter movements, while stable periods induce stronger shrinkage toward $\theta_{t-1}$. This contrasts with standard $g$-priors that scale variability by the statistical properties of the data ($X_t^{\prime} X_t$), which may not align with economically meaningful sources of parameter instability. By conditioning on $Z_t$, our prior variance directly reflects external information about when and why parameters should be allowed to change, thereby incorporating substantive economic knowledge into the model's flexibility.
	
	Second, from a reduced-form perspective, equation \eqref{eq:stacked_model} can be interpreted as a VAR with interaction terms between the regressors $X_t$ and the cumulative drivers $\sum_{j=1}^{t-1} Z_j$. This parallels the literature on structural VARs with interactions, where coefficients are allowed to vary with external regimes or indicators. For instance, \cite{Aastveitetal2017} study VARs with interactions between policy and uncertainty measures, while \cite{RameyZubairy2018} use local projections with interactions to allow for state-dependent effects of fiscal policy, and \cite{BachmannSims2012} examine how parameter stability varies with the level of uncertainty. Our framework shares the spirit of these approaches but provides a unified treatment in which all parameter changes are governed by a common set of observable drivers, facilitating interpretation while reducing dimensionality.
	
	Third, our model can be viewed as an alternative to TVP specifications in which the unknown state innovation $\eta_t$ is replaced by a linear combination of observable factors $\gamma Z_t$. Unlike reduced-rank TVP-VAR models that use latent factors to govern parameter dynamics \citep{Chanetal2020}, the AVP-VAR directly links time variation to interpretable economic drivers, thereby enhancing transparency and facilitating economic interpretation. This is particularly valuable when the goal is to understand how specific shocks or policy changes affect macroeconomic relationships.
	
	\paragraph{Parsimony and choice of drivers.} Notice that when $Z_{t}$ are random Gaussian variables, then $\gamma Z_{t}$ approximates arbitrarily well the random walk time variation. However, the AVP-VAR framework achieves greater parsimony when the number of relevant drivers is small relative to the sample size. In practical situations, it is fair to assume that we have a collection of drivers $Z_{t}$ that are informative about the state of the economy (e.g. breaks), but we are unsure about which exact drivers are informative for the TVPs. That is, the expectation is that each driver $j$ should affect coefficients only at specific time periods, and most drivers should be irrelevant. This sparsity structure is precisely what Bayesian shrinkage priors are designed to exploit.
	
	To achieve parsimony and select relevant drivers and time periods, we employ the horseshoe prior for $\text{vec}(\gamma)$ \citep{Carvalhoetal2010}. The general form of the prior, for the i-th element of $\gamma$ is
	\begin{eqnarray}
		\gamma_{i} | \psi_{i}, \tau & \sim & \mathcal{N}\left(0,\psi_{i}^{2}\tau^{2}\right), \\
		\psi_{i} & \sim & \text{Cauchy}_+(0,1), \\
		\tau & \sim & \text{Cauchy}_+(0,1).
	\end{eqnarray}
	The horseshoe prior is particularly well-suited for sparse signal recovery, as it combines strong shrinkage for irrelevant coefficients with minimal shrinkage for important ones. Importantly, the horseshoe prior has been shown to yield Bayes estimates that are consistent a posteriori and attain risk equivalent to the Bayes oracle under sparsity \citep{Armaganetal2013,Ghoshetal2016}. This theoretical property ensures that our model can effectively identify the subset of drivers and time periods that genuinely govern parameter variation, even when the candidate set is large.
	
	In this paper our benchmark for comparison is the random walk evolution of equation \eqref{eq:RW_theta}, which we will refer to simply as TVP. In order to maintain comparability and the effect of subjective prior choices, whenever estimating dynamics under this specification we also use the same horseshoe prior on $\eta_{t}$. That is, we assume $Q = diag(q_{1}^{2}\zeta^{2},...,q_{n(k+r)}^{2}\zeta^{2} )$ where $\zeta$ and $q_{i}$, $i=1,...,n(k+r)$ have half-Cauchy priors. This prior automatically adjusts the amount of time-variation in $\theta_{t}$ by shrinking $\eta_{t}$ more aggressively when any of the VAR dimensions $n$ or $k$ or $r$ increase.
	
	\paragraph{Relationship to prior tilting and external information.} Finally, our approach of conditioning parameter dynamics on $Z_t$ can be interpreted as incorporating prior information in a manner related to two strands of the Bayesian VAR literature. On one hand, the entropic tilting or relative entropy approach adjusts an ex-post predictive density to satisfy external moment conditions while remaining as close as possible, in the Kullback-Leibler sense, to the original model \citep{Robertsonetal2005,Cogleyetal2005,TallmanZaman2020}. On the other hand, the democratic prior of \cite{Wright2013} centers the VAR's long-run mean on survey expectations and shrinks heavily toward that level. Both approaches use external information to shape the forecast distribution. Relative-entropy tilting operates in the predictive space, whereas the democratic prior and AVP-VAR operate in the parameter space. In all three cases, external information disciplines the model without fundamentally altering the likelihood, thereby balancing data-driven flexibility with economically motivated restrictions.
	
	\subsection{Posterior Inference}
	\label{subsec:posterior}
	
	Estimation exploits the static-regression form in \eqref{eq:stacked_model}. We precompute cumulative drivers $C_t=\sum_{s=1}^{t-1} Z_s$ for $t=1,\ldots,T$. Partition $\theta=[\beta^\prime,\lambda^\prime]^\prime$ and $\gamma=[\gamma_\beta^\prime,\gamma_\lambda^\prime]^\prime$ conformably, where $\beta\in\mathbb{R}^{nk}$ contains baseline VAR coefficients, $\lambda\in\mathbb{R}^{nr}$ contains baseline factor loadings, $\gamma_\beta\in\mathbb{R}^{nk\times m}$, and $\gamma_\lambda\in\mathbb{R}^{nr\times m}$. For each equation $i=1,\ldots,n$, define augmented coefficient vectors
	\[
	\tilde\beta_i := \begin{bmatrix}\beta_i\\ \mathrm{vec}(\gamma_{\beta,i})\end{bmatrix}\in\mathbb{R}^{k+km},
	\qquad
	\tilde\lambda_i := \begin{bmatrix}\lambda_i\\ \mathrm{vec}(\gamma_{\lambda,i})\end{bmatrix}\in\mathbb{R}^{r+rm},
	\]
	where $\beta_i\in\mathbb{R}^{k}$ and $\lambda_i\in\mathbb{R}^{r}$ are the $i$-th equation's baseline coefficients, while $\gamma_{\beta,i}\in\mathbb{R}^{k\times m}$ and $\gamma_{\lambda,i}\in\mathbb{R}^{r\times m}$ map drivers to coefficient drift. Time-varying parameters are recovered deterministically as
	\begin{equation}
		\beta_{it} = \beta_i + \gamma_{\beta,i} C_t,
		\qquad
		\lambda_{it} = \lambda_i + \gamma_{\lambda,i} C_t,
		\qquad t=1,\ldots,T.
		\label{eq:recover_coef}
	\end{equation}
	
	Define augmented regressors $\tilde x_t := [x_t^\prime, (x_t\otimes C_t)^\prime]^\prime\in\mathbb{R}^{k+km}$ and $\tilde f_t := [f_t^\prime, (f_t\otimes C_t)^\prime]^\prime\in\mathbb{R}^{r+rm}$. Let $y_{it}$ denote the $i$-th element of $y_t$ and $\sigma_{it}^2$ the corresponding idiosyncratic variance. The Gibbs sampler iterates over the following steps, equation-by-equation.
	
	\paragraph{Step 1: Augmented VAR coefficients.}
	Given $\{\lambda_{it},f_t,\sigma_{it}^2\}$, define $\tilde y_{it}:=y_{it}-f_t^\prime \lambda_{it}$ so that
	\[
	\tilde y_{it} = \tilde x_t^\prime \tilde\beta_i + v_{it},
	\qquad v_{it}\sim\mathcal{N}(0,\sigma_{it}^2).
	\]
	Place a horseshoe prior on $\tilde\beta_i$: for $\ell=1,\ldots,k+km$,
	\begin{equation}
		\tilde\beta_{i\ell}\mid \psi_{i\ell},\tau_i \sim \mathcal{N}(0,\psi_{i\ell}^2\tau_i^2),\quad
		\psi_{i\ell}\sim \text{Cauchy}_+(0,1),\quad
		\tau_i\sim \text{Cauchy}_+(0,1).
		\label{eq:hs_beta}
	\end{equation}
	The conditional posterior is $\tilde\beta_i\mid \cdot \sim \mathcal{N}(\bar\mu_{\beta,i},\bar V_{\beta,i})$ with
	\[
	\bar V_{\beta,i}^{-1} = D_{\beta,i}^{-1} + \sum_{t=1}^T \sigma_{it}^{-2}\,\tilde x_t \tilde x_t^\prime,
	\qquad
	\bar\mu_{\beta,i} = \bar V_{\beta,i}\!\left(\sum_{t=1}^T \sigma_{it}^{-2}\,\tilde x_t\,\tilde y_{it}\right),
	\]
	where $D_{\beta,i}=\mathrm{diag}(\psi_{i1}^2\tau_i^2,\ldots,\psi_{i,k+km}^2\tau_i^2)$. Reconstruct $\beta_{it}$ via \eqref{eq:recover_coef}.
	
	\paragraph{Step 2: Augmented factor loadings.}
	Given $\{\beta_{it},f_t,\sigma_{it}^2\}$, define $\hat y_{it}:=y_{it}-x_t^\prime\beta_{it}$ so that
	\[
	\hat y_{it} = \tilde f_t^\prime \tilde\lambda_i + v_{it},
	\qquad v_{it}\sim\mathcal{N}(0,\sigma_{it}^2).
	\]
	Place a horseshoe prior on $\tilde\lambda_i$: for $\ell=1,\ldots,r+rm$,
	\begin{equation}
		\tilde\lambda_{i\ell}\mid \xi_{i\ell},\tau_{\lambda,i} \sim \mathcal{N}(0,\xi_{i\ell}^2\tau_{\lambda,i}^2),\quad
		\xi_{i\ell}\sim \text{Cauchy}_+(0,1),\quad
		\tau_{\lambda,i}\sim \text{Cauchy}_+(0,1).
		\label{eq:hs_lambda}
	\end{equation}
	The conditional posterior is $\tilde\lambda_i\mid \cdot \sim \mathcal{N}(\bar\mu_{\lambda,i},\bar V_{\lambda,i})$ with
	\[
	\bar V_{\lambda,i}^{-1} = D_{\lambda,i}^{-1} + \sum_{t=1}^T \sigma_{it}^{-2}\,\tilde f_t \tilde f_t^\prime,
	\qquad
	\bar\mu_{\lambda,i} = \bar V_{\lambda,i}\!\left(\sum_{t=1}^T \sigma_{it}^{-2}\,\tilde f_t\,\hat y_{it}\right),
	\]
	where $D_{\lambda,i}=\mathrm{diag}(\xi_{i1}^2\tau_{\lambda,i}^2,\ldots,\xi_{i,r+rm}^2\tau_{\lambda,i}^2)$. Reconstruct $\lambda_{it}$ via \eqref{eq:recover_coef} and form $\Lambda_t$.
	
	\paragraph{Step 3: Factors.}
	Conditional on $\{\beta_t,\Lambda_t,\Sigma_t\}$, define $\tilde y_t:=y_t - (I_n\otimes x_t^\prime)\beta_t$ and
	\[
	\bar G_t^{-1} = I_r + \Lambda_t^\prime \Sigma_t^{-1}\Lambda_t.
	\]
	Then $f_t \mid \cdot \sim \mathcal{N}(\bar G_t \Lambda_t^\prime \Sigma_t^{-1}\tilde y_t, \bar G_t)$.
	
	\paragraph{Step 4: Stochastic volatilities.}
	For each equation $i$, log-volatilities evolve as $\log\sigma_{it}=\log\sigma_{i,t-1}+\zeta_{it}$ with $\zeta_{it}\sim\mathcal{N}(0,\omega_i^2)$ from \eqref{eq:RW_sigma}. Update $\{\sigma_{it}^2\}_{t=1}^T$ using standard precision-based stochastic volatility algorithms conditional on residuals $v_{it}=y_{it}-x_t^\prime\beta_{it}-f_t^\prime\lambda_{it}$.
	
	\paragraph{Step 5: Horseshoe hyper-parameters.}
	Update local scales $\{\psi_{i,\ell}\},\{\xi_{i,\ell}\}$ and global scales $\tau_i,\tau_{\lambda,i}$ in \eqref{eq:hs_beta}--\eqref{eq:hs_lambda} via slice sampling or equivalent auxiliary-variable schemes. 
	
	\medskip
	\noindent
	Because time variation is encoded through observable $\{C_t\}$, all coefficient updates reduce to weighted least squares with no Kalman filtering required. Augmented dimensions scale as $k+km$ and $r+rm$, growing linearly in the number of drivers $m$ rather than sample size $T$, yielding substantial computational gains when $m\ll T$. More details about the Gibbs sampler are provided on the online supplement.

	
	\newpage	
	\section{Forecasting with macroeconomic AVP-VAR}
	\subsection{Forecast Setup}
	We assess the empirical performance of AVP-VAR using a representative three-variable system encompassing output, price inflation, and a short-term interest rate, and we use the first two variables to evaluate forecast performance. To ensure our findings generalize across time series with different properties, we consider two distinct datasets, a monthly dataset for the U.S. and a quarterly dataset for the euro area.\footnote{In the online supplement we give further evidence by repeating our exercise using quarterly U.S. time series from the \citet{mccracken2020fred} FRED-QD database.} Selection of the series to include as drivers of parameter variation is an important aspect of our model, and it is a data-specific decision. In general, following the evidence in the Monte Carlo exercises we select a moderate number of important variables that have captured key break events in the past and will probably continue to do so in the future, including (but not limited to): uncertainty, consumer and other survey expectations, credit risk, and financial stress.

	We evaluate AVP-VAR's forecasting performance through a standard pseudo out-of-sample analysis. Our benchmark set comprises seven competing specifications that span constant and time-varying parameter frameworks, as well as constant and time-varying volatility, namely:
	\begin{itemize}
		\item \textbf{Constant Parameter VAR (CP-VAR):} BVAR featuring time-invariant autoregressive coefficients and covariances and variances.
		\item \textbf{Constant Parameter VAR with Stochastic Volatility (CP-VAR-SV):} This is as above, but we allow for univariate stochastic volatility in each of the $n$ equations. Both covariances and variances are allowed to be time-varying.
		\item \textbf{Time-varying Parameter VAR with Empirical Bayes prior (TVP-VAR-EB):} A VAR with time-varying coefficients, covariances, and variances, all evolving as random walks. Following \citet{Primiceri2005}, we estimate a constant-parameter VAR on a training sample to inform the prior moments for the TVP-VAR. These moments are then further scaled using a shrinkage prior that limits excessive time variation in the parameters.
		\item \textbf{Time-varying Parameter VAR with Full Bayes prior (TVP-VAR):} Same model as above but with prior on $\beta_{t}$ tuned automatically by using a Horseshoe hierarchical (full Bayes) prior on the state variance of the random walk dynamics.
		\item \textbf{Stochastic Volatility, Outlier-robust VAR with Student-t errors (VAR-SVOt)} This is the best performing model in \cite{Carrieroetal2023} for forecasting the pandemic and other high-volatility periods.
		\item \textbf{Factor-augmented VAR (FAVAR):} This is a simple three-variable VAR with constant parameters, estimated with ordinary least squares, augmented with an additional factor extracted from the drivers $Z_{t}$.
		\item \textbf{Factor-augmented VAR (FAVAR-SV):} This is the three-variable CP-VAR-SV augmented with an additional factor extracted from the drivers $Z_{t}$.

	\end{itemize}
	
	The specifications above cover a wide range of VARs considered in the literature, spanning fully constant-parameter models, heteroskedastic specifications, and fully time-varying parameter frameworks. The two factor-augmented VARs in our comparison include only factors extracted from the driver set $Z_{t}$, allowing us to test whether AVP-VAR's superior forecasting performance is due solely to its broader information set, or also to its ability to model dynamic relationships more effectively. With the exception of the TVP-VAR-EB, which relies on semi-subjective tuning of prior hyper-parameters, and FAVAR, which is a simple ordinary least square estimation, all other models use the same horseshoe prior as AVP-VAR to select drivers. This prior enables tuning-free and adaptive shrinkage of both constant and time-varying parameters, reducing the influence of subjective prior choices. For the TVP-VAR-EB, we follow the exact recommendations of \citet{Primiceri2005}, who explores various hyper-parameter settings and ultimately adopts a conservative value that limits excessive time variation. Although these recommendations were based on data preceding major crises such as the global financial crisis (GFC), we adopt them as a proxy for how a less-experienced user might approach forecasting with a TVP-VAR, favouring stability over flexibility. The VAR-SVOt is estimated using the exact sampler and settings provided by \citet{Carrieroetal2023}. Finally, when estimating constant variances, we use diffuse priors; for stochastic volatility models, we employ weakly informative priors (not fully diffuse) to avoid numerical instability and explosive estimates. Full estimation details and parameter settings for all competing models are provided in the online supplement.
	
	All models are estimated on a three-variable VAR with two lags ($p = 2$), capturing economic activity, inflation, and the nominal interest rate. Results are qualitatively robust to alternative lag choices ($p \in \{1, 3, 4\}$), which are available in the online supplement. For all VARs that use the factor decomposition of the residuals variance matrix, the number of factors $r$ is set to one.\footnote{In unreported results, we also consider $r=2$ with qualitatively the same conclusions.} Forecasting performance is evaluated using a standard pseudo out-of-sample recursive scheme: estimation begins with the first 50\% of the sample, and the sample is then expanded one observation at a time, re-estimating the model and generating forecasts at each step until the end of the sample. We consider forecast horizons up to two years, corresponding to 24 months for monthly data and 8 quarters for quarterly data. Iterative VAR forecasting uses standard formulas; details are provided in the online supplement. Prior to estimation, all variables and drivers are standardized (demeaned and scaled to unit variance) to ensure numerical stability, particularly for models with time-varying parameters. The evaluation period spans multiple phases of the business cycle, including major economic disruptions such as the 2008–2009 financial crisis and the COVID-19 pandemic, providing a rigorous test of each model's ability to adapt to structural change.

	\subsection{Forecasts for the US: Monthly data and results}
	
	We estimate a three-variable VAR using monthly U.S. data spanning January 1985 to November 2023. The endogenous variables are: (i) inflation, measured as the monthly log-difference of the Personal Consumption Expenditures Price Index (PCEPI); (ii) real economic activity, proxied by the monthly log-difference of the Industrial Production Index (INDPRO); and (iii) the effective federal funds rate (FEDFUNDS), representing the monetary policy stance.
	
	The driver set $Z_t$ includes $m=12$ uncertainty-related indicators that capture diverse dimensions of economic and financial instability. These comprise macroeconomic and financial uncertainty \citep{Juradoetal2015, Ludvigsonetal2021}, geopolitical risk \citep{CaldaraIacoviello2022}, U.S. economic policy uncertainty \citep{Bakeretal2016}, global economic conditions \citep{Baumeisteretal2022}, consumer sentiment (University of Michigan), business conditions \citep{Aruobaetal2009}, equity market volatility \citep{Bakeretal2019}, and the global financial cycle \citep{MirandaRey2020}. These indicators span domestic and global factors, real and financial variables, and policy and market-based metrics, offering comprehensive coverage of potential sources of structural change. Table~\ref{tab:data_description} provides detailed descriptions, transformations, and sources for all variables.
	
	\begin{table}[htbp!]
		\singlespacing
		\centering
		\scriptsize
		\caption{Variables and Drivers U.S. monthly data} \label{tab:data_description}
		\begin{threeparttable}
			\begin{tabular}{@{}p{2cm}p{6cm}p{3cm}p{0.8cm}@{}}
				\toprule
				\textbf{Series ID} & \textbf{Description} & \textbf{Source} & \textbf{tcode} \\
				\midrule
				\multicolumn{4}{c}{\textbf{VAR Variables}} \\
				\midrule
				PCEPI & Personal Consumption Expenditures, seasonally adjusted. & FRED$^{1}$ & 5 \\
				FEDFUNDS & Federal Funds Effective Rate. & FRED$^{1}$ & 2 \\
				INDPRO & Industrial Production: Total Index, seasonally adjusted. & FRED$^{1}$ & 5 \\
				\addlinespace[0.5em]
				\midrule
				\multicolumn{4}{c}{\textbf{Drivers} ($Z_t$)} \\
				\midrule
				USEPU & US Economic Policy Uncertainty Index (news-based). & PolicyUncertainty.com & 1 \\
				GPR & Geopolitical Risk Index based on news coverage. & M. Iacoviello$^{2}$ & 1 \\
				GECON & Global economic activity index (16 components). & C. Baumeister$^{3}$ & 1 \\
				EBP, GZ & Excess Bond Premium and GZ Spread. & FRB$^{4}$ & 1 \\
				JLN12, JLNF12 & Macro and financial uncertainty indices. & S. Ludvigson$^{5}$ & 1 \\
				UMCSENT & University of Michigan Consumer Sentiment Index. & FRED$^{1}$ & 1 \\
				ISENT & Investor Sentiment Index & J. Wurgler$^{6}$ & 1 \\
				GFC & Global Financial Cycle Index. & S. Miranda-Agrippino$^{7}$ & 1 \\
				ADS & Aruoba-Diebold-Scotti Business Conditions Index. & Philadelphia Fed$^{8}$ & 1 \\
				EMV & Equity Market Volatility Index (newspaper-based). & PolicyUncertainty.com & 1 \\
				\midrule
				Frequency: & Monthly data & & \\
				Period: & 1985M1 -- 2023M11 & & \\
				\bottomrule
			\end{tabular}
			\begin{tablenotes}
				\scriptsize
				\item $^{1}$ Federal Reserve Economic Data
				\item $^{2}$ \href{https://www.matteoiacoviello.com/gpr.htm}{https://www.matteoiacoviello.com/gpr.htm}
				\item $^{3}$ \href{https://sites.google.com/site/cjsbaumeister/research}{https://sites.google.com/site/cjsbaumeister/research}
				\item $^{4}$ \href{https://www.federalreserve.gov/econres/notes/feds-notes/ebp\_csv.csv}{https://www.federalreserve.gov/econres/notes/feds-notes/ebp\_csv.csv}
				\item $^{5}$ \href{https://www.sydneyludvigson.com/macro-and-financial-uncertainty-indexes}{https://www.sydneyludvigson.com/macro-and-financial-uncertainty-indexes}
				\item $^{6}$ \href{https://pages.stern.nyu.edu/\~jwurgler/}{https://pages.stern.nyu.edu/\~jwurgler/}
				\item $^{7}$ \href{https://silviamirandaagrippino.com/code-data}{https://silviamirandaagrippino.com/code-data}
				\item $^{8}$ \href{https://www.philadelphiafed.org/surveys-and-data/real-time-data-research/ads}{https://www.philadelphiafed.org/surveys-and-data/real-time-data-research/ads}
			\end{tablenotes}
		\end{threeparttable}
	\end{table}
	
\begin{figure}[htbp!]  
		\centering
		\includegraphics[width=\textwidth]{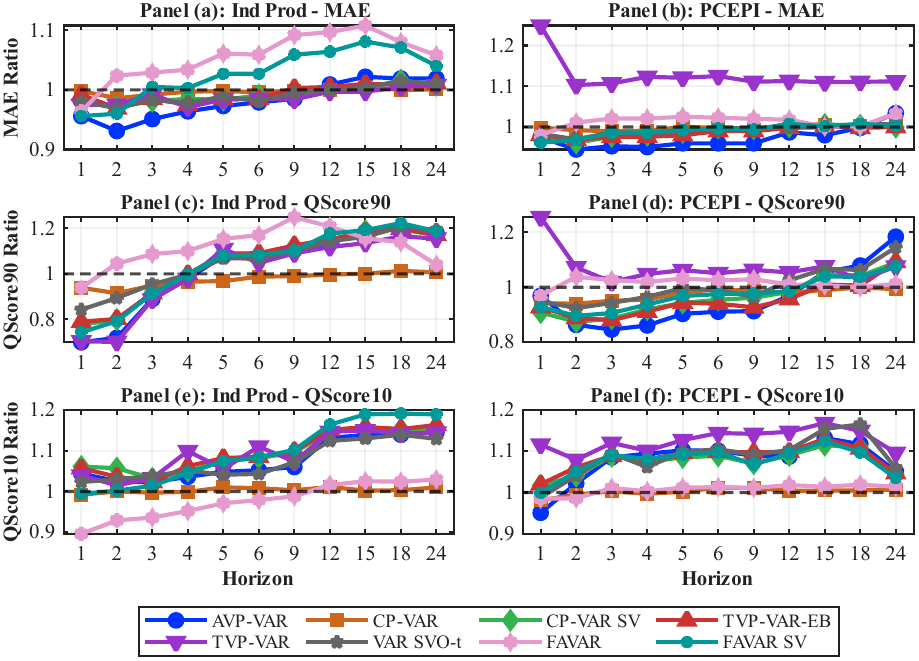}  
		\caption{Forecasting results using U.S. monthly data. The top row displays mean absolute forecast errors (MAE), while the middle and bottom rows show quantile scores for the upper (90th percentile) and lower (10th percentile) tails of each variable. All metrics are reported as ratios relative to a benchmark constant-parameter VAR($p$) estimated via least squares. A model is considered superior when its relative MAE, or relative quantile scores, are below one.} \label{fig:oos_US}
	\end{figure}

	Figure~\ref{fig:oos_US} and Table~\ref{tab:forecasting_mspeUS} summarize the forecasting results. AVP-VAR consistently outperforms all benchmark models for both industrial production and PCEPI inflation across many horizons. For industrial production, AVP-VAR achieves the lowest MSPE among all competitors from $h=3$ to $h=9$ months, with improvements up to 13\% relative to the OLS benchmark. For inflation, AVP-VAR delivers gains of 13\% at short horizons and remains as the top performer for most of forecasting horizons.
	
	Tail forecast evaluations reveal nuanced patterns: for lower tail risk (10th quantile), simpler models such as CP-VAR and FAVAR often outperform more complex specifications. However, for upper tail risk (90th quantile), AVP-VAR remains competitive, especially for inflation forecasts.
	
	Importantly, factor-augmented models that condition on $Z_t$ generally underperform AVP-VAR, suggesting that its superior performance stems not merely from access to additional information, but from its ability to adapt parameters dynamically. This distinction highlights the value of AVP-VAR’s architecture, which uses $Z_t$ to govern the evolution of parameters rather than treating them as additional regressors.
	
	\begin{table}[htbp!]
		\centering
		\caption{Forecasting Performance: Mean Squared Prediction Error Relative to Benchmark} \label{tab:forecasting_mspeUS}
		\scriptsize
		\begin{tabular}{@{}cS[table-format=1.3]S[table-format=1.3]S[table-format=1.3]S[table-format=1.3]S[table-format=1.3]S[table-format=1.3]S[table-format=1.3]S[table-format=1.3]@{}}
			\toprule
			{\textbf{h}} & {\textbf{AVP-VAR}} & {\textbf{CP-VAR}} & {\makecell{ \textbf{CP-VAR} \\ \textbf{SV}}} & {\makecell{ \textbf{TVP-VAR} \\ \textbf{EB}}} &{\makecell{\textbf{TVP-VAR}}} & {\makecell{ \textbf{VAR} \\ \textbf{SVOt}}} & {\textbf{FAVAR}} & {\makecell{ \textbf{FAVAR} \\ \textbf{SV}}}  \\
			\midrule
			\multicolumn{9}{c}{\textbf{Panel A: Industrial Production}} \\
			\midrule
			1  & \textbf{0.902} & \textbf{0.996} & \textbf{0.949} & \textbf{0.977} & \textbf{0.890} & \textbf{0.951} & \textbf{0.768}$^\dag$ & \textbf{0.849} \\
			2  & \textbf{0.805} & \textbf{0.949} & \textbf{0.863} & \textbf{0.853} & \textbf{0.784}$^\dag$ & \textbf{0.858} & \textbf{0.889} & \textbf{0.816} \\
			3  & \textbf{0.867}$^\dag$ & \textbf{0.967} & \textbf{0.879} & \textbf{0.881} & \textbf{0.886} & \textbf{0.891} & 1.017 & \textbf{0.928} \\
			4  & \textbf{0.938}$^\dag$ & \textbf{0.986} & \textbf{0.959} & \textbf{0.951} & \textbf{0.956} & \textbf{0.952} & 1.036 & \textbf{0.976} \\
			5  & \textbf{0.967}$^\dag$ & \textbf{0.991} & \textbf{0.988} & \textbf{0.986} & \textbf{0.991} & \textbf{0.984} & 1.040 & \textbf{0.999} \\
			6  & \textbf{0.972}$^\dag$ & \textbf{0.994} & \textbf{0.985} & \textbf{0.985} & \textbf{0.990} & \textbf{0.977} & 1.036 & \textbf{0.993} \\
			9  & \textbf{0.983}$^\dag$ & 1.000 & \textbf{0.996} & 1.005 & \textbf{0.994} & 1.004 & 1.063 & 1.028 \\
			12 & 1.001 & \textbf{0.997}$^\dag$ & 1.006 & 1.006 & 1.000 & 1.003 & 1.081 & 1.037 \\
			15 & 1.002 & 1.002 & 1.007 & 1.003 & \textbf{0.999}$^\dag$ & 1.006 & 1.059 & 1.031 \\
			18 & 1.001 & 1.000 & 1.010 & 1.006 & 1.002 & 1.011 & 1.054 & 1.038 \\
			24 & 1.001 & \textbf{0.999}$^\dag$ & 1.004 & 1.010 & 1.002 & 1.009 & 1.032 & 1.031 \\
			\midrule
			\multicolumn{9}{c}{\textbf{Panel B: PCEPI}} \\
			\midrule
			1  & \textbf{0.980} & \textbf{0.988} & \textbf{0.964} & \textbf{0.999} & 1.491 & \textbf{0.970} & \textbf{0.943}$^\dag$ & \textbf{0.950} \\
			2  & \textbf{0.881}$^\dag$ & \textbf{0.972} & \textbf{0.897} & \textbf{0.937} & 1.066 & \textbf{0.922} & \textbf{0.999} & \textbf{0.907} \\
			3  & \textbf{0.869}$^\dag$ & \textbf{0.971} & \textbf{0.926} & \textbf{0.930} & 1.055 & \textbf{0.944} & 1.017 & \textbf{0.946} \\
			4  & \textbf{0.886}$^\dag$ & \textbf{0.979} & \textbf{0.947} & \textbf{0.939} & 1.084 & \textbf{0.950} & 1.014 & \textbf{0.956} \\
			5  & \textbf{0.918}$^\dag$ & \textbf{0.997} & \textbf{0.972} & \textbf{0.973} & 1.112 & \textbf{0.978} & 1.034 & \textbf{0.991} \\
			6  & \textbf{0.912}$^\dag$ & \textbf{0.995} & 1.002 & \textbf{0.996} & 1.118 & 1.001 & 1.037 & 1.003 \\
			9  & \textbf{0.926}$^\dag$ & \textbf{0.990} & \textbf{0.993} & \textbf{0.993} & 1.103 & \textbf{0.992} & 1.028 & \textbf{0.986} \\
			12 & \textbf{0.964}$^\dag$ & 1.011 & \textbf{0.994} & \textbf{0.998} & 1.099 & \textbf{0.994} & 1.027 & 1.002 \\
			15 & \textbf{0.959}$^\dag$ & 1.006 & 1.005 & 1.006 & 1.108 & \textbf{0.996} & 1.009 & 1.002 \\
			18 & \textbf{0.989}$^\dag$ & \textbf{0.997} & 1.008 & 1.002 & 1.102 & 1.006 & 1.012 & 1.012 \\
			24 & 1.040 & 1.008 & 1.004 & 1.007 & 1.107 & 1.008 & 1.042 & 1.001 \\
			\midrule
			\multicolumn{9}{c}{\textbf{Panel C: Computational speed (total time relative to TVP-VAR-EB)}} \\
			\midrule
			Ratio & 0.294 & 0.171 & 0.567 & 1.000 & 1.789 & 0.131 & 0.001 & 0.712 \\
			\bottomrule
		\end{tabular}
		\vspace{0.5em}
		\par
		\begin{tabularx}{\textwidth}{X}
			{\footnotesize \textit{Notes:} The table reports mean squared prediction errors (MSPE) for each of the two forecasted variables, relative to a benchmark constant-parameter VAR($p$) estimated via least squares. Values below one (highlighted in bold) indicate superior forecasting performance compared to the benchmark. The forecast horizon is denoted by $h$ (in months), and the symbol $\dag$ marks the best-performing model at each horizon.}
		\end{tabularx}
	\end{table}
	
	\subsection{Forecasts for the EA: Quarterly data and results}
	
	We estimate a three-variable VAR using quarterly Euro Area data spanning 1975Q1 to 2024Q3. The endogenous variables are: (i) real economic activity, measured as the quarterly log-difference of real GDP (YER); (ii) inflation, measured as the quarterly log-difference of the Harmonized Index of Consumer Prices (HICP); and (iii) the nominal short-term interest rate (STN) in first differences, representing the monetary policy stance.
	
	The driver set $Z_t$ includes $m=10$ variables capturing external shocks and domestic conditions relevant to the Euro Area. These include commodity prices (oil and non-oil), labour market indicators (unemployment rate and productivity growth), global activity measures (world GDP growth, world demand, and trade-weighted world GDP), and exchange rate indicators (effective exchange rate and USD/EUR bilateral rate). This selection reflects the Euro Area’s sensitivity to global factors, distinguishing it from the U.S. application where uncertainty measures play a more central role. Table~\ref{tab:euro_area_variables} provides detailed descriptions, transformations, and sources for all variables.
	
	\begin{table}[htbp!]
		\singlespacing
		\centering
		\scriptsize
		\caption{Variables and Drivers Euro area quarterly data} \label{tab:euro_area_variables}
		\begin{threeparttable}
			\begin{tabular}{@{}llcc@{}}
				\toprule
				\textbf{Series ID} & \textbf{Description} & \textbf{Source} & \textbf{tcode} \\
				\midrule
				\multicolumn{4}{c}{ \textbf{VAR Variables}} \\
				\midrule
				YER & Real GDP & AWM & 5 \\
				HICP & HICP Overall Index & AWM & 5 \\
				STN & Nominal Short-Term Interest Rate & AWM & 2 \\
				\addlinespace[0.5em]
				\midrule
				\multicolumn{4}{c}{\textbf{Drivers} ($Z_t$)} \\
				\midrule
				POILU & Oil Prices & AWM & 5 \\
				PCOMU & Non-Oil Commodity Prices & AWM & 5 \\
				COMPR & Commodity Price Index & AWM & 5 \\
				URX & Unemployment Rate & AWM & 5 \\
				YWR & World GDP & AWM & 5 \\
				YWRX & World Demand & AWM & 5 \\
				TWGDP & Euro-Area-Trade-Weighted World GDP & AWM & 5 \\
				LPROD\_EMP & Real Labor Productivity per Person & AWM & 5 \\
				EEN & Effective Exchange Rate & AWM & 5 \\
				EXR AVG & USD/EUR Exchange Rate (Average) & AWM & 5 \\
				\midrule
				Frequency: & Quarterly data &  &  \\
				Period: & 1975Q1 -- 2024Q3  &  &  \\
				\bottomrule
			\end{tabular}
			\begin{tablenotes}
				\footnotesize
				\item Notes: tcode 1 = Levels, tcode 2 = first difference, tcode 5 = log-difference. AWM stands for the Area-Wide Model database, see \href{https://eabcn.org/data/area-wide-model}{https://eabcn.org/data/area-wide-model}
			\end{tablenotes}
		\end{threeparttable}
	\end{table}
	
	Table~\ref{tab:forecasting_mspe_EURO} and Figure~\ref{fig:oos_EURO} present the forecasting results. For real GDP growth, both AVP-VAR and TVP-VAR deliver the strongest performance, achieving up to 60\% improvement in MSPE at short horizons relative to the OLS benchmark. While the TVP-VAR has an edge over longer forecasting horizons, the differences with the AVP-VAR are negligible. For HICP inflation, there is no a clear winner, but the AVP-VAR remains among the top performers for $h>3$.
	
	Quantile forecast evaluations show AVP-VAR performs reasonably well in upper tail risk (90th quantile), especially for inflation, while performance for downside risk (10th quantile) is more mixed (good for inflation, but poor for growth). AVP-VAR also demonstrates strong computational efficiency, requiring only 43\% of TVP-VAR-EB’s estimation time while delivering superior real activity forecasts and competitive inflation performance.
	
	\begin{figure}[htbp!]  
		\centering
		\includegraphics[width=0.95\textwidth]{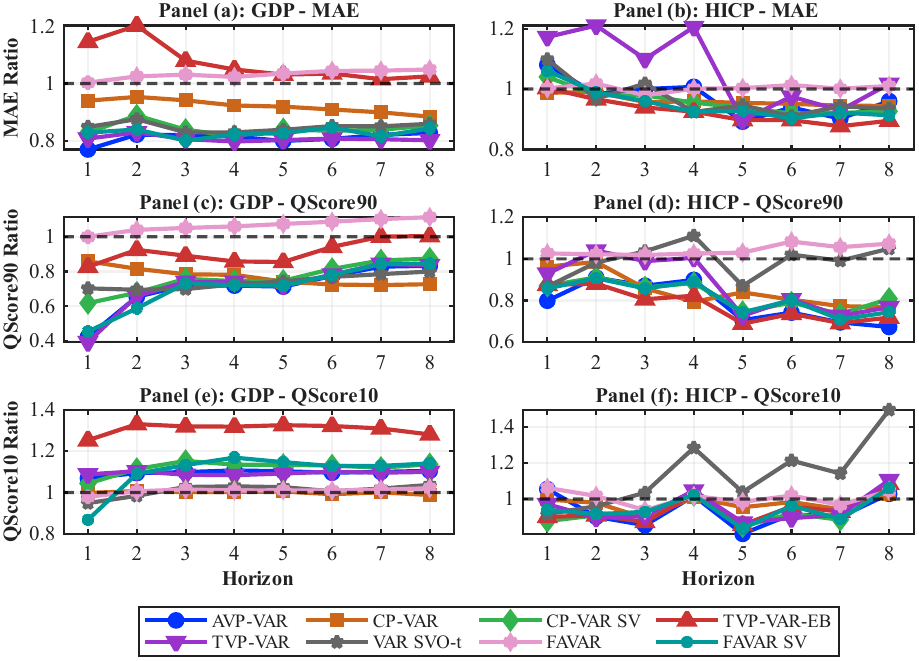}  
		\caption{Forecasting results using Euro Area quarterly data. The top row displays mean absolute forecast errors (MAE), while the middle and bottom rows show quantile scores for the upper (90th percentile) and lower (10th percentile) tails of each variable. All metrics are reported as ratios relative to a benchmark constant-parameter VAR($p$) estimated via least squares. A model is considered superior when its relative MAE, or relative quantile scores, are below one.} \label{fig:oos_EURO}
	\end{figure}
	
	\begin{table}[htbp!]
		\centering
		\caption{Forecasting Performance: Mean Squared Prediction Error Relative to Benchmark} \label{tab:forecasting_mspe_EURO}
		\scriptsize
		\begin{tabular}{@{}cS[table-format=1.3]S[table-format=1.3]S[table-format=1.3]S[table-format=1.3]S[table-format=1.3]S[table-format=1.3]S[table-format=1.3]S[table-format=1.3]@{}}
			\toprule
			{\textbf{h}} & {\textbf{AVP-VAR}} & {\textbf{CP-VAR}} & {\makecell{ \textbf{CP-VAR} \\ \textbf{SV}}} & {\makecell{ \textbf{TVP-VAR} \\ \textbf{EB}}} &{\textbf{TVP-VAR}} & {\makecell{ \textbf{VAR} \\ \textbf{SVOt}}} & {\textbf{FAVAR}} & {\makecell{ \textbf{FAVAR} \\ \textbf{SV}}}  \\
			\midrule
			\multicolumn{9}{c}{\textbf{Panel A: GDP}} \\
			\midrule
			1  & \textbf{0.391}$^\dag$ & \textbf{0.896} & \textbf{0.536} & \textbf{0.620} & \textbf{0.435} & \textbf{0.547} & 1.036 & \textbf{0.521} \\
			2  & \textbf{0.598}$^\dag$ & \textbf{0.857} & \textbf{0.665} & \textbf{0.813} & \textbf{0.605} & \textbf{0.661} & 1.074 & \textbf{0.634} \\
			3  & \textbf{0.613} & \textbf{0.803} & \textbf{0.624} & \textbf{0.708} & \textbf{0.613} & \textbf{0.619} & 1.111 & \textbf{0.609}$^\dag$ \\
			4  & \textbf{0.579} & \textbf{0.757} & \textbf{0.592} & \textbf{0.642} & \textbf{0.571}$^\dag$ & \textbf{0.587} & 1.131 & \textbf{0.587} \\
			5  & \textbf{0.602} & \textbf{0.740} & \textbf{0.614} & \textbf{0.656} & \textbf{0.599}$^\dag$ & \textbf{0.616} & 1.159 & \textbf{0.610} \\
			6  & \textbf{0.649} & \textbf{0.746} & \textbf{0.653} & \textbf{0.699} & \textbf{0.645}$^\dag$ & \textbf{0.659} & 1.181 & \textbf{0.656} \\
			7  & \textbf{0.672} & \textbf{0.742} & \textbf{0.677} & \textbf{0.713} & \textbf{0.667}$^\dag$ & \textbf{0.679} & 1.207 & \textbf{0.672} \\
			8  & \textbf{0.673} & \textbf{0.727} & \textbf{0.680} & \textbf{0.709} & \textbf{0.667}$^\dag$ & \textbf{0.681} & 1.245 & \textbf{0.679} \\
			\midrule
			\multicolumn{9}{c}{\textbf{Panel B: HICP}} \\
			\midrule
			1  & 1.682 & \textbf{0.981}$^\dag$ & 1.024 & \textbf{0.997} & 1.309 & 1.125 & 1.030 & 1.122 \\
			2  & \textbf{0.971} & \textbf{0.987} & \textbf{0.970} & \textbf{0.955}$^\dag$ & 1.383 & \textbf{0.957} & 1.027 & \textbf{0.996} \\
			3  & 1.018 & \textbf{0.894} & \textbf{0.922} & \textbf{0.838}$^\dag$ & 1.106 & \textbf{0.989} & 1.011 & \textbf{0.887} \\
			4  & \textbf{0.925} & \textbf{0.841} & \textbf{0.813} & \textbf{0.786} & 1.201 & \textbf{0.781}$^\dag$ & 1.096 & \textbf{0.784} \\
			5  & \textbf{0.753} & \textbf{0.863} & \textbf{0.824} & \textbf{0.751} & \textbf{0.740}$^\dag$ & \textbf{0.835} & 1.066 & \textbf{0.813} \\
			6  & \textbf{0.756} & \textbf{0.801} & \textbf{0.738} & \textbf{0.711}$^\dag$ & \textbf{0.782} & \textbf{0.716} & 1.162 & \textbf{0.715} \\
			7  & \textbf{0.672} & \textbf{0.750} & \textbf{0.688} & \textbf{0.629}$^\dag$ & \textbf{0.655} & \textbf{0.709} & 1.167 & \textbf{0.674} \\
			8  & \textbf{0.603} & \textbf{0.672} & \textbf{0.589} & \textbf{0.562} & \textbf{0.661} & \textbf{0.605} & 1.280 & \textbf{0.561}$^\dag$ \\
			\midrule
			\multicolumn{9}{c}{\textbf{Panel C: Computational speed (total time relative to TVP-VAR-EB)}} \\
			\midrule
			Ratio & 0.432 & 0.261 & 0.554 & 1.000 & 2.407  & 0.283 & 0.001 & 0.687 \\
			\bottomrule
		\end{tabular}	
		\vspace{0.5em}
		\par
		\begin{tabularx}{\textwidth}{X}
			{\footnotesize \textit{Notes:} The table reports mean squared prediction errors (MSPE) for each of the two forecasted variables, relative to a benchmark constant-parameter VAR($p$) estimated via least squares. Values below one (highlighted in bold) indicate superior forecasting performance compared to the benchmark. The forecast horizon is denoted by $h$ (in quarters), and the symbol $\dag$ marks the best-performing model at each horizon.}
		\end{tabularx}
	\end{table}
	
	\subsection{Robustness of results}
	
	\subsubsection{Different lag lengths ($p$).} 
	We assess the sensitivity of our results to the choice of lag order by estimating models with $p = 1, 3,$ and $4$, and evaluating their performance using the mean squared prediction error (MSPE), as well as quantile scores at the 90th (QS90) and 10th (QS10) percentiles. Detailed results are provided in the online supplement. The main conclusion remains robust: the AVP--VAR consistently delivers competitive predictive performance for both output growth and inflation, particularly at short and medium horizons, and ranks among the best-performing specifications across most variables and forecasting horizons. Overall, we find improvements in MSPE and QS90, while performance at the lower tail (QS10) remains comparatively weaker across all models, including AVP--VAR.

	

	\subsubsection{Different ordering for TVP-VAR-EB.} 
	
	The TVP-VAR-EB faces a well-known ordering dependency, as emphasized by \cite{Ariasetal2023}. The model relies on a triangular decomposition to identify time-varying covariances, which requires imposing a recursive structure. Under this identification scheme, the ordering of variables determines which shocks are assumed to have contemporaneous effects on others, directly affecting the model's forecasting distribution. Our baseline implementation follows the ordering: growth, inflation, and interest rates. 
	
	To assess whether alternative orderings materially alter our conclusions, we examine the reverse specification for the TVP-VAR-EB benchmark: interest rates, inflation, and growth. We conduct this exercise for both the U.S. and E.A. across different lag orders $p$. Full results appear in the online supplement. In terms of MSPE, the two orderings yield generally similar results across all databases, with some differences for higher order lags. That said, we do not observe a consistent winner between the two orderings. Notably, the alternative ordering does not overturn our main finding: AVP-VAR remains among the most competitive models across forecast horizons and evaluation metrics, regardless of the TVP-VAR-EB ordering used for comparison. 
	

	\subsubsection{FRED--QD database.} 
	We consider a three-variable VAR using the \citet{mccracken2020fred} FRED-QD database, spanning 1963Q1 to 2025Q2. The endogenous variables comprise: (i) real economic activity; (ii) inflation; and (iii) the effective federal funds rate.	We consider $m=9$ drivers capturing financial conditions and real activity dynamics that may signal parameter instability. These include: equity market indicators, real sector forward indicators, producer price inflation, credit market conditions, exchange rate movements, financial market volatility, and consumer sentiment. This set of drivers emphasizes financial market variables and real activity indicators readily available in FRED-QD, distinguishing it from our monthly U.S. specification that focuses more heavily on uncertainty indices. Details about variables, transformations, descriptions, drivers, and sources, are available in the online supplement.
	
	All results for $p=1,2,3,$ and $4$ are available in the online supplement. For real GDP growth, AVP-VAR shows again a good performance, consistently appearing among the top performers for most forecasting horizons. For PCE inflation, results are mixed. While AVP-VAR remains competitive, frequently ranking among the best specifications across horizons, TVP-VAR-EB and SVO-t seem to have an edge depending on the horizon. The quantile forecast evaluation mirrors these patterns: AVP-VAR performs well for upper tail risk (90th quantile), particularly for GDP, but shows weaker performance for extreme downside outcomes (10th quantile), where most models struggle. 
	
	While results are robust for GDP, they are not as impressive for inflation, compared to US monthly data. A possible explanation for this difference is that the driver set in this application consists entirely of standard FRED-QD macroeconomic aggregates, whereas our monthly analysis employed forward-looking uncertainty indices. 
	
\section{Stylized Facts on Time Variation}
\label{sec:stylized-facts}

This section examines the ability of the AVP--VAR to capture economically meaningful time variation. We combine evidence from controlled simulations with empirical diagnostics. The goal is twofold: to clarify the conditions under which the adaptive specification improves upon conventional random-walk TVPs, and to illustrate how these gains materialize in real data.

\subsection{Simulation Evidence}

Our Monte Carlo design isolates the behavior of time-varying parameters under diverse forms of instability. Rather than simulating full VAR systems, which would require imposing additional constraints to ensure stability and stationarity, we work with univariate regressions featuring four predictors. This choice reflects the equation-by-equation estimation strategy commonly used in practice and allows us to focus squarely on the dynamics of the coefficients without conflating results with identification or numerical stability issues. The predictors follow a VAR(1) process with spatial correlation, and the dependent variable evolves with stochastic volatility. This setup preserves realistic persistence and volatility patterns while avoiding the complexity of high-dimensional state spaces.

\subsubsection{Experimental Design}

We consider two data-generating processes (DGPs) that capture distinct forms of structural change. DGP~1 introduces a transitory break, that is, a short-lived but sharp jump in all coefficients, superimposed on a highly persistent component. This mimics episodes such as the COVID-19 shock, where abrupt disruptions coexist with underlying trends. The jump lasts for six observations and occurs at a randomly selected midpoint in the sample, emulating the sharp but brief pandemic-related recession observed in major economies. The persistent component follows a near-unit root process, while the jump component temporarily shifts all parameters by several standard deviations before reverting.

DGP~2 combines multiple nonlinear features: smooth regime-switching, threshold effects, and volatility co-movement. Parameters evolve based on a discrete regime state that transitions stochastically between two values, an interaction between the regime and predictor dynamics, and threshold-triggered shifts based on a stress indicator constructed from lagged dependent variables. These mechanisms generate heterogeneous and state-dependent parameter shifts, providing a stringent test of whether AVP--VAR can exploit external signals to track complex dynamics.

For each DGP, we simulate data across three sample sizes ($T \in \{50, 100, 200\}$) and three levels of predictor persistence ($\rho \in \{0, 0.5, 0.95\}$), conducting 1,000 Monte Carlo replications for each configuration. All models are estimated using Bayesian methods with 11,000 Gibbs iterations, discarding the first 1,000 as burn-in and thinning to retain 1,000 posterior draws.

We compare two types of external drivers: \emph{agnostic drivers} that contain no information about the DGP structure (simply random normal variates of varying dimensionality $m \in \{20, 40, 60\}$), and \emph{targeted drivers} that incorporate observable signals about structural features. In DGP~1, targeted drivers include an indicator for break periods and interactions with normal variates, allowing the model to learn break-specific parameter shifts without knowing their magnitude in advance. In DGP~2, targeted drivers progressively reveal information about the regime state, interaction signals between regimes and predictors, and threshold signals based on a stress indicator. Crucially, these signals reveal only the timing of potential breaks or regime shifts, not their magnitude or duration, which must still be estimated.

The full technical details of the simulation setup is provided in the accompanying online supplement. Model performance is assessed by computing the mean squared parameter error (MSPE), which averages the squared deviations between the estimated and true parameter paths across time and coefficients.

\subsubsection{Monte Carlo Results}

Table~\ref{tab:mc_results} presents MSPE ratios relative to the canonical random-walk TVP benchmark for $\rho = 0.95$. Values below unity indicate superior forecasting performance. Complete results for all specifications are available in the online supplement.

\begin{table}[htbp!]
\centering
\begin{threeparttable}
\caption{Monte Carlo Results: DGP1 (Transitory Jump) and DGP2 (Non-linear TVP and Breaks)}
\label{tab:mc_results}
\scriptsize
\begin{tabular}{@{}lcccc|cccc@{}}
\toprule
& \multicolumn{4}{c}{\textbf{DGP1: Transitory Jump}} & \multicolumn{4}{c}{\textbf{DGP2: Regimes \& Thresholds}} \\
\cmidrule(lr){2-5} \cmidrule(lr){6-9}
Model & $\beta_{1,t}$ & $\beta_{2,t}$ & $\beta_{3,t}$ & $\beta_{4,t}$ & $\beta_{1,t}$ & $\beta_{2,t}$ & $\beta_{3,t}$ & $\beta_{4,t}$ \\
\midrule
\multicolumn{9}{c}{\textbf{Panel A: $T = 50$, $\rho = 0.95$}} \\
\midrule
TVP & 1.00 & 1.00 & 1.00 & 1.00 & 1.00 & 1.00 & 1.00 & 1.00 \\
\addlinespace
\multicolumn{9}{l}{\emph{Agnostic Drivers}} \\
AVP ($m=20$) & 1.00 & 0.81 & 0.75 & 0.97 & 1.34 & 1.32 & 1.48 & 1.41 \\
AVP ($m=40$) & 0.99 & 0.81 & 0.75 & 0.96 & 1.18 & 1.16 & 1.22 & 1.19 \\
AVP ($m=60$) & 0.99 & 0.81 & 0.74 & 0.96 & 1.10 & 1.10 & 1.13 & 1.13 \\
\addlinespace
\multicolumn{9}{l}{\emph{Targeted Drivers}} \\
AVP ($m=20$) & 0.87 & 0.64 & 0.63 & 0.74 & 1.01 & 1.03 & 0.80 & 0.82 \\
AVP ($m=40$) & 0.75 & \textbf{0.58}$^\dag$ & 0.61 & 0.67 & \textbf{0.67}$^\dag$ & \textbf{0.74}$^\dag$ & \textbf{0.37}$^\dag$ & \textbf{0.54}$^\dag$ \\
AVP ($m=60$) & \textbf{0.71}$^\dag$ & 0.59 & \textbf{0.56}$^\dag$ & \textbf{0.65}$^\dag$ & 0.73 & 0.77 & 0.41 & 0.59 \\
\midrule
\multicolumn{9}{c}{\textbf{Panel B: $T = 100$, $\rho = 0.95$}} \\
\midrule
TVP & 1.00 & 1.00 & 1.00 & 1.00 & 1.00 & 1.00 & 1.00 & 1.00 \\
\addlinespace
\multicolumn{9}{l}{\emph{Agnostic Drivers}} \\
AVP ($m=20$) & 0.89 & 0.81 & 0.87 & 0.94 & 1.79 & 1.66 & 2.03 & 1.82 \\
AVP ($m=40$) & 0.89 & 0.80 & 0.87 & 0.93 & 1.42 & 1.34 & 1.50 & 1.40 \\
AVP ($m=60$) & 0.88 & 0.80 & 0.86 & 0.92 & 1.27 & 1.22 & 1.30 & 1.26 \\
\addlinespace
\multicolumn{9}{l}{\emph{Targeted Drivers}} \\
AVP ($m=20$) & 0.71 & 0.71 & 0.82 & 0.72 & 0.75 & 0.94 & 0.49 & 0.62 \\
AVP ($m=40$) & 0.72 & 0.66 & 0.69 & 0.74 & \textbf{0.43}$^\dag$ & \textbf{0.63}$^\dag$ & \textbf{0.23}$^\dag$ & \textbf{0.29}$^\dag$ \\
AVP ($m=60$) & \textbf{0.67}$^\dag$ & \textbf{0.62}$^\dag$ & \textbf{0.66}$^\dag$ & \textbf{0.69}$^\dag$ & 0.44 & \textbf{0.63}$^\dag$ & \textbf{0.23}$^\dag$ & 0.30 \\
\midrule
\multicolumn{9}{c}{\textbf{Panel C: $T = 200$, $\rho = 0.95$}} \\
\midrule
TVP & 1.00 & 1.00 & 1.00 & 1.00 & 1.00 & 1.00 & 1.00 & 1.00 \\
\addlinespace
\multicolumn{9}{l}{\emph{Agnostic Drivers}} \\
AVP ($m=20$) & 0.88 & 0.85 & 0.81 & 0.93 & 2.47 & 2.25 & 3.19 & 2.64 \\
AVP ($m=40$) & 0.87 & 0.82 & 0.80 & 0.92 & 1.79 & 1.69 & 2.09 & 1.85 \\
AVP ($m=60$) & 0.87 & 0.82 & 0.79 & 0.91 & 1.53 & 1.47 & 1.69 & 1.53 \\
\addlinespace
\multicolumn{9}{l}{\emph{Targeted Drivers}} \\
AVP ($m=20$) & 0.85 & 0.91 & 0.83 & 0.95 & 0.94 & 1.16 & 0.69 & 0.67 \\
AVP ($m=40$) & 0.82 & 0.80 & 0.81 & 0.87 & 0.42 & 0.65 & 0.24 & 0.27 \\
AVP ($m=60$) & \textbf{0.79}$^\dag$ & \textbf{0.78}$^\dag$ & \textbf{0.73}$^\dag$ & \textbf{0.83}$^\dag$ & \textbf{0.37}$^\dag$ & \textbf{0.59}$^\dag$ & \textbf{0.21}$^\dag$ & \textbf{0.24}$^\dag$ \\
\bottomrule
\end{tabular}
\begin{tablenotes}
\footnotesize
\item Notes: Each entry reports MSPE ratios relative to TVP benchmark across 1,000 Monte Carlo simulations. Bold entries indicate MSPE ratios below 1.00 (gains relative to benchmark). $^\dag$ highlights the best performing model for each coefficient.
\end{tablenotes}
\end{threeparttable}
\end{table}

Several patterns emerge from these simulations. First, for DGP~1, AVP relative performance generally improves with larger $m$, though this pattern is substantially more pronounced for targeted drivers. AVP with targeted drivers emerges as the best specification across 97\% of all entries (105 of 108 cases; see online supplement), with MSPE reductions frequently exceeding 20\% and reaching as high as 40\% relative to TVP when $\rho = 0.95$. Even AVP with agnostic drivers remains competitive with TVP, showing modest improvements despite having no information about the timing or magnitude of breaks. This result suggests that our proposed adaptive parameter variation does not suffer meaningful efficiency losses under conventional random-walk dynamics.

Second, for DGP~2, the contrast is starker. Agnostic AVP consistently underperforms the TVP benchmark across all configurations, with MSPE ratios sometimes exceeding 2.0. This underscores the importance of informative signals when parameter dynamics are complex and non-linear. In contrast, AVP with targeted drivers uniformly dominates TVP across all 72 entries when using $m = 40$ or $m = 60$ drivers, with MSPE reductions reaching 79\% in some cases (see online supplement). With $m = 20$, performance is more nuanced: AVP substantially outperforms TVP for $\beta_3$ and $\beta_4$ (improvements of 15--20\%), but modestly underperforms for $\beta_1$ and $\beta_2$ (see online supplement).

Third, the number of drivers $m$ plays a more critical role under DGP~2 than DGP~1, affecting both targeted and agnostic specifications. For targeted AVP, the gains can be dramatic: in some cases, moving from $m = 20$ to $m = 60$ transforms a 16\% underperformance into a 41\% improvement -- a swing of 57 percentage points. Under DGP~2, increasing $m$ simultaneously expands both the number of drivers and the richness of information about structural instabilities. Specifically, $m = 60$ entries include more information about regime behavior and threshold mechanisms than $m = 20$. The observed performance improvements therefore reflect both dimensions: the breadth of the information set and its alignment with the underlying sources of parameter variation.

\begin{figure}[htbp!]
\centering
\includegraphics[width=0.95\textwidth]{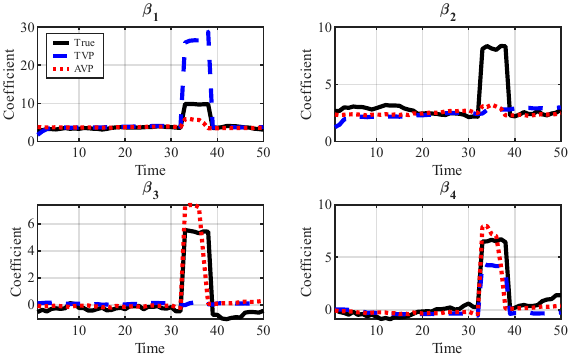}
\caption{Illustration of a Monte Carlo iteration for DGP1 with $T = 50$ and $\rho = 0.95$. The red line (AVP) considers $m = 60$ targeted drivers. Black line shows true parameter path; blue dashed line shows TVP estimates.}
\label{fig:mc_dgp1}
\end{figure}

\begin{figure}[htbp!]
\centering
\includegraphics[width=0.95\textwidth]{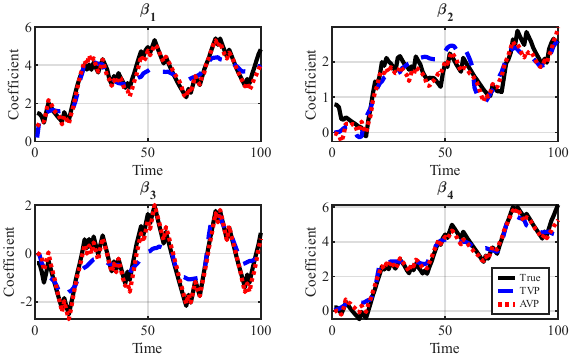}
\caption{Illustration of a Monte Carlo iteration for DGP2 with $T = 100$ and $\rho = 0.95$. The red line (AVP) considers $m = 60$ targeted drivers. Black line shows true parameter path; blue dashed line shows TVP estimates.}
\label{fig:mc_dgp2}
\end{figure}

Figures~\ref{fig:mc_dgp1} and \ref{fig:mc_dgp2} illustrate parameter estimates from single Monte Carlo draws, comparing TVP and AVP with $m = 60$ targeted drivers. For DGP~1 (Figure~\ref{fig:mc_dgp1}, left columns of Table~\ref{tab:mc_results}), TVP struggles to track the true parameters: it substantially overshoots the jump in $\beta_1$, fails to detect jumps in $\beta_2$ and $\beta_3$, and only partially captures the break in $\beta_4$. AVP demonstrates superior tracking ability, particularly for $\beta_3$ and $\beta_4$. For DGP~2 (Figure~\ref{fig:mc_dgp2}, right columns of Table~\ref{tab:mc_results}), the figure reveals that AVP tracks the true parameters considerably more accurately than TVP for most coefficients, successfully capturing the regime-switching and threshold dynamics that TVP largely misses.

Overall, these Monte Carlo exercises establish that when the random-walk evolution is contaminated with abrupt jumps, regime shifts, or threshold effects, the advantages of AVP become evident. In the presence of even mildly informative observables, the adaptive specification yields more accurate estimates of time variation than standard TVP methods. Complete details on functional forms, hyperparameters, prior settings, and robustness to different configurations appear in the online supplement.

\subsection{In-Sample Parameter Dynamics: U.S.\ Evidence}

\begin{figure}[htbp!]
\centering
\includegraphics[width=\textwidth]{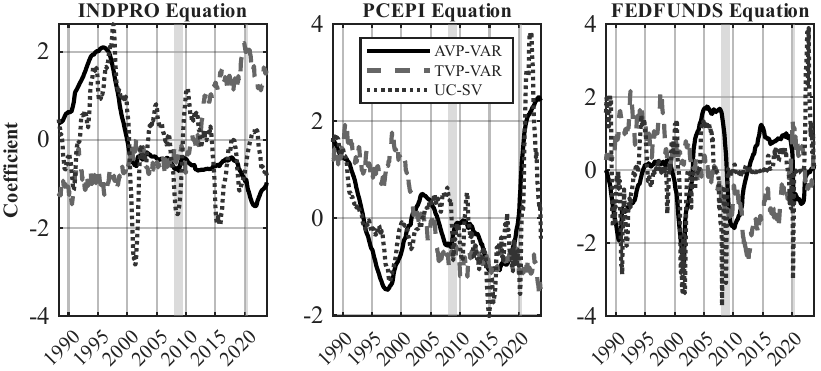}
\caption{Time-varying intercepts in a VAR(1) with U.S.\ monthly data. Solid line: AVP--VAR; dashed line: TVP--VAR (Full Bayes); dotted line: UC--SV. Shaded regions denote the GFC and COVID-19 pandemic.}
\label{fig:insample_main}
\end{figure}

To link these stylized facts to forecasting performance, we examine the evolution of intercepts in a parsimonious VAR(1) estimated on U.S.\ monthly data spanning January 1985 to November 2023. The system includes industrial production growth (INDPRO), PCE inflation (PCEPI), and the federal funds rate (FEDFUNDS). We focus on intercepts because they approximate time-varying trends and typically exhibit the clearest margin of variation in macroeconomic VARs. Figure~\ref{fig:insample_main} compares AVP--VAR to a TVP--VAR with hierarchical horseshoe priors (Full Bayes) and to the univariate UC--SV benchmark, which serves as a flexible reference known for strong forecasting performance.

Three observations are central. First, \emph{timeliness}: AVP--VAR intercepts adjust rapidly during crisis episodes, closely tracking UC--SV; TVP--VAR adjusts more slowly and often underestimates the magnitude of abrupt shifts. During the Global Financial Crisis (2008--2009) and the COVID-19 pandemic (2020), AVP--VAR intercepts exhibit sharp movements that mirror the UC--SV benchmark, capturing the dramatic changes in economic conditions. In contrast, TVP--VAR intercepts evolve more gradually, reflecting the random-walk prior's inability to accommodate sudden breaks without extensive lag.

Second, \emph{discipline}: outside of crises, AVP--VAR remains comparatively stable, reflecting the absence of strong signals in $Z_t$; TVP--VAR sometimes produces erratic local movements unrelated to economic conditions. This pattern is particularly evident in the industrial production and inflation equations during the mid-2000s and mid-2010s, where TVP--VAR exhibits volatility that bears little relation to observed macroeconomic developments, while AVP--VAR remains relatively smooth.

Third, \emph{heterogeneity}: while intercepts are the clearest margin of variation, the extent of time variation differs across equations; AVP--VAR accommodates this heterogeneity without overfitting, because the mapping from $Z_t$ to coefficients is sparse by construction. The federal funds rate equation displays more pronounced variation than the industrial production equation, reflecting differences in how monetary policy and real activity respond to structural shocks---a pattern that AVP--VAR captures naturally through driver selection.

These patterns clarify the division of labor between the likelihood and the prior: stochastic volatility and factor components absorb scale and co-movement in shocks, while the \emph{direction and timing} of coefficient movements are disciplined by $Z_t$. Substantively, the findings anticipate the forecasting gains documented in Section~3: the largest improvements occur during episodes of rapid structural change, precisely when random-walk TVPs adjust too slowly or too diffusely.

A potential concern is that the TVP--VAR specification with Full Bayes priors may still impose excessive regularization. To address this, the online supplement presents comparisons with TVP--VAR using empirical Bayes shrinkage (TVP--VAR--EB). The key insight remains unchanged: both TVP--VAR variants produce relatively flat intercepts compared to AVP--VAR and UC--SV. The online supplement also presents both standardized and unstandardised intercept paths, clarifying that differences in volatility scaling do not overturn our conclusions. In fact, when examining unstandardised coefficients, both TVP--VAR and TVP--VAR--EB display relatively stable parameters across most periods, including episodes of high volatility, reinforcing our claim that conventional TVP specifications tend to heavily regularize state innovations due to the high-dimensional parameter space.

The canonical TVP--VAR, faced with a large parameter space and a high-dimensional state covariance matrix, tends to heavily regularize the state innovations, leaving most parameters virtually time-invariant. In contrast, the AVP--VAR produces intercept dynamics that closely track the UC--SV estimates, while allowing the autoregressive coefficients to display heterogeneous patterns of variation: sometimes nearly constant, sometimes shifting more strongly. This ability to capture realistic time variation without being overburdened by excessive parametrization explains why the AVP--VAR delivers sharper estimates and systematically better forecasts.

Similar patterns emerge for the euro area quarterly data (1975Q1--2024Q3), as documented in the online supplement. Specifically, AVP--VAR successfully tracks the UC--SV benchmark during the post-COVID period (2020--2024), capturing the substantial upward shift in the inflation process, while both TVP--VAR variants fail to detect this shift and maintain relatively flat intercepts throughout this critical episode. The online supplement also contains cumulative residual plots for U.S.\ and euro area data, showing that AVP--VAR improves fit not only during crises but also across tranquil periods; robustness checks for different lag lengths, variable orderings in TVP--VAR--EB, and factor dimensions; and full in-sample figures for all three datasets. These materials reinforce the main message: externally driven parameter drift offers a transparent and replicable mechanism for timely adaptation, with gains concentrated in periods of structural stress and no meaningful efficiency loss in tranquil times.

	\section{Concluding Remarks}
	This paper introduces the Adaptively-Varying Parameter VAR (AVP-VAR), a new class of time-varying autoregressive models in which the coefficients evolve deterministically as linear functions of observable economic indicators. Unlike standard TVP-VAR models that rely on unobserved stochastic processes to capture parameter changes, the AVP-VAR framework assumes that all variation in parameters is driven by known external variables. This structure allows the state equation to be directly incorporated into the measurement equation, simplifying estimation to standard linear regression techniques. As a result, the model avoids the computational complexity of state-space filtering while retaining the flexibility of traditional TVP-VARs. The AVP-VAR offers three main advantages: it provides economically interpretable parameter dynamics, enables fully replicable estimation and forecasting, and significantly reduces computational time.

	Our Monte Carlo experiments demonstrate that this framework accurately tracks diverse parameter dynamics—including transitory jumps and regime switches. The empirical evaluation across three macroeconomic datasets reveals systematic patterns. The AVP-VAR consistently delivers competitive or superior forecasting accuracy for real economic activity. For inflation, performance varies across applications: the model shows strength in some settings, and in most cases it is one of the top competitors compared to established benchmarks.
	
	In-sample analysis reveals why conventional TVP-VARs underperform: confronted with high-dimensional state covariance matrices, they heavily regularize parameter evolution, leaving coefficients nearly constant even during obvious structural breaks. The AVP-VAR overcomes this limitation by disciplining time variation through observable drivers, producing parameter paths that closely track the flexible UC-SV benchmark.
	

	\bibliographystyle{apalike}
	\addcontentsline{toc}{section}{\refname}
	\bibliography{AVPVAR.bib}

\clearpage\newpage
\begin{appendix}
\begin{center}
{\LARGE Online Supplement to ``Learning from crises: A new class of time-varying parameter VARs with observable adaptation''}\\[1em]
{\large Nicolas Hardy and Dimitris Korobilis}\\[2em]
\end{center}

\tableofcontents
\clearpage

\section{Technical Details of the AVP--VAR Model}

\subsection{Model Specification}

We depart from standard TVPs by replacing the unobserved state innovation with an observable drift. Let $Z_t\in\mathbb{R}^m$ denote a vector of exogenous drivers (e.g., forward-looking indicators of uncertainty, stress, volatility, or expectations), and define the cumulative drivers $C_t := \sum_{s=1}^{t-1} Z_s$. The time-varying coefficients evolve according to
\begin{equation}
	\theta_t \;=\; \theta_{t-1} + \gamma Z_{t-1},
	\qquad\Rightarrow\qquad
	\theta_t \;=\; \theta + \gamma C_t,
	\label{eq:theta_evolve}
\end{equation}
where $\theta\in\mathbb{R}^{n(k+r)}$ collects baseline coefficients and $\gamma\in\mathbb{R}^{n(k+r)\times m}$ maps drivers to the drift of the coefficients. Unlike \citet{Fischeretal2023}, whose time-varying parameters combine pre-specified effect modifiers with stochastic innovations, AVP-VAR replaces state errors with linear combinations of observable economic drivers. This specification preserves the persistent dynamics of a random walk while ensuring all parameter drifts are explicitly determined by economic conditions. Let $x_t=[1,y_{t-1}^\prime,\ldots,y_{t-p}^\prime]^\prime\in\mathbb{R}^{k}$ with $k=np+1$, and $f_t\in\mathbb{R}^r$. Define $X_t := I_n\otimes[x_t^\prime,\,f_t^\prime]\in\mathbb{R}^{n\times n(k+r)}$ and $\theta_t=[\beta_t^\prime,\lambda_t^\prime]^\prime\in\mathbb{R}^{n(k+r)}$, where $\beta_t$ stacks the VAR coefficients and $\lambda_t=\mathrm{vec}(\Lambda_t)$ stacks the factor loadings. The measurement equation can then be written as
\begin{equation}
	y_t \;=\; X_t \theta_t + v_t \;=\; X_t\theta + X_t\gamma C_t + v_t,
	\quad v_t\sim\mathcal{N}(0,\Sigma_t),\quad \Sigma_t=\mathrm{diag}(\sigma_{1t}^2,\ldots,\sigma_{nt}^2).
	\label{eq:meas_avp}
\end{equation}
Using the mixed-product rule, $X_t\gamma C_t=(C_t^\prime\otimes X_t)\mathrm{vec}(\gamma)$, so that stacking $Y=[y_1^\prime,\ldots,y_T^\prime]^\prime$ and $V=[v_1^\prime,\ldots,v_T^\prime]^\prime$ yields the static linear regression
\begin{equation}
	Y \;=\; \tilde{X}(1_T\otimes \theta) +
	\tilde W \mathrm{vec}(\gamma) \;+\; V,
	\label{eq:stacked_model}
\end{equation}
with $\tilde X = I \otimes X \in\mathbb{R}^{Tn\times Tn(k+r)}$ and $\tilde W= \left[ C_1^\prime\!\otimes X_1, \hdots, C_T^\prime\!\otimes X_T \right]^{\prime} \in\mathbb{R}^{Tn\times mn(k+r)}$. This representation collapses the state equation into the measurement equation once $\{C_t\}_{t=1}^{T}$ are precomputed; no Kalman filtering is required.

\subsection{Posterior Inference}
\label{subsec:posterior}

Estimation exploits the static-regression form in \eqref{eq:stacked_model}. We precompute cumulative drivers $C_t=\sum_{s=1}^{t-1} Z_s$ for $t=1,\ldots,T$. Partition $\theta=[\beta^\prime,\lambda^\prime]^\prime$ and $\gamma=[\gamma_\beta^\prime,\gamma_\lambda^\prime]^\prime$ conformably, where $\beta\in\mathbb{R}^{nk}$ contains baseline VAR coefficients, $\lambda\in\mathbb{R}^{nr}$ contains baseline factor loadings, $\gamma_\beta\in\mathbb{R}^{nk\times m}$, and $\gamma_\lambda\in\mathbb{R}^{nr\times m}$. For each equation $i=1,\ldots,n$, define augmented coefficient vectors
\[
\tilde\beta_i := \begin{bmatrix}\beta_i\\ \mathrm{vec}(\gamma_{\beta,i})\end{bmatrix}\in\mathbb{R}^{k+km},
\qquad
\tilde\lambda_i := \begin{bmatrix}\lambda_i\\ \mathrm{vec}(\gamma_{\lambda,i})\end{bmatrix}\in\mathbb{R}^{r+rm},
\]
where $\beta_i\in\mathbb{R}^{k}$ and $\lambda_i\in\mathbb{R}^{r}$ are the $i$-th equation's baseline coefficients, while $\gamma_{\beta,i}\in\mathbb{R}^{k\times m}$ and $\gamma_{\lambda,i}\in\mathbb{R}^{r\times m}$ map drivers to coefficient drift. Time-varying parameters are recovered deterministically as
\begin{equation}
	\beta_{it} = \beta_i + \gamma_{\beta,i} C_t,
	\qquad
	\lambda_{it} = \lambda_i + \gamma_{\lambda,i} C_t,
	\qquad t=1,\ldots,T.
	\label{eq:recover_coef}
\end{equation}

Define augmented regressors $\tilde x_t := [x_t^\prime, (x_t\otimes C_t)^\prime]^\prime\in\mathbb{R}^{k+km}$ and $\tilde f_t := [f_t^\prime, (f_t\otimes C_t)^\prime]^\prime\in\mathbb{R}^{r+rm}$. Let $y_{it}$ denote the $i$-th element of $y_t$ and $\sigma_{it}^2$ the corresponding idiosyncratic variance. The Gibbs sampler is initialized using OLS estimates for $\beta_i$ and $\lambda_i$, and residual-based log-variance estimates for $\log\sigma_{it}$. Factor scores are initialized via principal components.

The Gibbs sampler iterates over the following steps.

\paragraph{Step 1: Augmented VAR coefficients.}
Given $\{\lambda_{it},f_t,\sigma_{it}^2\}$, define $\tilde y_{it}:=y_{it}-f_t^\prime \lambda_{it}$ so that
\[
\tilde y_{it} = \tilde x_t^\prime \tilde\beta_i + v_{it},
\qquad v_{it}\sim\mathcal{N}(0,\sigma_{it}^2).
\]
Place a horseshoe prior on $\tilde\beta_i$: for $\ell=1,\ldots,k+km$,
\begin{equation}
	\tilde\beta_{i\ell}\mid \psi_{i\ell},\tau_i \sim \mathcal{N}(0,\psi_{i\ell}^2\tau_i^2),\quad
	\psi_{i\ell}\sim \text{Cauchy}_+(0,1),\quad
	\tau_i\sim \text{Cauchy}_+(0,1).
	\label{eq:hs_beta}
\end{equation}
The conditional posterior is $\tilde\beta_i\mid \cdot \sim \mathcal{N}(\bar\mu_{\beta,i},\bar V_{\beta,i})$ with
\[
\bar V_{\beta,i}^{-1} = D_{\beta,i}^{-1} + \sum_{t=1}^T \sigma_{it}^{-2}\,\tilde x_t \tilde x_t^\prime,
\qquad
\bar\mu_{\beta,i} = \bar V_{\beta,i}\!\left(\sum_{t=1}^T \sigma_{it}^{-2}\,\tilde x_t\,\tilde y_{it}\right),
\]
where $D_{\beta,i}=\mathrm{diag}(\psi_{i1}^2\tau_i^2,\ldots,\psi_{i,k+km}^2\tau_i^2)$, and  $\bar V_{\beta,i}^{-1} \in \mathbb{R}^{(k+km)\times k+km }$. Finally, reconstruct $\beta_{it}$ via \eqref{eq:recover_coef}.

\paragraph{Step 2: Augmented factor loadings.}
Given $\{\beta_{it},f_t,\sigma_{it}^2\}$, define $\hat y_{it}:=y_{it}-x_t^\prime\beta_{it}$ so that
\[
\hat y_{it} = \tilde f_t^\prime \tilde\lambda_i + v_{it},
\qquad v_{it}\sim\mathcal{N}(0,\sigma_{it}^2).
\]
Place a horseshoe prior on $\tilde\lambda_i$: for $\ell=1,\ldots,r+rm$,
\begin{equation}
	\tilde\lambda_{i\ell}\mid \xi_{i\ell},\tau_{\lambda,i} \sim \mathcal{N}(0,\xi_{i\ell}^2\tau_{\lambda,i}^2),\quad
	\xi_{i\ell}\sim \text{Cauchy}_+(0,1),\quad
	\tau_{\lambda,i}\sim \text{Cauchy}_+(0,1).
	\label{eq:hs_lambda}
\end{equation}
The conditional posterior is $\tilde\lambda_i\mid \cdot \sim \mathcal{N}(\bar\mu_{\lambda,i},\bar V_{\lambda,i})$ with
\[
\bar V_{\lambda,i}^{-1} = D_{\lambda,i}^{-1} + \sum_{t=1}^T \sigma_{it}^{-2}\,\tilde f_t \tilde f_t^\prime,
\qquad
\bar\mu_{\lambda,i} = \bar V_{\lambda,i}\!\left(\sum_{t=1}^T \sigma_{it}^{-2}\,\tilde f_t\,\hat y_{it}\right),
\]
where $D_{\lambda,i}=\mathrm{diag}(\xi_{i1}^2\tau_{\lambda,i}^2,\ldots,\xi_{i,r+rm}^2\tau_{\lambda,i}^2)$ , and  $\bar V_{\lambda,i}^{-1} \in \mathbb{R}^{(r+rm)\times r+rm }$. Finally, reconstruct $\lambda_{it}$ via \eqref{eq:recover_coef} and form $\Lambda_t$.

\paragraph{Step 3: Factors.}
Conditional on $\{\beta_t,\Lambda_t,\Sigma_t\}$, 
\begin{equation}
	f_t \mid \cdot \sim \mathcal{N}(\bar G_t \Lambda_t^\prime \Sigma_t^{-1}\tilde y_t, \bar G_t),
\end{equation}
where  $\tilde y_t:=y_t - (I_n\otimes x_t^\prime)\beta_t$ and $\bar G_t^{-1} = I_r + \Lambda_t^\prime \Sigma_t^{-1}\Lambda_t.$
Because the factors are only used for forecasting in our model, identification is not essential, and rotations of $\Lambda_t$ and $f_t$ produce equivalent predictive densities.

\paragraph{Step 4: Stochastic volatilities.}
For each equation $i=1,\ldots,n$, let the log-volatility follow a random walk,
\begin{equation}
	\log\sigma_{it}=\log\sigma_{i,t-1}+\zeta_{it}, 
	\qquad 
	\zeta_{it}\sim\mathcal{N}(0,\omega_i^2),
	\qquad 
	\log\sigma_{i1}\sim\mathcal{N}(0,V_h),
	\label{eq:sv_rw_short}
\end{equation}
where $\omega_i^2$ controls the smoothness of volatility movements and $V_h=1$ sets the prior uncertainty of the initial state. Conditional on the residuals $v_{it}=y_{it}-x_t^\prime\beta_{it}-f_t^\prime\lambda_{it},$ define $y_{it}^*=\log(v_{it}^2+\varepsilon_0)$ with $\varepsilon_0=10^{-6}$ to avoid numerical issues. 
Following \citet{Kimetal1998}, the non-Gaussian $\log\chi_1^2$ distribution of $y_{it}^*$ is approximated by a seven-component normal mixture, introducing latent indicators $S_{it}\in\{1,\ldots,7\}$. Conditional on $S_{it}$, the vector of log-volatilities $(\log\sigma_{i1},\ldots,\log\sigma_{iT})'$ follows a Gaussian state-space model and is sampled using a precision-based algorithm, as outlined by \cite{Chan2012}.  

Finally, the innovation variance of the volatility process is updated as
\begin{equation}
	\omega_i^2 \mid \{\log\sigma_{it}\}_{t=1}^T \sim \mathrm{Inv\text{-}Gamma}\!\left(
	\frac{a_0 + T - 1}{2},\;
	\frac{b_0 + \sum_{t=2}^T (\log\sigma_{it}-\log\sigma_{i,t-1})^2}{2}
	\right),
\end{equation}
where we set $a_0 = 1$ and $b_0 = 0.01$.
\paragraph{Step 5: Horseshoe hyperparameters.}
For each equation $i=1,\ldots,n$, update local scales $\{\psi_{i\ell}\},\{\xi_{i\ell}\}$ and global scales $\tau_i,\tau_{\lambda,i}$ in \eqref{eq:hs_beta}--\eqref{eq:hs_lambda} via slice sampling. Local scales control element-wise shrinkage, while the global scales $\tau_i$ and $\tau_{\lambda,i}$ adapt overall shrinkage strength independently across equations. See the technical appendix in \cite{Korobilis2022a} for details on the slice sampling algorithm.

\medskip
\noindent
Because time variation is encoded through observable $\{C_t\}$, all coefficient updates reduce to weighted least squares with no Kalman filtering required. Augmented dimensions scale as $k+km$ and $r+rm$, growing linearly in the number of drivers $m$ rather than sample size $T$, yielding substantial computational gains when $m\ll T$.

\subsection{Forecasting Procedure}

Forecasts are generated recursively using the final posterior draws from the AVP--VAR model estimated up to time $t$. At each forecast origin $t$, the model parameters evolve one step ahead according to the deterministic mappings
\[
\beta_{i,t+1} = \beta_i + \gamma_{\beta,i} C_{t+1}, 
\qquad
\lambda_{i,t+1} = \lambda_i + \gamma_{\lambda,i} C_{t+1},
\]
for each equation $i=1,\ldots,n$. These updated coefficients are held fixed for all forecast horizons $h \ge 1$, i.e.
\[
\beta_{i,t+1} = \beta_{i,t+2} = \cdots = \beta_{i,t+h}, 
\qquad 
\lambda_{i,t+1} = \lambda_{i,t+2} = \cdots = \lambda_{i,t+h}.
\]
This assumption ensures that forecasts reflect the latest observed macroeconomic conditions encoded in $Z_t$ without requiring forecasts of the driver process itself.

Stochastic volatilities are treated analogously: the last estimated idiosyncratic variances $\Sigma_t = \mathrm{diag}(\sigma_{1t}^2,\ldots,\sigma_{nt}^2)$ are kept constant across horizons. Given $\Lambda_{t+1}$ and $\Sigma_t$, the reduced--form covariance matrix of shocks is
\[
\Omega_{t+1} = \Lambda_{t+1}\Lambda_{t+1}^\prime + \Sigma_t,
\]
which remains fixed for all simulated horizons.

Conditional on $\{\beta_{t+1},\Omega_{t+1}\}$, forecasts are simulated recursively from the reduced form
\[
y_{t+h} = X_{t+h-1}^\prime \beta_{t+1} + \varepsilon_{t+h},
\qquad 
\varepsilon_{t+h}\sim\mathcal{N}(0,\Omega_{t+1}),
\]
where $X_{t+h-1}$ collects the appropriate lagged values of simulated $y$’s. This recursion is repeated for $h=1,\ldots,H$ and for each posterior draw, yielding predictive distributions for every forecast horizon.

Point forecasts correspond to posterior medians, while predictive quantiles are computed directly from the simulated draws. Forecast evaluation is based on mean squared prediction error (MSPE) and mean absolute error (MAE) for central accuracy, and on quantile (pinball) loss at $\tau\in\{0.1,0.9\}$ to assess performance in the distribution tails.

\newpage

\section{Technical Details of Competing Models}

This appendix describes the benchmark models used in the empirical analysis. Posterior inference for all Bayesian models is conducted by Gibbs sampling, cycling through blocks whose conditional distributions are presented below.

\subsection{TVP--VAR--EB}

This specification follows the exact TVP-VAR likelihood specification and empirical Bayes (EB) prior used by \citet{Primiceri2005}, which is by now a long-established benchmark in the literature of time-varying parameter vector autoregressions.

Let $y_t$ be an $n\times1$ vector. The observation equation is
\begin{equation}
	y_t = B_{0,t} + B_{1,t} y_{t-1} + \cdots + B_{p,t} y_{t-p}
	+ A_t^{-1} H_t^{1/2} \varepsilon_t, 
	\qquad 
	\varepsilon_t \sim \mathcal{N}(0,I_n),
\end{equation}
where $A_t$ is lower unitriangular and
\[
H_t = \mathrm{diag}\!\left( \exp(h_{1t}),\ldots,\exp(h_{nt}) \right).
\]

Let
\[
\beta_t = \mathrm{vec}(B_{0,t},\ldots,B_{p,t}) \in \mathbb{R}^{K}, 
\qquad 
a_t = \mathrm{vech}(A_t) \in \mathbb{R}^{n(n-1)/2}.
\]
All states evolve as random walks:
\begin{align}
	\beta_t &= \beta_{t-1} + u_t, & u_t &\sim \mathcal{N}(0,Q), \\
	a_t &= a_{t-1} + v_t, & v_t &\sim \mathcal{N}(0,\Phi), \\
	h_t &= h_{t-1} + \eta_t, & \eta_t &\sim \mathcal{N}(0,\Omega).
\end{align}

\paragraph{Priors.}
Initial states are set using training-sample OLS. 
State innovation covariances follow inverse--Wishart priors:
\[
Q \sim \mathcal{IW}(\nu_Q,S_Q),\qquad
\Phi \sim \mathcal{IW}(\nu_\Phi,S_\Phi),\qquad
\Omega \sim \mathcal{IW}(\nu_\Omega,S_\Omega),
\]
with $\nu_Q = K+2$, $\nu_\Phi = n+2$, $\nu_\Omega = n+2$, 
and $S_Q=0.01I_K$, $S_\Phi=0.01I_{n(n-1)/2}$, $S_\Omega=0.01I_n$.

\paragraph{Posterior simulation.}

\begin{enumerate}
	\item[\textbf{1.}] \textbf{VAR coefficients $\beta_{1:T}$.}  
	Draw using the Carter--Kohn simulation smoother conditional on $(A_t,H_t,Q)$.
	
	\item[\textbf{2.}] \textbf{Contemporaneous relations $a_{1:T}$.}  
	Updated via a simulation smoother conditional on $(\Phi,H_t)$.
	
	\item[\textbf{3.}] \textbf{Stochastic volatilities $h_{1:T}$.}  
	Sampled using the mixture-of-normals method of \citet{Kimetal1998}.  
	Update $\Omega$ from its inverse--Wishart posterior.
	
	\item[\textbf{4.}] \textbf{Innovation variances.}  
	\[
	Q \mid \beta_{1:T} \sim 
	\mathcal{IW}\!\left(
	\nu_Q + T,\;
	S_Q + \sum_{t=2}^T (\beta_t - \beta_{t-1})(\beta_t - \beta_{t-1})'
	\right),
	\]
	with analogous updates for $\Phi$ and $\Omega$.
\end{enumerate}

\subsection{TVP--VAR}
This is a TVP-VAR with Full Bayes posterior. We follow a first-difference representation of the time-varying coefficients, similar to \cite{ChanJeliazkov2009}. Both autoregressive coefficients and covariances have horseshoe priors to discipline the amount of time variation. Let 
\[
y_t = X_t \beta_t + \varepsilon_t,
\qquad 
X_t = I_n \otimes (1,y_{t-1}',\ldots,y_{t-p}'),
\]
where $\beta_t = (\beta_{1,t}',\ldots,\beta_{n,t}')'$ and $\beta_{i,t}\in\mathbb{R}^{K}$. 

\paragraph{Factor stochastic volatility.}
The reduced-form innovations satisfy
\[
\varepsilon_t = L_t f_t + S_t^{1/2}\eta_t,
\qquad
f_t \sim \mathcal{N}(0,I_r),\quad 
\eta_t \sim \mathcal{N}(0,I_n),
\]
with
\[
L_t = (\lambda_{1,t},\ldots,\lambda_{n,t})',\qquad
S_t = \mathrm{diag}(\exp(h_{1t}),\ldots,\exp(h_{nt})).
\]
Thus 
\[
\Sigma_t = L_t L_t' + S_t.
\]

\paragraph{State evolution.}
All time-varying components follow random walks:
\begin{align}
	\beta_{i,t} &= \beta_{i,t-1} + u_{i,t}, & u_{i,t}\sim\mathcal{N}(0,Q_{\beta,i}),\\
	\lambda_{i,t} &= \lambda_{i,t-1} + \xi_{i,t}, & \xi_{i,t}\sim\mathcal{N}(0,Q_{\lambda,i}),\\
	h_{i,t} &= h_{i,t-1} + v_{i,t}, & v_{i,t}\sim\mathcal{N}(0,Q_{h,i}).
\end{align}

\paragraph{Difference representation.}
Define $T\times T$ first-difference matrices $\mathcal{D}_\beta$ and $\mathcal{D}_\lambda$:
\[
\gamma_i = \mathcal{D}_\beta \beta_i,\qquad 
\beta_i = \mathcal{D}_\beta^{-1}\gamma_i,
\]
\[
\delta_i = \mathcal{D}_\lambda \lambda_i,\qquad 
\lambda_i = \mathcal{D}_\lambda^{-1}\delta_i.
\]

\paragraph{Posterior simulation.}
Updates proceed equation-by-equation:

\begin{enumerate}
	\item[\textbf{1.}] 
	\textbf{Time-varying coefficients $\beta_{i,1:T}$.} Let $\tilde{y}_{i,t} = y_{i,t} - \lambda_{i,t}'f_t,$ and consider the differenced representation $\beta_i = \mathcal{D}_\beta^{-1}\gamma_i$. Stacking over time yields a Gaussian state-space model for $\gamma_i$.  
	Conditional posteriors are Gaussian, with Horseshoe priors on $\gamma_i$. 
	Recover $\beta_{i,t}$ by cumulative summation. See \cite{ChanJeliazkov2009} and the technical appendix in \cite{Korobilisetal2021} for details.
	
	\item[\textbf{2.}] 
	\textbf{Factors $f_t$.}
	Given $\tilde{y}_t = y_t - X_t\beta_t$,
	\[
	f_t \mid \cdot \sim 
	\mathcal{N}\!\left(
	V_t L_t'S_t^{-1}\tilde{y}_t,\;
	V_t
	\right),
	\qquad 
	V_t = (I_r + L_t'S_t^{-1}L_t)^{-1}.
	\]
	
	\item[\textbf{3.}] 
	\textbf{Factor loadings $\lambda_{i,1:T}$.}
	Using $u_{i,t} = y_{i,t} - x_t'\beta_{i,t}$ and the differenced representation 
	$\lambda_i = \mathcal{D}_\lambda^{-1}\delta_i$,  
	$\delta_i$ follows a Gaussian conditional posterior under Horseshoe priors.  
	Recover $\lambda_{i,t}$ by cumulative summation. This step mimics Step 1.
	
	\item[\textbf{4.}] 
	\textbf{Stochastic volatilities.}
	Residuals $\epsilon_{i,t} = y_{i,t}-x_t'\beta_{i,t}-\lambda_{i,t}'f_t$ 
	are used with the \citet{Kimetal1998} mixture approximation.  
	Sample $h_{i,1:T}$ via a precision-based FFBS algorithm, and update $Q_{h,i}$. See Step 4 from the AVP-VAR algorithm.
	
	\item[\textbf{5.}] 
	\textbf{Horseshoe hyperparameters.}
	Local and global shrinkage scales for $(\gamma_i,\delta_i)$ are updated via slice sampling. See the technical appendix in \cite{Korobilis2022a} for details.
\end{enumerate}

\subsection{VAR--SVO-t} 

The stochastic–volatility–with–outliers (SVO-$t$) model of 
\citet{Carrieroetal2023} augments a Gaussian VAR with a 
heavy–tailed volatility component driven by occasional discrete ``outlier'' 
states. The mechanism preserves conditional Gaussianity while allowing for 
large, infrequent jumps in volatility. Let $y_t=(y_{1t},\dots,y_{nt})'$ and  
$x_t = (1, y_{t-1}', \dots, y_{t-p}')'$.
The reduced–form VAR is
\begin{equation}
	y_t = B x_t + A^{-1}\Sigma_t^{1/2}\varepsilon_t,
	\qquad 
	\varepsilon_t \sim \mathcal{N}(0,I_n),
	\label{eq:SVO-eq}
\end{equation}
where $B$ collects autoregressive coefficients and $A$ is a lower–triangular 
contemporaneous matrix with ones on the diagonal.
Conditional on $(B,A)$, the reduced–form shocks satisfy
\[
u_t = A(y_t - Bx_t),
\qquad  
u_{i,t} \mid h_{i,t},\kappa_{i,t} \sim 
\mathcal{N}\!\bigl(0,\exp(h_{i,t})\kappa_{i,t}^2\bigr).
\]
Each log–volatility follows a random walk:
\[
h_{i,t} = h_{i,t-1} + \eta_{i,t}, 
\qquad 
\eta_{i,t} \sim \mathcal{N}(0,\phi_i^2).
\]
The SVO-$t$ mechanism introduces a discrete variance–inflation factor  
$\kappa_{i,t} \in \{1,2,\dots,S_{\max}\}$:
\[
\sigma_{i,t}^2 = \exp(h_{i,t})\kappa_{i,t}^2.
\]
Outliers occur with probability $\pi_i$:
\[
\Pr(\kappa_{i,t}=1)=1-\pi_i,
\qquad 
\Pr(\kappa_{i,t}=s)=\frac{\pi_i}{S_{\max}-1}, \quad s\ge2,
\]
with $S_{\max}=20$.

\paragraph{Priors.}

\begin{itemize}
	
	\item \textbf{VAR coefficients.}  
	A Minnesota prior is imposed equation by equation:
	\[
	\beta_i \sim \mathcal{N}(m_i,\,V_i),
	\]
	The prior variances are diagonal:
	\[
	\mathrm{Var}(\beta_{i,j,\ell})
	=
	\begin{cases}
		\dfrac{\lambda_1}{\ell^{\lambda_3}}, 
		& \text{own lag},\\[10pt]
		\dfrac{\lambda_1\lambda_2}{\ell^{\lambda_3}}
		\dfrac{s_i^2}{s_j^2}, 
		& \text{cross lag},
	\end{cases}
	\]
	with hyperparameters $(\lambda_1,\lambda_2,\lambda_3)=(0.05,0.5,2)$	and the intercept prior variance 
	\[
	\mathrm{Var}(\beta_{i,\mathrm{int}})=100\, s_i^2,
	\]
	where $s_i^2$ is the AR residual variance from the training sample.
	
	\item \textbf{Impact matrix $A$.}  
	For $i=2,\dots,n$, let $a_i=(a_{i1},\dots,a_{i,i-1})'$ denote the non-zero 
	elements in row $i$.  
	The prior is diffuse:
	\[
	a_i \sim \mathcal{N}(0,\,\Omega_{A,i}),
	\qquad
	\Omega_{A,i} = c_A I_{i-1},
	\qquad
	c_A = 10^6.
	\]
	
	\item \textbf{SV innovations.}  
	The variances of SV innovations are
	\[
	\Phi = \mathrm{diag}(\phi_1^2,\dots,\phi_n^2)
	\sim \mathcal{IW}(d_\Phi, S_\Phi),
	\]
	with
	\[
	d_\Phi = n+3,
	\qquad
	S_\Phi = 0.15\,I_n \times \frac{12}{np}.
	\]
	These values imply weakly informative shrinkage on the volatility process.
	
	\item \textbf{Outlier probabilities.}  
	For each series,
	\[
	\pi_i \sim \mathrm{Beta}(\alpha_\pi,\beta_\pi),
	\qquad
	\alpha_\pi = \frac{10\,np}{4},
	\qquad
	\beta_\pi  = 10\,np - \alpha_\pi.
	\]
	
	\item \textbf{Initial log-volatilities.}
	Initial states satisfy
	\[
	h_{i,1} \sim \mathcal{N}(0,\,100),
	\] 
	with the initial draw centered around log-squared AR residuals.

\end{itemize}

\paragraph{Posterior simulation.}

\begin{enumerate}
	
	\item[\textbf{1.}]
	\textbf{VAR coefficients $B$.}  
	Given $(A,\Sigma_t)$, each equation reduces to a weighted Gaussian regression.  
	Minnesota priors imply conjugate Gaussian posteriors.
	
	\item[\textbf{2.}]
	\textbf{Contemporaneous matrix $A$.}  
	Given shocks $u_t$, each row solves
	\[
	u_{i,t} = -a_i' u_{1:i-1,t} + \nu_{i,t},
	\qquad 
	\nu_{i,t}\sim\mathcal{N}(0,\sigma_{i,t}^2),
	\]
	yielding Gaussian posterior updates for $a_i$.
	
	\item[\textbf{3.}]
	\textbf{Mixture indicators for SV.}  
	Using the seven–component KSC approximation, 
	$\log(u_{i,t}^2)$ is represented as a mixture of Gaussian signals.  
	Mixture states are drawn from multinomial probabilities.
	
	\item[\textbf{4.}]
	\textbf{Outlier scales $\kappa_{i,t}$.}  
	For $s \in \{1,\dots,S_{\max}\}$,
	\[
	p(\kappa_{i,t}=s\mid\cdot)\propto 
	p(\kappa_{i,t}=s)\,
	\exp\!\left[
	-\tfrac12
	\Bigl(
	\frac{\log u_{i,t}^2 - h_{i,t} - \log s^2 - \mu_{s_{i,t}}}{\sigma_{s_{i,t}}}
	\Bigr)^2
	\right].
	\]
	
	\item[\textbf{5.}]
	\textbf{Outlier probabilities $\pi_i$.}  
	Let $I_{i,t}^{\mathrm{out}}=\mathbf{1}\{\kappa_{i,t}>1\}$.  
	The posterior is
	\[
	\pi_i \mid \kappa_{i,1:T}
	\sim
	\mathrm{Beta}\!\left(
	\alpha_\pi + \sum_t I_{i,t}^{\mathrm{out}},\;
	\beta_\pi  + \sum_t (1-I_{i,t}^{\mathrm{out}})
	\right).
	\]
	
	\item[\textbf{6.}]
	\textbf{Log–volatilities $h_{i,t}$.}  
	Conditional on $(s_{i,t},\kappa_{i,t})$, the transformed observations
	\[
	\log(u_{i,t}^2)=h_{i,t}+\log\kappa_{i,t}^2+\mu_{s_{i,t}}+\sigma_{s_{i,t}} z_{i,t}
	\]
	form a linear Gaussian state–space model.  
	Each $h_i$ is drawn using a precision–based FFBS algorithm.
	
	\item[\textbf{7.}]
	\textbf{SV innovation variances $\phi_i^2$.}  
	With $\eta_{i,t}=h_{i,t}-h_{i,t-1}$,
	\[
	\phi_i^2 \mid h_{i,1:T}
	\sim 
	\mathcal{IG}\!\left(
	d_\Phi + \tfrac{T}{2},\;
	S_\Phi + \tfrac12 \sum_{t=2}^T \eta_{i,t}^2
	\right).
	\]
	
\end{enumerate}

\subsection{FAVAR--SV}

A latent factor $F_t$ summarizing external information is extracted by recursive 
principal components using only data up to time $t$.  
The augmented system is
\[
\begin{pmatrix} F_t \\ y_t \end{pmatrix}
=
B_0 
+ B_1 \begin{pmatrix} F_{t-1} \\ y_{t-1} \end{pmatrix}
+ \cdots
+ B_p \begin{pmatrix} F_{t-p} \\ y_{t-p} \end{pmatrix}
+ A_t^{-1} H_t^{1/2} \varepsilon_t,
\]
with $A_t$ lower triangular and $H_t=\mathrm{diag}(\exp(h_{1t}),\ldots,\exp(h_{n+1,t}))$.  
Only the contemporaneous relations and volatilities evolve:
\[
a_t = a_{t-1} + \zeta_t,\qquad 
h_t = h_{t-1} + \eta_t.
\]
Estimation follows exactly the same approach as TVP--VAR, but this time just considering a fourth endogenous variable in the VAR.
\subsection{CP--VAR and CP--VAR--SV}

The CP--VAR imposes time-invariant parameters and homoskedastic Gaussian errors:
\[
y_t = B_0 + B_1 y_{t-1} + \cdots + B_p y_{t-p}
+ \Lambda f_t + \varepsilon_t,
\qquad \varepsilon_t \sim \mathcal{N}(0,\Sigma),
\]
where $f_t\sim\mathcal{N}(0,I_r)$, 
$\Lambda$ is fixed over time, and 
$\Sigma=\mathrm{diag}(\sigma_1^2,\ldots,\sigma_n^2)$.

The CP--VAR--SV variant allows $\Lambda_t$ and $h_t$ to follow random walks, while 
$(B_0,\ldots,B_p)$ remain constant. For this model, estimation is akin to TVP--VAR, but this time imposing constant autoregressive coefficients.

\subsection{FAVAR and OLS--VAR}

Both are VAR($p$) systems with constant parameters and homoskedastic errors:
\[
y_t = c + B_1 y_{t-1} + \cdots + B_p y_{t-p} + \varepsilon_t,
\qquad \varepsilon_t \sim \mathcal{N}(0,\Omega).
\]
OLS provides closed-form estimates
\[
B_{\mathrm{OLS}} = (X'X)^{-1}X'Y,
\qquad 
\Omega_{\mathrm{OLS}} = \frac{(Y-XB_{\mathrm{OLS}})'(Y-XB_{\mathrm{OLS}})}{T}.
\]

The FAVAR variant augments $y_t$ with a recursively estimated factor $F_t$.
\pagebreak

\newpage

\section{Simulations}
This is a detailed description of the Monte Carlo exercise we provide in the main paper. We conduct Monte Carlo experiments to evaluate the AVPs ability to track diverse parameter dynamics, such as persistent drift, temporary jumps, smooth regime changes, and threshold-induced discontinuities, and compare it to a standard random-walk time-varying-parameter benchmark. To this end, we consider two distinct data generating processes (DGPs): one with a transitory break, and one with regime switch and threshold effects. Rather than simulating full VAR systems, which would require imposing additional constraints to ensure stability and stationarity, we work with univariate regressions featuring four predictors. This choice reflects the equation-by-equation estimation strategy commonly used in practice and allows us to focus squarely on the dynamics of the coefficients without conflating results with identification or numerical stability issues. The predictors follow a VAR(1) process with spatial correlation, and the dependent variable evolves with stochastic volatility. This setup preserves realistic persistence and volatility patterns while avoiding the complexity of high-dimensional state spaces.

\subsection{Detailed Setup}
The generating model is a regression with time-varying parameters and stochastic volatility of the form
\begin{equation*}
	y_t = x_t' \beta_t + \sigma_t \varepsilon_t, \quad \varepsilon_t \sim \mathcal{N}(0, 1),
\end{equation*}
where $x_t \in \mathbb{R}^p$ is the vector of predictors, $\beta_t \in \mathbb{R}^p$ represents time-varying coefficients, and $\sigma_t > 0$ is the time-varying volatility. The predictors follow a VAR(1) with a spatial correlation structure:
\begin{equation*}
	x_t = \rho x_{t-1} + \eta_t, \quad \eta_t \sim \mathcal{N}(0, \Sigma_x),
\end{equation*}
where $\rho \in (-1, 1)$ is the autoregressive parameter, and the correlation matrix $\Sigma_x$ has elements $\left[\Sigma_x\right]_{i,j} = \rho^{|i-j|}$, $\quad i,j = 1, \ldots, p$. The process is initialized as the unconditional mean $x_0 = 0$, and a burn-in period of 100 observations is discarded. To allow for different degrees of persistence and correlation, we consider three different values for $\rho$:  $\rho \in \{0,0.5,0.95\}$. Finally, the log-volatility evolves as:
\begin{equation*}
	h_t = h_0 + 0.99(h_{t-1} - h_0) + T^{-1/2} \nu_t, \quad \nu_t \sim \mathcal{N}(0, 1),
\end{equation*}
where $h_0 = \log(\sigma_0)$ and $\sigma_t = \exp(h_t)$. We initialize with $\sigma_0=0.2$ and apply a burn-in of 100 observations to remove dependence on the initial condition.

The specification above is common among the two DGPs, their difference only being the dynamics of the TVPs $\beta_{t}$. In both cases $\beta_{t}$ will have a random walk component that is contaminated by some irregularity such as a jump or a non-linear function of some process. This setup allows us to test whether the AVP can recover parameter dynamics from explicit structural signals or alternatively approximate them using a sufficiently rich, but uninformative, noise basis. For each DGP, we simulate $T \in \{50,100,200\}$ observations (after burn-in), with $p=4$ predictors, and $N_{MC}=1,000$ Monte Carlo replications.

\paragraph{DGP1: Time-Varying Coefficients with Transitory Break (Jump)} In the first DGP the coefficients evolve as:
\begin{equation*}
	\beta_t = \theta_t + D_t \cdot 1_p,
\end{equation*}
where $\theta_t$ is a highly persistent component and $D_t$ is a scalar jump defined below. The persistent component follows a near-unit root process:
\begin{equation*}
	\theta_t = \mu + 0.99(\theta_{t-1} - \mu) + T^{-1/2} \nu_t, \quad \nu_t \sim \mathcal{N}(0, I_p),
\end{equation*}
where we define the mean $\mu \sim \mathcal{U}[-2, 2]$. The jump component introduces a transitory structural break lasting six observations, which roughly tries to emulate the sharp, but short, COVID-19-related recession in the U.S. and other major economies. Define the scalar jump indicator as:
\begin{equation}
	D_t = \begin{cases}
		w_t, & t \in \mathcal{H} \\
		0, & t \notin \mathcal{H}
	\end{cases}, \label{eq:jump}
\end{equation}
where $\mathcal{H} = \{\lfloor 2T/3 \rfloor, \ldots, \lfloor 2T/3 \rfloor + 5\}$ defines a six-period jump window, and $w_t \sim \mathcal{U}[4,8]$ is drawn independently at each $t \in \mathcal{H}$. All $p$ coefficients receive the same additive shock at each jump period, though the magnitude varies across the six periods. Put differently, a transitory structural break occurs approximately two-thirds through the sample and persists for six periods.

\paragraph{DGP2: Non-linear TVP with Regime Switches and Threshold Effects.}
The second DGP combines multiple sources of parameter instability: (1) a near-random walk component, (2) regime-switching dynamics, and (3) threshold effects. Formally, we decompose the coefficient evolution as:	
\begin{equation}
	\beta_{jt} = \theta_{jt}^{RW} + \theta_{jt}^{RS} + \theta_{jt}^{TE} + \eta_{jt},
\end{equation}
where $\eta_{jt} \sim \mathcal{N}(0, \sigma_{\eta,j}^2/T)$ are  parameter-specific innovations with values given in Table~\ref{tab:dgp2_params}. The first component is of the form
\begin{equation}
	\theta_{jt}^{RW} = \mu_j + \rho_j (\beta_{j,t-1} - \mu_j),
\end{equation}	
where $\rho_j = 0.95 + 0.04(j-1)/(p-1)$ varies deterministically from 0.95 to 0.99 across parameters, creating heterogeneous persistence in the baseline drift. The second component introduces smooth transitions between states based on the lagged dependent variable and is of the form
\begin{equation}
	\theta_{jt}^{RS} = \delta_j S_t + \alpha_j \tanh(\kappa_j x_{t-1,j}) S_t,
\end{equation}	
where $S_t = \frac{1}{1 + \exp(-2y_{t-1})}$ is a smooth transition function, $\delta_j$ captures direct regime effects and $\alpha_j$ governs how lagged predictors interact with the regime state through a bounded non-linear transformation. The third component captures discrete jumps of the form
\begin{equation}
	\theta_{jt}^{TE} = \phi_j \mathbb{I}(M_{jt} > \tau_j),
\end{equation}
whenever the function $M_{jt} = |y_{t-1}| + \omega_j|y_{t-2}|$ exceeds the threshold $\tau_j$, where $\phi_j$ defines the strength of the shift. Finally, we allow the error variance to also increase during regime transitions:	
\begin{equation}
	\text{var}(\varepsilon_t \vert S_t) = \sigma_t^2(1 + 0.5 S_t)
\end{equation}	
creating periods of heightened volatility that coincide with regime changes.

This richer DGP has several hyper-parameter values driving each of the four TVPs $\beta_{j}$, $j=1,..,4$. Table~\ref{tab:dgp2_params} summarizes all parameter-specific values. Parameter $\beta_1$ is primarily regime-driven with strong direct regime effects ($\delta_1 = 0.4$), $\beta_2$ emphasizes predictor-regime interactions ($\alpha_2 = 0.25, \kappa_2 = 2$), $\beta_3$ is dominated by threshold effects ($\phi_3 = 0.35$) with high stress sensitivity ($\omega_3 = 0.8$), and $\beta_4$ exhibits mixed dynamics with moderate values across all mechanisms.
\begin{table}[h]
	\centering
	\caption{TVP-Specific Hyper-parameters in DGP2} \label{tab:dgp2_params}
	\begin{tabular}{lcccccccc}
		\toprule
		Hyper-parameters	& $\rho_j$ & $\delta_j$ & $\alpha_j$ & $\kappa_j$ & $\phi_j$ & $\omega_j$ & $\tau_j$ & $\sigma_{\eta,j}$ \\
		\midrule
		values for $\beta_{1t}$ & 0.95 & 0.40 & 0.05 & 1 & 0.10 & 0.5 & 1.5 & 0.04 \\
		values for $\beta_{2t}$ & 0.96 & 0.15 & 0.25 & 2 & 0.05 & 0.5 & 1.2 & 0.05 \\
		values for $\beta_{3t}$ & 0.97 & 0.10 & 0.08 & 1 & 0.35 & 0.8 & 1.0 & 0.06 \\
		values for $\beta_{4t}$ & 0.99 & 0.20 & 0.12 & 1 & 0.15 & 0.3 & 1.3 & 0.07 \\
		\bottomrule
	\end{tabular}
\end{table}

While these DGPs appear complex, they serve a specific purpose. Simulating from a pure random walk would trivially favor the standard TVP estimator, while generating parameters as deterministic functions of $Z_t$ would favor our AVP by construction. Instead, we design DGPs that combine random walk dynamics with rich non-linear features (regime switches, interaction effects, and threshold mechanisms) to provide a neutral testing ground. We then compare how well each estimation approach recovers the true parameter paths when fed different sets of exogenous drivers. We compare two estimation approaches for the TVPs, namely the traditional random walk of the form 
\begin{equation}
	\beta_{t} = \beta_{t-1} + \eta_{t}, \eta_{t} \sim N(0,Q),
\end{equation}
and the adaptively-varying parameter scheme of the form
\begin{equation}
	\beta_{t} = \beta_{t-1} + \gamma Z_{t}^{j}. \label{eq:evol_MC}
\end{equation}
As already discussed in the previous Section, both $\gamma$ and $\eta_{t}$ have horseshoe priors as a default choice that regulates the amount of time-variation in the parameters.

For estimation under equation \eqref{eq:evol_MC}, we consider two types of drivers: \textbf{agnostic drivers} that contain no information about the DGP structure, and \textbf{targeted drivers} that include partial signals related to breaks or regimes. The agnostic drivers are simply 
\begin{equation}
	Z_t^j = [1, \mathcal{N}_m(0,I)] \quad \text{for } m \in \{20, 40, 60\} \text{ and } j \in \{1, 2, 3\}.
\end{equation}
Under these drivers, our proposed evolution collapses to a random walk with a reduced-rank state error. The targeted drivers, by contrast, incorporate observable signals---though imperfect---about the structural features governing parameter variation. Crucially, these signals reveal only the \emph{timing} of potential breaks or regime shifts, not their magnitude or duration, which must still be estimated.

In DGP1, the targeted drivers are 
\begin{equation}
	Z_t^j = [1, \mathcal{N}_9(0,I), \xi_t, \xi_t \times \mathcal{N}_m(0,I)] \quad \text{for } m \in \{10, 20, 30\} \text{ and } j \in \{4, 5, 6\},
\end{equation}
where $\xi_t = \mathbb{I}(t \in \mathcal{H})$ flags break periods. The interaction terms $\xi_t \cdot \mathcal{N}_m(0,I)$ allow the model to learn break-specific parameter shifts without knowing their intensity in advance. In DGP2, the targeted drivers progressively reveal more information: $Z_t^4$ includes the regime signal $\xi_t^{RS} = S_t$, $Z_t^5$ adds the interaction signal $\xi_t^{INT} = S_t \cdot p^{-1}\sum_{j=1}^p \tanh(x_{t-1,j})$, and $Z_t^6$ further incorporates the threshold signal $\xi_t^{TE} = \mathbb{I}(\tilde{M}_t > 1.0)$ where $\tilde{M}_t = |y_{t-1}| + 0.5|y_{t-2}|$ is a stress indicator. Each signal is again interacted with normal variates to capture heterogeneous effects across parameters.


\subsection{Monte Carlo results}

For each data generating process (DGP) and each sample size $T \in \{50, 100, 200\}$, we perform 1,000 independent Monte Carlo replications. In each replication, we begin by simulating a dataset according to the chosen DGP, which yields both the true time-varying parameter path and the corresponding observations.

Next, we estimate all models, that is, specification \eqref{eq:evol_MC} using the six different drivers outlined in the previous subsection and the standard random-walk benchmark, using Bayesian methods. Specifically, we implement Gibbs sampling with 11,000 iterations, discarding the first 1,000 as burn-in and thinning the chain by retaining every 10th draw. This results in 1,000 posterior samples per replication.

For each model and posterior draw, we reconstruct the estimated parameter path and compute the posterior mean trajectory. We then assess accuracy by calculating the mean squared parameter error (MSPE), which averages the squared deviations between the estimated and true parameter paths across time and coefficients. Finally, we report the average MSPE across all replications to summarize model performance.

\paragraph{DGP1: Transitory Breaks.}			
Table~\ref{tab:MonteCarloDGP1} presents Monte Carlo results for DGP1. Each entry of that table reports the MSPE ratio relative to the canonical TVP benchmark. There are four things worth highlighting. First, AVP relative performance generally improves with larger $m$. However, this pattern is substantially more pronounced for targeted drivers. With agnostic drivers, the gains from increasing $m$ are modest and often negligible. Second, the relative performance of AVP strengthens considerably with higher $\rho$. The most striking improvements occur at $\rho=0.95$, where MSPE reductions frequently exceed 20\% and reach as high as 40\% for targeted drivers. Third, even AVP with agnostic drivers remains competitive with TVP, though results vary systematically with $\rho$. When $\rho=0$, TVP dominates, outperforming agnostic AVP by approximately 2\% on average. At $\rho = 0.5$, results are mixed, with agnostic drivers showing advantages primarily in larger samples ($T=200$). Notably, at $\rho=0.95$, agnostic AVP consistently outperforms TVP, with improvements reaching up to 26\% in some entries. Fourth, as expected, AVP with targeted drivers substantially outperforms all alternatives. Across 105 of 108 entries (97.2\%), targeted AVP achieves lower MSPE than TVP. At $\rho=0.95$, gains sometimes exceed 40\% relative to TVP.  Moreover, 86 of 108 cases (79.6\%) yield improvements of at least 10\%.\footnote{Unreported LASSO results show substantially worse performance, with MSPE sometimes orders of magnitude higher. These results are available upon request.}

\begin{figure}[htbp!]  
	\centering
	\includegraphics[width=0.95\textwidth]{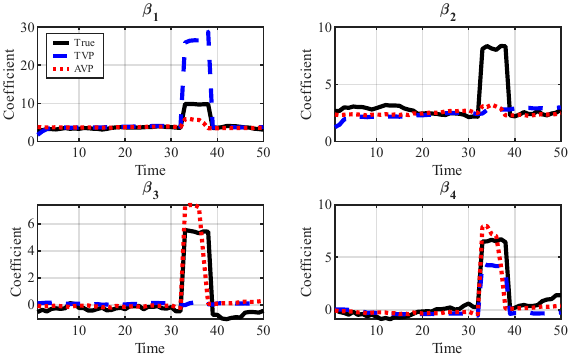}  
	\caption{Illustration of a Monte Carlo iteration for DGP1 with $T=50$ and $\rho=0.95$. The red line AVP considers $m=60$ targeted drivers.} \label{fig:plots_Montecarlo_jump}
\end{figure}

All in all, agnostic AVP tends to be competitive, despite not having any relevant information about the jumps, and show significant improvements for the case of high $\rho$. Moreover, targeted AVP is by far the best performing model, displaying large improvements compared to TVP, particularly for the case of large $m$. Figure~\ref{fig:plots_Montecarlo_jump} illustrates the DGP and parameter estimates from a single Monte Carlo draw, comparing TVP and AVP with $m=60$ targeted drivers. TVP struggles to track the true parameters: it substantially overshoots the jump in $\beta_1$ (panel 1), fails to detect jumps in $\beta_2$ and $\beta_3$ (panels 2-3), and only partially captures the break in $\beta_4$ (panel 4). AVP demonstrates superior tracking ability, particularly for $\beta_3$ and $\beta_4$, though it also misses the jump in $\beta_2$. Notably, in panel 1, AVP generates a more conservative adjustment that better approximates the true parameter trajectory.

\begin{table}[htbp!]
	\centering
	\caption{Monte Carlo results for DGP1 (Transitory Jump)}
	\label{tab:MonteCarloDGP1}
	\small
	\resizebox{0.95\textwidth}{!}{%
		\begin{tabular}{@{}l*{12}{S[table-format=1.2]}@{}}
			\toprule
			& \multicolumn{12}{c}{\textbf{T = 50}} \\
			\cmidrule(lr){1-13}
			& \multicolumn{4}{c}{\textbf{Panel A: $\rho = 0$}} & \multicolumn{4}{c}{\textbf{Panel B: $\rho = 0.50$}} & \multicolumn{4}{c}{\textbf{Panel C: $\rho = 0.95$}} \\
			\cmidrule(lr){2-5} \cmidrule(lr){6-9} \cmidrule(lr){10-13}
			\textbf{Model} &$\beta_{1,t}$&$\beta_{2,t}$&$\beta_{3,t}$&$\beta_{4,t}$& $\beta_{1,t}$&$\beta_{2,t}$&$\beta_{3,t}$&$\beta_{4,t}$ & $\beta_{1,t}$&$\beta_{2,t}$&$\beta_{3,t}$&$\beta_{4,t}$ \\
			\midrule
			TVP              &1.00  &1.00 &1.00  &1.00  &1.00  &1.00  &1.00 &1.00  &1.00  &1.00  &1.00  &1.00  \\			
			\textbf{Agnostic Drivers}& &&&&&&&&&&& \\
			AVP (m=20)  &1.03 &1.02  &1.04  &1.04  &1.04  &1.05  &1.00  &1.05  &\textbf{1.00}  &\textbf{0.81}  &\textbf{0.75} &\textbf{0.97}  \\
			AVP (m=40)  &1.02  &1.01  &1.03  &1.03  &1.02  &1.05  &\textbf{1.00}  &1.05  &\textbf{0.99}  &\textbf{0.81}  &\textbf{0.75}  &\textbf{0.96}  \\
			AVP (m=60)  &1.03  &1.01  &1.03  &1.03  &1.02  &1.04  &\textbf{0.99}  &1.04  &\textbf{0.99}  &\textbf{0.81}  &\textbf{0.74}  &\textbf{0.96} \\
			\textbf{Targeted Drivers}& &&&&&&&&&&& \\
			AVP (m=20)  &\textbf{0.82}  &\textbf{0.86}  &\textbf{0.86}  &\textbf{0.84}  &\textbf{0.80} &\textbf{0.85} &\textbf{0.78}  &\textbf{0.84}  &\textbf{0.87}  &\textbf{0.64}  &\textbf{0.63}  &\textbf{0.74} \\
			AVP (m=40)  &\textbf{0.78}  &\textbf{0.79}  &\textbf{0.80}  &\textbf{0.82}  &\textbf{0.77} &\textbf{0.77}  &\textbf{0.68}  &\textbf{0.78} &\textbf{0.75}  &$\textbf{0.58}^\dagger$  &\textbf{0.61}  &\textbf{0.67}  \\
			AVP (m=60)  &$\textbf{0.75}^\dagger$  &$\textbf{0.78}^\dagger$  &$\textbf{0.79}^\dagger$  &$\textbf{0.80}^\dagger$  &$\textbf{0.72}^\dagger$  &$\textbf{0.73}^\dagger$  &$\textbf{0.66}^\dagger$  &$\textbf{0.72}^\dagger$ &$\textbf{0.71}^\dagger$ &\textbf{0.59}  &$\textbf{0.56}^\dagger$  &$\textbf{0.65}^\dagger$  \\		
			\midrule
			& \multicolumn{12}{c}{\textbf{T = 100}} \\
			\cmidrule(lr){1-13}
			& \multicolumn{4}{c}{\textbf{Panel D: $\rho = 0$}} & \multicolumn{4}{c}{\textbf{Panel E: $\rho = 0.50$}} & \multicolumn{4}{c}{\textbf{Panel F: $\rho = 0.95$}} \\
			\cmidrule(lr){2-5} \cmidrule(lr){6-9} \cmidrule(lr){10-13}
			\textbf{Model} &$\beta_{1,t}$&$\beta_{2,t}$&$\beta_{3,t}$&$\beta_{4,t}$& $\beta_{1,t}$&$\beta_{2,t}$&$\beta_{3,t}$&$\beta_{4,t}$ & $\beta_{1,t}$&$\beta_{2,t}$&$\beta_{3,t}$&$\beta_{4,t}$ \\
			\midrule
			TVP             &1.00  &1.00 &1.00  &1.00  &1.00  &1.00  &1.00 &1.00  &1.00  &1.00  &1.00  &1.00  \\			
			\textbf{Agnostic Drivers}& &&&&&&&&&&& \\
			AVP (m=20)  &1.02&	1.02&	1.02&	1.03&1.00&	\textbf{0.98}&	1.02&	1.02&\textbf{0.89}&	\textbf{0.81}&	\textbf{0.87}&	\textbf{0.94}  \\
			AVP (m=40)  &1.01&	1.02&	1.02&	1.02&1.00&	\textbf{0.98}&	1.02&	1.02&\textbf{0.89}&	\textbf{0.80}&	\textbf{0.87}&	\textbf{0.93}  \\
			AVP (m=60)  &1.02&	1.02&	1.02&	1.02&1.00&	\textbf{0.98}&	1.02&	1.02&\textbf{0.88}&	\textbf{0.80}&	\textbf{0.86}&\textbf{	0.92}  \\		
			\textbf{Targeted Drivers}& &&&&&&&&&&& \\
			AVP (m=20)  &\textbf{0.90}&	\textbf{0.94}&	\textbf{0.87}&	\textbf{0.90}&\textbf{0.92}&	\textbf{0.89}&	\textbf{0.87}&	\textbf{0.95}&\textbf{0.71}&	\textbf{0.71}&	\textbf{0.82}&	\textbf{0.72}  \\
			AVP (m=40)  &$\textbf{0.83}^\dagger$ &	\textbf{0.89}&	\textbf{0.83}&	\textbf{0.88}& \textbf{0.83}&\textbf{0.85}&\textbf{0.85}&\textbf{0.86}&\textbf{0.72}&\textbf{0.66}&\textbf{0.69}&\textbf{0.74}  \\
			AVP (m=60)  &\textbf{0.86}  &$\textbf{0.85}^\dagger$  &$\textbf{0.81}^\dagger$  &$\textbf{0.82}^\dagger$  &$\textbf{0.82}^\dagger$  &$\textbf{0.77}^\dagger$  &$\textbf{0.83}^\dagger$  &$\textbf{0.84}^\dagger$  &$\textbf{0.67}^\dagger$  &$\textbf{0.62}^\dagger$  &$\textbf{0.66}^\dagger$  &$\textbf{0.69}^\dagger$  \\
			\midrule
			& \multicolumn{12}{c}{\textbf{T = 200}} \\
			\cmidrule(lr){1-13}
			& \multicolumn{4}{c}{\textbf{Panel G: $\rho = 0$}} & \multicolumn{4}{c}{\textbf{Panel H: $\rho = 0.50$}} & \multicolumn{4}{c}{\textbf{Panel I: $\rho = 0.95$}} \\
			\cmidrule(lr){2-5} \cmidrule(lr){6-9} \cmidrule(lr){10-13}
			\textbf{Model} &$\beta_{1,t}$&$\beta_{2,t}$&$\beta_{3,t}$&$\beta_{4,t}$& $\beta_{1,t}$&$\beta_{2,t}$&$\beta_{3,t}$&$\beta_{4,t}$ & $\beta_{1,t}$&$\beta_{2,t}$&$\beta_{3,t}$&$\beta_{4,t}$ \\
			\midrule
			TVP              &1.00  &1.00 &1.00  &1.00  &1.00  &1.00  &1.00 &1.00  &1.00  &1.00  &1.00  &1.00  \\			
			\textbf{Agnostic Drivers}& &&&&&&&&&&& \\
			AVP (m=20)  &1.01  &1.02  &1.03  &1.03  &1.00  &\textbf{0.98}   &\textbf{0.97}   &1.01  &\textbf{0.88}  &\textbf{0.85}  &\textbf{0.81}  &\textbf{0.93}  \\
			AVP (m=40)  &1.00  &1.02  &1.03  &1.03  &1.00  &\textbf{0.98}   &\textbf{0.96}  &1.01  &\textbf{0.87}  &\textbf{0.82}  &\textbf{0.80}  &\textbf{0.92}  \\
			AVP (m=60)  &1.00  &1.02  &1.02  &1.03  &\textbf{0.99}  &\textbf{0.97}   &\textbf{0.96}  &1.00  &\textbf{0.87}  &\textbf{0.82}  &\textbf{0.79}  &\textbf{0.91}  \\				
			\textbf{Targeted Drivers}& &&&&&&&&&&& \\
			AVP (m=20)  &1.00  &\textbf{0.94}  &\textbf{0.96}  &\textbf{0.99}  &\textbf{0.95}  &1.01  &1.04  &\textbf{0.99}  &\textbf{0.85} &\textbf{0.91}  &\textbf{0.83}  &\textbf{0.95} \\
			AVP (m=40)  &\textbf{0.89}  &\textbf{0.95}  &\textbf{0.91}  &\textbf{0.96}  &\textbf{0.92}  &\textbf{0.96}  &\textbf{0.91}  &\textbf{0.90}  &\textbf{0.82} &\textbf{0.80}  &\textbf{0.81}  &\textbf{0.87}  \\
			AVP (m=60)  &$\textbf{0.86}^\dagger$  &$\textbf{0.88}^\dagger$  &$\textbf{0.90}^\dagger$  &$\textbf{0.89}^\dagger$  &$\textbf{0.91}^\dagger$  &$\textbf{0.93}^\dagger$  &$\textbf{0.90}^\dagger$  &$\textbf{0.86}^\dagger$  &$\textbf{0.79}^\dagger$  &$\textbf{0.78}^\dagger$ &$\textbf{0.73}^\dagger$  &$\textbf{0.83}^\dagger$  \\				
			\bottomrule
		\end{tabular}%
	}
	\vspace{0.5em}
	\par
	\begin{tabularx}{\textwidth}{X}
		{\footnotesize \textit{Notes:} Each entry reports the MSPE ratios relative to TVP benchmark across the 1,000 Monte Carlo simulations. Bold entries indicates MSPE ratios lower than 1.00 indicates gains relative to the benchmark. $\dagger$ highlights the best performing model for each coefficient (lower MSPE ratio).}
	\end{tabularx}
\end{table}

\begin{figure}[htbp!] 
	\centering
	\includegraphics[width=0.95\textwidth]{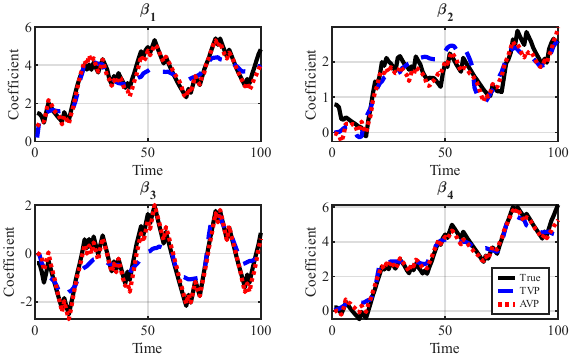}
	\caption{Illustration of a Monte Carlo iteration for DGP2 with $T=100$ and $\rho=0.95$. The red line AVP considers $m=60$ targeted drivers.}
	\label{fig:plots_Montecarlo_regimes}
\end{figure}

\paragraph{DGP2: Regime Switches and Threshold Effects.}	
Table~\ref{tab:MonteCarloDGP2} reports results for DGP2, which features multiple sources of instability including regime switches and threshold effects. Three things are worth mentioning. First, in stark contrast to DGP1, agnostic AVP consistently underperforms the TVP benchmark across all configurations. Not only does it fail to deliver any forecasting improvements, but in several cases it generates substantially higher MSPE. Second, AVP with targeted drivers again emerges as the best specification, though the gains are even more pronounced than in DGP1. In specific, using $m=40$ or $m=60$ drivers yields uniformly better performance than TVP across all 72 entries, with improvements reaching 79\% in some cases. With $m=20$, results are more nuanced: AVP consistently outperforms TVP for $\beta_3$ and $\beta_4$ by a substantial margin (around $15-20\%$), but underperforms for $\beta_1$ and $\beta_2$ by a relatively small margin ($1-3\%$). Third, the number of drivers $m$ plays a more critical role under DGP2 than DGP1, affecting both targeted and agnostic specifications. While agnostic AVP generally performs poorly, its relative performance improves markedly with larger $m$. For targeted AVP, the gains can be dramatic: in some cases, moving from $m=20$ to $m=60$ transforms a 16\% underperformance into a 41\% improvement—a swing of 57 percentage points. This result requires careful interpretation. Under DGP2, increasing $m$ simultaneously expands both the number of drivers and the richness of information about structural instabilities. Specifically, $m=60$ entries include more information about regime behavior and threshold mechanisms than $m=20$. The observed performance improvements therefore reflect both dimensions: the breadth of the information set and its alignment with the underlying sources of parameter variation. Figure~\ref{fig:plots_Montecarlo_regimes} illustrates parameter estimates from a single Monte Carlo draw under DGP2, comparing TVP and AVP with $m=60$ targeted drivers. The figure reveals that AVP tracks the true parameters considerably more accurately than TVP for most coefficients.

\begin{table}[htbp!]
	\centering
	\caption{Monte Carlo results for DGP2 (Non-linear TVP and breaks)}
	\label{tab:MonteCarloDGP2}
	\small
	\resizebox{0.95\textwidth}{!}{%
		\begin{tabular}{@{}l*{12}{S[table-format=1.2]}@{}}
			\toprule
			& \multicolumn{12}{c}{\textbf{T = 50}} \\
			\cmidrule(lr){1-13}
			& \multicolumn{4}{c}{\textbf{Panel A: $\rho = 0$}} & \multicolumn{4}{c}{\textbf{Panel B: $\rho = 0.50$}} & \multicolumn{4}{c}{\textbf{Panel C: $\rho = 0.95$}} \\
			\cmidrule(lr){2-5} \cmidrule(lr){6-9} \cmidrule(lr){10-13}
			\textbf{Model} &$\beta_{1,t}$&$\beta_{2,t}$&$\beta_{3,t}$&$\beta_{4,t}$& $\beta_{1,t}$&$\beta_{2,t}$&$\beta_{3,t}$&$\beta_{4,t}$ & $\beta_{1,t}$&$\beta_{2,t}$&$\beta_{3,t}$&$\beta_{4,t}$ \\
			\midrule
			TVP              &1.00  &1.00 &1.00  &1.00  &1.00  &1.00  &1.00 &1.00  &1.00  &1.00  &1.00  &1.00  \\			
			\textbf{Agnostic Drivers}& &&&&&&&&&&& \\
			AVP (m=20)  &1.38 &1.31  &1.45  &1.35  &1.35  &1.33  &1.45  &1.32  &1.34  &1.32  &1.48 &1.41  \\
			AVP (m=40)  &1.22  &1.16  &1.23  &1.18  &1.18  &1.14  &1.19  &1.17  &1.18  &1.16  &1.22  &1.19  \\
			AVP (m=60)  &1.13  &1.07  &1.12  &1.11  &1.15  &1.13  &1.12  &1.10  &1.10  &1.10  &1.13  &1.13 \\			
			\textbf{Targeted Drivers}& &&&&&&&&&&& \\
			AVP (m=20)  &1.03  &1.01  &\textbf{0.77}  &\textbf{0.85}  &1.01 &1.02 &\textbf{0.79}  &\textbf{0.81}  &1.01  &1.03  &\textbf{0.80}  &\textbf{0.82} \\
			AVP (m=40)  &$\textbf{0.69}^\dagger$  &$\textbf{0.73}^\dagger$  &$\textbf{0.38}^\dagger$  &$\textbf{0.53}^\dagger$  &$\textbf{0.68}^\dagger$ &$\textbf{0.75}^\dagger$  &$\textbf{0.37}^\dagger$  &$\textbf{0.51}^\dagger$ &$\textbf{0.67}^\dagger$ &$\textbf{0.74}^\dagger$  &$\textbf{0.37}^\dagger$  &$\textbf{0.54}^\dagger$  \\
			AVP (m=60)  &\textbf{0.75}  &\textbf{0.77}  &\textbf{0.42}  &\textbf{0.59}  &\textbf{0.75}  &\textbf{0.79}  &\textbf{0.42}  &\textbf{0.59} &\textbf{0.73} &\textbf{0.77}  &\textbf{0.41}  &\textbf{0.59}  \\			
			\midrule
			& \multicolumn{12}{c}{\textbf{T = 100}} \\
			\cmidrule(lr){1-13}
			& \multicolumn{4}{c}{\textbf{Panel D: $\rho = 0$}} & \multicolumn{4}{c}{\textbf{Panel E: $\rho = 0.50$}} & \multicolumn{4}{c}{\textbf{Panel F: $\rho = 0.95$}} \\
			\cmidrule(lr){2-5} \cmidrule(lr){6-9} \cmidrule(lr){10-13}
			\textbf{Model} &$\beta_{1,t}$&$\beta_{2,t}$&$\beta_{3,t}$&$\beta_{4,t}$& $\beta_{1,t}$&$\beta_{2,t}$&$\beta_{3,t}$&$\beta_{4,t}$ & $\beta_{1,t}$&$\beta_{2,t}$&$\beta_{3,t}$&$\beta_{4,t}$ \\
			\midrule
			TVP              &1.00  &1.00 &1.00  &1.00  &1.00  &1.00  &1.00 &1.00  &1.00  &1.00  &1.00  &1.00  \\			
			\textbf{Agnostic Drivers}& &&&&&&&&&&& \\
			AVP (m=20)  &1.79&	1.67&	2.03&	1.76&1.80&	1.62&	2.00&	1.75&1.79&	1.66&	2.03&	1.82  \\
			AVP (m=40)  &1.40&	1.37&	1.50&	1.36&1.40&	1.32&	1.48&	1.38&1.42&	1.34&	1.50&	1.40  \\
			AVP (m=60)  &1.27&	1.25&	1.32&	1.22&1.27&	1.23&	1.31&	1.27&1.27&	1.22&	1.30&	1.26  \\			
			\textbf{Targeted Drivers}& &&&&&&&&&&& \\
			AVP (m=20)  &\textbf{0.74}&	\textbf{0.98}&	\textbf{0.50}&	\textbf{0.60}&\textbf{0.77}&	\textbf{0.94}&	\textbf{0.48}&	\textbf{0.60}&\textbf{0.75}&	\textbf{0.94}&	\textbf{0.49}&	\textbf{0.62}  \\
			AVP (m=40)  &$\textbf{0.43}^\dagger$ &	$\textbf{0.63}^\dagger$&$\textbf{0.22}^\dagger$&	$\textbf{0.28}^\dagger$& $\textbf{0.42}^\dagger$&$\textbf{0.61}^\dagger$&$\textbf{0.23}^\dagger$&$\textbf{0.28}^\dagger$&$\textbf{0.43}^\dagger$&$\textbf{0.63}^\dagger$&$\textbf{0.23}^\dagger$&$\textbf{0.29}^\dagger$  \\
			AVP (m=60)  &\textbf{0.44}  &$\textbf{0.63}^\dagger$  &$\textbf{0.22}^\dagger$  &\textbf{0.30}  &\textbf{0.43}  &$\textbf{0.61}^\dagger$  &$\textbf{0.23}^\dagger$  &\textbf{0.29}  &\textbf{0.44}  &$\textbf{0.63}^\dagger$  &$\textbf{0.23}^\dagger$  &\textbf{0.30}  \\
			\midrule
			& \multicolumn{12}{c}{\textbf{T = 200}} \\
			\cmidrule(lr){1-13}
			& \multicolumn{4}{c}{\textbf{Panel G: $\rho = 0$}} & \multicolumn{4}{c}{\textbf{Panel H: $\rho = 0.50$}} & \multicolumn{4}{c}{\textbf{Panel I: $\rho = 0.95$}} \\
			\cmidrule(lr){2-5} \cmidrule(lr){6-9} \cmidrule(lr){10-13}
			\textbf{Model} &$\beta_{1,t}$&$\beta_{2,t}$&$\beta_{3,t}$&$\beta_{4,t}$& $\beta_{1,t}$&$\beta_{2,t}$&$\beta_{3,t}$&$\beta_{4,t}$ & $\beta_{1,t}$&$\beta_{2,t}$&$\beta_{3,t}$&$\beta_{4,t}$ \\
			\midrule
			TVP              &1.00  &1.00 &1.00  &1.00  &1.00  &1.00  &1.00 &1.00  &1.00  &1.00  &1.00  &1.00  \\			
			\textbf{Agnostic Drivers}& &&&&&&&&&&& \\
			AVP (m=20)  &2.54  &2.27  &3.21  &2.77  &2.51  &2.29   &3.27   &2.73  &2.47  &2.25  &3.19  &2.64  \\
			AVP (m=40)  &1.83  &1.69  &2.10  &1.90  &1.83  &1.69   &2.11  &1.84  &1.79  &1.69  &2.09  &1.85  \\
			AVP (m=60)  &1.56  &1.49  &1.72  &1.58  &1.56  &1.48   &1.74  &1.57  &1.53  &1.47  &1.69  &1.53  \\		
			\textbf{Targeted Drivers}& &&&&&&&&&&& \\
			AVP (m=20)  &\textbf{0.90}  &1.14  &\textbf{0.68}  &\textbf{0.67}  &\textbf{0.91}  &1.16  &\textbf{0.70}  &\textbf{0.67}  &\textbf{0.94} &1.16  &\textbf{0.69}  &\textbf{0.67} \\
			AVP (m=40)  &\textbf{0.42}  &\textbf{0.65}  &\textbf{0.25}  &\textbf{0.28}  &\textbf{0.42}  &\textbf{0.65}  &\textbf{0.25}  &\textbf{0.28}  &\textbf{0.42} &\textbf{0.65}  &\textbf{0.24}  &\textbf{0.27}  \\
			AVP (m=60)  &$\textbf{0.37}^\dagger$  &$\textbf{0.58}^\dagger$  &$\textbf{0.21}^\dagger$  &$\textbf{0.25}^\dagger$  &$\textbf{0.38}^\dagger$  &$\textbf{0.59}^\dagger$  &$\textbf{0.22}^\dagger$  &$\textbf{0.25}^\dagger$  &$\textbf{0.37}^\dagger$  &$\textbf{0.59}^\dagger$ &$\textbf{0.21}^\dagger$  &$\textbf{0.24}^\dagger$  \\			
			\bottomrule
		\end{tabular}%
	}
	\vspace{0.5em}
	\par
	\begin{tabularx}{\textwidth}{X}
		{\footnotesize \textit{Notes:} Each entry reports the MSPE ratios relative to TVP benchmark across the 1,000 Monte Carlo simulations. Bold entries indicates MSPE ratios lower than 1.00 indicates gains relative to the benchmark. $\dagger$ highlights the best performing model for each coefficient (lower MSPE ratio).}
	\end{tabularx}
\end{table}

\normalsize

\clearpage
\newpage


\section{Additional empirical results}	

\subsection{In-sample analysis and time-varying parameter estimates}

\begin{figure}[H]  
	\centering		
	\includegraphics[width=.95\textwidth,trim={0cm, 0cm, 0cm, 0cm}]{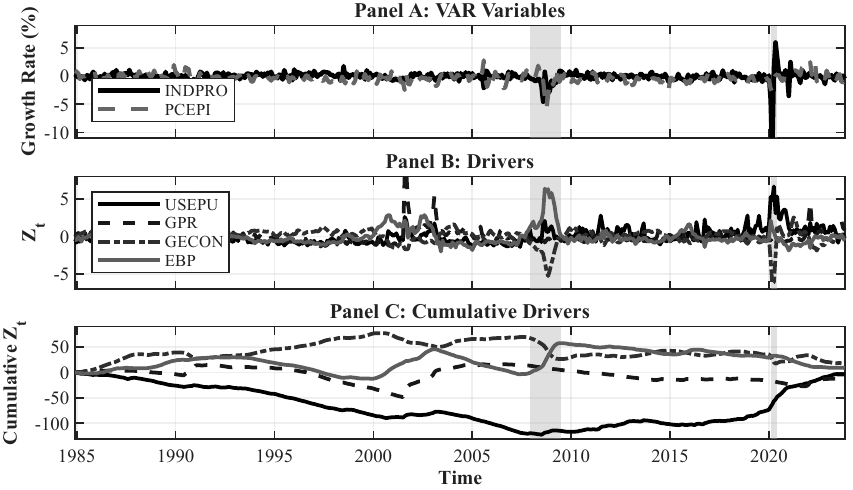}  
	\caption{VAR Variables and Drivers for the U.S. monthly data. Panel A displays the two target variables (INDPRO and PCEPI) in log-differences, from 1985 to 2024. Panel B shows four drivers $Z_t$: economic policy uncertainty (USEPU), geopolitical risk (GPR), global economic activity (GECON), and excess bond premium (EBP), all in levels.  Panel C presents the cumulative sum of these four drivers. Shaded regions denote the Global Financial Crisis (2008-2009) and the COVID--19 pandemic (2020). Variable definitions and data sources are provided in Table~1 of the manuscript.}
	\label{fig:plots_US}
\end{figure}

\begin{figure}[H]  
	\centering		
	\includegraphics[width=.95\textwidth,trim={0cm, 0cm, 0cm, 0cm}]{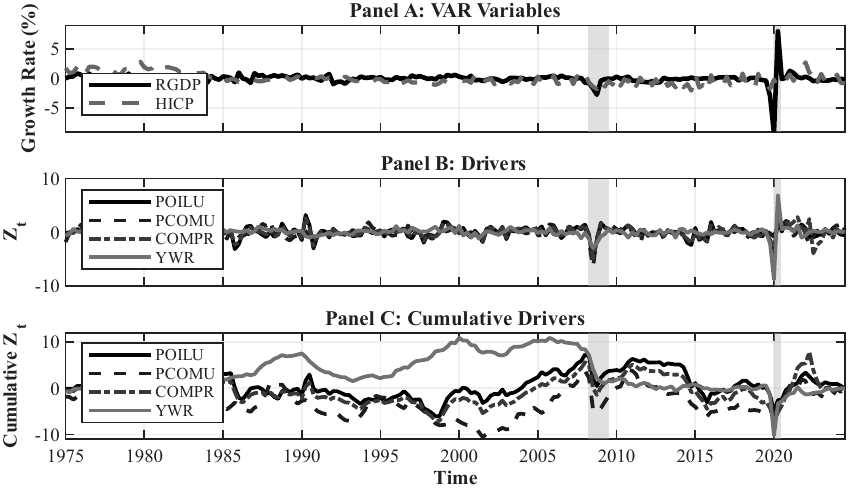}  
	\caption{VAR Variables and Drivers for the Euro Area quarterly data. Panel A displays the two target variables (RGDP and HICP) in log-differences, from 1975 to 2024. Panel B shows four drivers $Z_t$: oil prices (POILU), non-oil commodity prices (PCOMU), commodity price index (COMPR), and world GDP (YWR), all in log-differences.  Panel C presents the cumulative sum of these four drivers. Shaded regions denote the Global Financial Crisis (2008-2009) and the COVID--19 pandemic (2020). Variable definitions and data sources are provided in Table~3 of the manuscript.}
	\label{fig:plots_EURO}
\end{figure}

\paragraph{The relationship between VAR variables and drivers $Z_t$.} Figures~\ref{fig:plots_US}--\ref{fig:plots_EURO} illustrate the empirical motivation for our approach using U.S. and Euro Area data, respectively. Panel A displays the evolution of GDP growth and inflation, which exhibit pronounced volatility during major economic disruptions—most notably the 2008-2009 financial crisis and the 2020 COVID--19 pandemic. Panel B reveals the expected pattern: our drivers experience sharp increases precisely when macroeconomic stress intensifies, with the most dramatic spikes coinciding closely with target variable disruptions.\footnote{For visual clarity, we display only four drivers; the full set exhibits similar patterns.} Panel C reinforces this co-movement by illustrating how cumulative $Z$s capture the persistent nature of these shocks, displaying clear step-changes during crisis episodes. This correlation between target variables and $Z$ swings suggest that our AVP-VAR might successfully identify periods when VAR parameters require quick adjustments. Crucially, driver responses either precede or coincide with macroeconomic disruptions, indicating that the model can facilitate timely parameter adaptation rather than merely reacting to observed instability. This temporal alignment is central to AVP-VAR's forecasting improvements, as it enables the model to adjust its dynamics in anticipation of or contemporaneously with structural changes.

\paragraph{In-Sample Parameter Dynamics}
To connect the stylized facts from our simulations to the forecasting gains of AVP--VAR, we examine the evolution of intercepts in a parsimonious VAR(1) estimated on U.S. and Euro Area data. We focus on intercepts because they capture time-varying trends and typically exhibit the most pronounced variation in macroeconomic VARs. As a flexible benchmark known for strong forecasting performance \citep{StockWatson2007}, we consider the univariate UC--SV model, which represents a local-level process with stochastic volatility, formally given by:
\begin{align}
	y_{t+h} &= \tau_t + \sigma_t^{\varepsilon}\varepsilon_{t+h}, \label{eq:ucsv1}\\
	\tau_t &= \tau_{t-1} + \sigma_t^{\eta}\eta_t, \label{eq:ucsv2}\\
	\log \sigma_t^{\varepsilon} &= \log \sigma_{t-1}^{\varepsilon} + \zeta_t, \label{eq:ucsv3}\\
	\log \sigma_t^{\eta} &= \log \sigma_{t-1}^{\eta} + \xi_t, \label{eq:ucsv4}
\end{align}
where $\varepsilon_t$, $\eta_t$, $\zeta_t$, and $\xi_t$ are independent standard normal innovations. In this setup, $y_t$ evolves around a latent level $\tau_t$ that itself follows a random walk, while both the measurement error $\varepsilon_t$ and the state innovation $\eta_t$ exhibit stochastic volatility. The following subsections compare the intercepts estimated with AVP--VAR, TVP-VAR, TVP-VAR-EB, and UC--SV.

\subsubsection{In-Sample Parameter Dynamics: U.S. and Euro Area data}

\begin{figure}[H]  
	\centering		 
	\includegraphics[width=0.95\textwidth]{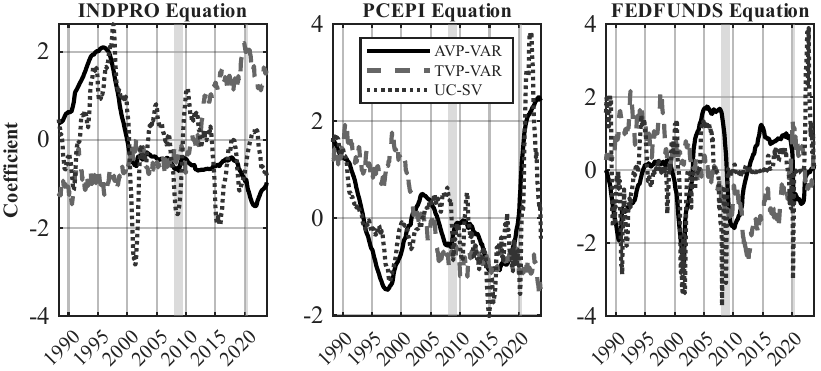}
	\caption{Comparison of Time-Varying Intercepts for U.S. monthly data. Standardized in-sample intercepts from three competing models: TVP-VAR (dashed line), AVP-VAR (solid line), and univariate unobserved components stochastic volatility model UC-SV (dotted line). The left panel displays coefficients from the Industrial Production equation, the center panel from the PCEPI inflation equation, and the right panel from the short-term interest rate equation. All models are estimated as VAR(1) systems. Shaded regions denote the Global Financial Crisis (2008–2009) and the COVID--19 pandemic (2020).} 
	\label{fig:intercepts_US_FB}
\end{figure}

\begin{figure}[H]  
	\centering
	\includegraphics[width=0.95\textwidth]{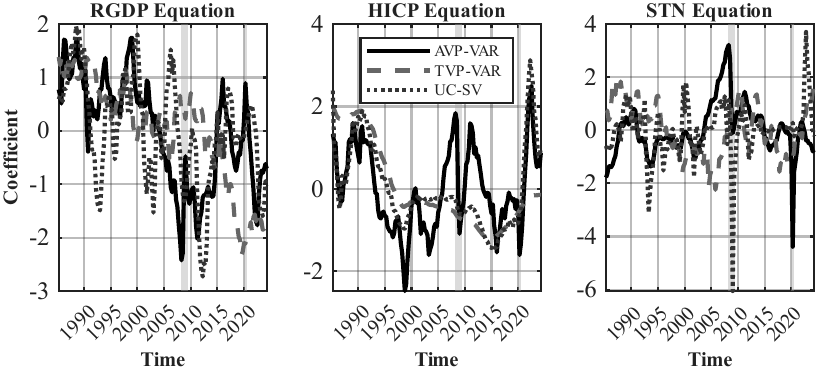}  
	\caption{Comparison of Time-Varying Intercepts for Euro Area quarterly data. Standardized in-sample intercepts from three competing models: TVP-VAR (dashed line), AVP-VAR (solid line), and univariate unobserved components stochastic volatility model UC-SV (dotted line). The left panel displays coefficients from the real GDP growth equation, the center panel from the HICP inflation equation, and the right panel from the short-term interest rate equation. All models are estimated as VAR(1) systems. Shaded regions denote the Global Financial Crisis (2008–2009) and the COVID--19 pandemic (2020).}	\label{fig:intercepts_EURO_FB}
\end{figure}

\begin{figure}[H]  
	\centering		 
	\includegraphics[width=0.95\textwidth]{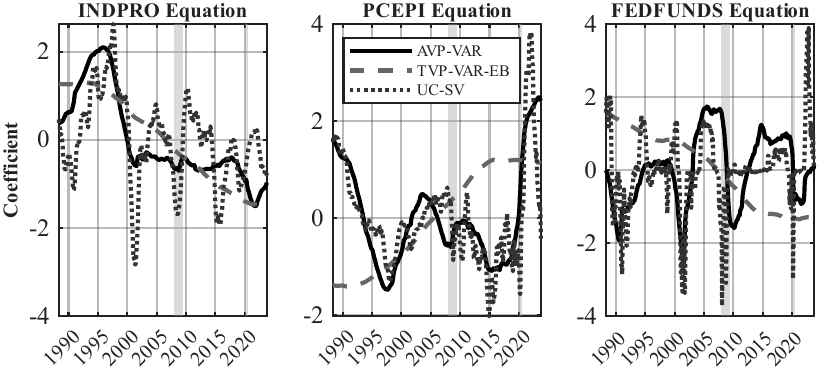}
	\caption{Comparison of Time-Varying Intercepts for U.S. monthly data. Standardized in-sample intercepts from three competing models: TVP-VAR-EB (dashed line), AVP-VAR (solid line), and univariate unobserved components stochastic volatility model UC-SV (dotted line). The left panel displays coefficients from the Industrial Production equation, the center panel from the PCEPI inflation equation, and the right panel from the short-term interest rate equation. All models are estimated as VAR(1) systems. Shaded regions denote the Global Financial Crisis (2008–2009) and the COVID--19 pandemic (2020).} \label{fig:intercepts_US_EB}
\end{figure}

\begin{figure}[H]  
	\centering
	\includegraphics[width=0.95\textwidth]{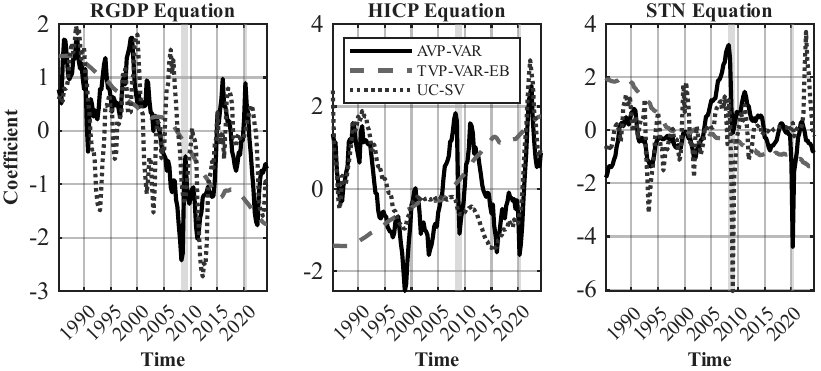}  
	\caption{Comparison of Time-Varying Intercepts for Euro Area quarterly data. Standardized in-sample intercepts from three competing models: TVP-VAR-EB (dashed line), AVP-VAR (solid line), and univariate unobserved components stochastic volatility model UC-SV (dotted line). The left panel displays coefficients from the real GDP growth equation, the center panel from the HICP inflation equation, and the right panel from the short-term interest rate equation. All models are estimated as VAR(1) systems. Shaded regions denote the Global Financial Crisis (2008–2009) and the COVID--19 pandemic (2020).}	
	\label{fig:intercepts_EURO_EB}
\end{figure}

\subsubsection{Unstandardised In-Sample Parameter Dynamics: U.S. and Euro Area data}

\begin{figure}[H]  
	\centering
	\includegraphics[width=0.95\textwidth]{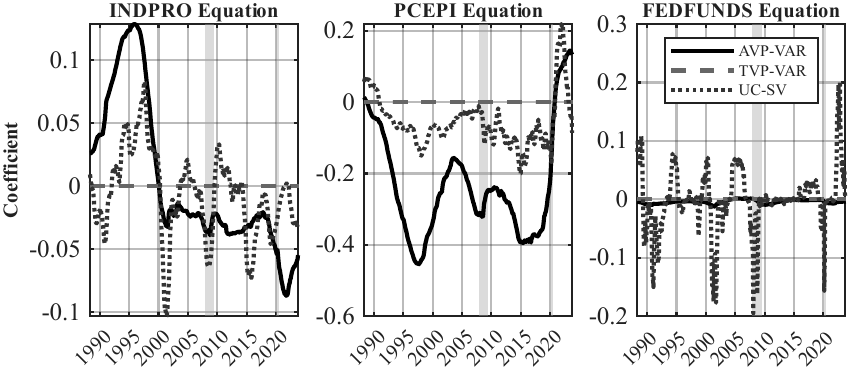}  
	\caption{Comparison of Time-Varying Intercepts for U.S. monthly data.  Unstandardised in-sample intercepts from three competing models: TVP-VAR (dashed line), AVP-VAR (solid line), and univariate unobserved components stochastic volatility model UC-SV (dotted line). The left panel displays coefficients from the Industrial Production equation, the center panel from the PCEPI inflation equation, and the right panel from the short-term interest rate equation. All models are estimated as VAR(1) systems. Shaded regions denote the Global Financial Crisis (2008–2009) and the COVID--19 pandemic (2020).}	\label{fig:intercepts_US_FB_nonstand}
\end{figure}

\begin{figure}[H]  
	\centering
	\includegraphics[width=0.95\textwidth]{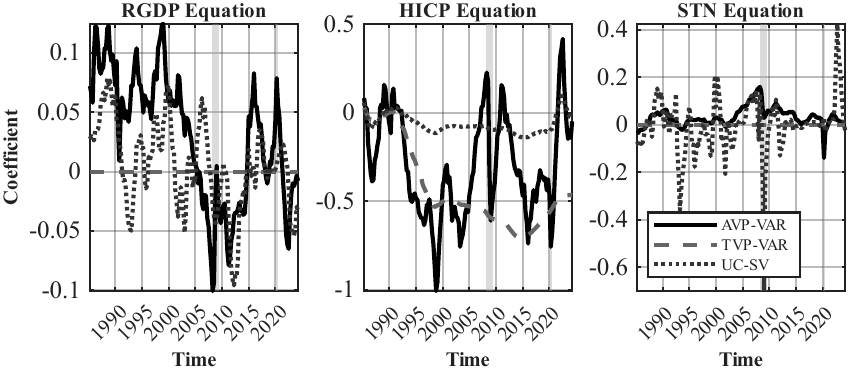}  
	\caption{Comparison of Time-Varying Intercepts for Euro Area quarterly data.  Unstandardised in-sample intercepts from three competing models: TVP-VAR (dashed line), AVP-VAR (solid line), and univariate unobserved components stochastic volatility model UC-SV (dotted line). The left panel displays coefficients from the real GDP growth equation, the center panel from the HICP inflation equation, and the right panel from the short-term interest rate equation. All models are estimated as VAR(1) systems. Shaded regions denote the Global Financial Crisis (2008–2009) and the COVID--19 pandemic (2020).}	\label{fig:intercepts_EURO_FB_nonstand}
\end{figure}

\begin{figure}[H]  
	\centering
	\includegraphics[width=0.95\textwidth]{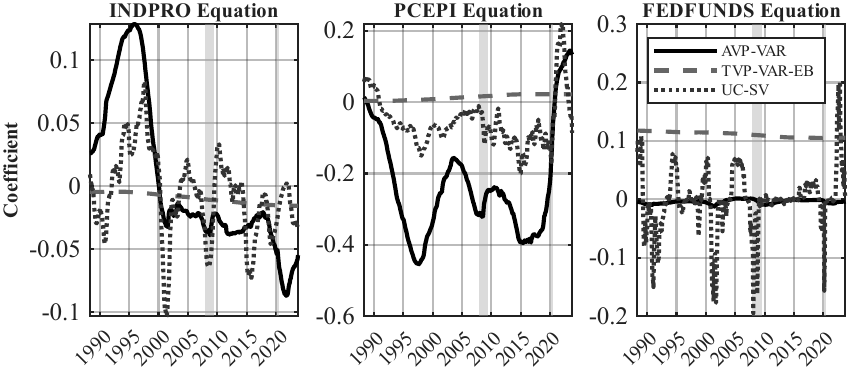}  
	\caption{Comparison of Time-Varying Intercepts for U.S. monthly data. Unstandardised in-sample intercepts from three competing models: TVP-VAR-EB (dashed line), AVP-VAR (solid line), and univariate unobserved components stochastic volatility model UC-SV (dotted line). The left panel displays coefficients from the Industrial Production equation, the center panel from the PCEPI inflation equation, and the right panel from the short-term interest rate equation. All models are estimated as VAR(1) systems. Shaded regions denote the Global Financial Crisis (2008–2009) and the COVID--19 pandemic (2020).}	\label{fig:intercepts_US_EB_nonstand}
\end{figure}

\begin{figure}[H]  
	\centering
	\includegraphics[width=0.95\textwidth]{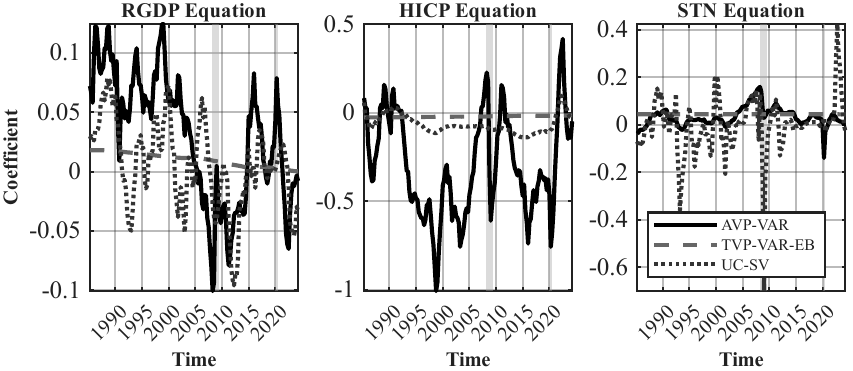}  
	\caption{Comparison of Time-Varying Intercepts for Euro Area quarterly data. Unstandardised in-sample intercepts from three competing models: TVP-VAR-EB (dashed line), AVP-VAR (solid line), and univariate unobserved components stochastic volatility model UC-SV (dotted line). The left panel displays coefficients from the real GDP growth equation, the center panel from the HICP inflation equation, and the right panel from the short-term interest rate equation. All models are estimated as VAR(1) systems. Shaded regions denote the Global Financial Crisis (2008–2009) and the COVID--19 pandemic (2020).}	\label{fig:intercepts_EURO_EB_nonstand}
\end{figure}

\subsubsection{In-Sample Fit: Cumulative residuals for U.S. and Euro Area}
\begin{figure}[H]  
	\centering
	\includegraphics[width=0.95\textwidth]{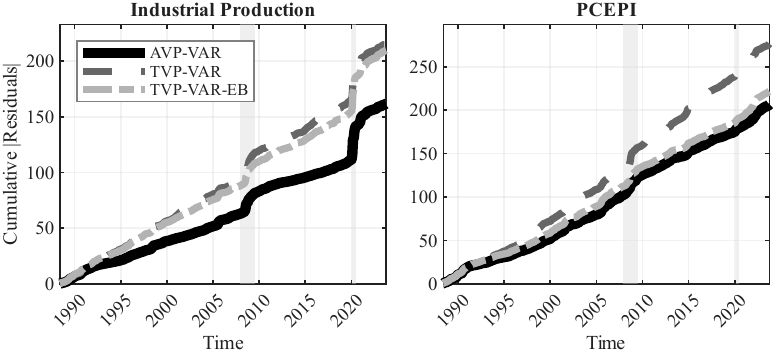}
	\caption{Absolute cumulative residuals for U.S. monthly Industrial Production (left) and PCEPI (right) for three models: the AVP–VAR, the TVP–VAR, and the TVP–VAR–EB. The figure summarizes in-sample fit from a VAR(2) estimated with 100,000 MCMC draws, discarding the first 5,000 as burn-in and retaining every 5th draw. Lower cumulative residuals indicate better in-sample tracking performance. Shaded regions denote the Global Financial Crisis (2008-2009) and the COVID--19 pandemic(2020). } \label{fig:residuals_US}
\end{figure}

\begin{figure}[H]  
	\centering
	\includegraphics[width=0.95\textwidth]{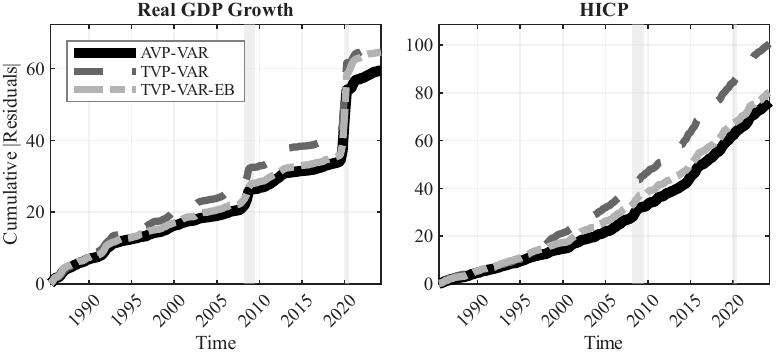}
	\caption{Absolute cumulative residuals for Euro Area quarterly GDP growth (left) and HICP (right) for three models: the AVP–VAR, the TVP–VAR, and the TVP–VAR–EB. The figure summarizes in-sample fit from a VAR(2) estimated with 100,000 MCMC draws, discarding the first 5,000 as burn-in and retaining every 5th draw. Lower cumulative residuals indicate better in-sample tracking performance. Shaded regions denote the Global Financial Crisis (2008-2009) and the COVID--19 pandemic(2020). } \label{fig:residuals_EURO}
\end{figure}

Having established that the AVP--VAR successfully tracks the UC--SV benchmark, a natural question arises: does this enhanced flexibility translate into a superior in-sample fit? Figures~\ref{fig:residuals_US}--\ref{fig:residuals_EURO} address this question by comparing absolute in-sample residuals across competing specifications for a VAR($2$) model. The results favour the AVP--VAR in both equations (output growth and inflation) and across sample periods.

For GDP growth, the AVP--VAR consistently outperforms both TVP--VAR--EB and TVP--VAR. In the U.S. case (Figure~\ref{fig:residuals_US}), the AVP--VAR exhibits smaller residuals throughout the entire sample, not only during crisis episodes. In contrast, for the Euro Area (Figure~\ref{fig:residuals_EURO}), the AVP--VAR and TVP--VAR--EB display similar performance for most of the sample, with the AVP--VAR gaining a clear advantage during the COVID--19 period. Overall, the TVP--VAR--EB and TVP--VAR produce broadly comparable results. Regarding inflation, the TVP--VAR tends to perform the worst across all periods, while the AVP--VAR and TVP--VAR--EB yield similar accuracy.

Taken together, these findings indicate that the benefits of the AVP--VAR extend beyond periods of turbulence. By allowing parameters to evolve in response to observable economic conditions that capture structural change, the model achieves a more consistent and superior fit overall.

\subsection{Robustness Exercise: Different Lag Lengths in the VAR($p$)}
This section extends all results in the paper to alternative lag lengths in the VAR, considering $p \in \{1,2,3,4\}$. For each specification, we present figures comparing MAE, QScore90, and QScore10, as well as tables reporting MSPE, QScore90, and QScore10. The analysis is conducted for both target variables (output growth and inflation) and for both datasets (U.S.\ monthly data and euro-area quarterly data). Results for the FRED--QD dataset are presented separately in Section~4.4.

\subsubsection{U.S. monthly data: VAR(1)}

\begin{figure}[htbp!]  
	\centering
	\includegraphics[width=\textwidth]{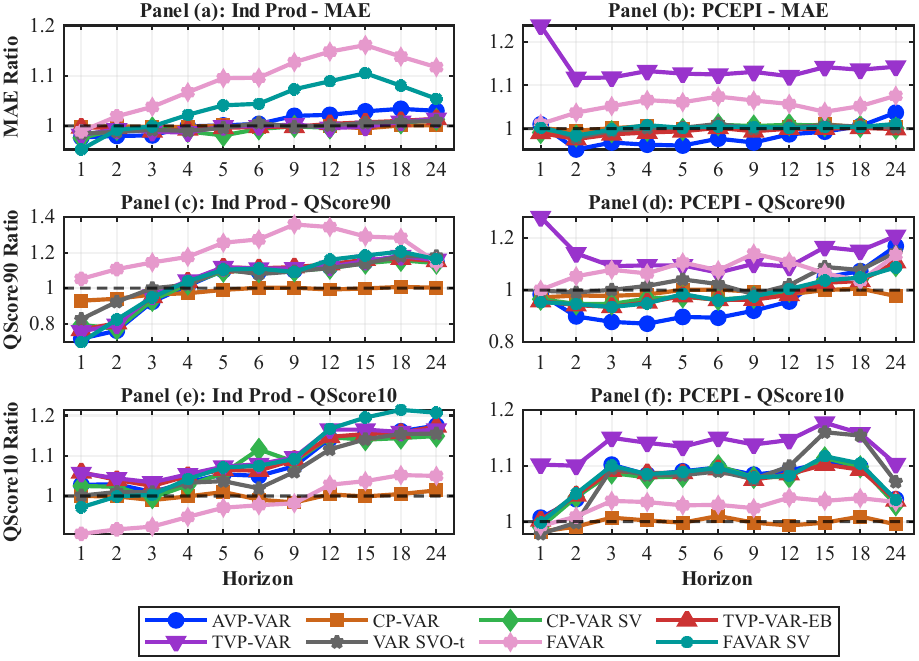}  
	\caption{Forecasting results using U.S. monthly data. The top row displays mean absolute forecast errors (MAE), while the middle and bottom rows show quantile scores for the upper (90th percentile) and lower (10th percentile) tails of each variable. All metrics are reported as ratios relative to a benchmark constant-parameter VAR($1$) estimated via least squares. A model is considered superior when its relative MAE, or relative quantile scores, are below one.} \label{fig:OOS_US_P1}
\end{figure}
\begin{table}[htbp!]
	\centering
	\caption{Forecasting Performance VAR(1): Mean Squared Prediction Error Relative to Benchmark} \label{tab:forecasting_mspe_US_p1}
	\scriptsize
	\begin{tabular}{@{}cS[table-format=1.3]S[table-format=1.3]S[table-format=1.3]S[table-format=1.3]S[table-format=1.3]S[table-format=1.3]S[table-format=1.3]S[table-format=1.3]@{}}
		\toprule
		{\textbf{h}} & {\textbf{AVP-VAR}} & {\textbf{CP-VAR}} & {\makecell{ \textbf{CP-VAR} \\ \textbf{SV}}} & {\makecell{ \textbf{{TVP-VAR}} \\ \textbf{EB}}} &{\textbf{TVP-VAR}} & {\makecell{ \textbf{VAR} \\ \textbf{SVOt}}} & {\textbf{FAVAR}} & {\makecell{ \textbf{FAVAR} \\ \textbf{SV}}}  \\
		\midrule
		\multicolumn{9}{c}{\textbf{Panel A: Industrial Production}} \\
		\midrule
		1 & \textbf{0.902} & 1.009 & \textbf{0.888} & \textbf{0.941} & \textbf{0.952} & \textbf{0.903} & \textbf{0.947} & \textbf{0.823}$^{\dag}$ \\
		2 & \textbf{0.851} & \textbf{0.987} & \textbf{0.859} & \textbf{0.848}$^{\dag}$ & \textbf{0.852} & \textbf{0.852} & \textbf{0.974} & \textbf{0.852} \\
		3 & \textbf{0.911}$^{\dag}$ & \textbf{0.989} & \textbf{0.934} & \textbf{0.926} & \textbf{0.923} & \textbf{0.935} & 1.049 & \textbf{0.955} \\
		4 & \textbf{0.972}$^{\dag}$ & \textbf{0.995} & \textbf{0.985} & \textbf{0.984} & \textbf{0.981} & \textbf{0.985} & 1.065 & \textbf{0.998} \\
		5 & \textbf{0.997} & \textbf{0.998} & \textbf{0.992}$^{\dag}$ & 1.000 & 1.007 & 1.004 & 1.070 & 1.016 \\
		6 & \textbf{0.991}$^{\dag}$ & \textbf{0.996} & \textbf{0.999} & 1.001 & \textbf{0.996} & 1.003 & 1.073 & 1.007 \\
		9 & 1.002 & 1.001 & 1.000 & 1.003 & 1.007 & 1.002 & 1.103 & 1.030 \\
		12 & 1.006 & \textbf{0.996}$^{\dag}$ & 1.002 & 1.006 & \textbf{0.999} & 1.009 & 1.133 & 1.054 \\
		15 & 1.010 & \textbf{0.997}$^{\dag}$ & 1.005 & 1.006 & \textbf{0.998} & 1.006 & 1.111 & 1.046 \\
		18 & 1.010 & 1.002 & 1.007 & 1.009 & 1.002 & 1.009 & 1.127 & 1.053 \\
		24 & 1.014 & \textbf{0.998}$^{\dag}$ & 1.009 & 1.011 & 1.005 & 1.011 & 1.089 & 1.037 \\
		\midrule
		\multicolumn{9}{c}{\textbf{Panel B: PCEPI}} \\
		\midrule
		1 & 1.072 & \textbf{0.993} & \textbf{0.983}$^{\dag}$ & 1.000 & 1.487 & 1.011 & 1.006 & 1.003 \\
		2 & \textbf{0.939}$^{\dag}$ & \textbf{0.994} & \textbf{0.983} & \textbf{0.981} & 1.197 & \textbf{0.984} & 1.046 & 1.000 \\
		3 & \textbf{0.925}$^{\dag}$ & \textbf{0.993} & \textbf{0.985} & \textbf{0.980} & 1.152 & \textbf{0.994} & 1.079 & 1.001 \\
		4 & \textbf{0.919}$^{\dag}$ & 1.006 & \textbf{0.988} & \textbf{0.988} & 1.149 & \textbf{0.997} & 1.088 & 1.009 \\
		5 & \textbf{0.921}$^{\dag}$ & \textbf{0.994} & \textbf{0.992} & \textbf{0.997} & 1.145 & 1.005 & 1.096 & 1.013 \\
		6 & \textbf{0.930}$^{\dag}$ & 1.008 & 1.012 & 1.007 & 1.140 & 1.015 & 1.108 & 1.012 \\
		9 & \textbf{0.934}$^{\dag}$ & 1.001 & 1.003 & \textbf{0.997} & 1.139 & \textbf{0.996} & 1.115 & 1.003 \\
		12 & \textbf{0.964}$^{\dag}$ & 1.001 & 1.004 & \textbf{0.999} & 1.132 & \textbf{0.996} & 1.112 & 1.006 \\
		15 & \textbf{0.965}$^{\dag}$ & 1.005 & 1.006 & 1.001 & 1.153 & 1.003 & 1.075 & 1.003 \\
		18 & \textbf{0.997}$^{\dag}$ & \textbf{0.998} & 1.003 & 1.007 & 1.157 & 1.003 & 1.090 & 1.009 \\
		24 & 1.046 & 1.002 & \textbf{0.996}$^{\dag}$ & 1.000 & 1.167 & 1.001 & 1.138 & 1.008 \\
		\bottomrule
	\end{tabular}	
	\vspace{0.5em}
	\par
	\begin{tabularx}{\textwidth}{X}
		{\footnotesize \textit{Notes:} The table reports mean squared prediction errors (MSPE) for each of the two forecasted variables, relative to a benchmark constant-parameter VAR($1$) estimated via least squares, for U.S. monthly data. Values below one (highlighted in bold) indicate superior forecasting performance compared to the benchmark. The forecast horizon is denoted by $h$ (in months), and the symbol $\dag$ marks the best-performing model at each horizon.}
	\end{tabularx}
\end{table}

\begin{table}[htbp!]
	\centering
	\caption{Forecasting Performance VAR(1): Quantile Scores 90\% Relative to Benchmark} \label{tab:forecasting_qs90_US_p1}
	\scriptsize
	\begin{tabular}{@{}cS[table-format=1.3]S[table-format=1.3]S[table-format=1.3]S[table-format=1.3]S[table-format=1.3]S[table-format=1.3]S[table-format=1.3]S[table-format=1.3]@{}}
		\toprule
		{\textbf{h}} & {\textbf{AVP-VAR}} & {\textbf{CP-VAR}} & {\makecell{ \textbf{CP-VAR} \\ \textbf{SV}}} & {\makecell{ \textbf{{TVP-VAR}} \\ \textbf{EB}}} &{\textbf{TVP-VAR}} & {\makecell{ \textbf{VAR} \\ \textbf{SVOt}}} & {\textbf{FAVAR}} & {\makecell{ \textbf{FAVAR} \\ \textbf{SV}}}  \\
		\midrule
		\multicolumn{9}{c}{\textbf{Panel A: Industrial Production}} \\
		\midrule
		1 & \textbf{0.715} & \textbf{0.929} & \textbf{0.793} & \textbf{0.765} & \textbf{0.756} & \textbf{0.825} & 1.052 & \textbf{0.695}$^{\dag}$ \\
		2 & \textbf{0.760}$^{\dag}$ & \textbf{0.940} & \textbf{0.770} & \textbf{0.802} & \textbf{0.793} & \textbf{0.923} & 1.107 & \textbf{0.823} \\
		3 & \textbf{0.923}$^{\dag}$ & \textbf{0.964} & \textbf{0.931} & \textbf{0.951} & \textbf{0.957} & \textbf{0.999} & 1.145 & \textbf{0.948} \\
		4 & 1.013 & \textbf{0.971}$^{\dag}$ & 1.023 & 1.037 & 1.042 & 1.019 & 1.176 & 1.039 \\
		5 & 1.100 & \textbf{0.990}$^{\dag}$ & 1.095 & 1.117 & 1.118 & 1.105 & 1.256 & 1.104 \\
		6 & 1.093 & 1.002 & 1.093 & 1.112 & 1.110 & 1.079 & 1.275 & 1.110 \\
		9 & 1.100 & 1.002 & 1.097 & 1.118 & 1.112 & 1.094 & 1.359 & 1.091 \\
		12 & 1.120 & \textbf{0.992}$^{\dag}$ & 1.112 & 1.135 & 1.120 & 1.114 & 1.344 & 1.160 \\
		15 & 1.149 & \textbf{0.998}$^{\dag}$ & 1.141 & 1.160 & 1.153 & 1.139 & 1.291 & 1.184 \\
		18 & 1.181 & 1.007 & 1.157 & 1.166 & 1.180 & 1.180 & 1.283 & 1.206 \\
		24 & 1.150 & 1.002 & 1.141 & 1.154 & 1.159 & 1.179 & 1.145 & 1.163 \\
		\midrule
		\multicolumn{9}{c}{\textbf{Panel B: PCEPI}} \\
		\midrule
		1 & \textbf{0.984} & \textbf{0.972} & \textbf{0.965} & \textbf{0.958} & 1.280 & \textbf{0.998} & 1.001 & \textbf{0.954}$^{\dag}$ \\
		2 & \textbf{0.899}$^{\dag}$ & \textbf{0.979} & \textbf{0.947} & \textbf{0.942} & 1.141 & \textbf{0.991} & 1.053 & \textbf{0.946} \\
		3 & \textbf{0.878}$^{\dag}$ & \textbf{0.978} & \textbf{0.947} & \textbf{0.936} & 1.092 & 1.001 & 1.078 & \textbf{0.935} \\
		4 & \textbf{0.871}$^{\dag}$ & \textbf{0.983} & \textbf{0.967} & \textbf{0.953} & 1.095 & 1.016 & 1.064 & \textbf{0.949} \\
		5 & \textbf{0.897}$^{\dag}$ & 1.001 & \textbf{0.974} & \textbf{0.977} & 1.095 & 1.041 & 1.106 & \textbf{0.985} \\
		6 & \textbf{0.894}$^{\dag}$ & 1.002 & \textbf{0.964} & \textbf{0.962} & 1.067 & 1.023 & 1.076 & \textbf{0.960} \\
		9 & \textbf{0.921}$^{\dag}$ & \textbf{0.994} & \textbf{0.972} & \textbf{0.963} & 1.101 & \textbf{0.984} & 1.140 & \textbf{0.976} \\
		12 & \textbf{0.956}$^{\dag}$ & \textbf{0.984} & \textbf{0.996} & \textbf{0.982} & 1.090 & 1.019 & 1.109 & 1.004 \\
		15 & 1.049 & 1.000 & 1.039 & 1.028 & 1.165 & 1.089 & 1.065 & 1.037 \\
		18 & 1.073 & 1.006 & 1.041 & 1.035 & 1.151 & 1.077 & 1.036 & 1.048 \\
		24 & 1.169 & \textbf{0.977}$^{\dag}$ & 1.094 & 1.107 & 1.208 & 1.145 & 1.134 & 1.087 \\
		\bottomrule
	\end{tabular}	
	\vspace{0.5em}
	\par
	\begin{tabularx}{\textwidth}{X}
		{\footnotesize \textit{Notes:} The table reports quantile scores 90\% (QS90) for each of the two forecasted variables, relative to a benchmark constant-parameter VAR($1$) estimated via least squares, for U.S. monthly data. Values below one (highlighted in bold) indicate superior forecasting performance compared to the benchmark. The forecast horizon is denoted by $h$ (in months), and the symbol $\dag$ marks the best-performing model at each horizon.}
	\end{tabularx}
\end{table}

\begin{table}[htbp!]
	\centering
	\caption{Forecasting Performance VAR(1): Quantile Scores 10\% Relative to Benchmark} \label{tab:forecasting_qs10_US_p1}
	\scriptsize
	\begin{tabular}{@{}cS[table-format=1.3]S[table-format=1.3]S[table-format=1.3]S[table-format=1.3]S[table-format=1.3]S[table-format=1.3]S[table-format=1.3]S[table-format=1.3]@{}}
		\toprule
		{\textbf{h}} & {\textbf{AVP-VAR}} & {\textbf{CP-VAR}} & {\makecell{ \textbf{CP-VAR} \\ \textbf{SV}}} & {\makecell{ \textbf{{TVP-VAR}} \\ \textbf{EB}}} &{\textbf{TVP-VAR}} & {\makecell{ \textbf{VAR} \\ \textbf{SVOt}}} & {\textbf{FAVAR}} & {\makecell{ \textbf{FAVAR} \\ \textbf{SV}}}  \\
		\midrule
		\multicolumn{9}{c}{\textbf{Panel A: GDP}} \\
		\midrule
		1 & 1.028 & 1.000 & 1.025 & 1.059 & 1.057 & 1.001 & \textbf{0.906}$^{\dag}$ & \textbf{0.972} \\
		2 & 1.029 & \textbf{0.999} & 1.023 & 1.040 & 1.044 & 1.010 & \textbf{0.917}$^{\dag}$ & \textbf{0.999} \\
		3 & 1.010 & \textbf{0.990} & \textbf{0.996} & 1.025 & 1.034 & 1.009 & \textbf{0.924}$^{\dag}$ & 1.003 \\
		4 & 1.034 & \textbf{0.999} & 1.027 & 1.050 & 1.054 & 1.033 & \textbf{0.948}$^{\dag}$ & 1.042 \\
		5 & 1.054 & 1.010 & 1.056 & 1.065 & 1.073 & 1.037 & \textbf{0.971}$^{\dag}$ & 1.071 \\
		6 & 1.050 & \textbf{0.991} & 1.116 & 1.063 & 1.079 & 1.019 & \textbf{0.977}$^{\dag}$ & 1.075 \\
		9 & 1.074 & \textbf{0.984} & 1.085 & 1.090 & 1.096 & 1.060 & \textbf{0.982}$^{\dag}$ & 1.092 \\
		12 & 1.143 & 1.004 & 1.146 & 1.147 & 1.165 & 1.116 & 1.028 & 1.167 \\
		15 & 1.154 & \textbf{0.999}$^{\dag}$ & 1.140 & 1.154 & 1.165 & 1.141 & 1.037 & 1.195 \\
		18 & 1.161 & 1.005 & 1.144 & 1.157 & 1.159 & 1.154 & 1.052 & 1.214 \\
		24 & 1.175 & 1.014 & 1.149 & 1.173 & 1.165 & 1.155 & 1.050 & 1.207 \\
		\midrule
		\multicolumn{9}{c}{\textbf{Panel B: HICP}} \\
		\midrule
		1 & 1.007 & \textbf{0.982} & \textbf{0.990} & 1.002 & 1.102 & \textbf{0.978}$^{\dag}$ & \textbf{0.992} & \textbf{0.998} \\
		2 & 1.040 & \textbf{0.991}$^{\dag}$ & 1.044 & 1.047 & 1.100 & \textbf{0.998} & 1.009 & 1.050 \\
		3 & 1.102 & 1.007 & 1.087 & 1.091 & 1.150 & 1.095 & 1.037 & 1.100 \\
		4 & 1.084 & 1.002 & 1.079 & 1.087 & 1.141 & 1.083 & 1.035 & 1.085 \\
		5 & 1.090 & \textbf{0.998}$^{\dag}$ & 1.081 & 1.087 & 1.134 & 1.083 & 1.029 & 1.089 \\
		6 & 1.097 & 1.010 & 1.100 & 1.094 & 1.150 & 1.090 & 1.030 & 1.096 \\
		9 & 1.084 & \textbf{0.998}$^{\dag}$ & 1.079 & 1.075 & 1.138 & 1.076 & 1.024 & 1.079 \\
		12 & 1.091 & \textbf{0.993}$^{\dag}$ & 1.080 & 1.085 & 1.146 & 1.099 & 1.043 & 1.082 \\
		15 & 1.106 & \textbf{0.998}$^{\dag}$ & 1.114 & 1.102 & 1.177 & 1.160 & 1.037 & 1.117 \\
		18 & 1.098 & 1.009 & 1.098 & 1.093 & 1.158 & 1.154 & 1.042 & 1.104 \\
		24 & 1.040 & \textbf{0.995}$^{\dag}$ & 1.030 & 1.038 & 1.103 & 1.071 & 1.035 & 1.039 \\
		\bottomrule
	\end{tabular}	
	\vspace{0.5em}
	\par
	\begin{tabularx}{\textwidth}{X}
		{\footnotesize \textit{Notes:} The table reports quantile scores 10\% (QS10) for each of the two forecasted variables, relative to a benchmark constant-parameter VAR($1$) estimated via least squares, for U.S. monthly data. Values below one (highlighted in bold) indicate superior forecasting performance compared to the benchmark. The forecast horizon is denoted by $h$ (in months), and the symbol $\dag$ marks the best-performing model at each horizon.}
	\end{tabularx}
\end{table}
\FloatBarrier

\subsubsection{U.S. monthly data: VAR(2)}
\begin{figure}[htbp!]  
	\centering
	\includegraphics[width=\textwidth]{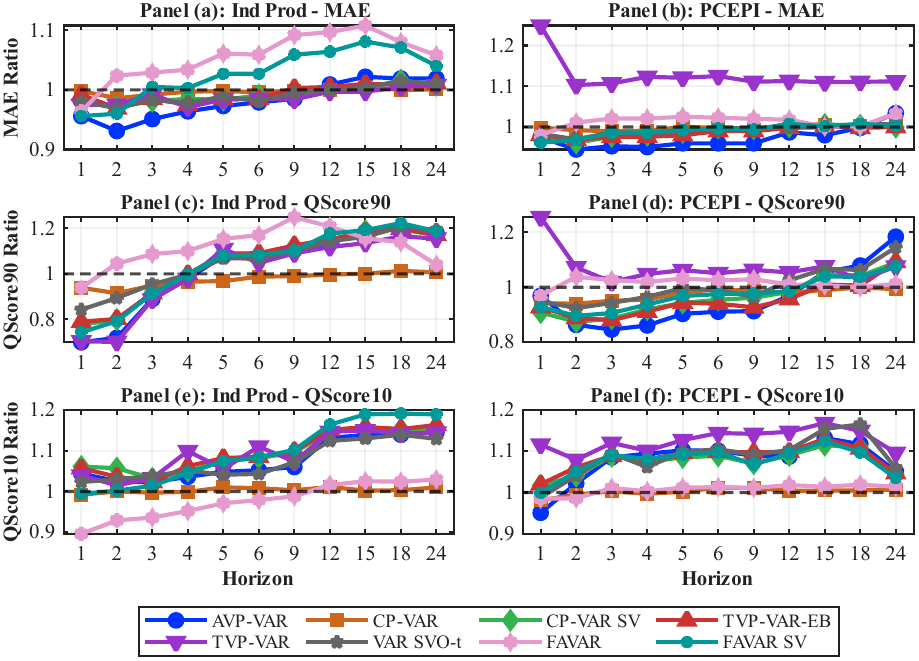}  
	\caption{Forecasting results using U.S. monthly data. The top row displays mean absolute forecast errors (MAE), while the middle and bottom rows show quantile scores for the upper (90th percentile) and lower (10th percentile) tails of each variable. All metrics are reported as ratios relative to a benchmark constant-parameter VAR($2$) estimated via least squares. A model is considered superior when its relative MAE, or relative quantile scores, are below one.} \label{fig:OOS_US_P2}
\end{figure}
\begin{table}[htbp!]
	\centering
	\caption{Forecasting Performance VAR(2): Mean Squared Prediction Error Relative to Benchmark} \label{tab:forecasting_mspe_US_p2}
	\scriptsize
	\begin{tabular}{@{}cS[table-format=1.3]S[table-format=1.3]S[table-format=1.3]S[table-format=1.3]S[table-format=1.3]S[table-format=1.3]S[table-format=1.3]S[table-format=1.3]@{}}
		\toprule
		{\textbf{h}} & {\textbf{AVP-VAR}} & {\textbf{CP-VAR}} & {\makecell{ \textbf{CP-VAR} \\ \textbf{SV}}} & {\makecell{ \textbf{{TVP-VAR}} \\ \textbf{EB}}} &{\textbf{TVP-VAR}} & {\makecell{ \textbf{VAR} \\ \textbf{SVOt}}} & {\textbf{FAVAR}} & {\makecell{ \textbf{FAVAR} \\ \textbf{SV}}}  \\
		\midrule
		\multicolumn{9}{c}{\textbf{Panel A: Industrial Production}} \\
		\midrule
		1 & \textbf{0.902} & \textbf{0.996} & \textbf{0.949} & \textbf{0.977} & \textbf{0.890} & \textbf{0.951} & \textbf{0.768}$^{\dag}$ & \textbf{0.849} \\
		2 & \textbf{0.805} & \textbf{0.949} & \textbf{0.863} & \textbf{0.853} & \textbf{0.784}$^{\dag}$ & \textbf{0.858} & \textbf{0.889} & \textbf{0.816} \\
		3 & \textbf{0.867}$^{\dag}$ & \textbf{0.967} & \textbf{0.879} & \textbf{0.881} & \textbf{0.886} & \textbf{0.891} & 1.017 & \textbf{0.928} \\
		4 & \textbf{0.938}$^{\dag}$ & \textbf{0.986} & \textbf{0.959} & \textbf{0.951} & \textbf{0.956} & \textbf{0.952} & 1.036 & \textbf{0.976} \\
		5 & \textbf{0.967}$^{\dag}$ & \textbf{0.991} & \textbf{0.988} & \textbf{0.986} & \textbf{0.991} & \textbf{0.984} & 1.040 & \textbf{0.999} \\
		6 & \textbf{0.972}$^{\dag}$ & \textbf{0.994} & \textbf{0.985} & \textbf{0.985} & \textbf{0.990} & \textbf{0.977} & 1.036 & \textbf{0.993} \\
		9 & \textbf{0.983}$^{\dag}$ & 1.000 & \textbf{0.996} & 1.005 & \textbf{0.994} & 1.004 & 1.063 & 1.028 \\
		12 & 1.001 & \textbf{0.997}$^{\dag}$ & 1.006 & 1.006 & 1.000 & 1.003 & 1.081 & 1.037 \\
		15 & 1.002 & 1.002 & 1.007 & 1.003 & \textbf{0.999}$^{\dag}$ & 1.006 & 1.059 & 1.031 \\
		18 & 1.001 & 1.000 & 1.010 & 1.006 & 1.002 & 1.011 & 1.054 & 1.038 \\
		24 & 1.001 & \textbf{0.999}$^{\dag}$ & 1.004 & 1.010 & 1.002 & 1.009 & 1.032 & 1.031 \\
		\midrule
		\multicolumn{9}{c}{\textbf{Panel B: PCEPI}} \\
		\midrule
		1 & \textbf{0.980} & \textbf{0.988} & \textbf{0.964} & \textbf{0.999} & 1.491 & \textbf{0.970} & \textbf{0.943}$^{\dag}$ & \textbf{0.950} \\
		2 & \textbf{0.881}$^{\dag}$ & \textbf{0.972} & \textbf{0.897} & \textbf{0.937} & 1.066 & \textbf{0.922} & \textbf{0.999} & \textbf{0.907} \\
		3 & \textbf{0.869}$^{\dag}$ & \textbf{0.971} & \textbf{0.926} & \textbf{0.930} & 1.055 & \textbf{0.944} & 1.017 & \textbf{0.946} \\
		4 & \textbf{0.886}$^{\dag}$ & \textbf{0.979} & \textbf{0.947} & \textbf{0.939} & 1.084 & \textbf{0.950} & 1.014 & \textbf{0.956} \\
		5 & \textbf{0.918}$^{\dag}$ & \textbf{0.997} & \textbf{0.972} & \textbf{0.973} & 1.112 & \textbf{0.978} & 1.034 & \textbf{0.991} \\
		6 & \textbf{0.912}$^{\dag}$ & \textbf{0.995} & 1.002 & \textbf{0.996} & 1.118 & 1.001 & 1.037 & 1.003 \\
		9 & \textbf{0.926}$^{\dag}$ & \textbf{0.990} & \textbf{0.993} & \textbf{0.993} & 1.103 & \textbf{0.992} & 1.028 & \textbf{0.986} \\
		12 & \textbf{0.964}$^{\dag}$ & 1.011 & \textbf{0.994} & \textbf{0.998} & 1.099 & \textbf{0.994} & 1.027 & 1.002 \\
		15 & \textbf{0.959}$^{\dag}$ & 1.006 & 1.005 & 1.006 & 1.108 & \textbf{0.996} & 1.009 & 1.002 \\
		18 & \textbf{0.989}$^{\dag}$ & \textbf{0.997} & 1.008 & 1.002 & 1.102 & 1.006 & 1.012 & 1.012 \\
		24 & 1.040 & 1.008 & 1.004 & 1.007 & 1.107 & 1.008 & 1.042 & 1.001 \\
		\bottomrule
	\end{tabular}	
	\vspace{0.5em}
	\par
	\begin{tabularx}{\textwidth}{X}
		{\footnotesize \textit{Notes:} The table reports mean squared prediction errors (MSPE) for each of the two forecasted variables, relative to a benchmark constant-parameter VAR($2$) estimated via least squares, for U.S. monthly data. Values below one (highlighted in bold) indicate superior forecasting performance compared to the benchmark. The forecast horizon is denoted by $h$ (in months), and the symbol $\dag$ marks the best-performing model at each horizon.}
	\end{tabularx}
\end{table}

\begin{table}[htbp!]
	\centering
	\caption{Forecasting Performance VAR(2): Quantile Scores 90\% Relative to Benchmark} \label{tab:forecasting_qs90_US_p2}
	\scriptsize
	\begin{tabular}{@{}cS[table-format=1.3]S[table-format=1.3]S[table-format=1.3]S[table-format=1.3]S[table-format=1.3]S[table-format=1.3]S[table-format=1.3]S[table-format=1.3]@{}}
		\toprule
		{\textbf{h}} & {\textbf{AVP-VAR}} & {\textbf{CP-VAR}} & {\makecell{ \textbf{CP-VAR} \\ \textbf{SV}}} & {\makecell{ \textbf{{TVP-VAR}} \\ \textbf{EB}}} &{\textbf{TVP-VAR}} & {\makecell{ \textbf{VAR} \\ \textbf{SVOt}}} & {\textbf{FAVAR}} & {\makecell{ \textbf{FAVAR} \\ \textbf{SV}}}  \\
		\midrule
		\multicolumn{9}{c}{\textbf{Panel A: Industrial Production}} \\
		\midrule
		1 & \textbf{0.700}$^{\dag}$ & \textbf{0.939} & \textbf{0.777} & \textbf{0.790} & \textbf{0.704} & \textbf{0.843} & \textbf{0.940} & \textbf{0.743} \\
		2 & \textbf{0.719} & \textbf{0.915} & \textbf{0.799} & \textbf{0.802} & \textbf{0.701}$^{\dag}$ & \textbf{0.894} & 1.045 & \textbf{0.792} \\
		3 & \textbf{0.889} & \textbf{0.949} & \textbf{0.918} & \textbf{0.915} & \textbf{0.887}$^{\dag}$ & \textbf{0.954} & 1.087 & \textbf{0.913} \\
		4 & \textbf{0.980} & \textbf{0.966}$^{\dag}$ & \textbf{0.991} & \textbf{0.999} & \textbf{0.978} & \textbf{0.997} & 1.100 & \textbf{0.998} \\
		5 & 1.078 & \textbf{0.968}$^{\dag}$ & 1.090 & 1.088 & 1.109 & 1.069 & 1.154 & 1.082 \\
		6 & 1.076 & \textbf{0.988}$^{\dag}$ & 1.086 & 1.091 & 1.048 & 1.067 & 1.170 & 1.083 \\
		9 & 1.105 & \textbf{0.990}$^{\dag}$ & 1.120 & 1.125 & 1.090 & 1.100 & 1.249 & 1.102 \\
		12 & 1.135 & \textbf{0.997}$^{\dag}$ & 1.147 & 1.152 & 1.118 & 1.143 & 1.208 & 1.177 \\
		15 & 1.185 & 1.001 & 1.190 & 1.184 & 1.135 & 1.163 & 1.156 & 1.192 \\
		18 & 1.201 & 1.013 & 1.195 & 1.199 & 1.165 & 1.206 & 1.139 & 1.223 \\
		24 & 1.179 & 1.007 & 1.170 & 1.171 & 1.155 & 1.193 & 1.040 & 1.187 \\
		\midrule
		\multicolumn{9}{c}{\textbf{Panel B: PCEPI}} \\
		\midrule
		1 & \textbf{0.969} & \textbf{0.942} & \textbf{0.908}$^{\dag}$ & \textbf{0.928} & 1.255 & \textbf{0.949} & \textbf{0.970} & \textbf{0.928} \\
		2 & \textbf{0.863}$^{\dag}$ & \textbf{0.940} & \textbf{0.874} & \textbf{0.888} & 1.072 & \textbf{0.924} & 1.038 & \textbf{0.896} \\
		3 & \textbf{0.846}$^{\dag}$ & \textbf{0.951} & \textbf{0.886} & \textbf{0.880} & 1.019 & \textbf{0.943} & 1.026 & \textbf{0.906} \\
		4 & \textbf{0.861}$^{\dag}$ & \textbf{0.955} & \textbf{0.919} & \textbf{0.911} & 1.048 & \textbf{0.965} & 1.019 & \textbf{0.937} \\
		5 & \textbf{0.903}$^{\dag}$ & \textbf{0.981} & \textbf{0.943} & \textbf{0.944} & 1.062 & \textbf{0.997} & 1.032 & \textbf{0.968} \\
		6 & \textbf{0.911}$^{\dag}$ & \textbf{0.993} & \textbf{0.954} & \textbf{0.940} & 1.051 & \textbf{0.989} & 1.025 & \textbf{0.975} \\
		9 & \textbf{0.913}$^{\dag}$ & \textbf{0.986} & \textbf{0.962} & \textbf{0.927} & 1.063 & \textbf{0.965} & 1.029 & \textbf{0.978} \\
		12 & \textbf{0.969} & \textbf{0.980} & \textbf{0.983} & \textbf{0.957}$^{\dag}$ & 1.054 & 1.025 & 1.010 & \textbf{0.985} \\
		15 & 1.055 & \textbf{0.990}$^{\dag}$ & 1.046 & 1.010 & 1.076 & 1.067 & 1.007 & 1.039 \\
		18 & 1.079 & \textbf{0.999}$^{\dag}$ & 1.041 & 1.006 & 1.028 & 1.063 & 1.000 & 1.038 \\
		24 & 1.184 & \textbf{0.993}$^{\dag}$ & 1.090 & 1.092 & 1.075 & 1.145 & 1.012 & 1.076 \\
		\bottomrule
	\end{tabular}	
	\vspace{0.5em}
	\par
	\begin{tabularx}{\textwidth}{X}
		{\footnotesize \textit{Notes:} The table reports quantile scores 90\% (QS90) for each of the two forecasted variables, relative to a benchmark constant-parameter VAR($2$) estimated via least squares, for U.S. monthly data. Values below one (highlighted in bold) indicate superior forecasting performance compared to the benchmark. The forecast horizon is denoted by $h$ (in months), and the symbol $\dag$ marks the best-performing model at each horizon.}
	\end{tabularx}
\end{table}

\begin{table}[htbp!]
	\centering
	\caption{Forecasting Performance VAR(2): Quantile Scores 10\% Relative to Benchmark} \label{tab:forecasting_qs10_US_p2}
	\scriptsize
	\begin{tabular}{@{}cS[table-format=1.3]S[table-format=1.3]S[table-format=1.3]S[table-format=1.3]S[table-format=1.3]S[table-format=1.3]S[table-format=1.3]S[table-format=1.3]@{}}
		\toprule
		{\textbf{h}} & {\textbf{AVP-VAR}} & {\textbf{CP-VAR}} & {\makecell{ \textbf{CP-VAR} \\ \textbf{SV}}} & {\makecell{ \textbf{{TVP-VAR}} \\ \textbf{EB}}} &{\textbf{TVP-VAR}} & {\makecell{ \textbf{VAR} \\ \textbf{SVOt}}} & {\textbf{FAVAR}} & {\makecell{ \textbf{FAVAR} \\ \textbf{SV}}}  \\
		\midrule
		\multicolumn{9}{c}{\textbf{Panel A: GDP}} \\
		\midrule
		1 & 1.044 & \textbf{0.992} & 1.061 & 1.057 & 1.039 & 1.023 & \textbf{0.896}$^{\dag}$ & \textbf{0.994} \\
		2 & 1.025 & 1.000 & 1.057 & 1.036 & 1.015 & 1.030 & \textbf{0.929}$^{\dag}$ & 1.002 \\
		3 & 1.022 & \textbf{0.997} & 1.032 & 1.023 & 1.032 & 1.038 & \textbf{0.936}$^{\dag}$ & 1.014 \\
		4 & 1.036 & \textbf{0.999} & 1.050 & 1.062 & 1.099 & 1.051 & \textbf{0.952}$^{\dag}$ & 1.045 \\
		5 & 1.048 & 1.010 & 1.074 & 1.082 & 1.055 & 1.041 & \textbf{0.971}$^{\dag}$ & 1.075 \\
		6 & 1.052 & 1.008 & 1.082 & 1.085 & 1.110 & 1.043 & \textbf{0.979}$^{\dag}$ & 1.083 \\
		9 & 1.060 & 1.003 & 1.095 & 1.099 & 1.073 & 1.071 & \textbf{0.988}$^{\dag}$ & 1.102 \\
		12 & 1.132 & 1.011 & 1.149 & 1.151 & 1.146 & 1.124 & 1.016 & 1.164 \\
		15 & 1.138 & 1.002 & 1.155 & 1.158 & 1.150 & 1.130 & 1.025 & 1.189 \\
		18 & 1.139 & 1.003 & 1.153 & 1.153 & 1.142 & 1.139 & 1.024 & 1.191 \\
		24 & 1.145 & 1.010 & 1.151 & 1.163 & 1.144 & 1.129 & 1.029 & 1.189 \\
		\midrule
		\multicolumn{9}{c}{\textbf{Panel B: HICP}} \\
		\midrule
		1 & \textbf{0.951}$^{\dag}$ & \textbf{0.981} & 1.017 & 1.020 & 1.116 & \textbf{0.988} & \textbf{0.985} & \textbf{0.999} \\
		2 & 1.022 & \textbf{0.995} & 1.040 & 1.061 & 1.079 & 1.041 & \textbf{0.986}$^{\dag}$ & 1.051 \\
		3 & 1.085 & 1.003 & 1.088 & 1.089 & 1.120 & 1.094 & 1.011 & 1.091 \\
		4 & 1.094 & \textbf{0.997}$^{\dag}$ & 1.075 & 1.080 & 1.101 & 1.062 & 1.003 & 1.081 \\
		5 & 1.102 & 1.002 & 1.085 & 1.096 & 1.127 & 1.098 & 1.012 & 1.092 \\
		6 & 1.101 & 1.012 & 1.089 & 1.099 & 1.144 & 1.105 & 1.013 & 1.097 \\
		9 & 1.089 & 1.011 & 1.070 & 1.097 & 1.142 & 1.091 & 1.011 & 1.068 \\
		12 & 1.087 & 1.004 & 1.090 & 1.099 & 1.147 & 1.103 & 1.018 & 1.094 \\
		15 & 1.132 & 1.005 & 1.115 & 1.130 & 1.168 & 1.154 & 1.014 & 1.124 \\
		18 & 1.117 & 1.005 & 1.108 & 1.107 & 1.149 & 1.165 & 1.019 & 1.097 \\
		24 & 1.051 & 1.007 & 1.044 & 1.047 & 1.096 & 1.059 & 1.014 & 1.035 \\
		\bottomrule
	\end{tabular}	
	\vspace{0.5em}
	\par
	\begin{tabularx}{\textwidth}{X}
		{\footnotesize \textit{Notes:} The table reports quantile scores 10\% (QS10) for each of the two forecasted variables, relative to a benchmark constant-parameter VAR($2$) estimated via least squares, for U.S. monthly data. Values below one (highlighted in bold) indicate superior forecasting performance compared to the benchmark. The forecast horizon is denoted by $h$ (in months), and the symbol $\dag$ marks the best-performing model at each horizon.}
	\end{tabularx}
\end{table}

\FloatBarrier


\subsubsection{U.S. monthly data: VAR(3)}
\begin{figure}[htbp!]  
	\centering
	\includegraphics[width=\textwidth]{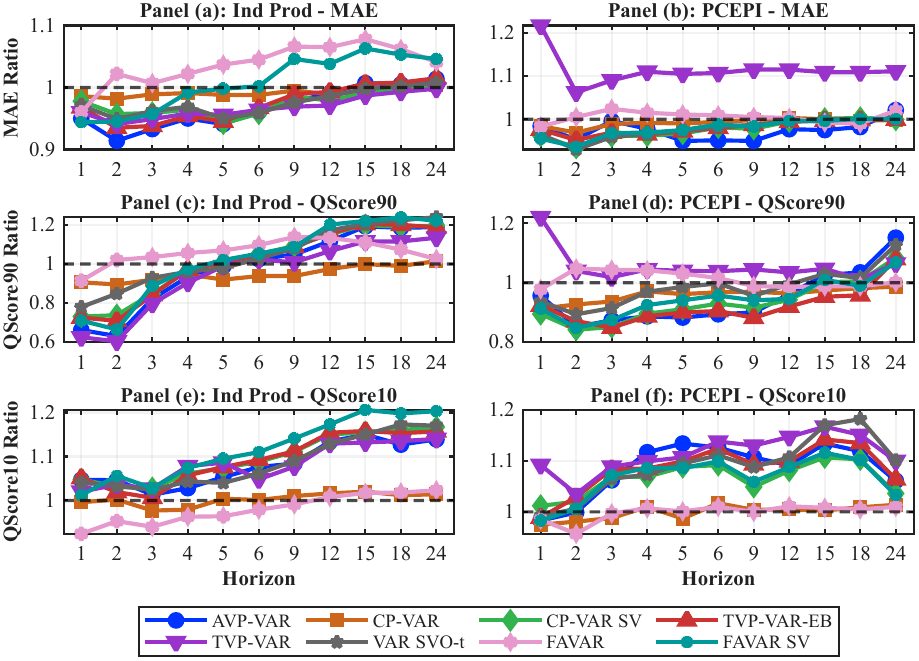}  
	\caption{Forecasting results using U.S. monthly data. The top row displays mean absolute forecast errors (MAE), while the middle and bottom rows show quantile scores for the upper (90th percentile) and lower (10th percentile) tails of each variable. All metrics are reported as ratios relative to a benchmark constant-parameter VAR($3$) estimated via least squares. A model is considered superior when its relative MAE, or relative quantile scores, are below one.} \label{fig:OOS_US_P3}
\end{figure}
\begin{table}[htbp!]
	\centering
	\caption{Forecasting Performance VAR(3): Mean Squared Prediction Error Relative to Benchmark} \label{tab:forecasting_mspe_US_p3}
	\scriptsize
	\begin{tabular}{@{}cS[table-format=1.3]S[table-format=1.3]S[table-format=1.3]S[table-format=1.3]S[table-format=1.3]S[table-format=1.3]S[table-format=1.3]S[table-format=1.3]@{}}
		\toprule
		{\textbf{h}} & {\textbf{AVP-VAR}} & {\textbf{CP-VAR}} & {\makecell{ \textbf{CP-VAR} \\ \textbf{SV}}} & {\makecell{ \textbf{{TVP-VAR}} \\ \textbf{EB}}} &{\textbf{TVP-VAR}} & {\makecell{ \textbf{VAR} \\ \textbf{SVOt}}} & {\textbf{FAVAR}} & {\makecell{ \textbf{FAVAR} \\ \textbf{SV}}}  \\
		\midrule
		\multicolumn{9}{c}{\textbf{Panel A: Industrial Production}} \\
		\midrule
		1 & \textbf{0.907} & \textbf{0.988} & \textbf{0.933} & \textbf{0.957} & \textbf{0.844} & \textbf{0.922} & \textbf{0.765}$^{\dag}$ & \textbf{0.837} \\
		2 & \textbf{0.770} & \textbf{0.943} & \textbf{0.814} & \textbf{0.806} & \textbf{0.711}$^{\dag}$ & \textbf{0.817} & \textbf{0.901} & \textbf{0.776} \\
		3 & \textbf{0.839} & \textbf{0.956} & \textbf{0.847} & \textbf{0.833} & \textbf{0.800}$^{\dag}$ & \textbf{0.848} & \textbf{0.990} & \textbf{0.859} \\
		4 & \textbf{0.891}$^{\dag}$ & \textbf{0.974} & \textbf{0.912} & \textbf{0.900} & \textbf{0.908} & \textbf{0.912} & 1.008 & \textbf{0.935} \\
		5 & \textbf{0.930}$^{\dag}$ & \textbf{0.982} & \textbf{0.932} & \textbf{0.930} & \textbf{0.945} & \textbf{0.937} & 1.004 & \textbf{0.955} \\
		6 & \textbf{0.944} & \textbf{0.980} & \textbf{0.951} & \textbf{0.948} & \textbf{0.951} & \textbf{0.940}$^{\dag}$ & 1.005 & \textbf{0.955} \\
		9 & \textbf{0.970}$^{\dag}$ & \textbf{0.991} & \textbf{0.985} & \textbf{0.978} & \textbf{0.976} & \textbf{0.974} & 1.026 & 1.006 \\
		12 & \textbf{0.982}$^{\dag}$ & \textbf{0.992} & \textbf{0.987} & \textbf{0.991} & \textbf{0.985} & \textbf{0.990} & 1.046 & 1.021 \\
		15 & \textbf{0.987}$^{\dag}$ & \textbf{0.994} & \textbf{0.996} & \textbf{0.994} & \textbf{0.992} & 1.000 & 1.038 & 1.023 \\
		18 & \textbf{0.987} & \textbf{0.997} & \textbf{0.997} & \textbf{0.997} & \textbf{0.985}$^{\dag}$ & 1.000 & 1.034 & 1.019 \\
		24 & \textbf{0.994}$^{\dag}$ & 1.000 & 1.002 & 1.009 & \textbf{0.995} & 1.006 & 1.023 & 1.028 \\
		\midrule
		\multicolumn{9}{c}{\textbf{Panel B: PCEPI}} \\
		\midrule
		1 & 1.007 & \textbf{0.981} & \textbf{0.949} & 1.021 & 1.444 & \textbf{0.951} & \textbf{0.960} & \textbf{0.942}$^{\dag}$ \\
		2 & \textbf{0.903} & \textbf{0.935} & \textbf{0.857}$^{\dag}$ & \textbf{0.928} & 1.001 & \textbf{0.868} & 1.007 & \textbf{0.858} \\
		3 & \textbf{0.943} & \textbf{0.953} & \textbf{0.879} & \textbf{0.903} & 1.015 & \textbf{0.875}$^{\dag}$ & 1.029 & \textbf{0.893} \\
		4 & \textbf{0.918} & \textbf{0.974} & \textbf{0.907}$^{\dag}$ & \textbf{0.916} & 1.064 & \textbf{0.914} & 1.018 & \textbf{0.933} \\
		5 & \textbf{0.915}$^{\dag}$ & \textbf{0.971} & \textbf{0.934} & \textbf{0.939} & 1.071 & \textbf{0.945} & 1.015 & \textbf{0.951} \\
		6 & \textbf{0.926}$^{\dag}$ & \textbf{0.994} & \textbf{0.975} & \textbf{0.983} & 1.092 & \textbf{0.981} & 1.018 & \textbf{0.986} \\
		9 & \textbf{0.912}$^{\dag}$ & \textbf{0.987} & \textbf{0.955} & \textbf{0.980} & 1.093 & \textbf{0.970} & \textbf{0.992} & \textbf{0.963} \\
		12 & \textbf{0.944}$^{\dag}$ & \textbf{0.998} & \textbf{0.979} & \textbf{0.989} & 1.093 & \textbf{0.979} & \textbf{0.992} & \textbf{0.977} \\
		15 & \textbf{0.954}$^{\dag}$ & \textbf{0.993} & \textbf{0.988} & \textbf{0.993} & 1.092 & \textbf{0.991} & \textbf{0.989} & \textbf{0.981} \\
		18 & \textbf{0.970}$^{\dag}$ & \textbf{0.994} & \textbf{0.989} & 1.002 & 1.101 & \textbf{0.992} & \textbf{0.991} & \textbf{0.991} \\
		24 & 1.026 & 1.000 & 1.002 & 1.008 & 1.108 & 1.006 & 1.024 & \textbf{0.997}$^{\dag}$ \\
		\bottomrule
	\end{tabular}	
	\vspace{0.5em}
	\par
	\begin{tabularx}{\textwidth}{X}
		{\footnotesize \textit{Notes:} The table reports mean squared prediction errors (MSPE) for each of the two forecasted variables, relative to a benchmark constant-parameter VAR($3$) estimated via least squares, for U.S. monthly data. Values below one (highlighted in bold) indicate superior forecasting performance compared to the benchmark. The forecast horizon is denoted by $h$ (in months), and the symbol $\dag$ marks the best-performing model at each horizon.}
	\end{tabularx}
\end{table}

\begin{table}[htbp!]
	\centering
	\caption{Forecasting Performance VAR(3): Quantile Scores 90\% Relative to Benchmark} \label{tab:forecasting_qs90_US_p3}
	\scriptsize
	\begin{tabular}{@{}cS[table-format=1.3]S[table-format=1.3]S[table-format=1.3]S[table-format=1.3]S[table-format=1.3]S[table-format=1.3]S[table-format=1.3]S[table-format=1.3]@{}}
		\toprule
		{\textbf{h}} & {\textbf{AVP-VAR}} & {\textbf{CP-VAR}} & {\makecell{ \textbf{CP-VAR} \\ \textbf{SV}}} & {\makecell{ \textbf{{TVP-VAR}} \\ \textbf{EB}}} &{\textbf{TVP-VAR}} & {\makecell{ \textbf{VAR} \\ \textbf{SVOt}}} & {\textbf{FAVAR}} & {\makecell{ \textbf{FAVAR} \\ \textbf{SV}}}  \\
		\midrule
		\multicolumn{9}{c}{\textbf{Panel A: Industrial Production}} \\
		\midrule
		1 & \textbf{0.662} & \textbf{0.906} & \textbf{0.731} & \textbf{0.726} & \textbf{0.625}$^{\dag}$ & \textbf{0.780} & \textbf{0.914} & \textbf{0.710} \\
		2 & \textbf{0.631} & \textbf{0.895} & \textbf{0.738} & \textbf{0.709} & \textbf{0.604}$^{\dag}$ & \textbf{0.848} & 1.020 & \textbf{0.665} \\
		3 & \textbf{0.821} & \textbf{0.919} & \textbf{0.848} & \textbf{0.841} & \textbf{0.796}$^{\dag}$ & \textbf{0.929} & 1.035 & \textbf{0.887} \\
		4 & \textbf{0.934} & \textbf{0.946} & \textbf{0.965} & \textbf{0.971} & \textbf{0.904}$^{\dag}$ & \textbf{0.960} & 1.056 & \textbf{0.973} \\
		5 & \textbf{0.988} & \textbf{0.920}$^{\dag}$ & 1.009 & 1.008 & \textbf{0.989} & \textbf{0.979} & 1.069 & 1.020 \\
		6 & 1.025 & \textbf{0.938}$^{\dag}$ & 1.050 & 1.048 & 1.023 & 1.028 & 1.096 & 1.053 \\
		9 & 1.048 & \textbf{0.938}$^{\dag}$ & 1.092 & 1.087 & 1.008 & 1.079 & 1.138 & 1.087 \\
		12 & 1.116 & \textbf{0.974}$^{\dag}$ & 1.152 & 1.156 & 1.066 & 1.172 & 1.133 & 1.201 \\
		15 & 1.192 & \textbf{0.999}$^{\dag}$ & 1.198 & 1.204 & 1.115 & 1.205 & 1.110 & 1.220 \\
		18 & 1.185 & \textbf{0.990}$^{\dag}$ & 1.191 & 1.200 & 1.115 & 1.223 & 1.073 & 1.238 \\
		24 & 1.188 & 1.014 & 1.196 & 1.187 & 1.134 & 1.239 & 1.025 & 1.221 \\
		\midrule
		\multicolumn{9}{c}{\textbf{Panel B: PCEPI}} \\
		\midrule
		1 & \textbf{0.957} & \textbf{0.913} & \textbf{0.895}$^{\dag}$ & \textbf{0.923} & 1.221 & \textbf{0.937} & \textbf{0.978} & \textbf{0.913} \\
		2 & \textbf{0.851} & \textbf{0.927} & \textbf{0.842}$^{\dag}$ & \textbf{0.871} & 1.040 & \textbf{0.894} & 1.048 & \textbf{0.848} \\
		3 & \textbf{0.874} & \textbf{0.937} & \textbf{0.851} & \textbf{0.849}$^{\dag}$ & 1.019 & \textbf{0.915} & 1.044 & \textbf{0.874} \\
		4 & \textbf{0.887}$^{\dag}$ & \textbf{0.971} & \textbf{0.896} & \textbf{0.887} & 1.045 & \textbf{0.970} & 1.042 & \textbf{0.925} \\
		5 & \textbf{0.883}$^{\dag}$ & \textbf{0.962} & \textbf{0.911} & \textbf{0.900} & 1.040 & \textbf{0.986} & 1.031 & \textbf{0.941} \\
		6 & \textbf{0.894}$^{\dag}$ & \textbf{0.968} & \textbf{0.930} & \textbf{0.905} & 1.039 & \textbf{0.998} & 1.016 & \textbf{0.957} \\
		9 & \textbf{0.900} & \textbf{0.970} & \textbf{0.916} & \textbf{0.882}$^{\dag}$ & 1.045 & \textbf{0.956} & \textbf{0.989} & \textbf{0.941} \\
		12 & \textbf{0.962} & \textbf{0.964} & \textbf{0.943} & \textbf{0.920}$^{\dag}$ & 1.035 & \textbf{0.995} & \textbf{0.986} & \textbf{0.946} \\
		15 & 1.026 & \textbf{0.974} & 1.002 & \textbf{0.953}$^{\dag}$ & 1.047 & 1.030 & \textbf{0.988} & 1.010 \\
		18 & 1.035 & \textbf{0.982} & 1.005 & \textbf{0.958}$^{\dag}$ & 1.012 & 1.016 & \textbf{0.988} & \textbf{0.990} \\
		24 & 1.152 & \textbf{0.987}$^{\dag}$ & 1.084 & 1.080 & 1.065 & 1.127 & 1.002 & 1.069 \\
		\bottomrule
	\end{tabular}	
	\vspace{0.5em}
	\par
	\begin{tabularx}{\textwidth}{X}
		{\footnotesize \textit{Notes:} The table reports quantile scores 90\% (QS90) for each of the two forecasted variables, relative to a benchmark constant-parameter VAR($3$) estimated via least squares, for U.S. monthly data. Values below one (highlighted in bold) indicate superior forecasting performance compared to the benchmark. The forecast horizon is denoted by $h$ (in months), and the symbol $\dag$ marks the best-performing model at each horizon.}
	\end{tabularx}
\end{table}

\begin{table}[htbp!]
	\centering
	\caption{Forecasting Performance VAR(3): Quantile Scores 10\% Relative to Benchmark} \label{tab:forecasting_qs10_US_p3}
	\scriptsize
	\begin{tabular}{@{}cS[table-format=1.3]S[table-format=1.3]S[table-format=1.3]S[table-format=1.3]S[table-format=1.3]S[table-format=1.3]S[table-format=1.3]S[table-format=1.3]@{}}
		\toprule
		{\textbf{h}} & {\textbf{AVP-VAR}} & {\textbf{CP-VAR}} & {\makecell{ \textbf{CP-VAR} \\ \textbf{SV}}} & {\makecell{ \textbf{{TVP-VAR}} \\ \textbf{EB}}} &{\textbf{TVP-VAR}} & {\makecell{ \textbf{VAR} \\ \textbf{SVOt}}} & {\textbf{FAVAR}} & {\makecell{ \textbf{FAVAR} \\ \textbf{SV}}}  \\
		\midrule
		\multicolumn{9}{c}{\textbf{Panel A: GDP}} \\
		\midrule
		1 & 1.047 & \textbf{0.996} & 1.048 & 1.051 & 1.020 & 1.041 & \textbf{0.924}$^{\dag}$ & 1.012 \\
		2 & 1.046 & 1.002 & 1.031 & 1.019 & 1.036 & 1.033 & \textbf{0.953}$^{\dag}$ & 1.056 \\
		3 & 1.013 & \textbf{0.977} & 1.028 & 1.006 & 1.027 & 1.024 & \textbf{0.940}$^{\dag}$ & 1.028 \\
		4 & 1.028 & \textbf{0.979} & 1.061 & 1.057 & 1.078 & 1.043 & \textbf{0.963}$^{\dag}$ & 1.074 \\
		5 & 1.053 & 1.004 & 1.074 & 1.076 & 1.086 & 1.040 & \textbf{0.964}$^{\dag}$ & 1.095 \\
		6 & 1.076 & 1.001 & 1.088 & 1.092 & 1.048 & 1.061 & \textbf{0.979}$^{\dag}$ & 1.110 \\
		9 & 1.084 & 1.010 & 1.111 & 1.112 & 1.077 & 1.090 & \textbf{0.991}$^{\dag}$ & 1.141 \\
		12 & 1.134 & 1.016 & 1.150 & 1.154 & 1.129 & 1.129 & 1.008 & 1.173 \\
		15 & 1.150 & 1.020 & 1.155 & 1.158 & 1.132 & 1.150 & 1.017 & 1.206 \\
		18 & 1.127 & 1.012 & 1.160 & 1.153 & 1.135 & 1.173 & 1.018 & 1.198 \\
		24 & 1.139 & 1.014 & 1.167 & 1.157 & 1.140 & 1.170 & 1.023 & 1.203 \\
		\midrule
		\multicolumn{9}{c}{\textbf{Panel B: HICP}} \\
		\midrule
		1 & \textbf{0.983} & \textbf{0.975}$^{\dag}$ & 1.012 & \textbf{0.989} & 1.092 & \textbf{0.992} & \textbf{0.985} & \textbf{0.983} \\
		2 & \textbf{0.999} & \textbf{0.981} & 1.020 & 1.029 & 1.035 & 1.006 & \textbf{0.957}$^{\dag}$ & 1.006 \\
		3 & 1.062 & \textbf{0.988}$^{\dag}$ & 1.069 & 1.084 & 1.089 & 1.065 & \textbf{0.997} & 1.072 \\
		4 & 1.117 & 1.010 & 1.069 & 1.085 & 1.099 & 1.073 & 1.007 & 1.086 \\
		5 & 1.134 & \textbf{0.987}$^{\dag}$ & 1.088 & 1.099 & 1.107 & 1.095 & 1.001 & 1.086 \\
		6 & 1.126 & 1.017 & 1.092 & 1.124 & 1.138 & 1.111 & 1.010 & 1.099 \\
		9 & 1.107 & 1.004 & 1.049 & 1.093 & 1.130 & 1.088 & 1.000 & 1.059 \\
		12 & 1.091 & 1.005 & 1.081 & 1.099 & 1.147 & 1.107 & 1.011 & 1.088 \\
		15 & 1.134 & 1.003 & 1.107 & 1.142 & 1.167 & 1.170 & 1.008 & 1.116 \\
		18 & 1.119 & 1.009 & 1.103 & 1.135 & 1.151 & 1.182 & 1.005 & 1.102 \\
		24 & 1.061 & 1.014 & 1.036 & 1.064 & 1.100 & 1.101 & 1.010 & 1.033 \\
		\bottomrule
	\end{tabular}	
	\vspace{0.5em}
	\par
	\begin{tabularx}{\textwidth}{X}
		{\footnotesize \textit{Notes:} The table reports quantile scores 10\% (QS10) for each of the two forecasted variables, relative to a benchmark constant-parameter VAR($3$) estimated via least squares, for U.S. monthly data. Values below one (highlighted in bold) indicate superior forecasting performance compared to the benchmark. The forecast horizon is denoted by $h$ (in months), and the symbol $\dag$ marks the best-performing model at each horizon.}
	\end{tabularx}
\end{table}

\FloatBarrier

\subsubsection{U.S. monthly data: VAR(4)}

\begin{figure}[htbp!]  
	\centering
	\includegraphics[width=\textwidth]{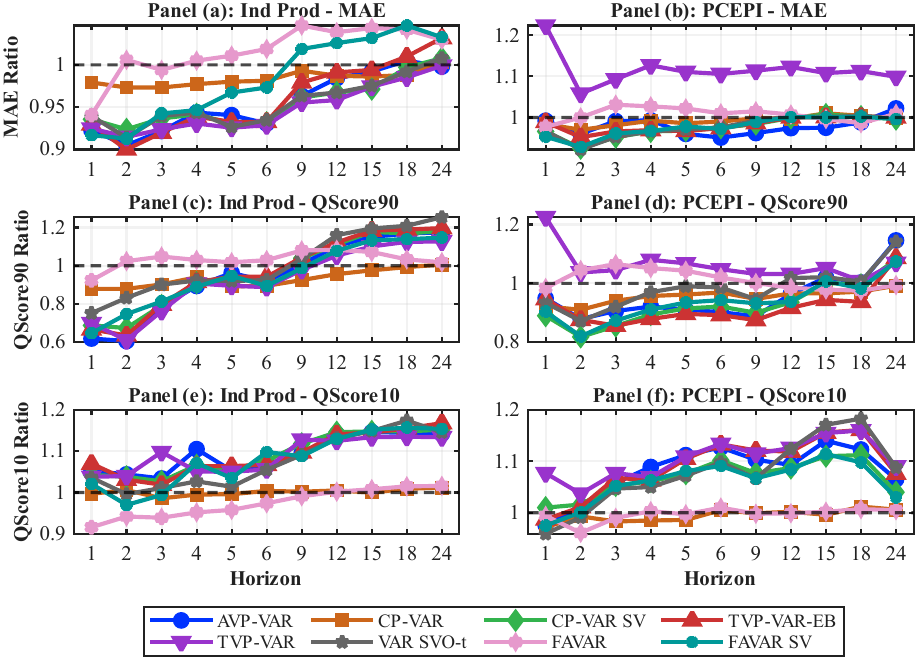}  
	\caption{Forecasting results using U.S. monthly data. The top row displays mean absolute forecast errors (MAE), while the middle and bottom rows show quantile scores for the upper (90th percentile) and lower (10th percentile) tails of each variable. All metrics are reported as ratios relative to a benchmark constant-parameter VAR($4$) estimated via least squares. A model is considered superior when its relative MAE, or relative quantile scores, are below one.} \label{fig:OOS_US_P4}
\end{figure}
\begin{table}[htbp!]
	\centering
	\caption{Forecasting Performance VAR(4): Mean Squared Prediction Error Relative to Benchmark} \label{tab:forecasting_mspe_US_p4}
	\scriptsize
	\begin{tabular}{@{}cS[table-format=1.3]S[table-format=1.3]S[table-format=1.3]S[table-format=1.3]S[table-format=1.3]S[table-format=1.3]S[table-format=1.3]S[table-format=1.3]@{}}
		\toprule
		{\textbf{h}} & {\textbf{AVP-VAR}} & {\textbf{CP-VAR}} & {\makecell{ \textbf{CP-VAR} \\ \textbf{SV}}} & {\makecell{ \textbf{{TVP-VAR}} \\ \textbf{EB}}} &{\textbf{TVP-VAR}} & {\makecell{ \textbf{VAR} \\ \textbf{SVOt}}} & {\textbf{FAVAR}} & {\makecell{ \textbf{FAVAR} \\ \textbf{SV}}}  \\
		\midrule
		\multicolumn{9}{c}{\textbf{Panel A: Industrial Production}} \\
		\midrule
		1 & \textbf{0.876} & \textbf{0.967} & \textbf{0.895} & \textbf{0.922} & \textbf{0.811} & \textbf{0.899} & \textbf{0.792}$^{\dag}$ & \textbf{0.845} \\
		2 & \textbf{0.744} & \textbf{0.918} & \textbf{0.798} & \textbf{0.754} & \textbf{0.680}$^{\dag}$ & \textbf{0.783} & \textbf{0.939} & \textbf{0.786} \\
		3 & \textbf{0.839} & \textbf{0.936} & \textbf{0.838} & \textbf{0.830} & \textbf{0.791}$^{\dag}$ & \textbf{0.845} & 1.045 & \textbf{0.888} \\
		4 & \textbf{0.884} & \textbf{0.955} & \textbf{0.888} & \textbf{0.882} & \textbf{0.877}$^{\dag}$ & \textbf{0.880} & 1.002 & \textbf{0.904} \\
		5 & \textbf{0.910}$^{\dag}$ & \textbf{0.966} & \textbf{0.921} & \textbf{0.911} & \textbf{0.920} & \textbf{0.917} & \textbf{0.987} & \textbf{0.936} \\
		6 & \textbf{0.900}$^{\dag}$ & \textbf{0.958} & \textbf{0.908} & \textbf{0.907} & \textbf{0.916} & \textbf{0.912} & \textbf{0.987} & \textbf{0.927} \\
		9 & \textbf{0.952} & \textbf{0.974} & \textbf{0.952} & \textbf{0.945}$^{\dag}$ & \textbf{0.955} & \textbf{0.948} & 1.003 & \textbf{0.972} \\
		12 & \textbf{0.977} & \textbf{0.987} & \textbf{0.978} & \textbf{0.982} & \textbf{0.970}$^{\dag}$ & \textbf{0.981} & 1.027 & 1.006 \\
		15 & \textbf{0.984} & \textbf{0.986} & \textbf{0.983} & \textbf{0.981}$^{\dag}$ & \textbf{0.986} & \textbf{0.987} & 1.025 & 1.007 \\
		18 & \textbf{0.986}$^{\dag}$ & \textbf{0.987} & \textbf{0.995} & \textbf{0.990} & \textbf{0.986} & \textbf{0.988} & 1.014 & 1.016 \\
		24 & \textbf{0.991}$^{\dag}$ & 1.002 & 1.002 & 1.011 & \textbf{0.998} & 1.006 & 1.017 & 1.022 \\
		\midrule
		\multicolumn{9}{c}{\textbf{Panel B: PCEPI}} \\
		\midrule
		1 & 1.003 & \textbf{0.966} & \textbf{0.941} & 1.046 & 1.411 & \textbf{0.942} & \textbf{0.954} & \textbf{0.923}$^{\dag}$ \\
		2 & \textbf{0.912} & \textbf{0.932} & \textbf{0.843} & \textbf{0.941} & \textbf{0.986} & \textbf{0.843} & \textbf{0.996} & \textbf{0.837}$^{\dag}$ \\
		3 & \textbf{0.948} & \textbf{0.938} & \textbf{0.863} & \textbf{0.895} & \textbf{0.999} & \textbf{0.860}$^{\dag}$ & 1.022 & \textbf{0.868} \\
		4 & \textbf{0.955} & \textbf{0.968} & \textbf{0.914} & \textbf{0.913} & 1.071 & \textbf{0.911}$^{\dag}$ & 1.037 & \textbf{0.925} \\
		5 & \textbf{0.925}$^{\dag}$ & \textbf{0.969} & \textbf{0.937} & \textbf{0.933} & 1.080 & \textbf{0.936} & 1.029 & \textbf{0.949} \\
		6 & \textbf{0.934}$^{\dag}$ & \textbf{0.990} & \textbf{0.969} & \textbf{0.972} & 1.104 & \textbf{0.972} & 1.025 & \textbf{0.969} \\
		9 & \textbf{0.930}$^{\dag}$ & \textbf{0.983} & \textbf{0.958} & \textbf{0.986} & 1.093 & \textbf{0.966} & 1.005 & \textbf{0.963} \\
		12 & \textbf{0.943}$^{\dag}$ & \textbf{0.989} & \textbf{0.980} & 1.007 & 1.102 & \textbf{0.985} & \textbf{0.993} & \textbf{0.973} \\
		15 & \textbf{0.959}$^{\dag}$ & \textbf{0.996} & \textbf{0.988} & \textbf{0.997} & 1.084 & \textbf{0.984} & \textbf{0.984} & \textbf{0.980} \\
		18 & \textbf{0.968}$^{\dag}$ & \textbf{0.994} & \textbf{0.982} & 1.003 & 1.088 & \textbf{0.981} & \textbf{0.976} & \textbf{0.979} \\
		24 & 1.034 & \textbf{0.995} & \textbf{0.997} & 1.014 & 1.093 & \textbf{0.994}$^{\dag}$ & 1.006 & \textbf{0.996} \\
		\bottomrule
	\end{tabular}	
	\vspace{0.5em}
	\par
	\begin{tabularx}{\textwidth}{X}
		{\footnotesize \textit{Notes:} The table reports mean squared prediction errors (MSPE) for each of the two forecasted variables, relative to a benchmark constant-parameter VAR($4$) estimated via least squares, for U.S. monthly data. Values below one (highlighted in bold) indicate superior forecasting performance compared to the benchmark. The forecast horizon is denoted by $h$ (in months), and the symbol $\dag$ marks the best-performing model at each horizon.}
	\end{tabularx}
\end{table}

\begin{table}[htbp!]
	\centering
	\caption{Forecasting Performance VAR(4): Quantile Scores 90\% Relative to Benchmark} \label{tab:forecasting_qs90_US_p4}
	\scriptsize
	\begin{tabular}{@{}cS[table-format=1.3]S[table-format=1.3]S[table-format=1.3]S[table-format=1.3]S[table-format=1.3]S[table-format=1.3]S[table-format=1.3]S[table-format=1.3]@{}}
		\toprule
		{\textbf{h}} & {\textbf{AVP-VAR}} & {\textbf{CP-VAR}} & {\makecell{ \textbf{CP-VAR} \\ \textbf{SV}}} & {\makecell{ \textbf{{TVP-VAR}} \\ \textbf{EB}}} &{\textbf{TVP-VAR}} & {\makecell{ \textbf{VAR} \\ \textbf{SVOt}}} & {\textbf{FAVAR}} & {\makecell{ \textbf{FAVAR} \\ \textbf{SV}}}  \\
		\midrule
		\multicolumn{9}{c}{\textbf{Panel A: Industrial Production}} \\
		\midrule
		1 & \textbf{0.619}$^{\dag}$ & \textbf{0.879} & \textbf{0.688} & \textbf{0.667} & \textbf{0.700} & \textbf{0.752} & \textbf{0.923} & \textbf{0.647} \\
		2 & \textbf{0.606}$^{\dag}$ & \textbf{0.879} & \textbf{0.674} & \textbf{0.635} & \textbf{0.610} & \textbf{0.832} & 1.024 & \textbf{0.747} \\
		3 & \textbf{0.784} & \textbf{0.902} & \textbf{0.792} & \textbf{0.797} & \textbf{0.763}$^{\dag}$ & \textbf{0.901} & 1.048 & \textbf{0.813} \\
		4 & \textbf{0.894} & \textbf{0.940} & \textbf{0.911} & \textbf{0.915} & \textbf{0.907} & \textbf{0.923} & 1.030 & \textbf{0.890}$^{\dag}$ \\
		5 & \textbf{0.962} & \textbf{0.892}$^{\dag}$ & \textbf{0.936} & \textbf{0.942} & \textbf{0.895} & \textbf{0.920} & 1.018 & \textbf{0.947} \\
		6 & \textbf{0.919} & \textbf{0.895} & \textbf{0.926} & \textbf{0.942} & \textbf{0.888}$^{\dag}$ & \textbf{0.925} & 1.029 & \textbf{0.894} \\
		9 & 1.008 & \textbf{0.925}$^{\dag}$ & 1.039 & 1.043 & \textbf{0.978} & 1.036 & 1.080 & \textbf{0.987} \\
		12 & 1.103 & \textbf{0.956}$^{\dag}$ & 1.112 & 1.122 & 1.053 & 1.160 & 1.083 & 1.078 \\
		15 & 1.151 & \textbf{0.976}$^{\dag}$ & 1.177 & 1.184 & 1.101 & 1.195 & 1.072 & 1.133 \\
		18 & 1.168 & \textbf{0.993}$^{\dag}$ & 1.169 & 1.189 & 1.125 & 1.209 & 1.033 & 1.139 \\
		24 & 1.175 & 1.006 & 1.181 & 1.197 & 1.127 & 1.254 & 1.018 & 1.147 \\
		\midrule
		\multicolumn{9}{c}{\textbf{Panel B: PCEPI}} \\
		\midrule
		1 & \textbf{0.949} & \textbf{0.922} & \textbf{0.890}$^{\dag}$ & \textbf{0.946} & 1.225 & \textbf{0.934} & \textbf{0.984} & \textbf{0.904} \\
		2 & \textbf{0.879} & \textbf{0.910} & \textbf{0.818}$^{\dag}$ & \textbf{0.876} & 1.037 & \textbf{0.872} & 1.045 & \textbf{0.820} \\
		3 & \textbf{0.906} & \textbf{0.941} & \textbf{0.855}$^{\dag}$ & \textbf{0.856} & 1.045 & \textbf{0.920} & 1.065 & \textbf{0.873} \\
		4 & \textbf{0.921} & \textbf{0.957} & \textbf{0.894} & \textbf{0.879}$^{\dag}$ & 1.080 & \textbf{0.970} & 1.052 & \textbf{0.911} \\
		5 & \textbf{0.911} & \textbf{0.962} & \textbf{0.915} & \textbf{0.896}$^{\dag}$ & 1.068 & \textbf{0.990} & 1.043 & \textbf{0.935} \\
		6 & \textbf{0.902} & \textbf{0.969} & \textbf{0.921} & \textbf{0.892}$^{\dag}$ & 1.048 & \textbf{0.986} & 1.020 & \textbf{0.943} \\
		9 & \textbf{0.888} & \textbf{0.947} & \textbf{0.900} & \textbf{0.875}$^{\dag}$ & 1.032 & \textbf{0.946} & 1.001 & \textbf{0.932} \\
		12 & \textbf{0.964} & \textbf{0.966} & \textbf{0.936} & \textbf{0.917}$^{\dag}$ & 1.032 & 1.018 & \textbf{0.985} & \textbf{0.937} \\
		15 & 1.020 & \textbf{0.966} & 1.002 & \textbf{0.944}$^{\dag}$ & 1.052 & 1.021 & \textbf{0.979} & 1.007 \\
		18 & 1.003 & \textbf{0.972} & \textbf{0.995} & \textbf{0.937}$^{\dag}$ & 1.008 & 1.006 & \textbf{0.975} & \textbf{0.983} \\
		24 & 1.146 & \textbf{0.992}$^{\dag}$ & 1.081 & 1.087 & 1.070 & 1.142 & \textbf{0.994} & 1.074 \\
		\bottomrule
	\end{tabular}	
	\vspace{0.5em}
	\par
	\begin{tabularx}{\textwidth}{X}
		{\footnotesize \textit{Notes:} The table reports quantile scores 90\% (QS90) for each of the two forecasted variables, relative to a benchmark constant-parameter VAR($4$) estimated via least squares, for U.S. monthly data. Values below one (highlighted in bold) indicate superior forecasting performance compared to the benchmark. The forecast horizon is denoted by $h$ (in months), and the symbol $\dag$ marks the best-performing model at each horizon.}
	\end{tabularx}
\end{table}

\begin{table}[htbp!]
	\centering
	\caption{Forecasting Performance VAR(4): Quantile Scores 10\% Relative to Benchmark} \label{tab:forecasting_qs10_US_p4}
	\scriptsize
	\begin{tabular}{@{}cS[table-format=1.3]S[table-format=1.3]S[table-format=1.3]S[table-format=1.3]S[table-format=1.3]S[table-format=1.3]S[table-format=1.3]S[table-format=1.3]@{}}
		\toprule
		{\textbf{h}} & {\textbf{AVP-VAR}} & {\textbf{CP-VAR}} & {\makecell{ \textbf{CP-VAR} \\ \textbf{SV}}} & {\makecell{ \textbf{{TVP-VAR}} \\ \textbf{EB}}} &{\textbf{TVP-VAR}} & {\makecell{ \textbf{VAR} \\ \textbf{SVOt}}} & {\textbf{FAVAR}} & {\makecell{ \textbf{FAVAR} \\ \textbf{SV}}}  \\
		\midrule
		\multicolumn{9}{c}{\textbf{Panel A: GDP}} \\
		\midrule
		1 & 1.052 & \textbf{0.997} & 1.043 & 1.070 & 1.038 & 1.036 & \textbf{0.916}$^{\dag}$ & 1.021 \\
		2 & 1.044 & 1.006 & 1.038 & 1.031 & 1.040 & \textbf{0.996} & \textbf{0.942}$^{\dag}$ & \textbf{0.970} \\
		3 & 1.034 & \textbf{0.986} & 1.027 & 1.019 & 1.098 & 1.011 & \textbf{0.939}$^{\dag}$ & \textbf{0.994} \\
		4 & 1.105 & \textbf{0.995} & 1.066 & 1.064 & 1.051 & 1.027 & \textbf{0.952}$^{\dag}$ & 1.071 \\
		5 & 1.053 & \textbf{0.996} & 1.054 & 1.064 & 1.050 & 1.014 & \textbf{0.958}$^{\dag}$ & 1.035 \\
		6 & 1.067 & 1.004 & 1.077 & 1.067 & 1.056 & 1.055 & \textbf{0.974}$^{\dag}$ & 1.098 \\
		9 & 1.095 & 1.002 & 1.118 & 1.097 & 1.128 & 1.091 & \textbf{0.991}$^{\dag}$ & 1.088 \\
		12 & 1.141 & 1.005 & 1.146 & 1.141 & 1.125 & 1.131 & 1.002 & 1.130 \\
		15 & 1.146 & 1.000 & 1.148 & 1.145 & 1.134 & 1.149 & 1.008 & 1.152 \\
		18 & 1.145 & 1.009 & 1.148 & 1.155 & 1.135 & 1.174 & 1.015 & 1.157 \\
		24 & 1.142 & 1.012 & 1.152 & 1.168 & 1.134 & 1.146 & 1.016 & 1.154 \\
		\midrule
		\multicolumn{9}{c}{\textbf{Panel B: HICP}} \\
		\midrule
		1 & \textbf{0.976} & \textbf{0.972} & 1.010 & \textbf{0.987} & 1.077 & \textbf{0.959}$^{\dag}$ & \textbf{0.994} & \textbf{0.975} \\
		2 & 1.010 & \textbf{0.993} & 1.016 & 1.011 & 1.037 & \textbf{0.991} & \textbf{0.961}$^{\dag}$ & 1.001 \\
		3 & 1.058 & \textbf{0.983}$^{\dag}$ & 1.053 & 1.068 & 1.077 & 1.047 & \textbf{0.990} & 1.052 \\
		4 & 1.089 & \textbf{0.985}$^{\dag}$ & 1.061 & 1.063 & 1.069 & 1.050 & 1.003 & 1.063 \\
		5 & 1.112 & \textbf{0.986}$^{\dag}$ & 1.076 & 1.105 & 1.109 & 1.072 & \textbf{0.997} & 1.081 \\
		6 & 1.127 & 1.005 & 1.102 & 1.133 & 1.134 & 1.098 & 1.010 & 1.091 \\
		9 & 1.103 & \textbf{0.999} & 1.077 & 1.121 & 1.115 & 1.067 & \textbf{0.998}$^{\dag}$ & 1.067 \\
		12 & 1.093 & 1.002 & 1.087 & 1.119 & 1.126 & 1.124 & \textbf{0.999}$^{\dag}$ & 1.084 \\
		15 & 1.140 & \textbf{0.996}$^{\dag}$ & 1.110 & 1.156 & 1.155 & 1.171 & 1.002 & 1.114 \\
		18 & 1.123 & 1.012 & 1.112 & 1.161 & 1.160 & 1.183 & 1.007 & 1.098 \\
		24 & 1.065 & 1.006 & 1.040 & 1.076 & 1.090 & 1.088 & 1.005 & 1.030 \\
		\bottomrule
	\end{tabular}	
	\vspace{0.5em}
	\par
	\begin{tabularx}{\textwidth}{X}
		{\footnotesize \textit{Notes:} The table reports quantile scores 10\% (QS10) for each of the two forecasted variables, relative to a benchmark constant-parameter VAR($4$) estimated via least squares, for U.S. monthly data. Values below one (highlighted in bold) indicate superior forecasting performance compared to the benchmark. The forecast horizon is denoted by $h$ (in months), and the symbol $\dag$ marks the best-performing model at each horizon.}
	\end{tabularx}
\end{table}

\FloatBarrier

\subsubsection{Euro Area quarterly data: VAR(1)}


\begin{figure}[htbp!]  
	\centering
	\includegraphics[width=0.95\textwidth]{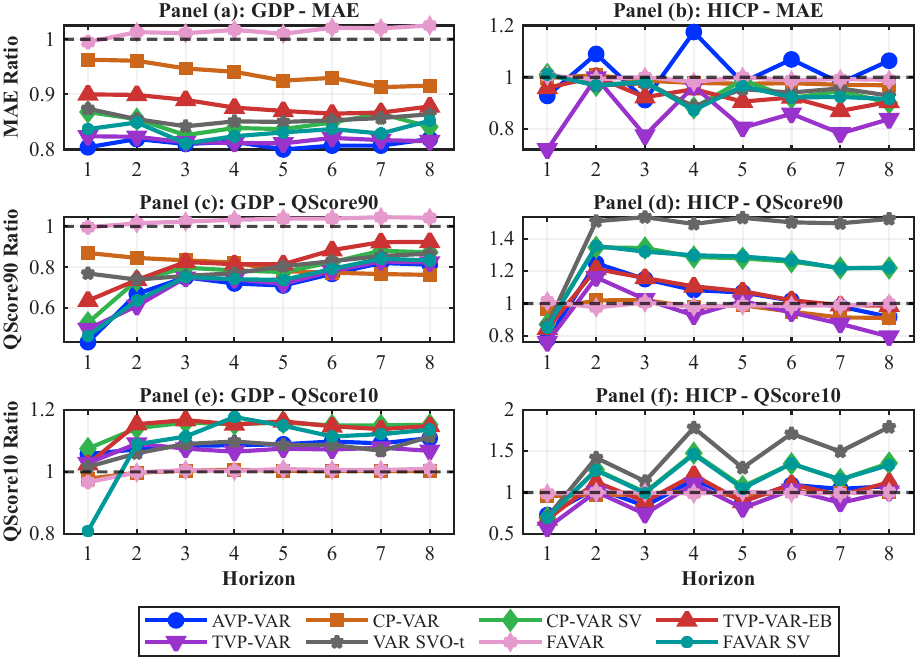}  
	\caption{Forecasting results using Euro Area quarterly data. The top row displays mean absolute forecast errors (MAE), while the middle and bottom rows show quantile scores for the upper (90th percentile) and lower (10th percentile) tails of each variable. All metrics are reported as ratios relative to a benchmark constant-parameter VAR($1$) estimated via least squares. A model is considered superior when its relative MAE, or relative quantile scores, are below one.} \label{fig:oos_EURO}
\end{figure}

\begin{table}[htbp!]
	\centering
	\caption{Forecasting Performance VAR(1): Mean Squared Prediction Error Relative to Benchmark} \label{tab:forecasting_mspe_EURO_p1}
	\scriptsize
	\begin{tabular}{@{}cS[table-format=1.3]S[table-format=1.3]S[table-format=1.3]S[table-format=1.3]S[table-format=1.3]S[table-format=1.3]S[table-format=1.3]S[table-format=1.3]@{}}
		\toprule
		{\textbf{h}} & {\textbf{AVP-VAR}} & {\textbf{CP-VAR}} & {\makecell{ \textbf{CP-VAR} \\ \textbf{SV}}} & {\makecell{ \textbf{{TVP-VAR}} \\ \textbf{EB}}} &{\textbf{TVP-VAR}} & {\makecell{ \textbf{VAR} \\ \textbf{SVOt}}} & {\textbf{FAVAR}} & {\makecell{ \textbf{FAVAR} \\ \textbf{SV}}}  \\
		\midrule
		\multicolumn{9}{c}{\textbf{Panel A: GDP}} \\
		\midrule
		1 & \textbf{0.425}$^{\dag}$ & \textbf{0.919} & \textbf{0.590} & \textbf{0.547} & \textbf{0.440} & \textbf{0.564} & \textbf{0.997} & \textbf{0.545} \\
		2 & \textbf{0.636}$^{\dag}$ & \textbf{0.888} & \textbf{0.668} & \textbf{0.682} & \textbf{0.645} & \textbf{0.663} & 1.034 & \textbf{0.683} \\
		3 & \textbf{0.652}$^{\dag}$ & \textbf{0.850} & \textbf{0.655} & \textbf{0.673} & \textbf{0.652} & \textbf{0.658} & 1.041 & \textbf{0.652} \\
		4 & \textbf{0.606} & \textbf{0.813} & \textbf{0.612} & \textbf{0.623} & \textbf{0.604}$^{\dag}$ & \textbf{0.617} & 1.057 & \textbf{0.611} \\
		5 & \textbf{0.625} & \textbf{0.789} & \textbf{0.634} & \textbf{0.639} & \textbf{0.623}$^{\dag}$ & \textbf{0.634} & 1.064 & \textbf{0.631} \\
		6 & \textbf{0.657}$^{\dag}$ & \textbf{0.788} & \textbf{0.666} & \textbf{0.671} & \textbf{0.660} & \textbf{0.667} & 1.068 & \textbf{0.665} \\
		7 & \textbf{0.672}$^{\dag}$ & \textbf{0.772} & \textbf{0.686} & \textbf{0.687} & \textbf{0.674} & \textbf{0.685} & 1.076 & \textbf{0.678} \\
		8 & \textbf{0.670} & \textbf{0.759} & \textbf{0.674} & \textbf{0.681} & \textbf{0.665}$^{\dag}$ & \textbf{0.681} & 1.085 & \textbf{0.677} \\
		\midrule
		\multicolumn{9}{c}{\textbf{Panel B: HICP}} \\
		\midrule
		1 & 1.263 & \textbf{0.987} & 1.028 & \textbf{0.928} & \textbf{0.705}$^{\dag}$ & 1.033 & \textbf{0.998} & 1.021 \\
		2 & 1.211 & 1.002 & \textbf{0.953} & \textbf{0.999} & 1.121 & \textbf{0.953} & \textbf{0.980} & \textbf{0.946}$^{\dag}$ \\
		3 & \textbf{0.921} & \textbf{0.984} & 1.015 & \textbf{0.886} & \textbf{0.732}$^{\dag}$ & \textbf{0.998} & 1.003 & 1.009 \\
		4 & 1.337 & \textbf{0.951} & \textbf{0.793} & \textbf{0.938} & 1.025 & \textbf{0.808} & \textbf{0.971} & \textbf{0.773}$^{\dag}$ \\
		5 & 1.025 & \textbf{0.955} & \textbf{0.973} & \textbf{0.840} & \textbf{0.765}$^{\dag}$ & \textbf{0.942} & \textbf{0.993} & \textbf{0.933} \\
		6 & 1.193 & \textbf{0.932} & \textbf{0.841}$^{\dag}$ & \textbf{0.865} & \textbf{0.873} & \textbf{0.873} & \textbf{0.980} & \textbf{0.843} \\
		7 & 1.030 & \textbf{0.921} & \textbf{0.882} & \textbf{0.795} & \textbf{0.737}$^{\dag}$ & \textbf{0.905} & \textbf{0.988} & \textbf{0.838} \\
		8 & 1.121 & \textbf{0.888} & \textbf{0.761}$^{\dag}$ & \textbf{0.796} & \textbf{0.786} & \textbf{0.805} & \textbf{0.984} & \textbf{0.781} \\
		\bottomrule
	\end{tabular}	
	\vspace{0.5em}
	\par
	\begin{tabularx}{\textwidth}{X}
		{\footnotesize \textit{Notes:} The table reports mean squared prediction errors (MSPE) for each of the two forecasted variables, relative to a benchmark constant-parameter VAR($1$) estimated via least squares, for Euro Area quarterly data. Values below one (highlighted in bold) indicate superior forecasting performance compared to the benchmark. The forecast horizon is denoted by $h$ (in quarters), and the symbol $\dag$ marks the best-performing model at each horizon.}
	\end{tabularx}
\end{table}

\begin{table}[htbp!]
	\centering
	\caption{Forecasting Performance VAR(1): Quantile Scores 90\% Relative to Benchmark} \label{tab:forecasting_qs90_EURO_p1}
	\scriptsize
	\begin{tabular}{@{}cS[table-format=1.3]S[table-format=1.3]S[table-format=1.3]S[table-format=1.3]S[table-format=1.3]S[table-format=1.3]S[table-format=1.3]S[table-format=1.3]@{}}
		\toprule
		{\textbf{h}} & {\textbf{AVP-VAR}} & {\textbf{CP-VAR}} & {\makecell{ \textbf{CP-VAR} \\ \textbf{SV}}} & {\makecell{ \textbf{{TVP-VAR}} \\ \textbf{EB}}} &{\textbf{TVP-VAR}} & {\makecell{ \textbf{VAR} \\ \textbf{SVOt}}} & {\textbf{FAVAR}} & {\makecell{ \textbf{FAVAR} \\ \textbf{SV}}}  \\
		\midrule
		\multicolumn{9}{c}{\textbf{Panel A: GDP}} \\
		\midrule
		1 & \textbf{0.434}$^{\dag}$ & \textbf{0.870} & \textbf{0.529} & \textbf{0.637} & \textbf{0.498} & \textbf{0.770} & \textbf{0.996} & \textbf{0.464} \\
		2 & \textbf{0.670} & \textbf{0.846} & \textbf{0.736} & \textbf{0.738} & \textbf{0.611}$^{\dag}$ & \textbf{0.740} & 1.015 & \textbf{0.638} \\
		3 & \textbf{0.751} & \textbf{0.834} & \textbf{0.798} & \textbf{0.826} & \textbf{0.749}$^{\dag}$ & \textbf{0.758} & 1.024 & \textbf{0.753} \\
		4 & \textbf{0.721}$^{\dag}$ & \textbf{0.823} & \textbf{0.785} & \textbf{0.815} & \textbf{0.741} & \textbf{0.771} & 1.032 & \textbf{0.742} \\
		5 & \textbf{0.711}$^{\dag}$ & \textbf{0.781} & \textbf{0.777} & \textbf{0.815} & \textbf{0.719} & \textbf{0.805} & 1.038 & \textbf{0.738} \\
		6 & \textbf{0.768}$^{\dag}$ & \textbf{0.774} & \textbf{0.825} & \textbf{0.884} & \textbf{0.781} & \textbf{0.830} & 1.038 & \textbf{0.792} \\
		7 & \textbf{0.821} & \textbf{0.767}$^{\dag}$ & \textbf{0.881} & \textbf{0.923} & \textbf{0.829} & \textbf{0.855} & 1.045 & \textbf{0.842} \\
		8 & \textbf{0.813} & \textbf{0.763}$^{\dag}$ & \textbf{0.873} & \textbf{0.924} & \textbf{0.823} & \textbf{0.867} & 1.042 & \textbf{0.834} \\
		\midrule
		\multicolumn{9}{c}{\textbf{Panel B: HICP}} \\
		\midrule
		1 & \textbf{0.777} & \textbf{0.965} & \textbf{0.871} & \textbf{0.844} & \textbf{0.760}$^{\dag}$ & \textbf{0.865} & 1.013 & \textbf{0.852} \\
		2 & 1.249 & 1.019 & 1.348 & 1.214 & 1.162 & 1.510 & \textbf{0.975}$^{\dag}$ & 1.357 \\
		3 & 1.151 & 1.021 & 1.344 & 1.158 & 1.026 & 1.535 & 1.005$^{\dag}$ & 1.321 \\
		4 & 1.083 & \textbf{0.976} & 1.288 & 1.106 & \textbf{0.926}$^{\dag}$ & 1.493 & \textbf{0.981} & 1.299 \\
		5 & 1.068 & \textbf{0.989} & 1.280 & 1.076 & 1.015 & 1.531 & \textbf{0.988}$^{\dag}$ & 1.291 \\
		6 & 1.018 & \textbf{0.949} & 1.259 & 1.020 & \textbf{0.944}$^{\dag}$ & 1.503 & \textbf{0.986} & 1.268 \\
		7 & \textbf{0.981} & \textbf{0.913} & 1.220 & \textbf{0.989} & \textbf{0.874}$^{\dag}$ & 1.497 & \textbf{0.988} & 1.218 \\
		8 & \textbf{0.917} & \textbf{0.911} & 1.220 & \textbf{0.987} & \textbf{0.795}$^{\dag}$ & 1.524 & \textbf{0.998} & 1.222 \\
		\bottomrule
	\end{tabular}	
	\vspace{0.5em}
	\par
	\begin{tabularx}{\textwidth}{X}
		{\footnotesize \textit{Notes:} The table reports quantile scores 90\% (QS90) for each of the two forecasted variables, relative to a benchmark constant-parameter VAR($1$) estimated via least squares. Values below one (highlighted in bold) indicate superior forecasting performance compared to the benchmark. The forecast horizon is denoted by $h$ (in quarters), and the symbol $\dag$ marks the best-performing model at each horizon.}
	\end{tabularx}
\end{table}

\begin{table}[htbp!]
	\centering
	\caption{Forecasting Performance VAR(1): Quantile Scores 10\% Relative to Benchmark} \label{tab:forecasting_qs10_EURO_p1}
	\scriptsize
	\begin{tabular}{@{}cS[table-format=1.3]S[table-format=1.3]S[table-format=1.3]S[table-format=1.3]S[table-format=1.3]S[table-format=1.3]S[table-format=1.3]S[table-format=1.3]@{}}
		\toprule
		{\textbf{h}} & {\textbf{AVP-VAR}} & {\textbf{CP-VAR}} & {\makecell{ \textbf{CP-VAR} \\ \textbf{SV}}} & {\makecell{ \textbf{{TVP-VAR}} \\ \textbf{EB}}} &{\textbf{TVP-VAR}} & {\makecell{ \textbf{VAR} \\ \textbf{SVOt}}} & {\textbf{FAVAR}} & {\makecell{ \textbf{FAVAR} \\ \textbf{SV}}}  \\
		\midrule
		\multicolumn{9}{c}{\textbf{Panel A: GDP}} \\
		\midrule
		1 & 1.057 & \textbf{0.978} & 1.075 & 1.025 & 1.028 & 1.016 & \textbf{0.967} & \textbf{0.809}$^{\dag}$ \\
		2 & 1.073 & \textbf{0.997}$^{\dag}$ & 1.141 & 1.154 & 1.091 & 1.060 & \textbf{0.998} & 1.089 \\
		3 & 1.084 & 1.004 & 1.159 & 1.167 & 1.075 & 1.090 & 1.006 & 1.114 \\
		4 & 1.086 & 1.007 & 1.154 & 1.153 & 1.065 & 1.098 & 1.003 & 1.177 \\
		5 & 1.089 & 1.003 & 1.159 & 1.163 & 1.075 & 1.087 & 1.009 & 1.149 \\
		6 & 1.098 & 1.004 & 1.150 & 1.147 & 1.072 & 1.087 & 1.006 & 1.114 \\
		7 & 1.092 & 1.003 & 1.150 & 1.138 & 1.078 & 1.068 & 1.006 & 1.122 \\
		8 & 1.109 & 1.003 & 1.152 & 1.148 & 1.068 & 1.113 & 1.010 & 1.136 \\
		\midrule
		\multicolumn{9}{c}{\textbf{Panel B: HICP}} \\
		\midrule
		1 & \textbf{0.727} & \textbf{0.956} & \textbf{0.694} & \textbf{0.676} & \textbf{0.573}$^{\dag}$ & \textbf{0.703} & \textbf{0.987} & \textbf{0.698} \\
		2 & 1.053 & \textbf{0.972}$^{\dag}$ & 1.276 & 1.125 & 1.009 & 1.420 & 1.003 & 1.269 \\
		3 & \textbf{0.837} & \textbf{0.975} & 1.003 & \textbf{0.892} & \textbf{0.740}$^{\dag}$ & 1.136 & \textbf{0.999} & \textbf{0.994} \\
		4 & 1.154 & 1.025 & 1.476 & 1.211 & 1.105 & 1.784 & \textbf{0.987}$^{\dag}$ & 1.463 \\
		5 & \textbf{0.902} & \textbf{0.947} & 1.072 & \textbf{0.887} & \textbf{0.811}$^{\dag}$ & 1.298 & 1.009 & 1.050 \\
		6 & 1.092 & 1.016 & 1.348 & 1.102 & 1.032 & 1.713 & \textbf{0.993}$^{\dag}$ & 1.350 \\
		7 & 1.044 & \textbf{0.995} & 1.156 & \textbf{0.965} & \textbf{0.875}$^{\dag}$ & 1.496 & \textbf{0.991} & 1.157 \\
		8 & 1.077 & 1.003 & 1.356 & 1.124 & 1.007 & 1.792 & 1.004 & 1.333 \\
		\bottomrule
	\end{tabular}	
	\vspace{0.5em}
	\par
	\begin{tabularx}{\textwidth}{X}
		{\footnotesize \textit{Notes:} The table reports quantile scores 10\% (QS10) for each of the two forecasted variables, relative to a benchmark constant-parameter VAR($1$) estimated via least squares, for Euro Area quarterly data. Values below one (highlighted in bold) indicate superior forecasting performance compared to the benchmark. The forecast horizon is denoted by $h$ (in quarters), and the symbol $\dag$ marks the best-performing model at each horizon.}
	\end{tabularx}
\end{table}

\FloatBarrier

\subsubsection{Euro Area quarterly data: VAR(2)}

\begin{figure}[htbp!]  
	\centering
	\includegraphics[width=0.95\textwidth]{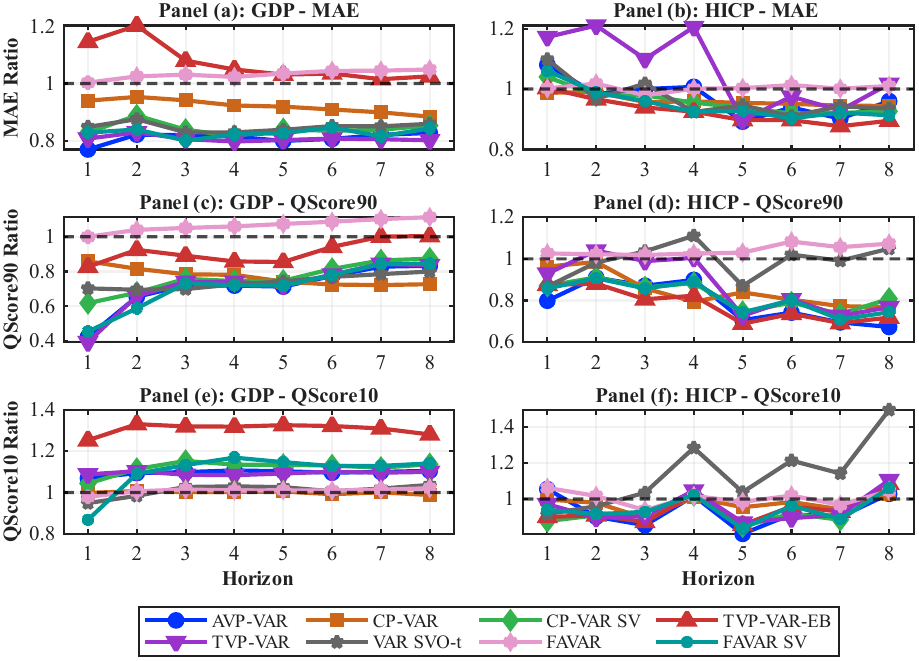}  
	\caption{Forecasting results using Euro Area quarterly data. The top row displays mean absolute forecast errors (MAE), while the middle and bottom rows show quantile scores for the upper (90th percentile) and lower (10th percentile) tails of each variable. All metrics are reported as ratios relative to a benchmark constant-parameter VAR($2$) estimated via least squares. A model is considered superior when its relative MAE, or relative quantile scores, are below one.} \label{fig:oos_EURO_p2}
\end{figure}

\begin{table}[htbp!]
	\centering
	\caption{Forecasting Performance VAR(2): Mean Squared Prediction Error Relative to Benchmark} \label{tab:forecasting_mspe_EURO_p2}
	\scriptsize
	\begin{tabular}{@{}cS[table-format=1.3]S[table-format=1.3]S[table-format=1.3]S[table-format=1.3]S[table-format=1.3]S[table-format=1.3]S[table-format=1.3]S[table-format=1.3]@{}}
		\toprule
		{\textbf{h}} & {\textbf{AVP-VAR}} & {\textbf{CP-VAR}} & {\makecell{ \textbf{CP-VAR} \\ \textbf{SV}}} & {\makecell{ \textbf{{TVP-VAR}} \\ \textbf{EB}}} &{\textbf{TVP-VAR}} & {\makecell{ \textbf{VAR} \\ \textbf{SVOt}}} & {\textbf{FAVAR}} & {\makecell{ \textbf{FAVAR} \\ \textbf{SV}}}  \\
		\midrule
		\multicolumn{9}{c}{\textbf{Panel A: GDP}} \\
		\midrule
		1 & \textbf{0.391}$^{\dag}$ & \textbf{0.896} & \textbf{0.536} & \textbf{0.620} & \textbf{0.435} & \textbf{0.547} & 1.036 & \textbf{0.521} \\
		2 & \textbf{0.598}$^{\dag}$ & \textbf{0.857} & \textbf{0.665} & \textbf{0.813} & \textbf{0.605} & \textbf{0.661} & 1.074 & \textbf{0.634} \\
		3 & \textbf{0.613} & \textbf{0.803} & \textbf{0.624} & \textbf{0.708} & \textbf{0.613} & \textbf{0.619} & 1.111 & \textbf{0.609}$^{\dag}$ \\
		4 & \textbf{0.579} & \textbf{0.757} & \textbf{0.592} & \textbf{0.642} & \textbf{0.571}$^{\dag}$ & \textbf{0.587} & 1.131 & \textbf{0.587} \\
		5 & \textbf{0.602} & \textbf{0.740} & \textbf{0.614} & \textbf{0.656} & \textbf{0.599}$^{\dag}$ & \textbf{0.616} & 1.159 & \textbf{0.610} \\
		6 & \textbf{0.649} & \textbf{0.746} & \textbf{0.653} & \textbf{0.699} & \textbf{0.645}$^{\dag}$ & \textbf{0.659} & 1.181 & \textbf{0.656} \\
		7 & \textbf{0.672} & \textbf{0.742} & \textbf{0.677} & \textbf{0.713} & \textbf{0.667}$^{\dag}$ & \textbf{0.679} & 1.207 & \textbf{0.672} \\
		8 & \textbf{0.673} & \textbf{0.727} & \textbf{0.680} & \textbf{0.709} & \textbf{0.667}$^{\dag}$ & \textbf{0.681} & 1.245 & \textbf{0.679} \\
		\midrule
		\multicolumn{9}{c}{\textbf{Panel B: HICP}} \\
		\midrule
		1 & 1.682 & \textbf{0.981}$^{\dag}$ & 1.024 & \textbf{0.997} & 1.309 & 1.125 & 1.030 & 1.122 \\
		2 & \textbf{0.971} & \textbf{0.987} & \textbf{0.970} & \textbf{0.955}$^{\dag}$ & 1.383 & \textbf{0.957} & 1.027 & \textbf{0.996} \\
		3 & 1.018 & \textbf{0.894} & \textbf{0.922} & \textbf{0.838}$^{\dag}$ & 1.106 & \textbf{0.989} & 1.011 & \textbf{0.887} \\
		4 & \textbf{0.925} & \textbf{0.841} & \textbf{0.813} & \textbf{0.786} & 1.201 & \textbf{0.781}$^{\dag}$ & 1.096 & \textbf{0.784} \\
		5 & \textbf{0.753} & \textbf{0.863} & \textbf{0.824} & \textbf{0.751} & \textbf{0.740}$^{\dag}$ & \textbf{0.835} & 1.066 & \textbf{0.813} \\
		6 & \textbf{0.756} & \textbf{0.801} & \textbf{0.738} & \textbf{0.711}$^{\dag}$ & \textbf{0.782} & \textbf{0.716} & 1.162 & \textbf{0.715} \\
		7 & \textbf{0.672} & \textbf{0.750} & \textbf{0.688} & \textbf{0.629}$^{\dag}$ & \textbf{0.655} & \textbf{0.709} & 1.167 & \textbf{0.674} \\
		8 & \textbf{0.603} & \textbf{0.672} & \textbf{0.589} & \textbf{0.562} & \textbf{0.661} & \textbf{0.605} & 1.280 & \textbf{0.561}$^{\dag}$ \\
		\bottomrule
	\end{tabular}	
	\vspace{0.5em}
	\par
	\begin{tabularx}{\textwidth}{X}
		{\footnotesize \textit{Notes:} The table reports mean squared prediction errors (MSPE) for each of the two forecasted variables, relative to a benchmark constant-parameter VAR($2$) estimated via least squares, for Euro Area quarterly data. Values below one (highlighted in bold) indicate superior forecasting performance compared to the benchmark. The forecast horizon is denoted by $h$ (in quarters), and the symbol $\dag$ marks the best-performing model at each horizon.}
	\end{tabularx}
\end{table}

\begin{table}[htbp!]
	\centering
	\caption{Forecasting Performance VAR(2): Quantile Scores 90\% Relative to Benchmark} \label{tab:forecasting_qs90_EURO_p2}
	\scriptsize
	\begin{tabular}{@{}cS[table-format=1.3]S[table-format=1.3]S[table-format=1.3]S[table-format=1.3]S[table-format=1.3]S[table-format=1.3]S[table-format=1.3]S[table-format=1.3]@{}}
		\toprule
		{\textbf{h}} & {\textbf{AVP-VAR}} & {\textbf{CP-VAR}} & {\makecell{ \textbf{CP-VAR} \\ \textbf{SV}}} & {\makecell{ \textbf{{TVP-VAR}} \\ \textbf{EB}}} &{\textbf{TVP-VAR}} & {\makecell{ \textbf{VAR} \\ \textbf{SVOt}}} & {\textbf{FAVAR}} & {\makecell{ \textbf{FAVAR} \\ \textbf{SV}}}  \\
		\midrule
		\multicolumn{9}{c}{\textbf{Panel A: GDP}} \\
		\midrule
		1 & \textbf{0.423} & \textbf{0.859} & \textbf{0.618} & \textbf{0.827} & \textbf{0.393}$^{\dag}$ & \textbf{0.703} & 1.000 & \textbf{0.454} \\
		2 & \textbf{0.654} & \textbf{0.816} & \textbf{0.678} & \textbf{0.925} & \textbf{0.659} & \textbf{0.695} & 1.040 & \textbf{0.588}$^{\dag}$ \\
		3 & \textbf{0.722} & \textbf{0.784} & \textbf{0.758} & \textbf{0.891} & \textbf{0.740} & \textbf{0.699}$^{\dag}$ & 1.051 & \textbf{0.732} \\
		4 & \textbf{0.719}$^{\dag}$ & \textbf{0.781} & \textbf{0.745} & \textbf{0.858} & \textbf{0.741} & \textbf{0.726} & 1.060 & \textbf{0.721} \\
		5 & \textbf{0.713}$^{\dag}$ & \textbf{0.742} & \textbf{0.747} & \textbf{0.855} & \textbf{0.719} & \textbf{0.743} & 1.074 & \textbf{0.718} \\
		6 & \textbf{0.774} & \textbf{0.725}$^{\dag}$ & \textbf{0.815} & \textbf{0.942} & \textbf{0.782} & \textbf{0.769} & 1.087 & \textbf{0.780} \\
		7 & \textbf{0.828} & \textbf{0.722}$^{\dag}$ & \textbf{0.864} & 1.001 & \textbf{0.844} & \textbf{0.785} & 1.102 & \textbf{0.843} \\
		8 & \textbf{0.833} & \textbf{0.727}$^{\dag}$ & \textbf{0.874} & 1.004 & \textbf{0.836} & \textbf{0.800} & 1.113 & \textbf{0.845} \\
		\midrule
		\multicolumn{9}{c}{\textbf{Panel B: HICP}} \\
		\midrule
		1 & \textbf{0.799}$^{\dag}$ & \textbf{0.963} & \textbf{0.871} & \textbf{0.875} & \textbf{0.929} & \textbf{0.867} & 1.026 & \textbf{0.858} \\
		2 & \textbf{0.910} & \textbf{0.985} & \textbf{0.911} & \textbf{0.881}$^{\dag}$ & 1.040 & \textbf{0.982} & 1.022 & \textbf{0.907} \\
		3 & \textbf{0.871} & \textbf{0.858} & \textbf{0.863} & \textbf{0.806}$^{\dag}$ & \textbf{0.989} & 1.036 & 1.019 & \textbf{0.858} \\
		4 & \textbf{0.899} & \textbf{0.794}$^{\dag}$ & \textbf{0.884} & \textbf{0.823} & 1.005 & 1.111 & 1.025 & \textbf{0.890} \\
		5 & \textbf{0.706} & \textbf{0.839} & \textbf{0.745} & \textbf{0.689}$^{\dag}$ & \textbf{0.727} & \textbf{0.870} & 1.030 & \textbf{0.743} \\
		6 & \textbf{0.742} & \textbf{0.804} & \textbf{0.790} & \textbf{0.736}$^{\dag}$ & \textbf{0.806} & 1.020 & 1.083 & \textbf{0.801} \\
		7 & \textbf{0.695} & \textbf{0.773} & \textbf{0.735} & \textbf{0.692}$^{\dag}$ & \textbf{0.726} & \textbf{0.991} & 1.056 & \textbf{0.709} \\
		8 & \textbf{0.673}$^{\dag}$ & \textbf{0.768} & \textbf{0.808} & \textbf{0.718} & \textbf{0.768} & 1.049 & 1.073 & \textbf{0.744} \\
		\bottomrule
	\end{tabular}	
	\vspace{0.5em}
	\par
	\begin{tabularx}{\textwidth}{X}
		{\footnotesize \textit{Notes:} The table reports quantile scores 90\% (QS90) for each of the two forecasted variables, relative to a benchmark constant-parameter VAR($2$) estimated via least squares, for Euro Area quarterly data. Values below one (highlighted in bold) indicate superior forecasting performance compared to the benchmark. The forecast horizon is denoted by $h$ (in quarters), and the symbol $\dag$ marks the best-performing model at each horizon.}
	\end{tabularx}
\end{table}

\begin{table}[htbp!]
	\centering
	\caption{Forecasting Performance VAR(2): Quantile Scores 10\% Relative to Benchmark} \label{tab:forecasting_qs10_EURO_p2}
	\scriptsize
	\begin{tabular}{@{}cS[table-format=1.3]S[table-format=1.3]S[table-format=1.3]S[table-format=1.3]S[table-format=1.3]S[table-format=1.3]S[table-format=1.3]S[table-format=1.3]@{}}
		\toprule
		{\textbf{h}} & {\textbf{AVP-VAR}} & {\textbf{CP-VAR}} & {\makecell{ \textbf{CP-VAR} \\ \textbf{SV}}} & {\makecell{ \textbf{{TVP-VAR}} \\ \textbf{EB}}} &{\textbf{TVP-VAR}} & {\makecell{ \textbf{VAR} \\ \textbf{SVOt}}} & {\textbf{FAVAR}} & {\makecell{ \textbf{FAVAR} \\ \textbf{SV}}}  \\
		\midrule
		\multicolumn{9}{c}{\textbf{Panel A: GDP}} \\
		\midrule
		1 & 1.069 & \textbf{0.998} & 1.044 & 1.253 & 1.088 & \textbf{0.949} & \textbf{0.979} & \textbf{0.868}$^{\dag}$ \\
		2 & 1.094 & 1.007 & 1.114 & 1.331 & 1.103 & \textbf{0.984}$^{\dag}$ & 1.006 & 1.091 \\
		3 & 1.099 & 1.002 & 1.153 & 1.320 & 1.086 & 1.027 & 1.015 & 1.133 \\
		4 & 1.107 & 1.003 & 1.135 & 1.319 & 1.084 & 1.030 & 1.010 & 1.169 \\
		5 & 1.103 & 1.005 & 1.134 & 1.326 & 1.092 & 1.026 & 1.011 & 1.147 \\
		6 & 1.096 & \textbf{0.992}$^{\dag}$ & 1.131 & 1.322 & 1.101 & 1.006 & 1.009 & 1.128 \\
		7 & 1.100 & 1.001 & 1.120 & 1.310 & 1.096 & 1.019 & 1.014 & 1.129 \\
		8 & 1.105 & \textbf{0.988}$^{\dag}$ & 1.137 & 1.281 & 1.100 & 1.037 & 1.021 & 1.141 \\
		\midrule
		\multicolumn{9}{c}{\textbf{Panel B: HICP}} \\
		\midrule
		1 & 1.055 & \textbf{0.997} & \textbf{0.877}$^{\dag}$ & \textbf{0.900} & \textbf{0.971} & \textbf{0.916} & 1.062 & \textbf{0.939} \\
		2 & \textbf{0.901} & \textbf{0.979} & \textbf{0.909} & \textbf{0.909} & \textbf{0.895}$^{\dag}$ & \textbf{0.960} & 1.017 & \textbf{0.916} \\
		3 & \textbf{0.854}$^{\dag}$ & \textbf{0.894} & \textbf{0.905} & \textbf{0.872} & \textbf{0.904} & 1.035 & \textbf{0.938} & \textbf{0.928} \\
		4 & 1.017 & 1.009 & 1.022 & 1.024 & 1.047 & 1.285 & 1.017 & 1.022 \\
		5 & \textbf{0.806}$^{\dag}$ & \textbf{0.954} & \textbf{0.847} & \textbf{0.856} & \textbf{0.869} & 1.038 & \textbf{0.984} & \textbf{0.835} \\
		6 & \textbf{0.915} & \textbf{0.983} & \textbf{0.919} & \textbf{0.958} & \textbf{0.893}$^{\dag}$ & 1.214 & 1.019 & \textbf{0.962} \\
		7 & \textbf{0.900} & \textbf{0.943} & \textbf{0.884}$^{\dag}$ & \textbf{0.930} & \textbf{0.913} & 1.142 & \textbf{0.967} & \textbf{0.889} \\
		8 & 1.029 & 1.030 & 1.067 & 1.084 & 1.105 & 1.496 & 1.045 & 1.060 \\
		\bottomrule
	\end{tabular}	
	\vspace{0.5em}
	\par
	\begin{tabularx}{\textwidth}{X}
		{\footnotesize \textit{Notes:} The table reports quantile scores 10\% (QS10) for each of the two forecasted variables, relative to a benchmark constant-parameter VAR($2$) estimated via least squares, for Euro Area quarterly data. Values below one (highlighted in bold) indicate superior forecasting performance compared to the benchmark. The forecast horizon is denoted by $h$ (in quarters), and the symbol $\dag$ marks the best-performing model at each horizon.}
	\end{tabularx}
\end{table}

\FloatBarrier

\subsubsection{Euro Area quarterly data: VAR(3)}

\begin{figure}[htbp!]  
	\centering
	\includegraphics[width=0.95\textwidth]{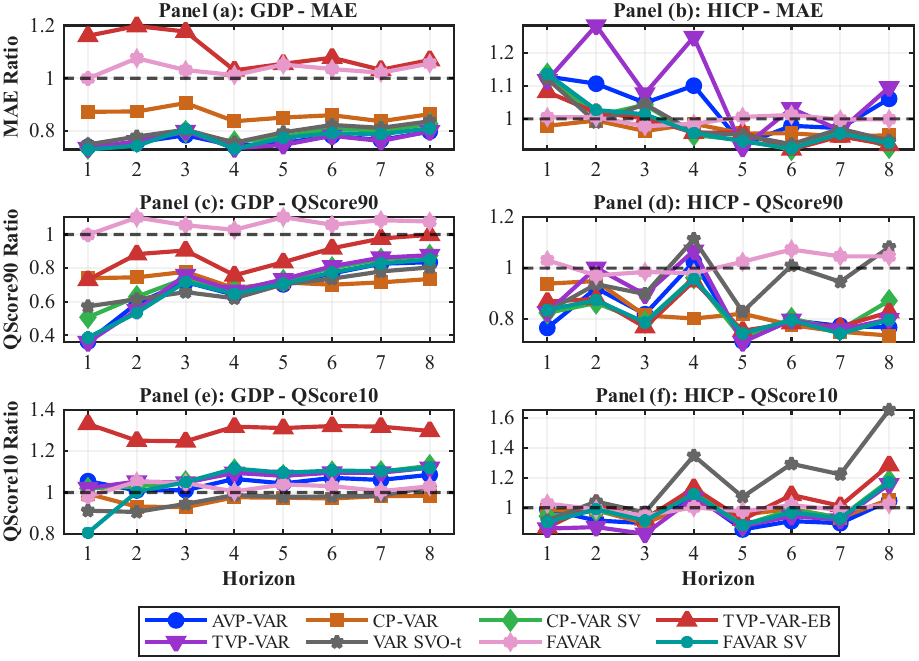}  
	\caption{Forecasting results using Euro Area quarterly data. The top row displays mean absolute forecast errors (MAE), while the middle and bottom rows show quantile scores for the upper (90th percentile) and lower (10th percentile) tails of each variable. All metrics are reported as ratios relative to a benchmark constant-parameter VAR($3$) estimated via least squares. A model is considered superior when its relative MAE, or relative quantile scores, are below one.} \label{fig:oos_EURO_p3}
\end{figure}

\begin{table}[htbp!]
	\centering
	\caption{Forecasting Performance VAR(3): Mean Squared Prediction Error Relative to Benchmark} \label{tab:forecasting_mspe_EURO_p3}
	\scriptsize
	\begin{tabular}{@{}cS[table-format=1.3]S[table-format=1.3]S[table-format=1.3]S[table-format=1.3]S[table-format=1.3]S[table-format=1.3]S[table-format=1.3]S[table-format=1.3]@{}}
		\toprule
		{\textbf{h}} & {\textbf{AVP-VAR}} & {\textbf{CP-VAR}} & {\makecell{ \textbf{CP-VAR} \\ \textbf{SV}}} & {\makecell{ \textbf{{TVP-VAR}} \\ \textbf{EB}}} &{\textbf{TVP-VAR}} & {\makecell{ \textbf{VAR} \\ \textbf{SVOt}}} & {\textbf{FAVAR}} & {\makecell{ \textbf{FAVAR} \\ \textbf{SV}}}  \\
		\midrule
		\multicolumn{9}{c}{\textbf{Panel A: GDP}} \\
		\midrule
		1 & \textbf{0.391} & \textbf{0.835} & \textbf{0.475} & \textbf{0.584} & \textbf{0.388}$^{\dag}$ & \textbf{0.487} & 1.048 & \textbf{0.464} \\
		2 & \textbf{0.565}$^{\dag}$ & \textbf{0.793} & \textbf{0.602} & \textbf{0.807} & \textbf{0.577} & \textbf{0.613} & 1.156 & \textbf{0.596} \\
		3 & \textbf{0.592}$^{\dag}$ & \textbf{0.768} & \textbf{0.601} & \textbf{0.736} & \textbf{0.595} & \textbf{0.604} & 1.122 & \textbf{0.606} \\
		4 & \textbf{0.538} & \textbf{0.692} & \textbf{0.551} & \textbf{0.623} & \textbf{0.536}$^{\dag}$ & \textbf{0.550} & 1.085 & \textbf{0.543} \\
		5 & \textbf{0.585} & \textbf{0.708} & \textbf{0.598} & \textbf{0.655} & \textbf{0.580}$^{\dag}$ & \textbf{0.602} & 1.136 & \textbf{0.592} \\
		6 & \textbf{0.640}$^{\dag}$ & \textbf{0.718} & \textbf{0.650} & \textbf{0.710} & \textbf{0.640} & \textbf{0.652} & 1.131 & \textbf{0.644} \\
		7 & \textbf{0.661}$^{\dag}$ & \textbf{0.712} & \textbf{0.671} & \textbf{0.722} & \textbf{0.661} & \textbf{0.674} & 1.135 & \textbf{0.670} \\
		8 & \textbf{0.679}$^{\dag}$ & \textbf{0.727} & \textbf{0.690} & \textbf{0.733} & \textbf{0.680} & \textbf{0.691} & 1.176 & \textbf{0.684} \\
		\midrule
		\multicolumn{9}{c}{\textbf{Panel B: HICP}} \\
		\midrule
		1 & 1.375 & \textbf{0.971}$^{\dag}$ & 1.219 & 1.134 & 1.200 & 1.212 & 1.021 & 1.245 \\
		2 & 1.106 & \textbf{0.978}$^{\dag}$ & 1.017 & 1.038 & 1.403 & \textbf{0.995} & \textbf{0.984} & 1.029 \\
		3 & \textbf{0.995} & \textbf{0.890}$^{\dag}$ & 1.032 & \textbf{0.949} & 1.021 & 1.033 & \textbf{0.994} & \textbf{0.988} \\
		4 & 1.051 & \textbf{0.863} & \textbf{0.847}$^{\dag}$ & \textbf{0.877} & 1.236 & \textbf{0.854} & 1.037 & \textbf{0.847} \\
		5 & \textbf{0.763} & \textbf{0.853} & \textbf{0.837} & \textbf{0.847} & \textbf{0.742}$^{\dag}$ & \textbf{0.868} & 1.047 & \textbf{0.830} \\
		6 & \textbf{0.774} & \textbf{0.792} & \textbf{0.715}$^{\dag}$ & \textbf{0.751} & \textbf{0.822} & \textbf{0.744} & 1.163 & \textbf{0.716} \\
		7 & \textbf{0.745} & \textbf{0.760} & \textbf{0.758} & \textbf{0.728} & \textbf{0.699}$^{\dag}$ & \textbf{0.770} & 1.147 & \textbf{0.738} \\
		8 & \textbf{0.719} & \textbf{0.671} & \textbf{0.592}$^{\dag}$ & \textbf{0.608} & \textbf{0.727} & \textbf{0.633} & 1.232 & \textbf{0.609} \\
		\bottomrule
	\end{tabular}	
	\vspace{0.5em}
	\par
	\begin{tabularx}{\textwidth}{X}
		{\footnotesize \textit{Notes:} The table reports mean squared prediction errors (MSPE) for each of the two forecasted variables, relative to a benchmark constant-parameter VAR($3$) estimated via least squares, for Euro Area quarterly data. Values below one (highlighted in bold) indicate superior forecasting performance compared to the benchmark. The forecast horizon is denoted by $h$ (in quarters), and the symbol $\dag$ marks the best-performing model at each horizon.}
	\end{tabularx}
\end{table}


\begin{table}[htbp!]
	\centering
	\caption{Forecasting Performance VAR(3): Quantile Scores 90\% Relative to Benchmark} \label{tab:forecasting_qs90_EURO_p3}
	\scriptsize
	\begin{tabular}{@{}cS[table-format=1.3]S[table-format=1.3]S[table-format=1.3]S[table-format=1.3]S[table-format=1.3]S[table-format=1.3]S[table-format=1.3]S[table-format=1.3]@{}}
		\toprule
		{\textbf{h}} & {\textbf{AVP-VAR}} & {\textbf{CP-VAR}} & {\makecell{ \textbf{CP-VAR} \\ \textbf{SV}}} & {\makecell{ \textbf{{TVP-VAR}} \\ \textbf{EB}}} &{\textbf{TVP-VAR}} & {\makecell{ \textbf{VAR} \\ \textbf{SVOt}}} & {\textbf{FAVAR}} & {\makecell{ \textbf{FAVAR} \\ \textbf{SV}}}  \\
		\midrule
		\multicolumn{9}{c}{\textbf{Panel A: GDP}} \\
		\midrule
		1 & \textbf{0.363} & \textbf{0.740} & \textbf{0.507} & \textbf{0.730} & \textbf{0.359}$^{\dag}$ & \textbf{0.572} & \textbf{0.998} & \textbf{0.386} \\
		2 & \textbf{0.590} & \textbf{0.745} & \textbf{0.630} & \textbf{0.883} & \textbf{0.550} & \textbf{0.615} & 1.101 & \textbf{0.534}$^{\dag}$ \\
		3 & \textbf{0.717} & \textbf{0.778} & \textbf{0.739} & \textbf{0.905} & \textbf{0.758} & \textbf{0.657}$^{\dag}$ & 1.055 & \textbf{0.718} \\
		4 & \textbf{0.641} & \textbf{0.680} & \textbf{0.670} & \textbf{0.757} & \textbf{0.665} & \textbf{0.620}$^{\dag}$ & 1.028 & \textbf{0.647} \\
		5 & \textbf{0.703}$^{\dag}$ & \textbf{0.715} & \textbf{0.726} & \textbf{0.835} & \textbf{0.733} & \textbf{0.703} & 1.103 & \textbf{0.707} \\
		6 & \textbf{0.768} & \textbf{0.701}$^{\dag}$ & \textbf{0.802} & \textbf{0.918} & \textbf{0.811} & \textbf{0.737} & 1.058 & \textbf{0.774} \\
		7 & \textbf{0.826} & \textbf{0.715}$^{\dag}$ & \textbf{0.852} & \textbf{0.976} & \textbf{0.862} & \textbf{0.780} & 1.085 & \textbf{0.828} \\
		8 & \textbf{0.835} & \textbf{0.735}$^{\dag}$ & \textbf{0.875} & \textbf{0.998} & \textbf{0.877} & \textbf{0.804} & 1.077 & \textbf{0.851} \\
		\midrule
		\multicolumn{9}{c}{\textbf{Panel B: HICP}} \\
		\midrule
		1 & \textbf{0.765}$^{\dag}$ & \textbf{0.939} & \textbf{0.824} & \textbf{0.868} & \textbf{0.826} & \textbf{0.830} & 1.031 & \textbf{0.836} \\
		2 & \textbf{0.921} & \textbf{0.949} & \textbf{0.863}$^{\dag}$ & \textbf{0.879} & 1.001 & \textbf{0.937} & \textbf{0.971} & \textbf{0.874} \\
		3 & \textbf{0.819} & \textbf{0.814} & \textbf{0.788} & \textbf{0.768}$^{\dag}$ & \textbf{0.896} & \textbf{0.898} & \textbf{0.985} & \textbf{0.786} \\
		4 & 1.025 & \textbf{0.802}$^{\dag}$ & \textbf{0.962} & \textbf{0.949} & 1.068 & 1.114 & \textbf{0.980} & \textbf{0.958} \\
		5 & \textbf{0.713} & \textbf{0.821} & \textbf{0.722} & \textbf{0.754} & \textbf{0.709}$^{\dag}$ & \textbf{0.828} & 1.025 & \textbf{0.744} \\
		6 & \textbf{0.795} & \textbf{0.776}$^{\dag}$ & \textbf{0.802} & \textbf{0.784} & \textbf{0.797} & 1.010 & 1.073 & \textbf{0.795} \\
		7 & \textbf{0.773} & \textbf{0.750} & \textbf{0.758} & \textbf{0.770} & \textbf{0.758} & \textbf{0.946} & 1.046 & \textbf{0.743}$^{\dag}$ \\
		8 & \textbf{0.767} & \textbf{0.735}$^{\dag}$ & \textbf{0.872} & \textbf{0.827} & \textbf{0.799} & 1.082 & 1.046 & \textbf{0.797} \\
		\bottomrule
	\end{tabular}	
	\vspace{0.5em}
	\par
	\begin{tabularx}{\textwidth}{X}
		{\footnotesize \textit{Notes:} The table reports quantile scores 90\% (QS90) for each of the two forecasted variables, relative to a benchmark constant-parameter VAR($3$) estimated via least squares, for Euro Area quarterly data. Values below one (highlighted in bold) indicate superior forecasting performance compared to the benchmark. The forecast horizon is denoted by $h$ (in quarters), and the symbol $\dag$ marks the best-performing model at each horizon.}
	\end{tabularx}
\end{table}


\begin{table}[htbp!]
	\centering
	\caption{Forecasting Performance VAR(3): Quantile Scores 10\% Relative to Benchmark} \label{tab:forecasting_qs10_EURO_p3}
	\scriptsize
	\begin{tabular}{@{}cS[table-format=1.3]S[table-format=1.3]S[table-format=1.3]S[table-format=1.3]S[table-format=1.3]S[table-format=1.3]S[table-format=1.3]S[table-format=1.3]@{}}
		\toprule
		{\textbf{h}} & {\textbf{AVP-VAR}} & {\textbf{CP-VAR}} & {\makecell{ \textbf{CP-VAR} \\ \textbf{SV}}} & {\makecell{ \textbf{{TVP-VAR}} \\ \textbf{EB}}} &{\textbf{TVP-VAR}} & {\makecell{ \textbf{VAR} \\ \textbf{SVOt}}} & {\textbf{FAVAR}} & {\makecell{ \textbf{FAVAR} \\ \textbf{SV}}}  \\
		\midrule
		\multicolumn{9}{c}{\textbf{Panel A: GDP}} \\
		\midrule
		1 & 1.055 & \textbf{0.994} & 1.020 & 1.332 & 1.022 & \textbf{0.912} & \textbf{0.980} & \textbf{0.805}$^{\dag}$ \\
		2 & 1.010 & \textbf{0.932} & 1.030 & 1.251 & 1.053 & \textbf{0.906}$^{\dag}$ & 1.053 & 1.003 \\
		3 & 1.011 & \textbf{0.929}$^{\dag}$ & 1.052 & 1.248 & 1.048 & \textbf{0.945} & 1.046 & 1.050 \\
		4 & 1.066 & \textbf{0.977}$^{\dag}$ & 1.109 & 1.319 & 1.095 & \textbf{0.988} & 1.005 & 1.116 \\
		5 & 1.044 & \textbf{0.975}$^{\dag}$ & 1.082 & 1.312 & 1.081 & \textbf{0.979} & 1.039 & 1.096 \\
		6 & 1.071 & \textbf{0.971}$^{\dag}$ & 1.099 & 1.322 & 1.096 & \textbf{0.981} & 1.031 & 1.106 \\
		7 & 1.061 & \textbf{0.983}$^{\dag}$ & 1.100 & 1.319 & 1.096 & \textbf{0.985} & 1.007 & 1.104 \\
		8 & 1.086 & \textbf{0.984}$^{\dag}$ & 1.130 & 1.298 & 1.120 & 1.026 & 1.030 & 1.120 \\
		\midrule
		\multicolumn{9}{c}{\textbf{Panel B: HICP}} \\
		\midrule
		1 & \textbf{0.989} & \textbf{0.982} & \textbf{0.944} & \textbf{0.869} & \textbf{0.861}$^{\dag}$ & \textbf{0.915} & 1.027 & \textbf{0.904} \\
		2 & \textbf{0.915} & \textbf{0.983} & \textbf{0.978} & 1.027 & \textbf{0.871}$^{\dag}$ & 1.041 & \textbf{0.989} & \textbf{0.990} \\
		3 & \textbf{0.897} & \textbf{0.896} & \textbf{0.930} & \textbf{0.923} & \textbf{0.826}$^{\dag}$ & \textbf{0.968} & \textbf{0.947} & \textbf{0.914} \\
		4 & 1.085 & 1.029 & 1.115 & 1.131 & 1.066 & 1.346 & 1.002$^{\dag}$ & 1.090 \\
		5 & \textbf{0.855}$^{\dag}$ & \textbf{0.954} & \textbf{0.884} & \textbf{0.921} & \textbf{0.877} & 1.072 & \textbf{0.976} & \textbf{0.884} \\
		6 & \textbf{0.911}$^{\dag}$ & \textbf{0.992} & \textbf{0.986} & 1.085 & \textbf{0.944} & 1.292 & 1.017 & \textbf{0.969} \\
		7 & \textbf{0.897}$^{\dag}$ & \textbf{0.933} & \textbf{0.935} & 1.012 & \textbf{0.920} & 1.222 & \textbf{0.985} & \textbf{0.929} \\
		8 & 1.046 & 1.054 & 1.180 & 1.283 & 1.156 & 1.651 & 1.023 & 1.169 \\
		\bottomrule
	\end{tabular}	
	\vspace{0.5em}
	\par
	\begin{tabularx}{\textwidth}{X}
		{\footnotesize \textit{Notes:} The table reports quantile scores 10\% (QS10) for each of the two forecasted variables, relative to a benchmark constant-parameter VAR($3$) estimated via least squares, for Euro Area quarterly data. Values below one (highlighted in bold) indicate superior forecasting performance compared to the benchmark. The forecast horizon is denoted by $h$ (in quarters), and the symbol $\dag$ marks the best-performing model at each horizon.}
	\end{tabularx}
\end{table}

\FloatBarrier

\subsubsection{Euro Area quarterly data: VAR(4)}

\begin{figure}[htbp!]  
	\centering
	\includegraphics[width=0.95\textwidth]{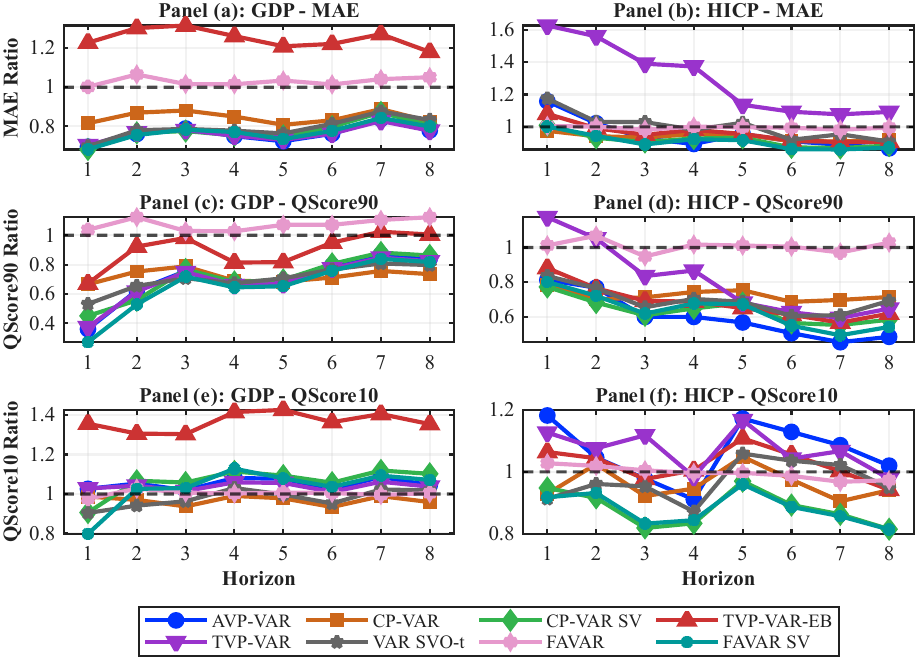}  
	\caption{Forecasting results using Euro Area quarterly data. The top row displays mean absolute forecast errors (MAE), while the middle and bottom rows show quantile scores for the upper (90th percentile) and lower (10th percentile) tails of each variable. All metrics are reported as ratios relative to a benchmark constant-parameter VAR($4$) estimated via least squares. A model is considered superior when its relative MAE, or relative quantile scores, are below one.} \label{fig:oos_EURO_p4}
\end{figure}
\begin{table}[htbp!]
	\centering
	\caption{Forecasting Performance VAR(4): Mean Squared Prediction Error Relative to Benchmark} \label{tab:forecasting_mspe_EURO_p4}
	\scriptsize
	\begin{tabular}{@{}cS[table-format=1.3]S[table-format=1.3]S[table-format=1.3]S[table-format=1.3]S[table-format=1.3]S[table-format=1.3]S[table-format=1.3]S[table-format=1.3]@{}}
		\toprule
		{\textbf{h}} & {\textbf{AVP-VAR}} & {\textbf{CP-VAR}} & {\makecell{ \textbf{CP-VAR} \\ \textbf{SV}}} & {\makecell{ \textbf{{TVP-VAR}} \\ \textbf{EB}}} &{\textbf{TVP-VAR}} & {\makecell{ \textbf{VAR} \\ \textbf{SVOt}}} & {\textbf{FAVAR}} & {\makecell{ \textbf{FAVAR} \\ \textbf{SV}}}  \\
		\midrule
		\multicolumn{9}{c}{\textbf{Panel A: GDP}} \\
		\midrule
		1 & \textbf{0.394} & \textbf{0.786} & \textbf{0.442} & \textbf{0.583} & \textbf{0.392}$^{\dag}$ & \textbf{0.467} & 1.065 & \textbf{0.475} \\
		2 & \textbf{0.609}$^{\dag}$ & \textbf{0.800} & \textbf{0.638} & \textbf{0.829} & \textbf{0.620} & \textbf{0.653} & 1.137 & \textbf{0.623} \\
		3 & \textbf{0.632} & \textbf{0.772} & \textbf{0.633} & \textbf{0.793} & \textbf{0.629}$^{\dag}$ & \textbf{0.638} & 1.050 & \textbf{0.635} \\
		4 & \textbf{0.564}$^{\dag}$ & \textbf{0.700} & \textbf{0.577} & \textbf{0.698} & \textbf{0.565} & \textbf{0.589} & 1.066 & \textbf{0.583} \\
		5 & \textbf{0.605}$^{\dag}$ & \textbf{0.694} & \textbf{0.615} & \textbf{0.708} & \textbf{0.605} & \textbf{0.622} & 1.124 & \textbf{0.607} \\
		6 & \textbf{0.657} & \textbf{0.706} & \textbf{0.666} & \textbf{0.746} & \textbf{0.656}$^{\dag}$ & \textbf{0.670} & 1.138 & \textbf{0.658} \\
		7 & \textbf{0.716} & \textbf{0.743} & \textbf{0.724} & \textbf{0.792} & \textbf{0.715}$^{\dag}$ & \textbf{0.727} & 1.183 & \textbf{0.721} \\
		8 & \textbf{0.696}$^{\dag}$ & \textbf{0.718} & \textbf{0.705} & \textbf{0.765} & \textbf{0.697} & \textbf{0.714} & 1.238 & \textbf{0.702} \\
		\midrule
		\multicolumn{9}{c}{\textbf{Panel B: HICP}} \\
		\midrule
		1 & 1.653 & \textbf{0.957}$^{\dag}$ & 1.012 & 1.189 & 2.023 & 1.155 & 1.082 & 1.009 \\
		2 & \textbf{0.931} & \textbf{0.862} & \textbf{0.835}$^{\dag}$ & \textbf{0.975} & 1.748 & \textbf{0.915} & 1.143 & \textbf{0.895} \\
		3 & \textbf{0.628}$^{\dag}$ & \textbf{0.762} & \textbf{0.646} & \textbf{0.752} & 1.190 & \textbf{0.755} & \textbf{0.997} & \textbf{0.649} \\
		4 & \textbf{0.617}$^{\dag}$ & \textbf{0.798} & \textbf{0.712} & \textbf{0.833} & 1.200 & \textbf{0.735} & 1.054 & \textbf{0.708} \\
		5 & \textbf{0.563}$^{\dag}$ & \textbf{0.739} & \textbf{0.654} & \textbf{0.663} & \textbf{0.781} & \textbf{0.729} & 1.020 & \textbf{0.635} \\
		6 & \textbf{0.480}$^{\dag}$ & \textbf{0.634} & \textbf{0.535} & \textbf{0.559} & \textbf{0.663} & \textbf{0.568} & 1.081 & \textbf{0.507} \\
		7 & \textbf{0.398}$^{\dag}$ & \textbf{0.591} & \textbf{0.459} & \textbf{0.471} & \textbf{0.547} & \textbf{0.509} & 1.055 & \textbf{0.441} \\
		8 & \textbf{0.367}$^{\dag}$ & \textbf{0.583} & \textbf{0.471} & \textbf{0.482} & \textbf{0.555} & \textbf{0.484} & 1.122 & \textbf{0.456} \\
		\bottomrule
	\end{tabular}	
	\vspace{0.5em}
	\par
	\begin{tabularx}{\textwidth}{X}
		{\footnotesize \textit{Notes:} The table reports mean squared prediction errors (MSPE) for each of the two forecasted variables, relative to a benchmark constant-parameter VAR($4$) estimated via least squares, for Euro Area quarterly data. Values below one (highlighted in bold) indicate superior forecasting performance compared to the benchmark. The forecast horizon is denoted by $h$ (in quarters), and the symbol $\dag$ marks the best-performing model at each horizon.}
	\end{tabularx}
\end{table}

\begin{table}[htbp!]
	\centering
	\caption{Forecasting Performance VAR(4): Quantile Scores 90\% Relative to Benchmark} \label{tab:forecasting_qs90_EURO_p4}
	\scriptsize
	\begin{tabular}{@{}cS[table-format=1.3]S[table-format=1.3]S[table-format=1.3]S[table-format=1.3]S[table-format=1.3]S[table-format=1.3]S[table-format=1.3]S[table-format=1.3]@{}}
		\toprule
		{\textbf{h}} & {\textbf{AVP-VAR}} & {\textbf{CP-VAR}} & {\makecell{ \textbf{CP-VAR} \\ \textbf{SV}}} & {\makecell{ \textbf{{TVP-VAR}} \\ \textbf{EB}}} &{\textbf{TVP-VAR}} & {\makecell{ \textbf{VAR} \\ \textbf{SVOt}}} & {\textbf{FAVAR}} & {\makecell{ \textbf{FAVAR} \\ \textbf{SV}}}  \\
		\midrule
		\multicolumn{9}{c}{\textbf{Panel A: GDP}} \\
		\midrule
		1 & \textbf{0.361} & \textbf{0.666} & \textbf{0.451} & \textbf{0.671} & \textbf{0.374} & \textbf{0.532} & 1.037 & \textbf{0.272}$^{\dag}$ \\
		2 & \textbf{0.624} & \textbf{0.753} & \textbf{0.557} & \textbf{0.923} & \textbf{0.617} & \textbf{0.655} & 1.123 & \textbf{0.529}$^{\dag}$ \\
		3 & \textbf{0.756} & \textbf{0.788} & \textbf{0.767} & \textbf{0.982} & \textbf{0.749} & \textbf{0.707}$^{\dag}$ & 1.029 & \textbf{0.718} \\
		4 & \textbf{0.677} & \textbf{0.690} & \textbf{0.682} & \textbf{0.814} & \textbf{0.661} & \textbf{0.673} & 1.027 & \textbf{0.645}$^{\dag}$ \\
		5 & \textbf{0.683} & \textbf{0.684} & \textbf{0.696} & \textbf{0.817} & \textbf{0.668} & \textbf{0.707} & 1.071 & \textbf{0.653}$^{\dag}$ \\
		6 & \textbf{0.793} & \textbf{0.712}$^{\dag}$ & \textbf{0.805} & \textbf{0.950} & \textbf{0.774} & \textbf{0.767} & 1.071 & \textbf{0.758} \\
		7 & \textbf{0.868} & \textbf{0.757}$^{\dag}$ & \textbf{0.882} & 1.024 & \textbf{0.848} & \textbf{0.808} & 1.104 & \textbf{0.839} \\
		8 & \textbf{0.843} & \textbf{0.735}$^{\dag}$ & \textbf{0.862} & 1.005 & \textbf{0.823} & \textbf{0.797} & 1.123 & \textbf{0.818} \\
		\midrule
		\multicolumn{9}{c}{\textbf{Panel B: HICP}} \\
		\midrule
		1 & \textbf{0.812} & \textbf{0.784} & \textbf{0.772}$^{\dag}$ & \textbf{0.881} & 1.175 & \textbf{0.835} & 1.012 & \textbf{0.801} \\
		2 & \textbf{0.768} & \textbf{0.700} & \textbf{0.682}$^{\dag}$ & \textbf{0.764} & 1.052 & \textbf{0.765} & 1.068 & \textbf{0.721} \\
		3 & \textbf{0.599}$^{\dag}$ & \textbf{0.712} & \textbf{0.608} & \textbf{0.692} & \textbf{0.835} & \textbf{0.655} & \textbf{0.948} & \textbf{0.618} \\
		4 & \textbf{0.599}$^{\dag}$ & \textbf{0.743} & \textbf{0.648} & \textbf{0.687} & \textbf{0.867} & \textbf{0.703} & 1.017 & \textbf{0.677} \\
		5 & \textbf{0.567}$^{\dag}$ & \textbf{0.753} & \textbf{0.679} & \textbf{0.650} & \textbf{0.681} & \textbf{0.686} & 1.011 & \textbf{0.672} \\
		6 & \textbf{0.505}$^{\dag}$ & \textbf{0.686} & \textbf{0.564} & \textbf{0.611} & \textbf{0.626} & \textbf{0.605} & 1.005 & \textbf{0.548} \\
		7 & \textbf{0.453}$^{\dag}$ & \textbf{0.697} & \textbf{0.552} & \textbf{0.566} & \textbf{0.593} & \textbf{0.609} & \textbf{0.971} & \textbf{0.493} \\
		8 & \textbf{0.484}$^{\dag}$ & \textbf{0.714} & \textbf{0.583} & \textbf{0.618} & \textbf{0.647} & \textbf{0.692} & 1.028 & \textbf{0.541} \\
		
		\bottomrule
	\end{tabular}	
	\vspace{0.5em}
	\par
	\begin{tabularx}{\textwidth}{X}
		{\footnotesize \textit{Notes:} The table reports quantile scores 90\% (QS90) for each of the two forecasted variables, relative to a benchmark constant-parameter VAR($4$) estimated via least squares, for Euro Area quarterly data. Values below one (highlighted in bold) indicate superior forecasting performance compared to the benchmark. The forecast horizon is denoted by $h$ (in quarters), and the symbol $\dag$ marks the best-performing model at each horizon.}
	\end{tabularx}
\end{table}


\begin{table}[htbp!]
	\centering
	\caption{Forecasting Performance VAR(4): Quantile Scores 10\% Relative to Benchmark} \label{tab:forecasting_qs10_EURO_p4}
	\scriptsize
	\begin{tabular}{@{}cS[table-format=1.3]S[table-format=1.3]S[table-format=1.3]S[table-format=1.3]S[table-format=1.3]S[table-format=1.3]S[table-format=1.3]S[table-format=1.3]@{}}
		\toprule
		{\textbf{h}} & {\textbf{AVP-VAR}} & {\textbf{CP-VAR}} & {\makecell{ \textbf{CP-VAR} \\ \textbf{SV}}} & {\makecell{ \textbf{{TVP-VAR}} \\ \textbf{EB}}} &{\textbf{TVP-VAR}} & {\makecell{ \textbf{VAR} \\ \textbf{SVOt}}} & {\textbf{FAVAR}} & {\makecell{ \textbf{FAVAR} \\ \textbf{SV}}}  \\
		\midrule
		\multicolumn{9}{c}{\textbf{Panel A: GDP}} \\
		\midrule
		1 & 1.026 & \textbf{0.987} & \textbf{0.906} & 1.356 & 1.028 & \textbf{0.904} & \textbf{0.976} & \textbf{0.798}$^{\dag}$ \\
		2 & 1.053 & \textbf{0.970} & 1.067 & 1.306 & 1.042 & \textbf{0.941}$^{\dag}$ & 1.027 & 1.026 \\
		3 & 1.019 & \textbf{0.937}$^{\dag}$ & 1.058 & 1.303 & 1.015 & \textbf{0.963} & 1.007 & 1.030 \\
		4 & 1.081 & \textbf{0.990}$^{\dag}$ & 1.114 & 1.415 & 1.059 & 1.030 & 1.002 & 1.128 \\
		5 & 1.077 & \textbf{0.978}$^{\dag}$ & 1.093 & 1.426 & 1.056 & \textbf{0.997} & 1.007 & 1.079 \\
		6 & 1.025 & \textbf{0.934}$^{\dag}$ & 1.057 & 1.364 & 1.010 & \textbf{0.952} & 1.005 & 1.036 \\
		7 & 1.086 & \textbf{0.989}$^{\dag}$ & 1.119 & 1.405 & 1.062 & 1.023 & \textbf{0.995} & 1.093 \\
		8 & 1.045 & \textbf{0.959}$^{\dag}$ & 1.102 & 1.354 & 1.037 & 1.018 & 1.004 & 1.071 \\
		\midrule
		\multicolumn{9}{c}{\textbf{Panel B: HICP}} \\
		\midrule
		1 & 1.182 & \textbf{0.929} & \textbf{0.948} & 1.064 & 1.127 & \textbf{0.913}$^{\dag}$ & 1.028 & \textbf{0.918} \\
		2 & 1.046 & 1.025 & \textbf{0.916}$^{\dag}$ & 1.044 & 1.076 & \textbf{0.961} & 1.019 & \textbf{0.932} \\
		3 & \textbf{0.978} & \textbf{0.922} & \textbf{0.819}$^{\dag}$ & \textbf{0.977} & 1.118 & \textbf{0.953} & 1.005 & \textbf{0.833} \\
		4 & \textbf{0.911} & \textbf{0.946} & \textbf{0.834}$^{\dag}$ & 1.005 & \textbf{0.984} & \textbf{0.872} & \textbf{0.996} & \textbf{0.844} \\
		5 & 1.172 & 1.049 & \textbf{0.970} & 1.108 & 1.167 & 1.059 & \textbf{0.997} & \textbf{0.960}$^{\dag}$ \\
		6 & 1.129 & \textbf{0.975} & \textbf{0.893} & 1.056 & 1.043 & 1.036 & \textbf{0.987} & \textbf{0.887}$^{\dag}$ \\
		7 & 1.086 & \textbf{0.905} & \textbf{0.863} & 1.002 & 1.067 & 1.022 & \textbf{0.968} & \textbf{0.857}$^{\dag}$ \\
		8 & 1.020 & \textbf{0.941} & \textbf{0.815} & \textbf{0.942} & \textbf{0.978} & \textbf{0.949} & \textbf{0.973} & \textbf{0.814}$^{\dag}$ \\
		\bottomrule
	\end{tabular}	
	\vspace{0.5em}
	\par
	\begin{tabularx}{\textwidth}{X}
		{\footnotesize \textit{Notes:} The table reports quantile scores 10\% (QS10) for each of the two forecasted variables, relative to a benchmark constant-parameter VAR($4$) estimated via least squares, for Euro Area quarterly data. Values below one (highlighted in bold) indicate superior forecasting performance compared to the benchmark. The forecast horizon is denoted by $h$ (in quarters), and the symbol $\dag$ marks the best-performing model at each horizon.}
	\end{tabularx}
\end{table}

\FloatBarrier


\subsection{Robustness exercise: Different TVP-VAR-EB orderings.}	

\subsubsection{U.S. monthly data}

\begin{table}[htbp!]
	\centering
	\caption{Mean squared prediction error results comparing different TVP-VAR-EB orderings: U.S. quarterly data.}
	\label{tab:forecast_accuracy_us}
	\scriptsize
	\begin{tabular}{@{}clccc ccc ccc@{}}
		\toprule
		& & \multicolumn{3}{c}{$p=1$} & \multicolumn{3}{c}{$p=2$} & \multicolumn{3}{c}{$p=3$} \\
		\cmidrule(lr){3-5} \cmidrule(lr){6-8} \cmidrule(lr){9-11}
		&h & {AVP-VAR} & \makecell{TVP-VAR\\EB} & \makecell{TVP-VAR\\EB Alt} 
		& {AVP-VAR} & \makecell{TVP-VAR\\EB} & \makecell{TVP-VAR\\EB Alt} 
		& {AVP-VAR} & \makecell{TVP-VAR\\EB} & \makecell{TVP-VAR\\EB Alt} \\
		\midrule
		\multicolumn{11}{@{}l}{Industrial Production} \\
		\addlinespace[0.1cm]
		& 1 & \textbf{0.902} & \textbf{0.941} & \textbf{0.959} & \textbf{0.902} & \textbf{0.977} & \textbf{0.976} & \textbf{0.907} & \textbf{0.957} & \textbf{0.929} \\
		& 2 & \textbf{0.851} & \textbf{0.848} & \textbf{0.854} & \textbf{0.805} & \textbf{0.853} & \textbf{0.845} & \textbf{0.770} & \textbf{0.806} & \textbf{0.798} \\
		& 3 & \textbf{0.911} & \textbf{0.926} & \textbf{0.930} & \textbf{0.867} & \textbf{0.881} & \textbf{0.885} & \textbf{0.839} & \textbf{0.833} & \textbf{0.849} \\
		& 4 & \textbf{0.972} & \textbf{0.984} & \textbf{0.986} & \textbf{0.938} & \textbf{0.951} & \textbf{0.953} & \textbf{0.891} & \textbf{0.900} & \textbf{0.902} \\
		& 5 & \textbf{0.997} & 1.000 & 1.008 & \textbf{0.967} & \textbf{0.986} & \textbf{0.989} & \textbf{0.930} & \textbf{0.930} & \textbf{0.935} \\
		& 6 & \textbf{0.991} & 1.001 & 1.001 & \textbf{0.972} & \textbf{0.985} & \textbf{0.986} & \textbf{0.944} & \textbf{0.948} & \textbf{0.944} \\
		& 9 & 1.002 & 1.003 & 1.003 & \textbf{0.983} & 1.005 & \textbf{0.998} & \textbf{0.970} & \textbf{0.978} & \textbf{0.973} \\
		& 12 & 1.006 & 1.006 & 1.006 & 1.001 & 1.006 & 1.007 & \textbf{0.982} & \textbf{0.991} & \textbf{0.996} \\
		& 15 & 1.010 & 1.006 & 1.008 & 1.002 & 1.003 & 1.005 & \textbf{0.987} & \textbf{0.994} & \textbf{0.998} \\
		& 18 & 1.010 & 1.009 & 1.010 & 1.001 & 1.006 & 1.007 & \textbf{0.987} & \textbf{0.997} & 1.008 \\
		& 24 & 1.014 & 1.011 & 1.011 & 1.001 & 1.010 & 1.011 & \textbf{0.994} & 1.009 & 1.007 \\
		\addlinespace[0.2cm]
		\multicolumn{11}{@{}l}{PCEPI} \\
		\addlinespace[0.1cm]
		& 1 & 1.072 & 1.000 & 1.004 & \textbf{0.980} & \textbf{0.999} & \textbf{0.997} & 1.007 & 1.021 & \textbf{0.995} \\
		& 2 & \textbf{0.939} & \textbf{0.981} & \textbf{0.974} & \textbf{0.881} & \textbf{0.937} & \textbf{0.927} & \textbf{0.903} & \textbf{0.928} & \textbf{0.934} \\
		& 3 & \textbf{0.925} & \textbf{0.980} & \textbf{0.983} & \textbf{0.869} & \textbf{0.930} & \textbf{0.934} & \textbf{0.943} & \textbf{0.903} & \textbf{0.907} \\
		& 4 & \textbf{0.919} & \textbf{0.988} & \textbf{0.986} & \textbf{0.886} & \textbf{0.939} & \textbf{0.946} & \textbf{0.918} & \textbf{0.916} & \textbf{0.943} \\
		& 5 & \textbf{0.921} & \textbf{0.997} & \textbf{0.996} & \textbf{0.918} & \textbf{0.973} & \textbf{0.978} & \textbf{0.915} & \textbf{0.939} & \textbf{0.958} \\
		& 6 & \textbf{0.930} & 1.007 & 1.002 & \textbf{0.912} & \textbf{0.996} & \textbf{0.993} & \textbf{0.926} & \textbf{0.983} & 1.031 \\
		& 9 & \textbf{0.934} & \textbf{0.997} & \textbf{0.992} & \textbf{0.926} & \textbf{0.993} & \textbf{0.991} & \textbf{0.912} & \textbf{0.980} & 1.018 \\
		& 12 & \textbf{0.964} & \textbf{0.999} & \textbf{0.996} & \textbf{0.964} & \textbf{0.998} & \textbf{0.996} & \textbf{0.944} & \textbf{0.989} & \textbf{0.990} \\
		& 15 & \textbf{0.965} & 1.001 & \textbf{0.999} & \textbf{0.959} & 1.006 & 1.001 & \textbf{0.954} & \textbf{0.993} & \textbf{0.985} \\
		& 18 & \textbf{0.997} & 1.007 & 1.000 & \textbf{0.989} & 1.002 & 1.003 & \textbf{0.970} & 1.002 & \textbf{0.999} \\
		& 24 & 1.046 & 1.000 & 1.001 & 1.040 & 1.007 & 1.003 & 1.026 & 1.008 & 1.014 \\
		\bottomrule
	\end{tabular}
	\begin{minipage}{\textwidth}
		\small
		\vspace{0.2cm}
		\textit{Notes:} The table reports mean squared prediction errors (MSPE) relative to a benchmark constant-parameter VAR($p$) estimated by ordinary least squares. Values below one (in bold) indicate superior performance relative to the benchmark. The forecast horizon is denoted by $h$, and $p$ denotes the lag order. TVP-VAR-EB refers to the baseline specification with the ordering: growth, inflation, and interest rates. TVP-VAR-EB Alt uses the alternative ordering: interest rates, inflation, and growth.
	\end{minipage}
\end{table}

\begin{table}[htbp!]
	\centering
	\caption{Quantile Score (90\%) results comparing different TVP-VAR-EB orderings: U.S. quarterly data.}
	\label{tab:quantile_score_90_us}
	\scriptsize
	\begin{tabular}{@{}clccc ccc ccc@{}}
		\toprule
		& & \multicolumn{3}{c}{$p=1$} & \multicolumn{3}{c}{$p=2$} & \multicolumn{3}{c}{$p=3$} \\
		\cmidrule(lr){3-5} \cmidrule(lr){6-8} \cmidrule(lr){9-11}
		&h & {AVP-VAR} & \makecell{TVP-VAR\\EB} & \makecell{TVP-VAR\\EB Alt} 
		& {AVP-VAR} & \makecell{TVP-VAR\\EB} & \makecell{TVP-VAR\\EB Alt} 
		& {AVP-VAR} & \makecell{TVP-VAR\\EB} & \makecell{TVP-VAR\\EB Alt} \\
		\midrule
		\multicolumn{11}{@{}l}{Industrial Production} \\
		\addlinespace[0.1cm]
		& 1 & \textbf{0.715} & \textbf{0.765} & \textbf{0.772} & \textbf{0.700} & \textbf{0.790} & \textbf{0.795} & \textbf{0.662} & \textbf{0.726} & \textbf{0.711} \\
		& 2 & \textbf{0.760} & \textbf{0.802} & \textbf{0.802} & \textbf{0.719} & \textbf{0.802} & \textbf{0.799} & \textbf{0.631} & \textbf{0.709} & \textbf{0.696} \\
		& 3 & \textbf{0.923} & \textbf{0.951} & \textbf{0.955} & \textbf{0.889} & \textbf{0.915} & \textbf{0.922} & \textbf{0.821} & \textbf{0.841} & \textbf{0.843} \\
		& 4 & 1.013 & 1.037 & 1.045 & \textbf{0.980} & \textbf{0.999} & 1.010 & \textbf{0.934} & \textbf{0.971} & \textbf{0.966} \\
		& 5 & 1.100 & 1.117 & 1.125 & 1.078 & 1.088 & 1.096 & \textbf{0.988} & 1.008 & 1.017 \\
		& 6 & 1.093 & 1.112 & 1.112 & 1.076 & 1.091 & 1.100 & 1.025 & 1.048 & 1.055 \\
		& 9 & 1.100 & 1.118 & 1.119 & 1.105 & 1.125 & 1.136 & 1.048 & 1.087 & 1.104 \\
		& 12 & 1.120 & 1.135 & 1.136 & 1.135 & 1.152 & 1.162 & 1.116 & 1.156 & 1.171 \\
		& 15 & 1.149 & 1.160 & 1.160 & 1.185 & 1.184 & 1.186 & 1.192 & 1.204 & 1.221 \\
		& 18 & 1.181 & 1.166 & 1.170 & 1.201 & 1.199 & 1.202 & 1.185 & 1.200 & 1.200 \\
		& 24 & 1.150 & 1.154 & 1.164 & 1.179 & 1.171 & 1.189 & 1.188 & 1.187 & 1.204 \\
		\addlinespace[0.2cm]
		\multicolumn{11}{@{}l}{PCEPI} \\
		\addlinespace[0.1cm]
		& 1 & \textbf{0.984} & \textbf{0.958} & \textbf{0.962} & \textbf{0.969} & \textbf{0.928} & \textbf{0.931} & \textbf{0.957} & \textbf{0.923} & \textbf{0.931} \\
		& 2 & \textbf{0.899} & \textbf{0.942} & \textbf{0.963} & \textbf{0.863} & \textbf{0.888} & \textbf{0.918} & \textbf{0.851} & \textbf{0.871} & \textbf{0.886} \\
		& 3 & \textbf{0.878} & \textbf{0.936} & \textbf{0.967} & \textbf{0.846} & \textbf{0.880} & \textbf{0.928} & \textbf{0.874} & \textbf{0.849} & \textbf{0.876} \\
		& 4 & \textbf{0.871} & \textbf{0.953} & \textbf{0.982} & \textbf{0.861} & \textbf{0.911} & \textbf{0.950} & \textbf{0.887} & \textbf{0.887} & \textbf{0.920} \\
		& 5 & \textbf{0.897} & \textbf{0.977} & 1.007 & \textbf{0.903} & \textbf{0.944} & \textbf{0.985} & \textbf{0.883} & \textbf{0.900} & \textbf{0.941} \\
		& 6 & \textbf{0.894} & \textbf{0.962} & \textbf{0.989} & \textbf{0.911} & \textbf{0.940} & \textbf{0.995} & \textbf{0.894} & \textbf{0.905} & \textbf{0.962} \\
		& 9 & \textbf{0.921} & \textbf{0.963} & 1.007 & \textbf{0.913} & \textbf{0.927} & \textbf{0.999} & \textbf{0.900} & \textbf{0.882} & \textbf{0.942} \\
		& 12 & \textbf{0.956} & \textbf{0.982} & 1.010 & \textbf{0.969} & \textbf{0.957} & \textbf{0.993} & \textbf{0.962} & \textbf{0.920} & \textbf{0.969} \\
		& 15 & 1.049 & 1.028 & 1.060 & 1.055 & 1.010 & 1.064 & 1.026 & \textbf{0.953} & 1.002 \\
		& 18 & 1.073 & 1.035 & 1.060 & 1.079 & 1.006 & 1.044 & 1.035 & \textbf{0.958} & \textbf{0.989} \\
		& 24 & 1.169 & 1.107 & 1.091 & 1.184 & 1.092 & 1.090 & 1.152 & 1.080 & 1.151 \\
		\bottomrule
	\end{tabular}
	\begin{minipage}{\textwidth}
		\small
		\vspace{0.2cm}
		\textit{Notes:} The table reports quantile scores at the 90\% level relative to a benchmark constant-parameter VAR($p$) estimated by ordinary least squares. Values below one (in bold) indicate superior performance relative to the benchmark. The forecast horizon is denoted by $h$, and $p$ denotes the lag order. TVP-VAR-EB refers to the baseline specification with the ordering: growth, inflation, and interest rates. TVP-VAR-EB Alt uses the alternative ordering: interest rates, inflation, and growth.
	\end{minipage}
\end{table}

\begin{table}[htbp]
	\centering
	\caption{Quantile Score (10\%) results comparing different TVP-VAR-EB orderings: U.S. quarterly data.}
	\label{tab:quantile_score_10_us}
	\scriptsize
	\begin{tabular}{@{}clccc ccc ccc@{}}
		\toprule
		& & \multicolumn{3}{c}{$p=1$} & \multicolumn{3}{c}{$p=2$} & \multicolumn{3}{c}{$p=3$} \\
		\cmidrule(lr){3-5} \cmidrule(lr){6-8} \cmidrule(lr){9-11}
		&h & {AVP-VAR} & \makecell{TVP-VAR\\EB} & \makecell{TVP-VAR\\EB Alt} 
		& {AVP-VAR} & \makecell{TVP-VAR\\EB} & \makecell{TVP-VAR\\EB Alt} 
		& {AVP-VAR} & \makecell{TVP-VAR\\EB} & \makecell{TVP-VAR\\EB Alt} \\
		\midrule
		\multicolumn{11}{@{}l}{Industrial Production} \\
		\addlinespace[0.1cm]
		& 1 & 1.028 & 1.059 & 1.069 & 1.044 & 1.057 & 1.059 & 1.047 & 1.051 & 1.041 \\
		& 2 & 1.029 & 1.040 & 1.038 & 1.025 & 1.036 & 1.034 & 1.046 & 1.019 & 1.018 \\
		& 3 & 1.010 & 1.025 & 1.027 & 1.022 & 1.023 & 1.031 & 1.013 & 1.006 & 1.013 \\
		& 4 & 1.034 & 1.050 & 1.051 & 1.036 & 1.062 & 1.054 & 1.028 & 1.057 & 1.055 \\
		& 5 & 1.054 & 1.065 & 1.059 & 1.048 & 1.082 & 1.082 & 1.053 & 1.076 & 1.068 \\
		& 6 & 1.050 & 1.063 & 1.064 & 1.052 & 1.085 & 1.082 & 1.076 & 1.092 & 1.087 \\
		& 9 & 1.074 & 1.090 & 1.087 & 1.060 & 1.099 & 1.100 & 1.084 & 1.112 & 1.111 \\
		& 12 & 1.143 & 1.147 & 1.149 & 1.132 & 1.151 & 1.158 & 1.134 & 1.154 & 1.152 \\
		& 15 & 1.154 & 1.154 & 1.160 & 1.138 & 1.158 & 1.166 & 1.150 & 1.158 & 1.173 \\
		& 18 & 1.161 & 1.157 & 1.163 & 1.139 & 1.153 & 1.162 & 1.127 & 1.153 & 1.184 \\
		& 24 & 1.175 & 1.173 & 1.167 & 1.145 & 1.163 & 1.164 & 1.139 & 1.157 & 1.189 \\
		\addlinespace[0.2cm]
		\multicolumn{11}{@{}l}{PCEPI} \\
		\addlinespace[0.1cm]
		& 1 & 1.007 & 1.002 & \textbf{0.986} & \textbf{0.951} & 1.020 & \textbf{0.999} & \textbf{0.983} & \textbf{0.989} & \textbf{0.988} \\
		& 2 & 1.040 & 1.047 & 1.039 & 1.022 & 1.061 & 1.038 & \textbf{0.999} & 1.029 & 1.026 \\
		& 3 & 1.102 & 1.091 & 1.090 & 1.085 & 1.089 & 1.086 & 1.062 & 1.084 & 1.073 \\
		& 4 & 1.084 & 1.087 & 1.077 & 1.094 & 1.080 & 1.072 & 1.117 & 1.085 & 1.081 \\
		& 5 & 1.090 & 1.087 & 1.080 & 1.102 & 1.096 & 1.088 & 1.134 & 1.099 & 1.097 \\
		& 6 & 1.097 & 1.094 & 1.086 & 1.101 & 1.099 & 1.086 & 1.126 & 1.124 & 1.103 \\
		& 9 & 1.084 & 1.075 & 1.063 & 1.089 & 1.097 & 1.082 & 1.107 & 1.093 & 1.072 \\
		& 12 & 1.091 & 1.085 & 1.073 & 1.087 & 1.099 & 1.087 & 1.091 & 1.099 & 1.093 \\
		& 15 & 1.106 & 1.102 & 1.110 & 1.132 & 1.130 & 1.129 & 1.134 & 1.142 & 1.182 \\
		& 18 & 1.098 & 1.093 & 1.095 & 1.117 & 1.107 & 1.110 & 1.119 & 1.135 & 1.196 \\
		& 24 & 1.040 & 1.038 & 1.027 & 1.051 & 1.047 & 1.045 & 1.061 & 1.064 & 1.094 \\
		\bottomrule
	\end{tabular}
	\begin{minipage}{\textwidth}
		\small
		\vspace{0.2cm}
		\textit{Notes:} The table reports quantile scores at the 10\% level relative to a benchmark constant-parameter VAR($p$) estimated by ordinary least squares. Values below one (in bold) indicate superior performance relative to the benchmark. The forecast horizon is denoted by $h$, and $p$ denotes the lag order. TVP-VAR-EB refers to the baseline specification with the ordering: growth, inflation, and interest rates. TVP-VAR-EB Alt uses the alternative ordering: interest rates, inflation, and growth.
	\end{minipage}
\end{table}

\FloatBarrier

\subsubsection{Euro Area quarterly data}

\begin{table}[htbp!]
	\centering
	\caption{Mean squared prediction error results comparing different TVP-VAR-EB orderings: Euro Area quarterly data.}
	\label{tab:forecast_accuracy}
	\scriptsize
	\begin{tabular}{@{}clccc ccc ccc@{}}
		\toprule
		& & \multicolumn{3}{c}{$p=1$} & \multicolumn{3}{c}{$p=2$} & \multicolumn{3}{c}{$p=3$} \\
		\cmidrule(lr){3-5} \cmidrule(lr){6-8} \cmidrule(lr){9-11}
		&h & {AVP-VAR} & \makecell{TVP-VAR\\EB} & \makecell{TVP-VAR\\EB Alt} 
		& {AVP-VAR} & \makecell{TVP-VAR\\EB} & \makecell{TVP-VAR\\EB Alt} 
		& {AVP-VAR} & \makecell{TVP-VAR\\EB} & \makecell{TVP-VAR\\EB Alt} \\
		\midrule
		\multicolumn{11}{@{}l}{GDPC1} \\
		\addlinespace[0.1cm]
		& 1 & \textbf{0.425} & \textbf{0.547} & \textbf{0.582} & \textbf{0.391} & \textbf{0.620} & \textbf{0.631} & \textbf{0.391} & \textbf{0.584} & \textbf{0.598} \\
		& 2 & \textbf{0.636} & \textbf{0.682} & \textbf{0.700} & \textbf{0.598} & \textbf{0.813} & \textbf{0.856} & \textbf{0.565} & \textbf{0.807} & \textbf{0.824} \\
		& 3 & \textbf{0.652} & \textbf{0.673} & \textbf{0.678} & \textbf{0.613} & \textbf{0.708} & \textbf{0.748} & \textbf{0.592} & \textbf{0.736} & \textbf{0.755} \\
		& 4 & \textbf{0.606} & \textbf{0.623} & \textbf{0.624} & \textbf{0.579} & \textbf{0.642} & \textbf{0.673} & \textbf{0.538} & \textbf{0.623} & \textbf{0.640} \\
		& 5 & \textbf{0.625} & \textbf{0.639} & \textbf{0.640} & \textbf{0.602} & \textbf{0.656} & \textbf{0.689} & \textbf{0.585} & \textbf{0.655} & \textbf{0.677} \\
		& 6 & \textbf{0.657} & \textbf{0.671} & \textbf{0.671} & \textbf{0.649} & \textbf{0.699} & \textbf{0.723} & \textbf{0.640} & \textbf{0.710} & \textbf{0.729} \\
		& 7 & \textbf{0.672} & \textbf{0.687} & \textbf{0.685} & \textbf{0.672} & \textbf{0.713} & \textbf{0.738} & \textbf{0.661} & \textbf{0.722} & \textbf{0.742} \\
		& 8 & \textbf{0.670} & \textbf{0.681} & \textbf{0.680} & \textbf{0.673} & \textbf{0.709} & \textbf{0.733} & \textbf{0.679} & \textbf{0.733} & \textbf{0.753} \\
		\addlinespace[0.2cm]
		\multicolumn{11}{@{}l}{HICP} \\
		\addlinespace[0.1cm]
		& 1 & 1.263 & \textbf{0.928} & \textbf{0.925} & 1.682 & \textbf{0.997} & \textbf{0.990} & 1.375 & 1.134 & 1.122 \\
		& 2 & 1.211 & \textbf{0.999} & \textbf{0.990} & \textbf{0.971} & \textbf{0.955} & \textbf{0.954} & 1.106 & 1.038 & 1.032 \\
		& 3 & \textbf{0.921} & \textbf{0.886} & \textbf{0.870} & 1.018 & \textbf{0.838} & \textbf{0.838} & \textbf{0.995} & \textbf{0.949} & \textbf{0.951} \\
		& 4 & 1.337 & \textbf{0.938} & \textbf{0.915} & \textbf{0.925} & \textbf{0.786} & \textbf{0.779} & 1.051 & \textbf{0.877} & \textbf{0.861} \\
		& 5 & 1.025 & \textbf{0.840} & \textbf{0.804} & 0.753 & \textbf{0.751} & \textbf{0.751} & 0.763 & \textbf{0.847} & \textbf{0.832} \\
		& 6 & 1.193 & \textbf{0.865} & \textbf{0.831} & 0.756 & \textbf{0.711} & \textbf{0.709} & 0.774 & \textbf{0.751} & \textbf{0.740} \\
		& 7 & 1.03 & \textbf{0.795} & \textbf{0.757} & 0.672 & \textbf{0.629} & \textbf{0.629} & 0.745 & \textbf{0.728} & \textbf{0.723} \\
		& 8 & 1.121 & \textbf{0.796} & \textbf{0.766} & 0.603 & \textbf{0.562} & \textbf{0.561} & 0.719 & \textbf{0.608} & \textbf{0.591} \\
		\bottomrule
	\end{tabular}
	\begin{minipage}{\textwidth}
		\small
		\vspace{0.2cm}
		\textit{Notes:} The table reports mean squared prediction errors (MSPE) relative to a benchmark constant-parameter VAR($p$) estimated by ordinary least squares. Values below one (in bold) indicate superior performance relative to the benchmark. The forecast horizon is denoted by $h$, and $p$ denotes the lag order. TVP-VAR-EB refers to the baseline specification with the ordering: growth, inflation, and interest rates. TVP-VAR-EB Alt uses the alternative ordering: interest rates, inflation, and growth.
	\end{minipage}
\end{table}

\begin{table}[htbp!]
	\centering
	\caption{Quantile Score (90\%) results comparing different TVP-VAR-EB orderings: Euro Area quarterly data.}
	\label{tab:quantile_score_90}
	\scriptsize
	\begin{tabular}{@{}clccc ccc ccc@{}}
		\toprule
		& & \multicolumn{3}{c}{$p=1$} & \multicolumn{3}{c}{$p=2$} & \multicolumn{3}{c}{$p=3$} \\
		\cmidrule(lr){3-5} \cmidrule(lr){6-8} \cmidrule(lr){9-11}
		&h & {AVP-VAR} & \makecell{TVP-VAR\\EB} & \makecell{TVP-VAR\\EB Alt} 
		& {AVP-VAR} & \makecell{TVP-VAR\\EB} & \makecell{TVP-VAR\\EB Alt} 
		& {AVP-VAR} & \makecell{TVP-VAR\\EB} & \makecell{TVP-VAR\\EB Alt} \\
		\midrule
		\multicolumn{11}{@{}l}{ GDPC1} \\
		\addlinespace[0.1cm]
		& 1 & \textbf{0.434} & \textbf{0.637} & \textbf{0.681} & \textbf{0.423} & \textbf{0.827} & \textbf{0.848} & \textbf{0.363} & \textbf{0.730} & \textbf{0.748} \\
		& 2 & \textbf{0.670} & \textbf{0.738} & \textbf{0.762} & \textbf{0.654} & \textbf{0.925} & \textbf{0.975} & \textbf{0.590} & \textbf{0.883} & \textbf{0.892} \\
		& 3 & \textbf{0.751} & \textbf{0.826} & \textbf{0.824} & \textbf{0.722} & \textbf{0.891} & \textbf{0.947} & \textbf{0.717} & \textbf{0.905} & \textbf{0.932} \\
		& 4 & \textbf{0.721} & \textbf{0.815} & \textbf{0.815} & \textbf{0.719} & \textbf{0.858} & \textbf{0.888} & \textbf{0.641} & \textbf{0.757} & \textbf{0.760} \\
		& 5 & \textbf{0.711} & \textbf{0.815} & \textbf{0.812} & \textbf{0.713} & \textbf{0.855} & \textbf{0.872} & \textbf{0.703} & \textbf{0.835} & \textbf{0.836} \\
		& 6 & \textbf{0.768} & \textbf{0.884} & \textbf{0.873} & \textbf{0.774} & \textbf{0.942} & \textbf{0.957} & \textbf{0.768} & \textbf{0.918} & \textbf{0.919} \\
		& 7 & \textbf{0.821} & \textbf{0.923} & \textbf{0.920} & \textbf{0.828} & 1.001 & 1.025 & \textbf{0.826} & \textbf{0.976} & \textbf{0.991} \\
		& 8 & \textbf{0.813} & \textbf{0.924} & \textbf{0.918} & \textbf{0.833} & 1.004 & 1.029 & \textbf{0.835} & \textbf{0.998} & 1.012 \\
		\addlinespace[0.2cm]
		\multicolumn{11}{@{}l}{ HICP} \\
		\addlinespace[0.1cm]
		& 1 & \textbf{0.777} & \textbf{0.844} & \textbf{0.830} & \textbf{0.799} & \textbf{0.875} & \textbf{0.891} & \textbf{0.765} & \textbf{0.868} & \textbf{0.871} \\
		& 2 & 1.249 & 1.214 & 1.178 & \textbf{0.910} & \textbf{0.881} & \textbf{0.895} & \textbf{0.921} & \textbf{0.879} & \textbf{0.885} \\
		& 3 & 1.151 & 1.158 & 1.120 & \textbf{0.871} & \textbf{0.806} & \textbf{0.795} & \textbf{0.819} & \textbf{0.768} & \textbf{0.751} \\
		& 4 & 1.083 & 1.106 & 1.063 & \textbf{0.899} & \textbf{0.823} & \textbf{0.824} & 1.025 & \textbf{0.949} & \textbf{0.966} \\
		& 5 & 1.068 & 1.076 & 1.045 & \textbf{0.706} & \textbf{0.689} & \textbf{0.709} & \textbf{0.713} & \textbf{0.754} & \textbf{0.765} \\
		& 6 & 1.018 & 1.020 & \textbf{0.989} & \textbf{0.742} & \textbf{0.736} & \textbf{0.749} & \textbf{0.795} & \textbf{0.784} & \textbf{0.788} \\
		& 7 & \textbf{0.981} & \textbf{0.989} & \textbf{0.959} & \textbf{0.695} & \textbf{0.692} & \textbf{0.702} & \textbf{0.773} & \textbf{0.770} & \textbf{0.775} \\
		& 8 & \textbf{0.917} & \textbf{0.987} & \textbf{0.949} & \textbf{0.673} & \textbf{0.718} & \textbf{0.727} & \textbf{0.767} & \textbf{0.827} & \textbf{0.833} \\
		\bottomrule
	\end{tabular}
	\begin{minipage}{\textwidth}
		\small
		\vspace{0.2cm}
		\textit{Notes:} The table reports quantile scores at the 90\% level relative to a benchmark constant-parameter VAR($p$) estimated by ordinary least squares. Values below one (in bold) indicate superior performance relative to the benchmark. The forecast horizon is denoted by $h$, and $p$ denotes the lag order. TVP-VAR-EB refers to the baseline specification with the ordering: growth, inflation, and interest rates. TVP-VAR-EB Alt uses the alternative ordering: interest rates, inflation, and growth.
	\end{minipage}
\end{table}

\begin{table}[htbp!]
	\centering
	\caption{Quantile Score (10\%) results comparing different TVP-VAR-EB orderings: Euro Area quarterly data.}
	\label{tab:quantile_score_10}
	\scriptsize
	\begin{tabular}{@{}clccc ccc ccc@{}}
		\toprule
		& & \multicolumn{3}{c}{$p=1$} & \multicolumn{3}{c}{$p=2$} & \multicolumn{3}{c}{$p=3$} \\
		\cmidrule(lr){3-5} \cmidrule(lr){6-8} \cmidrule(lr){9-11}
		&h & {AVP-VAR} & \makecell{TVP-VAR\\EB} & \makecell{TVP-VAR\\EB Alt} 
		& {AVP-VAR} & \makecell{TVP-VAR\\EB} & \makecell{TVP-VAR\\EB Alt} 
		& {AVP-VAR} & \makecell{TVP-VAR\\EB} & \makecell{TVP-VAR\\EB Alt} \\
		\midrule
		\multicolumn{11}{@{}l}{GDPC1} \\
		\addlinespace[0.1cm]
		& 1 & 1.057 & 1.025 & 1.011 & 1.069 & 1.253 & 1.238 & 1.055 & 1.332 & 1.294 \\
		& 2 & 1.073 & 1.154 & 1.169 & 1.094 & 1.331 & 1.367 & 1.010 & 1.251 & 1.273 \\
		& 3 & 1.084 & 1.167 & 1.182 & 1.099 & 1.320 & 1.362 & 1.011 & 1.248 & 1.272 \\
		& 4 & 1.086 & 1.153 & 1.179 & 1.107 & 1.319 & 1.372 & 1.066 & 1.319 & 1.348 \\
		& 5 & 1.089 & 1.163 & 1.182 & 1.103 & 1.326 & 1.393 & 1.044 & 1.312 & 1.346 \\
		& 6 & 1.098 & 1.147 & 1.170 & 1.096 & 1.322 & 1.395 & 1.071 & 1.322 & 1.361 \\
		& 7 & 1.092 & 1.138 & 1.162 & 1.100 & 1.310 & 1.375 & 1.061 & 1.319 & 1.354 \\
		& 8 & 1.109 & 1.148 & 1.161 & 1.105 & 1.281 & 1.337 & 1.086 & 1.298 & 1.327 \\
		\addlinespace[0.2cm]
		\multicolumn{11}{@{}l}{HICP} \\
		\addlinespace[0.1cm]
		& 1 & \textbf{0.727} & \textbf{0.676} & \textbf{0.664} & 1.055 & \textbf{0.900} & \textbf{0.892} & \textbf{0.989} & \textbf{0.869} & \textbf{0.852} \\
		& 2 & 1.053 & 1.125 & 1.097 & \textbf{0.901} & \textbf{0.909} & \textbf{0.889} & \textbf{0.915} & 1.027 & \textbf{0.997} \\
		& 3 & \textbf{0.837} & \textbf{0.892} & \textbf{0.859} & \textbf{0.854} & \textbf{0.872} & \textbf{0.856} & \textbf{0.897} & \textbf{0.923} & \textbf{0.904} \\
		& 4 & 1.154 & 1.211 & 1.181 & 1.017 & 1.024 & 1.001 & 1.085 & 1.131 & 1.087 \\
		& 5 & \textbf{0.902} & \textbf{0.887} & \textbf{0.870} & \textbf{0.806} & \textbf{0.856} & \textbf{0.851} & \textbf{0.855} & \textbf{0.921} & \textbf{0.904} \\
		& 6 & 1.092 & 1.102 & 1.081 & \textbf{0.915} & \textbf{0.958} & \textbf{0.942} & \textbf{0.911} & 1.085 & 1.061 \\
		& 7 & 1.044 & \textbf{0.965} & \textbf{0.950} & \textbf{0.900} & \textbf{0.930} & \textbf{0.924} & \textbf{0.897} & 1.012 & \textbf{0.993} \\
		& 8 & 1.077 & 1.124 & 1.103 & 1.029 & 1.084 & 1.073 & 1.046 & 1.283 & 1.254 \\
		\bottomrule
	\end{tabular}
	\begin{minipage}{\textwidth}
		\small
		\vspace{0.2cm}
		\textit{Notes:} The table reports quantile scores at the 10\% level relative to a benchmark constant-parameter VAR($p$) estimated by ordinary least squares. Values below one (in bold) indicate superior performance relative to the benchmark. The forecast horizon is denoted by $h$, and $p$ denotes the lag order. TVP-VAR-EB refers to the baseline specification with the ordering: growth, inflation, and interest rates. TVP-VAR-EB Alt uses the alternative ordering: interest rates, inflation, and growth.
	\end{minipage}
\end{table}

\FloatBarrier


\subsection{Robustness exercise: FRED--QD data}			

\begin{table}[htbp!]
	\centering
	\singlespacing
	\small
	\caption{Variables and Drivers FRED-QD quarterly data}
	\label{tab:fred_qd_variables}
	\begin{threeparttable}
		\begin{tabular}{@{}llc@{}}
			\toprule
			\textbf{Series ID} & \textbf{Description}  & \textbf{tcode} \\
			\midrule
			\multicolumn{3}{c}{{ VAR Variables}} \\
			\midrule
			GDPC1 & Real Gross Domestic Product  & 5 \\
			PCECTPI & Personal Consumption Expenditures &  5 \\
			FEDFUNDS & Effective Federal Funds Rate &  2 \\
			\addlinespace[0.5em]
			\midrule
			\multicolumn{3}{c}{{ Drivers ($Z_t$)}} \\
			\midrule
			S\&P 500 & S\&P's Common Stock Price Index: Composite & 5 \\
			S\&P PE ratio & S\&P's Price-Earnings Ratio &  5 \\
			PERMITS & New Private Housing Units  & 5 \\
			PPIIDC & Producer Price Index by Ind. Commodities &  6 \\
			BAA10YM & Baa Corp. Bond Yield Relative to 10Y Tbill &  1 \\
			TOTALSLx & Total Consumer Credit Outstanding &  5 \\
			EXJPUSx & Japan / U.S. Foreign Exchange Rate &  5 \\
			VXOCLSx & CBOE S\&P 100 Volatility Index & 1 \\
			UMCSENTx & University of Michigan: Consumer Sentiment &  1 \\
			\midrule
			Frequency: & Quarterly data &   \\
			Period: & 1963Q1 -- 2025Q2 &  \\
			\bottomrule
		\end{tabular}
		\begin{tablenotes}
			\footnotesize
			\item \textit{Note:} Data are from \citet*{mccracken2020fred}, see this paper for definitions of transformation codes (column \emph{tcode}). 				
		\end{tablenotes}
	\end{threeparttable}
\end{table}

\subsubsection{FRED--QD: VAR(1)}	
\begin{table}[htbp!]
	\centering
	\caption{Forecasting Performance VAR(1): Mean Squared Prediction Error Relative to Benchmark} \label{tab:forecasting_mspe_FRED_p1}
	\scriptsize
	\begin{tabular}{@{}cS[table-format=1.3]S[table-format=1.3]S[table-format=1.3]S[table-format=1.3]S[table-format=1.3]S[table-format=1.3]S[table-format=1.3]S[table-format=1.3]@{}}
		\toprule
		{\textbf{h}} & {\textbf{AVP-VAR}} & {\textbf{CP-VAR}} & {\makecell{ \textbf{CP-VAR} \\ \textbf{SV}}} & {\makecell{ \textbf{{TVP-VAR}} \\ \textbf{EB}}} &{\textbf{TVP-VAR}} & {\makecell{ \textbf{VAR} \\ \textbf{SVOt}}} & {\textbf{FAVAR}} & {\makecell{ \textbf{FAVAR} \\ \textbf{SV}}}  \\
		\midrule
		\multicolumn{9}{c}{\textbf{Panel A: GDPC1}} \\
		\midrule
		1 & \textbf{0.747}$^{\dag}$ & \textbf{0.974} & \textbf{0.892} & \textbf{0.875} & \textbf{0.758} & \textbf{0.887} & \textbf{0.930} & \textbf{0.808} \\
		2 & \textbf{0.919}$^{\dag}$ & \textbf{0.986} & \textbf{0.952} & \textbf{0.923} & \textbf{0.923} & \textbf{0.963} & \textbf{0.994} & \textbf{0.949} \\
		3 & \textbf{0.944}$^{\dag}$ & \textbf{0.982} & \textbf{0.972} & \textbf{0.947} & \textbf{0.949} & \textbf{0.975} & \textbf{0.994} & \textbf{0.972} \\
		4 & \textbf{0.961}$^{\dag}$ & \textbf{0.989} & \textbf{0.982} & \textbf{0.965} & \textbf{0.963} & \textbf{0.985} & 1.014 & \textbf{0.986} \\
		5 & \textbf{0.956}$^{\dag}$ & \textbf{0.993} & \textbf{0.977} & \textbf{0.962} & \textbf{0.966} & \textbf{0.985} & 1.009 & \textbf{0.988} \\
		6 & \textbf{0.970} & \textbf{0.986} & \textbf{0.979} & \textbf{0.983} & \textbf{0.967}$^{\dag}$ & \textbf{0.994} & 1.011 & \textbf{0.991} \\
		7 & \textbf{0.975}$^{\dag}$ & \textbf{0.993} & \textbf{0.991} & \textbf{0.986} & \textbf{0.978} & 1.003 & 1.016 & 1.011 \\
		8 & \textbf{0.974} & \textbf{0.988} & \textbf{0.997} & \textbf{0.978} & \textbf{0.971}$^{\dag}$ & \textbf{0.995} & 1.015 & 1.002 \\
		\midrule
		\multicolumn{9}{c}{\textbf{Panel B: PCECTPI}} \\
		\midrule
		1 & 1.150 & \textbf{0.980} & \textbf{0.974} & \textbf{0.980} & 1.242 & \textbf{0.954}$^{\dag}$ & 1.044 & \textbf{0.976} \\
		2 & \textbf{0.851} & \textbf{0.971} & \textbf{0.945} & \textbf{0.934} & \textbf{0.821}$^{\dag}$ & \textbf{0.903} & 1.111 & \textbf{0.931} \\
		3 & \textbf{0.844} & \textbf{0.962} & \textbf{0.904} & \textbf{0.904} & \textbf{0.830}$^{\dag}$ & \textbf{0.886} & 1.136 & \textbf{0.900} \\
		4 & \textbf{0.790} & \textbf{0.953} & \textbf{0.881} & \textbf{0.884} & \textbf{0.752}$^{\dag}$ & \textbf{0.869} & 1.127 & \textbf{0.883} \\
		5 & \textbf{0.809} & \textbf{0.943} & \textbf{0.844} & \textbf{0.851} & \textbf{0.766}$^{\dag}$ & \textbf{0.837} & 1.084 & \textbf{0.834} \\
		6 & \textbf{0.802} & \textbf{0.938} & \textbf{0.834} & \textbf{0.847} & \textbf{0.736}$^{\dag}$ & \textbf{0.809} & 1.047 & \textbf{0.833} \\
		7 & \textbf{0.760} & \textbf{0.946} & \textbf{0.821} & \textbf{0.845} & \textbf{0.698}$^{\dag}$ & \textbf{0.781} & 1.042 & \textbf{0.817} \\
		8 & \textbf{0.746} & \textbf{0.948} & \textbf{0.801} & \textbf{0.844} & \textbf{0.691}$^{\dag}$ & \textbf{0.769} & 1.047 & \textbf{0.800} \\
		\bottomrule
	\end{tabular}	
	\vspace{0.5em}
	\par
	\begin{tabularx}{\textwidth}{X}
		{\footnotesize \textit{Notes:} The table reports mean squared prediction errors (MSPE) for each of the two forecasted variables, relative to a benchmark constant-parameter VAR($1$) estimated via least squares. Values below one (highlighted in bold) indicate superior forecasting performance compared to the benchmark. The forecast horizon is denoted by $h$ (in quarters), and the symbol $\dag$ marks the best-performing model at each horizon.}
	\end{tabularx}
\end{table}

\begin{table}[htbp!]
	\centering
	\caption{Forecasting Performance VAR(1): Quantile Scores 90\% Relative to Benchmark} \label{tab:forecasting_qs90_FRED_p1}
	\scriptsize
	\begin{tabular}{@{}cS[table-format=1.3]S[table-format=1.3]S[table-format=1.3]S[table-format=1.3]S[table-format=1.3]S[table-format=1.3]S[table-format=1.3]S[table-format=1.3]@{}}
		\toprule
		{\textbf{h}} & {\textbf{AVP-VAR}} & {\textbf{CP-VAR}} & {\makecell{ \textbf{CP-VAR} \\ \textbf{SV}}} & {\makecell{ \textbf{{TVP-VAR}} \\ \textbf{EB}}} &{\textbf{TVP-VAR}} & {\makecell{ \textbf{VAR} \\ \textbf{SVOt}}} & {\textbf{FAVAR}} & {\makecell{ \textbf{FAVAR} \\ \textbf{SV}}}  \\
		\midrule
		\multicolumn{9}{c}{\textbf{Panel A: GDPC1}} \\
		\midrule
		1 & \textbf{0.648} & \textbf{0.909} & \textbf{0.591} & \textbf{0.778} & \textbf{0.575} & \textbf{0.868} & \textbf{0.903} & \textbf{0.549}$^{\dag}$ \\
		2 & \textbf{0.851} & \textbf{0.910} & \textbf{0.899} & \textbf{0.912} & \textbf{0.838}$^{\dag}$ & \textbf{0.892} & \textbf{0.942} & \textbf{0.864} \\
		3 & \textbf{0.880}$^{\dag}$ & \textbf{0.974} & \textbf{0.938} & \textbf{0.955} & \textbf{0.881} & \textbf{0.908} & \textbf{0.968} & \textbf{0.937} \\
		4 & \textbf{0.885}$^{\dag}$ & \textbf{0.975} & \textbf{0.944} & \textbf{0.958} & \textbf{0.890} & \textbf{0.922} & \textbf{0.992} & \textbf{0.955} \\
		5 & \textbf{0.880}$^{\dag}$ & \textbf{0.977} & \textbf{0.959} & \textbf{0.979} & \textbf{0.898} & \textbf{0.959} & 1.001 & \textbf{0.959} \\
		6 & \textbf{0.909}$^{\dag}$ & \textbf{0.975} & \textbf{0.965} & \textbf{0.993} & \textbf{0.916} & \textbf{0.973} & 1.010 & \textbf{0.985} \\
		7 & \textbf{0.925}$^{\dag}$ & \textbf{0.980} & \textbf{0.981} & 1.009 & \textbf{0.925} & \textbf{0.999} & 1.013 & \textbf{0.991} \\
		8 & \textbf{0.952} & \textbf{0.980} & 1.023 & 1.037 & \textbf{0.951}$^{\dag}$ & 1.012 & 1.011 & 1.018 \\
		\midrule
		\multicolumn{9}{c}{\textbf{Panel B: PCECTPI}} \\
		\midrule
		1 & 1.023 & \textbf{0.943} & \textbf{0.905} & \textbf{0.923} & \textbf{0.954} & \textbf{0.854}$^{\dag}$ & 1.053 & \textbf{0.904} \\
		2 & \textbf{0.857} & \textbf{0.950} & \textbf{0.807} & \textbf{0.869} & \textbf{0.779}$^{\dag}$ & \textbf{0.794} & 1.067 & \textbf{0.810} \\
		3 & \textbf{0.825} & \textbf{0.910} & \textbf{0.838} & \textbf{0.875} & \textbf{0.721}$^{\dag}$ & \textbf{0.840} & 1.075 & \textbf{0.825} \\
		4 & \textbf{0.805} & \textbf{0.928} & \textbf{0.858} & \textbf{0.898} & \textbf{0.692}$^{\dag}$ & \textbf{0.880} & 1.067 & \textbf{0.857} \\
		5 & \textbf{0.820} & \textbf{0.940} & \textbf{0.904} & \textbf{0.952} & \textbf{0.732}$^{\dag}$ & \textbf{0.950} & 1.059 & \textbf{0.902} \\
		6 & \textbf{0.806} & \textbf{0.934} & \textbf{0.907} & \textbf{0.968} & \textbf{0.719}$^{\dag}$ & \textbf{0.942} & 1.014 & \textbf{0.918} \\
		7 & \textbf{0.813} & \textbf{0.937} & \textbf{0.943} & 1.007 & \textbf{0.707}$^{\dag}$ & \textbf{0.982} & \textbf{0.998} & \textbf{0.944} \\
		8 & \textbf{0.830} & \textbf{0.958} & \textbf{0.979} & 1.049 & \textbf{0.724}$^{\dag}$ & 1.006 & 1.011 & \textbf{0.982} \\
		\bottomrule
	\end{tabular}	
	\vspace{0.5em}
	\par
	\begin{tabularx}{\textwidth}{X}
		{\footnotesize \textit{Notes:} The table reports quantile scores 90\% (QS90) for each of the two forecasted variables, relative to a benchmark constant-parameter VAR($1$) estimated via least squares. Values below one (highlighted in bold) indicate superior forecasting performance compared to the benchmark. The forecast horizon is denoted by $h$ (in quarters), and the symbol $\dag$ marks the best-performing model at each horizon.}
	\end{tabularx}
\end{table}

\begin{table}[htbp!]
	\centering
	\caption{Forecasting Performance VAR(1): Quantile Scores 10\% Relative to Benchmark} \label{tab:forecasting_qs10_FRED_p1}
	\scriptsize
	\begin{tabular}{@{}cS[table-format=1.3]S[table-format=1.3]S[table-format=1.3]S[table-format=1.3]S[table-format=1.3]S[table-format=1.3]S[table-format=1.3]S[table-format=1.3]@{}}
		\toprule
		{\textbf{h}} & {\textbf{AVP-VAR}} & {\textbf{CP-VAR}} & {\makecell{ \textbf{CP-VAR} \\ \textbf{SV}}} & {\makecell{ \textbf{{TVP-VAR}} \\ \textbf{EB}}} &{\textbf{TVP-VAR}} & {\makecell{ \textbf{VAR} \\ \textbf{SVOt}}} & {\textbf{FAVAR}} & {\makecell{ \textbf{FAVAR} \\ \textbf{SV}}}  \\
		\midrule
		\multicolumn{9}{c}{\textbf{Panel A: GDPC1}} \\
		\midrule
		1 & 1.066 & \textbf{0.952} & 1.056 & 1.065 & 1.037 & \textbf{0.994} & \textbf{0.944}$^{\dag}$ & 1.016 \\
		2 & 1.037 & \textbf{0.973} & 1.074 & 1.074 & 1.074 & \textbf{0.981} & \textbf{0.957}$^{\dag}$ & 1.052 \\
		3 & 1.097 & \textbf{0.984}$^{\dag}$ & 1.095 & 1.122 & 1.078 & 1.034 & \textbf{0.998} & 1.084 \\
		4 & 1.118 & 1.005 & 1.116 & 1.151 & 1.101 & 1.049 & 1.004$^{\dag}$ & 1.103 \\
		5 & 1.107 & 1.013 & 1.116 & 1.148 & 1.099 & 1.057 & \textbf{0.992}$^{\dag}$ & 1.076 \\
		6 & 1.145 & 1.015 & 1.116 & 1.170 & 1.089 & 1.070 & \textbf{0.991}$^{\dag}$ & 1.101 \\
		7 & 1.129 & 1.002 & 1.114 & 1.161 & 1.091 & 1.077 & \textbf{0.991}$^{\dag}$ & 1.097 \\
		8 & 1.154 & 1.010 & 1.132 & 1.153 & 1.084 & 1.071 & 1.003 & 1.084 \\
		\midrule
		\multicolumn{9}{c}{\textbf{Panel B: PCECTPI}} \\
		\midrule
		1 & 1.164 & 1.014 & 1.118 & 1.146 & 1.110 & 1.118 & 1.006 & 1.116 \\
		2 & 1.003 & 1.014 & 1.098 & 1.125 & 1.052 & 1.028 & 1.003 & 1.101 \\
		3 & 1.053 & 1.002 & 1.119 & 1.159 & 1.102 & 1.047 & 1.003 & 1.111 \\
		4 & 1.040 & \textbf{0.996}$^{\dag}$ & 1.141 & 1.163 & 1.070 & 1.049 & 1.002 & 1.136 \\
		5 & 1.055 & 1.021 & 1.124 & 1.120 & 1.101 & 1.107 & 1.017 & 1.125 \\
		6 & 1.080 & 1.033 & 1.169 & 1.140 & 1.130 & 1.134 & 1.006 & 1.163 \\
		7 & 1.074 & 1.031 & 1.132 & 1.120 & 1.106 & 1.109 & 1.025 & 1.136 \\
		8 & 1.094 & 1.032 & 1.158 & 1.128 & 1.159 & 1.142 & 1.031 & 1.191 \\
		\bottomrule
	\end{tabular}	
	\vspace{0.5em}
	\par
	\begin{tabularx}{\textwidth}{X}
		{\footnotesize \textit{Notes:} The table reports quantile scores 10\% (QS10) for each of the two forecasted variables, relative to a benchmark constant-parameter VAR($1$) estimated via least squares. Values below one (highlighted in bold) indicate superior forecasting performance compared to the benchmark. The forecast horizon is denoted by $h$ (in quarters), and the symbol $\dag$ marks the best-performing model at each horizon.}
	\end{tabularx}
\end{table}

\begin{figure}[htbp!]  
	\centering
	\includegraphics[width=0.95\textwidth]{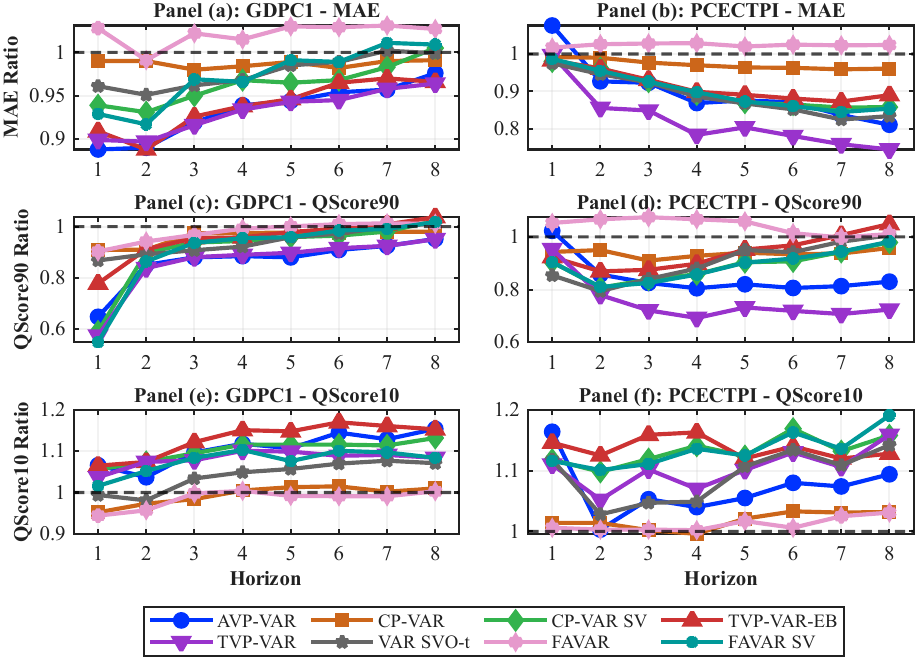}  
	\caption{Forecasting results using FRED quarterly data. The top row displays mean absolute forecast errors (MAE), while the middle and bottom rows show quantile scores for the upper (90th percentile) and lower (10th percentile) tails of each variable. All metrics are reported as ratios relative to a benchmark constant-parameter VAR($1$) estimated via least squares. A model is considered superior when its relative MAE, or relative quantile scores, are below one.} \label{fig:oos_FRED_p1}
\end{figure}

\FloatBarrier
\subsubsection{FRED--QD: VAR(2)}	

\begin{table}[htbp!]
	\centering
	\caption{Forecasting Performance VAR(2): Mean Squared Prediction Error Relative to Benchmark} \label{tab:forecasting_mspe_FRED_p2}
	\scriptsize
	\begin{tabular}{@{}cS[table-format=1.3]S[table-format=1.3]S[table-format=1.3]S[table-format=1.3]S[table-format=1.3]S[table-format=1.3]S[table-format=1.3]S[table-format=1.3]@{}}
		\toprule
		{\textbf{h}} & {\textbf{AVP-VAR}} & {\textbf{CP-VAR}} & {\makecell{ \textbf{CP-VAR} \\ \textbf{SV}}} & {\makecell{ \textbf{{TVP-VAR}} \\ \textbf{EB}}} &{\textbf{TVP-VAR}} & {\makecell{ \textbf{VAR} \\ \textbf{SVOt}}} & {\textbf{FAVAR}} & {\makecell{ \textbf{FAVAR} \\ \textbf{SV}}}  \\
		\midrule
		\multicolumn{9}{c}{\textbf{Panel A: GDPC1}} \\
		\midrule
		1 & \textbf{0.717} & \textbf{0.937} & \textbf{0.826} & \textbf{0.848} & \textbf{0.698}$^{\dag}$ & \textbf{0.846} & \textbf{0.943} & \textbf{0.857} \\
		2 & \textbf{0.830}$^{\dag}$ & \textbf{0.968} & \textbf{0.930} & \textbf{0.940} & \textbf{0.848} & \textbf{0.934} & \textbf{0.982} & \textbf{0.941} \\
		3 & \textbf{0.918}$^{\dag}$ & \textbf{0.962} & \textbf{0.956} & \textbf{0.956} & \textbf{0.918} & \textbf{0.954} & \textbf{0.980} & \textbf{0.941} \\
		4 & \textbf{0.944}$^{\dag}$ & \textbf{0.973} & \textbf{0.966} & \textbf{0.975} & \textbf{0.946} & \textbf{0.968} & \textbf{0.996} & \textbf{0.973} \\
		5 & \textbf{0.945} & \textbf{0.977} & \textbf{0.960} & \textbf{0.955} & \textbf{0.944}$^{\dag}$ & \textbf{0.968} & \textbf{0.989} & \textbf{0.965} \\
		6 & \textbf{0.958}$^{\dag}$ & \textbf{0.990} & \textbf{0.975} & \textbf{0.989} & \textbf{0.962} & \textbf{0.977} & 1.001 & \textbf{0.998} \\
		7 & \textbf{0.961}$^{\dag}$ & \textbf{0.985} & \textbf{0.974} & \textbf{0.976} & \textbf{0.965} & \textbf{0.990} & 1.005 & \textbf{0.994} \\
		8 & \textbf{0.956}$^{\dag}$ & \textbf{0.976} & \textbf{0.980} & \textbf{0.964} & \textbf{0.959} & \textbf{0.990} & \textbf{0.998} & \textbf{0.993} \\
		\midrule
		\multicolumn{9}{c}{\textbf{Panel B: PCECTPI}} \\
		\midrule
		1 & 1.118 & \textbf{0.969} & \textbf{0.962} & \textbf{0.978} & 1.000 & \textbf{0.937}$^{\dag}$ & 1.043 & \textbf{0.968} \\
		2 & \textbf{0.960} & \textbf{0.972} & \textbf{0.939} & \textbf{0.934}$^{\dag}$ & \textbf{0.937} & \textbf{0.942} & 1.120 & \textbf{0.960} \\
		3 & \textbf{0.927} & \textbf{0.953} & \textbf{0.925} & \textbf{0.920} & \textbf{0.849}$^{\dag}$ & \textbf{0.917} & 1.171 & \textbf{0.932} \\
		4 & \textbf{0.880} & \textbf{0.952} & \textbf{0.916} & \textbf{0.895} & \textbf{0.804}$^{\dag}$ & \textbf{0.904} & 1.183 & \textbf{0.922} \\
		5 & \textbf{0.932} & \textbf{0.948} & \textbf{0.900} & \textbf{0.875} & \textbf{0.818}$^{\dag}$ & \textbf{0.884} & 1.158 & \textbf{0.909} \\
		6 & \textbf{0.902} & \textbf{0.956} & \textbf{0.888} & \textbf{0.879} & \textbf{0.775}$^{\dag}$ & \textbf{0.878} & 1.142 & \textbf{0.895} \\
		7 & \textbf{0.874} & \textbf{0.943} & \textbf{0.885} & \textbf{0.868} & \textbf{0.770}$^{\dag}$ & \textbf{0.867} & 1.140 & \textbf{0.876} \\
		8 & \textbf{0.842} & \textbf{0.950} & \textbf{0.868} & \textbf{0.843} & \textbf{0.741}$^{\dag}$ & \textbf{0.860} & 1.161 & \textbf{0.873} \\
		\bottomrule
	\end{tabular}	
	\vspace{0.5em}
	\par
	\begin{tabularx}{\textwidth}{X}
		{\footnotesize \textit{Notes:} The table reports mean squared prediction errors (MSPE) for each of the two forecasted variables, relative to a benchmark constant-parameter VAR($2$) estimated via least squares. Values below one (highlighted in bold) indicate superior forecasting performance compared to the benchmark. The forecast horizon is denoted by $h$ (in quarters), and the symbol $\dag$ marks the best-performing model at each horizon.}
	\end{tabularx}
\end{table}

\begin{table}[htbp!]
	\centering
	\caption{Forecasting Performance VAR(2): Quantile Scores 90\% Relative to Benchmark} \label{tab:forecasting_qs90_FRED_p2}
	\scriptsize
	\begin{tabular}{@{}cS[table-format=1.3]S[table-format=1.3]S[table-format=1.3]S[table-format=1.3]S[table-format=1.3]S[table-format=1.3]S[table-format=1.3]S[table-format=1.3]@{}}
		\toprule
		{\textbf{h}} & {\textbf{AVP-VAR}} & {\textbf{CP-VAR}} & {\makecell{ \textbf{CP-VAR} \\ \textbf{SV}}} & {\makecell{ \textbf{{TVP-VAR}} \\ \textbf{EB}}} &{\textbf{TVP-VAR}} & {\makecell{ \textbf{VAR} \\ \textbf{SVOt}}} & {\textbf{FAVAR}} & {\makecell{ \textbf{FAVAR} \\ \textbf{SV}}}  \\
		\midrule
		\multicolumn{9}{c}{\textbf{Panel A: GDPC1}} \\
		\midrule
		1 & \textbf{0.663} & \textbf{0.887} & \textbf{0.573}$^{\dag}$ & \textbf{0.780} & \textbf{0.626} & \textbf{0.850} & \textbf{0.916} & \textbf{0.575} \\
		2 & \textbf{0.835} & \textbf{0.905} & \textbf{0.880} & \textbf{0.892} & \textbf{0.876} & \textbf{0.916} & \textbf{0.971} & \textbf{0.799}$^{\dag}$ \\
		3 & \textbf{0.878}$^{\dag}$ & \textbf{0.951} & \textbf{0.891} & \textbf{0.934} & \textbf{0.904} & \textbf{0.886} & \textbf{0.974} & \textbf{0.909} \\
		4 & \textbf{0.878}$^{\dag}$ & \textbf{0.964} & \textbf{0.912} & \textbf{0.941} & \textbf{0.897} & \textbf{0.902} & \textbf{0.971} & \textbf{0.934} \\
		5 & \textbf{0.885}$^{\dag}$ & \textbf{0.981} & \textbf{0.928} & \textbf{0.972} & \textbf{0.919} & \textbf{0.926} & \textbf{0.984} & \textbf{0.951} \\
		6 & \textbf{0.894}$^{\dag}$ & \textbf{0.981} & \textbf{0.952} & \textbf{0.996} & \textbf{0.933} & \textbf{0.946} & \textbf{0.988} & \textbf{0.979} \\
		7 & \textbf{0.915}$^{\dag}$ & \textbf{0.981} & \textbf{0.964} & 1.013 & \textbf{0.950} & \textbf{0.978} & \textbf{0.998} & \textbf{0.981} \\
		8 & \textbf{0.944}$^{\dag}$ & \textbf{0.988} & \textbf{0.990} & 1.052 & \textbf{0.980} & 1.001 & \textbf{0.998} & 1.013 \\
		\midrule
		\multicolumn{9}{c}{\textbf{Panel B: PCECTPI}} \\
		\midrule
		1 & \textbf{0.943} & \textbf{0.924} & \textbf{0.850} & \textbf{0.816}$^{\dag}$ & \textbf{0.928} & \textbf{0.899} & 1.052 & \textbf{0.853} \\
		2 & \textbf{0.930} & \textbf{0.943} & \textbf{0.783} & \textbf{0.790} & \textbf{0.856} & \textbf{0.873} & 1.094 & \textbf{0.763}$^{\dag}$ \\
		3 & \textbf{0.866} & \textbf{0.882} & \textbf{0.766} & \textbf{0.789} & \textbf{0.803} & \textbf{0.852} & 1.108 & \textbf{0.762}$^{\dag}$ \\
		4 & \textbf{0.862} & \textbf{0.870} & \textbf{0.847} & \textbf{0.859} & \textbf{0.797}$^{\dag}$ & \textbf{0.898} & 1.116 & \textbf{0.820} \\
		5 & \textbf{0.905} & \textbf{0.938} & \textbf{0.905} & \textbf{0.937} & \textbf{0.835}$^{\dag}$ & \textbf{0.958} & 1.114 & \textbf{0.900} \\
		6 & \textbf{0.900} & \textbf{0.927} & \textbf{0.934} & \textbf{0.981} & \textbf{0.823}$^{\dag}$ & \textbf{0.985} & 1.089 & \textbf{0.923} \\
		7 & \textbf{0.923} & \textbf{0.928} & \textbf{0.971} & 1.016 & \textbf{0.829}$^{\dag}$ & 1.028 & 1.105 & \textbf{0.969} \\
		8 & \textbf{0.922} & \textbf{0.947} & 1.003 & 1.035 & \textbf{0.808}$^{\dag}$ & 1.058 & 1.122 & 1.005 \\
		\bottomrule
	\end{tabular}	
	\vspace{0.5em}
	\par
	\begin{tabularx}{\textwidth}{X}
		{\footnotesize \textit{Notes:} The table reports quantile scores 90\% (QS90) for each of the two forecasted variables, relative to a benchmark constant-parameter VAR($2$) estimated via least squares. Values below one (highlighted in bold) indicate superior forecasting performance compared to the benchmark. The forecast horizon is denoted by $h$ (in quarters), and the symbol $\dag$ marks the best-performing model at each horizon.}
	\end{tabularx}
\end{table}

\begin{table}[htbp!]
	\centering
	\caption{Forecasting Performance VAR(2): Quantile Scores 10\% Relative to Benchmark} \label{tab:forecasting_qs10_FRED_p2}
	\scriptsize
	\begin{tabular}{@{}cS[table-format=1.3]S[table-format=1.3]S[table-format=1.3]S[table-format=1.3]S[table-format=1.3]S[table-format=1.3]S[table-format=1.3]S[table-format=1.3]@{}}
		\toprule
		{\textbf{h}} & {\textbf{AVP-VAR}} & {\textbf{CP-VAR}} & {\makecell{ \textbf{CP-VAR} \\ \textbf{SV}}} & {\makecell{ \textbf{{TVP-VAR}} \\ \textbf{EB}}} &{\textbf{TVP-VAR}} & {\makecell{ \textbf{VAR} \\ \textbf{SVOt}}} & {\textbf{FAVAR}} & {\makecell{ \textbf{FAVAR} \\ \textbf{SV}}}  \\
		\midrule
		\multicolumn{9}{c}{\textbf{Panel A: GDPC1}} \\
		\midrule
		1 & 1.006 & \textbf{0.984} & 1.003 & 1.050 & 1.043 & \textbf{0.952} & \textbf{0.955} & \textbf{0.939}$^{\dag}$ \\
		2 & 1.011 & 1.005 & 1.053 & 1.060 & 1.079 & \textbf{0.956}$^{\dag}$ & \textbf{0.970} & 1.046 \\
		3 & 1.069 & \textbf{0.993} & 1.092 & 1.134 & 1.101 & 1.013 & \textbf{0.991}$^{\dag}$ & 1.098 \\
		4 & 1.118 & \textbf{0.994}$^{\dag}$ & 1.123 & 1.175 & 1.124 & 1.026 & 1.005 & 1.134 \\
		5 & 1.113 & 1.024 & 1.120 & 1.196 & 1.136 & 1.031 & 1.001& 1.121 \\
		6 & 1.135 & 1.011 & 1.127 & 1.217 & 1.143 & 1.070 & 1.006 & 1.122 \\
		7 & 1.129 & 1.010 & 1.120 & 1.199 & 1.126 & 1.059 & \textbf{0.998}$^{\dag}$ & 1.129 \\
		8 & 1.122 & 1.015 & 1.108 & 1.174 & 1.115 & 1.069 & \textbf{0.996}$^{\dag}$ & 1.122 \\
		\midrule
		\multicolumn{9}{c}{\textbf{Panel B: PCECTPI}} \\
		\midrule
		1 & 1.172 & 1.007 & 1.136 & 1.180 & 1.179 & 1.098 & \textbf{0.994}$^{\dag}$ & 1.115 \\
		2 & 1.053 & 1.022 & 1.116 & 1.114 & 1.103 & 1.092 & \textbf{0.993}$^{\dag}$ & 1.092 \\
		3 & 1.063 & 1.021 & 1.132 & 1.146 & 1.095 & 1.123 & 1.007 & 1.133 \\
		4 & 1.066 & 1.024 & 1.147 & 1.155 & 1.066 & 1.104 & 1.016 & 1.153 \\
		5 & 1.107 & 1.033 & 1.167 & 1.128 & 1.093 & 1.121 & 1.053 & 1.158 \\
		6 & 1.090 & 1.024 & 1.177 & 1.145 & 1.090 & 1.124 & 1.024 & 1.184 \\
		7 & 1.064 & 1.013 & 1.104 & 1.117 & 1.108 & 1.115 & 1.009 & 1.129 \\
		8 & 1.097 & 1.032 & 1.153 & 1.097 & 1.129 & 1.134 & \textbf{0.997}$^{\dag}$ & 1.174 \\
		\bottomrule
	\end{tabular}	
	\vspace{0.5em}
	\par
	\begin{tabularx}{\textwidth}{X}
		{\footnotesize \textit{Notes:} The table reports quantile scores 10\% (QS10) for each of the two forecasted variables, relative to a benchmark constant-parameter VAR($2$) estimated via least squares. Values below one (highlighted in bold) indicate superior forecasting performance compared to the benchmark. The forecast horizon is denoted by $h$ (in quarters), and the symbol $\dag$ marks the best-performing model at each horizon.}
	\end{tabularx}
\end{table}

\begin{figure}[htbp!]  
	\centering
	\includegraphics[width=0.95\textwidth]{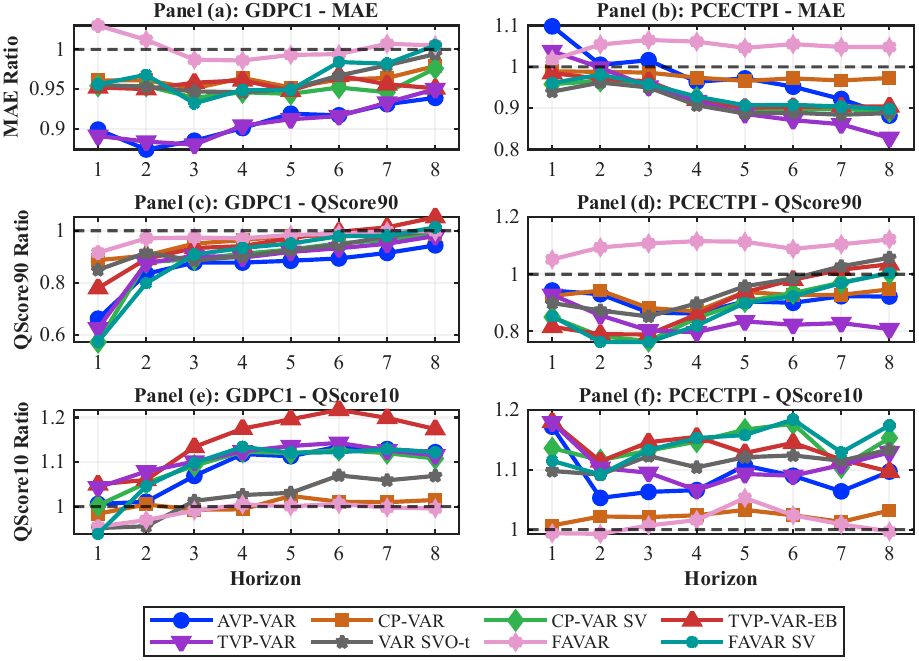}  
	\caption{Forecasting results using FRED quarterly data. The top row displays mean absolute forecast errors (MAE), while the middle and bottom rows show quantile scores for the upper (90th percentile) and lower (10th percentile) tails of each variable. All metrics are reported as ratios relative to a benchmark constant-parameter VAR($2$) estimated via least squares. A model is considered superior when its relative MAE, or relative quantile scores, are below one.} \label{fig:oos_FRED_p2}
\end{figure}

\FloatBarrier

\subsubsection{FRED--QD: VAR(3)}			
\begin{table}[htbp!]
	\centering
	\caption{Forecasting Performance VAR(3): Mean Squared Prediction Error Relative to Benchmark} \label{tab:forecasting_mspe_FRED_p3}
	\scriptsize
	\begin{tabular}{@{}cS[table-format=1.3]S[table-format=1.3]S[table-format=1.3]S[table-format=1.3]S[table-format=1.3]S[table-format=1.3]S[table-format=1.3]S[table-format=1.3]@{}}
		\toprule
		{\textbf{h}} & {\textbf{AVP-VAR}} & {\textbf{CP-VAR}} & {\makecell{ \textbf{CP-VAR} \\ \textbf{SV}}} & {\makecell{ \textbf{{TVP-VAR}} \\ \textbf{EB}}} &{\textbf{TVP-VAR}} & {\makecell{ \textbf{VAR} \\ \textbf{SVOt}}} & {\textbf{FAVAR}} & {\makecell{ \textbf{FAVAR} \\ \textbf{SV}}}  \\
		\midrule
		\multicolumn{9}{c}{\textbf{Panel A: GDPC1}} \\
		\midrule
		1 & \textbf{0.685}$^{\dag}$ & \textbf{0.909} & \textbf{0.779} & \textbf{0.819} & \textbf{0.686} & \textbf{0.807} & \textbf{0.932} & \textbf{0.766} \\
		2 & \textbf{0.841}$^{\dag}$ & \textbf{0.953} & \textbf{0.936} & \textbf{0.954} & \textbf{0.843} & \textbf{0.938} & \textbf{0.999} & \textbf{0.922} \\
		3 & \textbf{0.937} & \textbf{0.982} & \textbf{0.959} & \textbf{0.965} & \textbf{0.932}$^{\dag}$ & \textbf{0.970} & \textbf{0.988} & \textbf{0.950} \\
		4 & \textbf{0.945}$^{\dag}$ & \textbf{0.979} & \textbf{0.969} & \textbf{0.991} & \textbf{0.947} & \textbf{0.976} & \textbf{0.996} & \textbf{0.973} \\
		5 & \textbf{0.957}$^{\dag}$ & \textbf{0.988} & \textbf{0.970} & \textbf{0.974} & \textbf{0.957} & \textbf{0.981} & \textbf{0.996} & \textbf{0.972} \\
		6 & \textbf{0.962} & \textbf{0.982} & \textbf{0.971} & \textbf{0.996} & \textbf{0.953}$^{\dag}$ & \textbf{0.984} & 1.003 & \textbf{0.979} \\
		7 & \textbf{0.956} & \textbf{0.977} & \textbf{0.968} & \textbf{0.973} & \textbf{0.954}$^{\dag}$ & \textbf{0.979} & 1.001 & \textbf{0.987} \\
		8 & \textbf{0.948} & \textbf{0.977} & \textbf{0.973} & \textbf{0.956} & \textbf{0.944}$^{\dag}$ & \textbf{0.987} & \textbf{0.989} & \textbf{0.972} \\
		\midrule
		\multicolumn{9}{c}{\textbf{Panel B: PCECTPI}} \\
		\midrule
		1 & 1.136 & \textbf{0.966} & \textbf{0.946} & \textbf{0.940} & 1.522 & \textbf{0.920}$^{\dag}$ & 1.066 & \textbf{0.950} \\
		2 & \textbf{0.972} & \textbf{0.957} & \textbf{0.932} & \textbf{0.909}$^{\dag}$ & 1.152 & \textbf{0.922} & 1.194 & \textbf{0.953} \\
		3 & \textbf{0.941} & \textbf{0.955} & \textbf{0.916} & \textbf{0.904} & 1.094 & \textbf{0.900}$^{\dag}$ & 1.248 & \textbf{0.948} \\
		4 & \textbf{0.910} & \textbf{0.934} & \textbf{0.892} & \textbf{0.868}$^{\dag}$ & \textbf{0.995} & \textbf{0.887} & 1.250 & \textbf{0.924} \\
		5 & \textbf{0.925} & \textbf{0.937} & \textbf{0.873} & \textbf{0.853}$^{\dag}$ & \textbf{0.985} & \textbf{0.869} & 1.273 & \textbf{0.908} \\
		6 & \textbf{0.898} & \textbf{0.940} & \textbf{0.872} & \textbf{0.867} & \textbf{0.937} & \textbf{0.856}$^{\dag}$ & 1.312 & \textbf{0.902} \\
		7 & \textbf{0.861} & \textbf{0.939} & \textbf{0.871} & \textbf{0.852}$^{\dag}$ & \textbf{0.892} & \textbf{0.862} & 1.342 & \textbf{0.901} \\
		8 & \textbf{0.829}$^{\dag}$ & \textbf{0.931} & \textbf{0.857} & \textbf{0.830} & \textbf{0.857} & \textbf{0.846} & 1.407 & \textbf{0.916} \\
		\bottomrule
	\end{tabular}	
	\vspace{0.5em}
	\par
	\begin{tabularx}{\textwidth}{X}
		{\footnotesize \textit{Notes:} The table reports mean squared prediction errors (MSPE) for each of the two forecasted variables, relative to a benchmark constant-parameter VAR($3$) estimated via least squares. Values below one (highlighted in bold) indicate superior forecasting performance compared to the benchmark. The forecast horizon is denoted by $h$ (in quarters), and the symbol $\dag$ marks the best-performing model at each horizon.}
	\end{tabularx}
\end{table}

\begin{table}[htbp!]
	\centering
	\caption{Forecasting Performance VAR(3): Quantile Scores 90\% Relative to Benchmark} \label{tab:forecasting_qs90_FRED_p3}
	\scriptsize
	\begin{tabular}{@{}cS[table-format=1.3]S[table-format=1.3]S[table-format=1.3]S[table-format=1.3]S[table-format=1.3]S[table-format=1.3]S[table-format=1.3]S[table-format=1.3]@{}}
		\toprule
		{\textbf{h}} & {\textbf{AVP-VAR}} & {\textbf{CP-VAR}} & {\makecell{ \textbf{CP-VAR} \\ \textbf{SV}}} & {\makecell{ \textbf{{TVP-VAR}} \\ \textbf{EB}}} &{\textbf{TVP-VAR}} & {\makecell{ \textbf{VAR} \\ \textbf{SVOt}}} & {\textbf{FAVAR}} & {\makecell{ \textbf{FAVAR} \\ \textbf{SV}}}  \\
		\midrule
		\multicolumn{9}{c}{\textbf{Panel A: GDPC1}} \\
		\midrule
		1 & \textbf{0.663} & \textbf{0.892} & \textbf{0.737} & \textbf{0.766} & \textbf{0.593}$^{\dag}$ & \textbf{0.825} & \textbf{0.905} & \textbf{0.648} \\
		2 & \textbf{0.848} & \textbf{0.934} & \textbf{0.852} & \textbf{0.902} & \textbf{0.844} & \textbf{0.893} & \textbf{0.979} & \textbf{0.759}$^{\dag}$ \\
		3 & \textbf{0.886} & \textbf{0.957} & \textbf{0.895} & \textbf{0.945} & \textbf{0.880} & \textbf{0.871}$^{\dag}$ & \textbf{0.955} & \textbf{0.875} \\
		4 & \textbf{0.874}$^{\dag}$ & \textbf{0.970} & \textbf{0.895} & \textbf{0.949} & \textbf{0.897} & \textbf{0.891} & \textbf{0.972} & \textbf{0.889} \\
		5 & \textbf{0.885}$^{\dag}$ & \textbf{0.985} & \textbf{0.915} & \textbf{0.985} & \textbf{0.893} & \textbf{0.910} & \textbf{0.985} & \textbf{0.904} \\
		6 & \textbf{0.910}$^{\dag}$ & \textbf{0.979} & \textbf{0.926} & 1.004 & \textbf{0.910} & \textbf{0.921} & \textbf{0.992} & \textbf{0.925} \\
		7 & \textbf{0.907}$^{\dag}$ & \textbf{0.976} & \textbf{0.931} & 1.006 & \textbf{0.912} & \textbf{0.940} & \textbf{0.994} & \textbf{0.920} \\
		8 & \textbf{0.951} & \textbf{0.979} & \textbf{0.974} & 1.037 & \textbf{0.950}$^{\dag}$ & \textbf{0.974} & \textbf{0.994} & \textbf{0.959} \\
		\midrule
		\multicolumn{9}{c}{\textbf{Panel B: PCECTPI}} \\
		\midrule
		1 & \textbf{0.931} & \textbf{0.865} & \textbf{0.791} & \textbf{0.739}$^{\dag}$ & 1.255 & \textbf{0.845} & 1.057 & \textbf{0.786} \\
		2 & \textbf{0.881} & \textbf{0.904} & \textbf{0.718} & \textbf{0.721} & 1.106 & \textbf{0.800} & 1.166 & \textbf{0.709}$^{\dag}$ \\
		3 & \textbf{0.818} & \textbf{0.818} & \textbf{0.700}$^{\dag}$ & \textbf{0.700} & 1.019 & \textbf{0.826} & 1.172 & \textbf{0.717} \\
		4 & \textbf{0.845} & \textbf{0.835} & \textbf{0.809} & \textbf{0.775}$^{\dag}$ & \textbf{0.940} & \textbf{0.862} & 1.145 & \textbf{0.797} \\
		5 & \textbf{0.876} & \textbf{0.875} & \textbf{0.863} & \textbf{0.847}$^{\dag}$ & \textbf{0.925} & \textbf{0.942} & 1.180 & \textbf{0.858} \\
		6 & \textbf{0.883} & \textbf{0.881} & \textbf{0.884} & \textbf{0.897} & \textbf{0.880}$^{\dag}$ & \textbf{0.991} & 1.211 & \textbf{0.893} \\
		7 & \textbf{0.884} & \textbf{0.866} & \textbf{0.895} & \textbf{0.918} & \textbf{0.839}$^{\dag}$ & \textbf{0.973} & 1.225 & \textbf{0.904} \\
		8 & \textbf{0.885} & \textbf{0.895} & \textbf{0.929} & \textbf{0.947} & \textbf{0.813}$^{\dag}$ & \textbf{0.985} & 1.244 & \textbf{0.932} \\
		\bottomrule
	\end{tabular}	
	\vspace{0.5em}
	\par
	\begin{tabularx}{\textwidth}{X}
		{\footnotesize \textit{Notes:} The table reports quantile scores 90\% (QS90) for each of the two forecasted variables, relative to a benchmark constant-parameter VAR($3$) estimated via least squares. Values below one (highlighted in bold) indicate superior forecasting performance compared to the benchmark. The forecast horizon is denoted by $h$ (in quarters), and the symbol $\dag$ marks the best-performing model at each horizon.}
	\end{tabularx}
\end{table}

\begin{table}[htbp!]
	\centering
	\caption{Forecasting Performance VAR(3): Quantile Scores 10\% Relative to Benchmark} \label{tab:forecasting_qs10_FRED_p3}
	\scriptsize
	\begin{tabular}{@{}cS[table-format=1.3]S[table-format=1.3]S[table-format=1.3]S[table-format=1.3]S[table-format=1.3]S[table-format=1.3]S[table-format=1.3]S[table-format=1.3]@{}}
		\toprule
		{\textbf{h}} & {\textbf{AVP-VAR}} & {\textbf{CP-VAR}} & {\makecell{ \textbf{CP-VAR} \\ \textbf{SV}}} & {\makecell{ \textbf{{TVP-VAR}} \\ \textbf{EB}}} &{\textbf{TVP-VAR}} & {\makecell{ \textbf{VAR} \\ \textbf{SVOt}}} & {\textbf{FAVAR}} & {\makecell{ \textbf{FAVAR} \\ \textbf{SV}}}  \\
		\midrule
		\multicolumn{9}{c}{\textbf{Panel A: GDPC1}} \\
		\midrule
		1 & 1.028 & \textbf{0.993} & \textbf{0.999} & 1.067 & 1.028 & \textbf{0.947} & \textbf{0.968} & \textbf{0.927}$^{\dag}$ \\
		2 & 1.024 & 1.008 & 1.048 & 1.067 & 1.056 & \textbf{0.972}$^{\dag}$ & \textbf{0.985} & 1.032 \\
		3 & 1.108 & 1.021 & 1.098 & 1.160 & 1.097 & 1.023 & 1.005 & 1.103 \\
		4 & 1.126 & 1.024 & 1.112 & 1.203 & 1.118 & 1.031 & 1.012 & 1.134 \\
		5 & 1.122 & 1.018 & 1.112 & 1.215 & 1.123 & 1.040 & 1.003 & 1.098 \\
		6 & 1.150 & 1.014 & 1.120 & 1.249 & 1.128 & 1.060 & 1.005 & 1.118 \\
		7 & 1.133 & 1.006 & 1.100 & 1.218 & 1.096 & 1.063 & 1.005 & 1.093 \\
		8 & 1.123 & \textbf{0.998} & 1.086 & 1.201 & 1.096 & 1.062 & \textbf{0.992}$^{\dag}$ & 1.075 \\
		\midrule
		\multicolumn{9}{c}{\textbf{Panel B: PCECTPI}} \\
		\midrule
		1 & 1.175 & 1.011 & 1.124 & 1.165 & 1.207 & 1.107 & 1.012 & 1.136 \\
		2 & 1.073 & 1.022 & 1.117 & 1.114 & 1.105 & 1.071 & 1.001 & 1.118 \\
		3 & 1.084 & 1.013 & 1.141 & 1.130 & 1.146 & 1.084 & 1.009 & 1.143 \\
		4 & 1.057 & 1.010 & 1.114 & 1.117 & 1.071 & 1.081 & 1.021 & 1.141 \\
		5 & 1.081 & 1.023 & 1.103 & 1.078 & 1.070 & 1.076 & 1.043 & 1.125 \\
		6 & 1.100 & 1.046 & 1.168 & 1.122 & 1.093 & 1.108 & 1.063 & 1.188 \\
		7 & 1.124 & 1.000 & 1.122 & 1.098 & 1.140 & 1.068 & 1.059 & 1.160 \\
		8 & 1.112 & 1.019 & 1.151 & 1.099 & 1.217 & 1.078 & 1.054 & 1.200 \\
		\bottomrule
	\end{tabular}	
	\vspace{0.5em}
	\par
	\begin{tabularx}{\textwidth}{X}
		{\footnotesize \textit{Notes:} The table reports quantile scores 10\% (QS10) for each of the two forecasted variables, relative to a benchmark constant-parameter VAR($3$) estimated via least squares. Values below one (highlighted in bold) indicate superior forecasting performance compared to the benchmark. The forecast horizon is denoted by $h$ (in quarters), and the symbol $\dag$ marks the best-performing model at each horizon.}
	\end{tabularx}
\end{table}

\begin{figure}[htbp!]  
	\centering
	\includegraphics[width=0.95\textwidth]{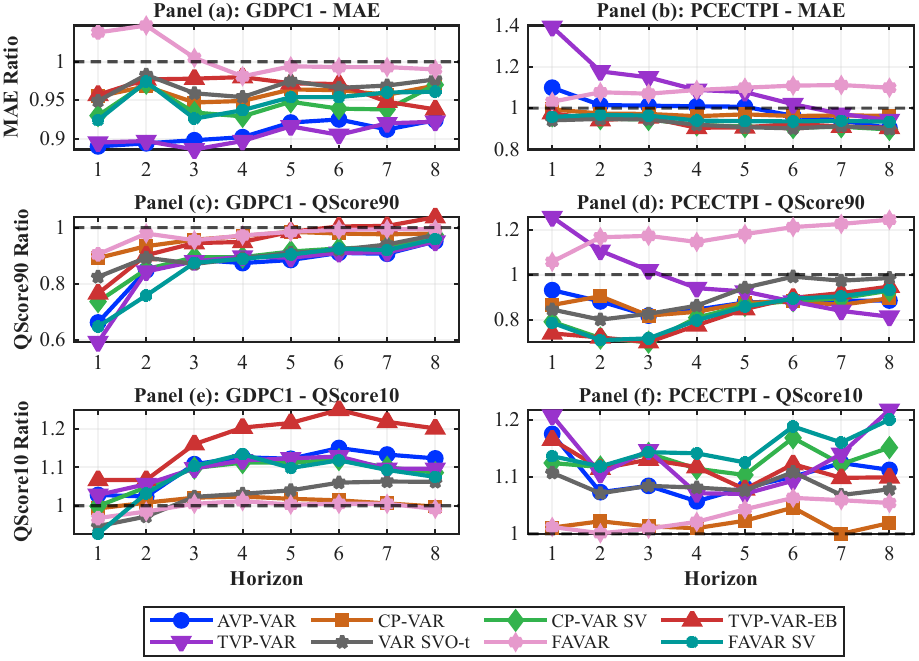}  
	\caption{Forecasting results using FRED quarterly data. The top row displays mean absolute forecast errors (MAE), while the middle and bottom rows show quantile scores for the upper (90th percentile) and lower (10th percentile) tails of each variable. All metrics are reported as ratios relative to a benchmark constant-parameter VAR($2$) estimated via least squares. A model is considered superior when its relative MAE, or relative quantile scores, are below one.} \label{fig:oos_FRED_p3}
\end{figure}

\FloatBarrier

\subsubsection{FRED--QD: VAR(4)}	

\begin{table}[htbp!]
	\centering
	\caption{Forecasting Performance VAR(4): Mean Squared Prediction Error Relative to Benchmark} \label{tab:forecasting_mspe_FRED_p4}
	\scriptsize
	\begin{tabular}{@{}cS[table-format=1.3]S[table-format=1.3]S[table-format=1.3]S[table-format=1.3]S[table-format=1.3]S[table-format=1.3]S[table-format=1.3]S[table-format=1.3]@{}}
		\toprule
		{\textbf{h}} & {\textbf{AVP-VAR}} & {\textbf{CP-VAR}} & {\makecell{ \textbf{CP-VAR} \\ \textbf{SV}}} & {\makecell{ \textbf{{TVP-VAR}} \\ \textbf{EB}}} &{\textbf{TVP-VAR}} & {\makecell{ \textbf{VAR} \\ \textbf{SVOt}}} & {\textbf{FAVAR}} & {\makecell{ \textbf{FAVAR} \\ \textbf{SV}}}  \\
		\midrule
		\multicolumn{9}{c}{\textbf{Panel A: GDPC1}} \\
		\midrule
		1 & \textbf{0.696} & \textbf{0.895} & \textbf{0.781} & \textbf{0.841} & \textbf{0.684}$^{\dag}$ & \textbf{0.807} & \textbf{0.937} & \textbf{0.753} \\
		2 & \textbf{0.847} & \textbf{0.934} & \textbf{0.935} & \textbf{0.966} & \textbf{0.839}$^{\dag}$ & \textbf{0.928} & \textbf{0.984} & \textbf{0.921} \\
		3 & \textbf{0.940} & \textbf{0.955} & \textbf{0.981} & 1.031 & \textbf{0.933}$^{\dag}$ & \textbf{0.971} & 1.000 & \textbf{0.945} \\
		4 & \textbf{0.941}$^{\dag}$ & \textbf{0.974} & \textbf{0.960} & 1.053 & \textbf{0.945} & \textbf{0.973} & 1.008 & \textbf{0.974} \\
		5 & \textbf{0.945} & \textbf{0.978} & \textbf{0.962} & 1.013 & \textbf{0.937}$^{\dag}$ & \textbf{0.969} & 1.009 & \textbf{0.956} \\
		6 & \textbf{0.957} & \textbf{0.988} & \textbf{0.982} & 1.034 & \textbf{0.955}$^{\dag}$ & \textbf{0.984} & 1.001 & \textbf{0.995} \\
		7 & \textbf{0.944}$^{\dag}$ & \textbf{0.986} & \textbf{0.962} & 1.000 & \textbf{0.949} & \textbf{0.973} & \textbf{0.992} & \textbf{0.977} \\
		8 & \textbf{0.945}$^{\dag}$ & \textbf{0.987} & \textbf{0.973} & \textbf{0.988} & \textbf{0.952} & \textbf{0.981} & \textbf{0.987} & \textbf{0.976} \\
		\midrule
		\multicolumn{9}{c}{\textbf{Panel B: PCECTPI}} \\
		\midrule
		1 & \textbf{0.981} & \textbf{0.925} & \textbf{0.831} & \textbf{0.801} & 1.375 & \textbf{0.782}$^{\dag}$ & 1.058 & \textbf{0.850} \\
		2 & \textbf{0.832} & \textbf{0.904} & \textbf{0.832} & \textbf{0.737}$^{\dag}$ & 1.014 & \textbf{0.774} & 1.151 & \textbf{0.852} \\
		3 & \textbf{0.821} & \textbf{0.911} & \textbf{0.847} & \textbf{0.723}$^{\dag}$ & \textbf{0.953} & \textbf{0.776} & 1.166 & \textbf{0.849} \\
		4 & \textbf{0.820} & \textbf{0.933} & \textbf{0.870} & \textbf{0.710}$^{\dag}$ & \textbf{0.915} & \textbf{0.815} & 1.172 & \textbf{0.868} \\
		5 & \textbf{0.786} & \textbf{0.889} & \textbf{0.819} & \textbf{0.680}$^{\dag}$ & \textbf{0.890} & \textbf{0.760} & 1.197 & \textbf{0.828} \\
		6 & \textbf{0.758} & \textbf{0.875} & \textbf{0.810} & \textbf{0.724}$^{\dag}$ & \textbf{0.851} & \textbf{0.786} & 1.209 & \textbf{0.825} \\
		7 & \textbf{0.759} & \textbf{0.895} & \textbf{0.847} & \textbf{0.729}$^{\dag}$ & \textbf{0.841} & \textbf{0.813} & 1.197 & \textbf{0.850} \\
		8 & \textbf{0.733} & \textbf{0.892} & \textbf{0.846} & \textbf{0.721}$^{\dag}$ & \textbf{0.821} & \textbf{0.813} & 1.244 & \textbf{0.856} \\
		\bottomrule
	\end{tabular}	
	\vspace{0.5em}
	\par
	\begin{tabularx}{\textwidth}{X}
		{\footnotesize \textit{Notes:} The table reports mean squared prediction errors (MSPE) for each of the two forecasted variables, relative to a benchmark constant-parameter VAR($4$) estimated via least squares. Values below one (highlighted in bold) indicate superior forecasting performance compared to the benchmark. The forecast horizon is denoted by $h$ (in quarters), and the symbol $\dag$ marks the best-performing model at each horizon.}
	\end{tabularx}
\end{table}

\begin{table}[htbp!]
	\centering
	\caption{Forecasting Performance VAR(4): Quantile Scores 90\% Relative to Benchmark} \label{tab:forecasting_qs90_FRED_p4}
	\scriptsize
	\begin{tabular}{@{}cS[table-format=1.3]S[table-format=1.3]S[table-format=1.3]S[table-format=1.3]S[table-format=1.3]S[table-format=1.3]S[table-format=1.3]S[table-format=1.3]@{}}
		\toprule
		{\textbf{h}} & {\textbf{AVP-VAR}} & {\textbf{CP-VAR}} & {\makecell{ \textbf{CP-VAR} \\ \textbf{SV}}} & {\makecell{ \textbf{{TVP-VAR}} \\ \textbf{EB}}} &{\textbf{TVP-VAR}} & {\makecell{ \textbf{VAR} \\ \textbf{SVOt}}} & {\textbf{FAVAR}} & {\makecell{ \textbf{FAVAR} \\ \textbf{SV}}}  \\
		\midrule
		\multicolumn{9}{c}{\textbf{Panel A: GDPC1}} \\
		\midrule
		1 & \textbf{0.717} & \textbf{0.921} & \textbf{0.610} & \textbf{0.833} & \textbf{0.622} & \textbf{0.860} & \textbf{0.981} & \textbf{0.584}$^{\dag}$ \\
		2 & \textbf{0.911} & \textbf{0.938} & \textbf{0.945} & \textbf{0.996} & \textbf{0.907} & \textbf{0.921} & \textbf{0.994} & \textbf{0.843}$^{\dag}$ \\
		3 & \textbf{0.927} & \textbf{0.973} & \textbf{0.965} & 1.007 & \textbf{0.924} & \textbf{0.885}$^{\dag}$ & \textbf{0.972} & \textbf{0.939} \\
		4 & \textbf{0.908} & \textbf{0.968} & \textbf{0.959} & \textbf{0.993} & \textbf{0.924} & \textbf{0.894}$^{\dag}$ & \textbf{0.982} & \textbf{0.949} \\
		5 & \textbf{0.905} & \textbf{0.961} & \textbf{0.959} & \textbf{0.993} & \textbf{0.905} & \textbf{0.903}$^{\dag}$ & \textbf{0.990} & \textbf{0.929} \\
		6 & \textbf{0.908}$^{\dag}$ & \textbf{0.982} & \textbf{0.973} & 1.032 & \textbf{0.918} & \textbf{0.922} & \textbf{0.984} & \textbf{0.968} \\
		7 & \textbf{0.905}$^{\dag}$ & \textbf{0.980} & \textbf{0.971} & 1.023 & \textbf{0.915} & \textbf{0.951} & \textbf{0.985} & \textbf{0.955} \\
		8 & \textbf{0.959} & \textbf{0.980} & 1.022 & 1.064 & \textbf{0.950}$^{\dag}$ & \textbf{0.977} & \textbf{0.985} & 1.002 \\
		\midrule
		\multicolumn{9}{c}{\textbf{Panel B: PCECTPI}} \\
		\midrule
		1 & \textbf{0.817} & \textbf{0.830} & \textbf{0.765} & \textbf{0.670}$^{\dag}$ & 1.204 & \textbf{0.726} & 1.068 & \textbf{0.760} \\
		2 & \textbf{0.799} & \textbf{0.858} & \textbf{0.693} & \textbf{0.634}$^{\dag}$ & 1.060 & \textbf{0.728} & 1.130 & \textbf{0.707} \\
		3 & \textbf{0.739} & \textbf{0.799} & \textbf{0.684} & \textbf{0.629}$^{\dag}$ & \textbf{0.965} & \textbf{0.798} & 1.097 & \textbf{0.688} \\
		4 & \textbf{0.794} & \textbf{0.832} & \textbf{0.782} & \textbf{0.708}$^{\dag}$ & \textbf{0.919} & \textbf{0.868} & 1.096 & \textbf{0.785} \\
		5 & \textbf{0.768} & \textbf{0.817} & \textbf{0.782} & \textbf{0.737}$^{\dag}$ & \textbf{0.872} & \textbf{0.859} & 1.136 & \textbf{0.772} \\
		6 & \textbf{0.774}$^{\dag}$ & \textbf{0.782} & \textbf{0.784} & \textbf{0.781} & \textbf{0.831} & \textbf{0.883} & 1.132 & \textbf{0.796} \\
		7 & \textbf{0.792}$^{\dag}$ & \textbf{0.833} & \textbf{0.823} & \textbf{0.852} & \textbf{0.820} & \textbf{0.896} & 1.103 & \textbf{0.821} \\
		8 & \textbf{0.806} & \textbf{0.864} & \textbf{0.871} & \textbf{0.898} & \textbf{0.802}$^{\dag}$ & \textbf{0.911} & 1.118 & \textbf{0.859} \\
		\bottomrule
	\end{tabular}	
	\vspace{0.5em}
	\par
	\begin{tabularx}{\textwidth}{X}
		{\footnotesize \textit{Notes:} The table reports quantile scores 90\% (QS90) for each of the two forecasted variables, relative to a benchmark constant-parameter VAR($4$) estimated via least squares. Values below one (highlighted in bold) indicate superior forecasting performance compared to the benchmark. The forecast horizon is denoted by $h$ (in quarters), and the symbol $\dag$ marks the best-performing model at each horizon.}
	\end{tabularx}
\end{table}

\begin{table}[htbp!]
	\centering
	\caption{Forecasting Performance VAR(4): Quantile Scores 10\% Relative to Benchmark} \label{tab:forecasting_qs10_FRED_p4}
	\scriptsize
	\begin{tabular}{@{}cS[table-format=1.3]S[table-format=1.3]S[table-format=1.3]S[table-format=1.3]S[table-format=1.3]S[table-format=1.3]S[table-format=1.3]S[table-format=1.3]@{}}
		\toprule
		{\textbf{h}} & {\textbf{AVP-VAR}} & {\textbf{CP-VAR}} & {\makecell{ \textbf{CP-VAR} \\ \textbf{SV}}} & {\makecell{ \textbf{{TVP-VAR}} \\ \textbf{EB}}} &{\makecell{ \textbf{{\textbf{TVP-VAR}}} \\ \textbf{FB}}} & {\makecell{ \textbf{VAR} \\ \textbf{SVOt}}} & {\textbf{FAVAR}} & {\makecell{ \textbf{FAVAR} \\ \textbf{SV}}}  \\
		\midrule
		\multicolumn{9}{c}{\textbf{Panel A: GDPC1}} \\
		\midrule
		1 & \textbf{0.997} & \textbf{0.959} & 1.016 & 1.048 & 1.003 & \textbf{0.929}$^{\dag}$ & \textbf{0.953} & \textbf{0.929} \\
		2 & \textbf{0.999} & \textbf{0.978} & 1.046 & 1.066 & 1.016 & \textbf{0.924}$^{\dag}$ & \textbf{0.971} & 1.023 \\
		3 & 1.072 & \textbf{0.997}$^{\dag}$ & 1.134 & 1.160 & 1.084 & 1.006 & 1.008 & 1.111 \\
		4 & 1.105 & 1.018$^{\dag}$ & 1.145 & 1.226 & 1.104 & 1.042 & 1.018 & 1.143 \\
		5 & 1.082 & \textbf{0.994}$^{\dag}$ & 1.127 & 1.228 & 1.080 & 1.008 & 1.027 & 1.115 \\
		6 & 1.146 & 1.029 & 1.170 & 1.305 & 1.138 & 1.064 & 1.014 & 1.153 \\
		7 & 1.119 & 1.012 & 1.162 & 1.305 & 1.115 & 1.052 & \textbf{0.997}$^{\dag}$ & 1.136 \\
		8 & 1.103 & \textbf{0.997} & 1.125 & 1.284 & 1.091 & 1.050 & \textbf{0.982}$^{\dag}$ & 1.114 \\
		\midrule
		\multicolumn{9}{c}{\textbf{Panel B: PCECTPI}} \\
		\midrule
		1 & 1.129 & \textbf{0.970}$^{\dag}$ & 1.079 & 1.123 & 1.155 & 1.036 & \textbf{0.992} & 1.080 \\
		2 & 1.033 & \textbf{0.988}$^{\dag}$ & 1.089 & 1.067 & 1.058 & 1.011 & \textbf{0.999} & 1.079 \\
		3 & 1.035 & 1.017 & 1.119 & 1.112 & 1.106 & 1.054 & 1.012 & 1.085 \\
		4 & 1.030 & 1.026 & 1.121 & 1.095 & 1.081 & 1.046 & 1.022 & 1.117 \\
		5 & 1.034 & 1.013 & 1.095 & 1.082 & 1.009 & 1.008 & 1.036 & 1.096 \\
		6 & 1.023 & 1.006 & 1.116 & 1.103 & 1.036 & 1.000 & 1.011 & 1.112 \\
		7 & 1.067 & 1.017 & 1.129 & 1.158 & 1.131 & 1.010 & 1.017 & 1.127 \\
		8 & 1.043 & \textbf{0.992}$^{\dag}$ & 1.147 & 1.198 & 1.189 & 1.076 & 1.024 & 1.159 \\
		\bottomrule
	\end{tabular}	
	\vspace{0.5em}
	\par
	\begin{tabularx}{\textwidth}{X}
		{\footnotesize \textit{Notes:} The table reports quantile scores 10\% (QS10) for each of the two forecasted variables, relative to a benchmark constant-parameter VAR($4$) estimated via least squares. Values below one (highlighted in bold) indicate superior forecasting performance compared to the benchmark. The forecast horizon is denoted by $h$ (in quarters), and the symbol $\dag$ marks the best-performing model at each horizon.}
	\end{tabularx}
\end{table}

\begin{figure}[htbp!]  
	\centering
	\includegraphics[width=0.95\textwidth]{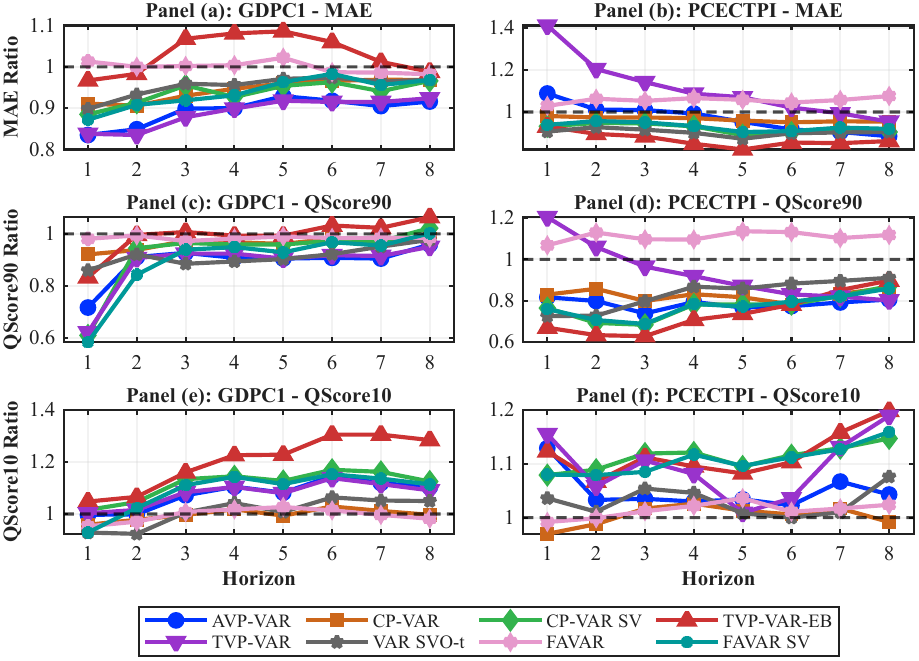}  
	\caption{Forecasting results using FRED quarterly data. The top row displays mean absolute forecast errors (MAE), while the middle and bottom rows show quantile scores for the upper (90th percentile) and lower (10th percentile) tails of each variable. All metrics are reported as ratios relative to a benchmark constant-parameter VAR($4$) estimated via least squares. A model is considered superior when its relative MAE, or relative quantile scores, are below one.} \label{fig:oos_FRED_p4}
\end{figure}

\FloatBarrier

\newpage
\bibliographystyle{apalike}
\addcontentsline{toc}{section}{\refname}
\bibliography{APVAR.bib}

\end{appendix}
	
\end{document}